%% file: friction.tex
\documentclass[11pt]{article}
\usepackage{verbatim,amsmath,amssymb}
\usepackage{epsfig,float,color}
\usepackage[font=small,labelfont=bf]{caption}
\usepackage{geometry}
\usepackage{setspace}
\usepackage{natbib}

\usepackage{amsthm,mathrsfs,amsfonts,dsfont} 
\usepackage{hyperref}
\usepackage[labelfont=bf]{caption}
\usepackage{subfig}
\usepackage{array}
\usepackage{makecell}
\usepackage{multirow}

\definecolor{darkblue}{rgb}{0,0,1}
\definecolor{col1}{rgb}{1,0,1}
\definecolor{col2}{rgb}{0,0.5,0}
\definecolor{col3}{rgb}{0.5,0,1}
\definecolor{col4}{rgb}{0.1,.75,0}

\hypersetup{pdftex=true, colorlinks=true, breaklinks=true, linkcolor=darkblue, menucolor=darkblue, pagecolor=darkblue, citecolor=darkblue, urlcolor=darkblue}

\newtheoremstyle{rem}
{6pt}
{6pt}
{\small}
{}
{\bf}
{:}
{.5em}
{}

\theoremstyle{rem}
\newtheorem{remark}{Remark}[section]

\input{neco_updated.tex}

\newcommand{\gn}{g_\mrn}
\newcommand{\ga}{g_\mra}
\newcommand{\gz}{g_\mrz}
\newcommand{\xs}{\bx_\mrs}
\newcommand{\Psic}{\Psi_\mrc}
\newcommand{\Psit}{\bar\Psi_\mrt}

\newcommand{\Sm}{\bar S}

\graphicspath{ {images/} }

\begin{document}

\begin{center}
\Large{\bf{A continuum contact model for friction between graphene sheets that accounts for surface anisotropy and curvature}}\\

\end{center}

\renewcommand{\thefootnote}{\fnsymbol{footnote}}

\begin{center}
\large{Aningi Mokhalingam$^{\mra}$, Shakti S.~Gupta$^{\mra}$ and Roger A.~Sauer$^{\mrb,\mrc,\mrd}$\footnote[1]{corresponding author, email: roger.sauer@pg.edu.pl, sauer@aices.rwth-aachen.de}
}\\
\vspace{4mm}

\small{\textit{
$^\mra$Department of Mechanical Engineering, Indian Institute of Technology Kanpur, UP 208016, India \\[1.1mm]
$^\mrb$Aachen Institute for Advanced Study in Computational Engineering Science (AICES), \\ 
RWTH Aachen University, Templergraben 55, 52056 Aachen, Germany \\[1.1mm]
$^\mrc$Faculty of Civil and Environmental Engineering, Gda\'{n}sk University of Technology, ul.~Narutowicza 11/12, 80-233 Gda\'{n}sk, Poland \\[1.1mm]
$^\mrd$Department of Mechanical Engineering, Indian Institute of Technology Guwahati, Assam 781039, India \\[1.1mm]
}}

\end{center}

\vspace{-1mm}

\renewcommand{\thefootnote}{\arabic{footnote}}

\begin{center}

\small{Published\footnote{This pdf is the personal version of an article whose journal version is available at \href{http://dx.doi.org/10.1103/PhysRevB.109.035435}{https://journals.aps.org/prb/}} 
in \textit{Phys. Rev. B}, \href{http://dx.doi.org/10.1103/PhysRevB.109.035435}{DOI: 10.1103/PhysRevB.109.035435} \\
Submitted on 15 September 2023; Revised on 22 December 2023; Accepted on 3 January 2024} 

\end{center}

\vspace{-3mm}


\rule{\linewidth}{.15mm}
{\bf Abstract:}
Understanding the interaction mechanics between graphene layers and co-axial carbon nanotubes (CNTs) is essential for modeling graphene and CNT-based nanoelectromechanical systems. This work proposes a new continuum contact model to study interlayer interactions between curved graphene sheets. The continuum model is calibrated and validated using molecular dynamics (MD) simulations. These are carried out employing the reactive empirical bond order (REBO)+Lennard-Jones (LJ) potential to model the interactions within a sheet, while the LJ,  Kolmogorov-Crespi (KC), and Lebedeva potentials are used to model the interactions between sheets. The continuum contact model is formulated for separation distances greater than 0.29 nm, when sliding contact becomes non-dissipative and can be described by a potential. In this regime, sheet deformations are sufficiently small and do not affect the sheet interactions substantially. This allows to treat the master contact surface as rigid, thus simplifying the contact formulation greatly. The model calibration is conducted systematically for a sequence of different stackings using existing and newly proposed ansatz functions.  
The calibrated continuum model is then implemented in a curvilinear finite element (FE) shell formulation to investigate the pull-out and twisting interactions between co-axial CNTs. The resisting pull-out forces and torques depend strongly on the chirality of the considered CNTs. The absolute differences between FE and MD results are very small, and can be attributed to model assumptions and loading conditions.

\vspace{-5mm}

\rule{\linewidth}{.15mm}
\section{Introduction}\label{s:intro}
Graphene is a 2D material with tightly packed carbon atoms in a hexagonal lattice structure that can be isolated from bulk graphite through micromechanical exfoliation \citep{Novoselov2004}. Experimental studies have shown that bi- and multi-layer graphene have remarkable thermomechanical and tribological properties \citep{Ver2004, Zheng2008, Berman2014_01}. The strong in-plane covalent bonds and weak nonbonded forces between the layers offer tunability of properties through different stackings. In particular, interlayer twisting or stretching of bilayer graphene can result in superlubricity and superconductivity \citep{Dienwiebel2004a, Feng2013, Cao2018, shen2020}. Due to these tunable properties, multilayered graphene has the potential to be used in various engineering applications \citep{Bunch2007, Bae2010, Liu2016}. 

The interaction mechanics of graphene layers has been studied using different techniques viz., experiments \citep{Dienwiebel2004a, Dienwiebel2005, Zheng2008}, theory \citep{Ver2004, Reguzzoni2012, Lopes2012, Liu_2014, Kumar2016, Wang2017b, Xue2022}, and atomistic simulations \citep{Guo2007, Xu_2011, Popov2011, ZHANG2015, Wang2017, LI2020, AFSHARIRAD2021}. For instance, Dienwiebel et al. \citep{Dienwiebel2004a} measured ultra-low friction or superlubricity between graphene layers due to incommensurability obtained through the relative rotation between the graphene layers using a frictional force microscope. The superlubricity in twisted bilayer graphene and graphene heterojunctions is governed by Moir\'e patterns formed between the mismatched layers \citep{Mele2010, Lopes2012, Koshino2020}. In another work, Dienwiebel et al. \citep{Dienwiebel2005} reported anisotropic friction with an angular periodicity of ~60$^\circ$. Verhoeven et al. \citep{Ver2004} investigated rigid graphene flake interactions over a graphene surface employing the modified Prandtl-Tomlinson model \citep{Tomlinson1929_01} and reported that the frictional forces depend upon the flake size and relative rotation between the graphene flake and substrate. Further, they have approximated the interaction energy using only the first Fourier components with the wavelengths $\sqrt{3}a_{\text{cc}}$, $1.5a_{\text{cc}}$ and $3a_{\text{cc}}$, where $a_{\text{cc}}$ is the covalent C-C bond length. Using the Lennard-Jones (LJ) potential, Xu et al.~\citep{Xu_2011} investigated the influence of the number of graphene layers on stick-slip friction and reported that these forces reduce with a decrease in the number of layers. Wang et al.~\citep{Wang2017b} studied the size effect on the interlayer shear behavior in bilayer graphene, accounting for elastic deformation in the graphene sheets employing a nonlinear shear-lag model \citep{Cox_1952}. They reported that the maximum shear force depends on the length and width of the sheet. However, for a length beyond 20 nm, the shear force is constant due to non-uniform relative displacement between the sheets. Using first-principle calculations, Sun et al. \citep{Sun2018} reported that bilayer graphene sliding friction reduces with increasing sheet contact pressure\footnote{The common term \textit{contact pressure} is used to denote the pressure between two graphene sheets even though they always remain at a nanoscale distance.} and becomes zero at a critical point due to the transition of the potential energy surface from a corrugated to a flattened and to a counter-corrugated state. Using the finite element method, Xue et al.~\citep{Xue2022} studied the dynamics of peeling and sliding graphene nanoribbons on a graphene substrate and reported that adhesive and shear interactions of graphene sheets influence the sliding behavior. Afsharirad et al. \citep{AFSHARIRAD2021} studied the inter-layer interactions between the walls of double-walled carbon nanotubes (DWCNT) using the LJ potential and reported that zigzag CNTs\footnote{Zigzag CNTs have zigzag circumference and armchair axis, while it is the opposite for armchair CNTs.} show larger axial sliding resistance than other kinds of CNTs. Arroyo and Belytschko \citep{Arroyo2004b, Arroyo2004a, arroyo2005} formulated a continuum contact theory for curved monolayer lattices via the exponential Cauchy-Born rule and implemented it in the finite element method to investigate the mechanics of CNTs.   

The variation of the interaction energy with the relative displacement of two graphene sheets is dominated by their $\pi-$orbital overlap at lower separation distances \citep{Kolmogorov2005}. The corrugation amplitude of the potential relief at these lower separation distances obtained from the LJ potential underestimates the interactions. This lead to the development of various new interaction potentials: The KC potential \citep{Kolmogorov2005}, a registry-dependent interlayer potential, the potential of Lebedeva et al. \citep{Lebedeva2011_01}, a potential enriched with density-functional theory (DFT) data, the potential of Jiang and Park \citep{Jiang2015_01}, a modification of the LJ potential by introducing Gaussian terms, the potential of Wen et al. \citep{Wen2018}, a modification of the KC potential by adding a dihedral-angle-dependent term to the repulsive part, and the potential of Leven and Maaravi et al. \citep{Leven2016, Maaravi2017}, a potential considering many-body dispersion effects. These potentials are successful in predicting the bulk properties of graphitic systems \citep{Ouyang2020}.  

The general continuum description of anisotropic friction based on frame-invariant tensors goes back to the works of Zmitrowicz \citep{Zmitrowicz1981,Zmitrowicz1989,Zmitrowicz1992} -- covering both centrosymmetric (forward/backward equivalent) and non-centrosymmetric (forward/backward different) anisotropic friction. Tensorial descriptions have become the basis for general nonlinear finite element (FE) formulations for frictional contact \citep{laursen,wriggers-contact}. The first nonlinear FE formulations for centrosymmetric anisotropic friction go back to the works of Park and Kwak \citep{park94} and  Buczkowski and Kleiber \citep{buczkowski97}. Subsequently, these formulations have been extended to non-centrosymmetric friction \citep{jones06}, anisotropic sticking \citep{Konyukhov2006_01,Konyukhov2006_02}, boundary element methods \citep{rodriguez13} and isogeometric analysis \citep{temizer14}, and they have been used in the computational study of various applications, such as wear \citep{RODRIGUEZ20121}, contact homogenization \citep{stupkiewicz14, temizer14}, and droplet sliding \citep{sauer16}. Recent works have also proposed general coupling models for friction and adhesion, both for isotropic friction \citep{mergel19,mergel21} and anisotropic friction \citep{hu22}.

Here, we develop a new continuum contact model for simulating and studying the non-dissipative anisotropic interaction of curved graphene bilayers. The continuum model is calibrated from near-zero Kelvin molecular dynamics (MD) simulations within the range of their validity. The MD simulations employ the reactive empirical bond order (REBO)+LJ potential to model the strong covalent interactions of carbon atoms within the sheets and employ various long-range interaction potentials to model the interactions between the two sheets. A nonlinear finite element contact formulation is then implemented using the calibrated continuum contact model. The proposed model is validated from the pull-out and twisting of DWCNTs. In summary, the novelties of the current work are: \\ [-7mm]
\begin{itemize}
\item Formulation of a continuum contact model for curved commensurate graphene sheets.\\[-5mm]
\item Calibration of the model from MD data across a wide range of contact pressures. \\[-5mm]
\item Nonlinear finite element implementation of the model.\\[-5mm]
\item Application of the model to the pull-out and twisting of CNTs from/within DWCNTs. \\[-5mm]
\item Validation and verification of the model using MD data and analytical results. \\[-5mm]
\end{itemize}

The remainder of the paper is organized as follows: The atomic simulation procedure and its interatomic potentials are described in Sec~\ref{atomic_simulations}. The description of the continuum interaction model, contact kinematics and tractions are given in Sec.~\ref{continuum_model}. The model calibration and behavior for flat bilayer graphene sliding are presented in Sec.~\ref{s:CCex}. The finite element formulation and the numerical results of CNT pull-out and twisting are then described in Secs.~\ref{s:fem_sec} and \ref{FEM}, respectively, followed by conclusions in Sec.~\ref{conclusion}. 
\section{Molecular simulations} \label{atomic_simulations}
In order to calibrate the proposed continuum model, molecular simulations of the interaction between two graphene sheets are used. The simulated sheets are approximately square with size 10.08 nm $\times$ 10.16 nm. The covalent and long-range bond interactions between the carbon atoms within graphene are modeled using the second generation REBO+LJ \citep{Stuart2000} potential, while the long-range bond interactions between the sheets are modeled using the LJ \citep{Lj1924}, KC \citep{Kolmogorov2005}, or Lebedeva \citep{Lebedeva2011_01} potentials (see Appendix~\ref{comp_pot} for details).

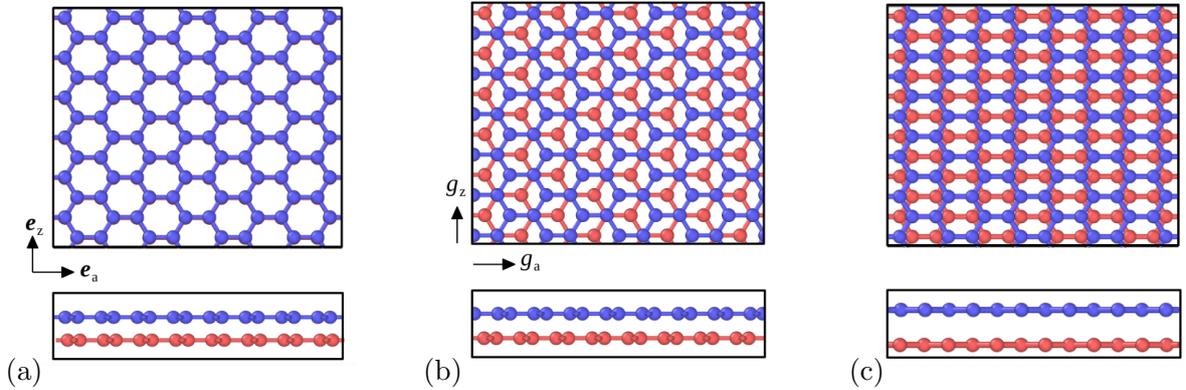
\begin{figure}[!htbp]
\begin{center} \unitlength1cm
\begin{picture}(0,4.6)
\put(-7.5,-0.3){\includegraphics[width=45mm, trim={0 0 3mm 0},clip]{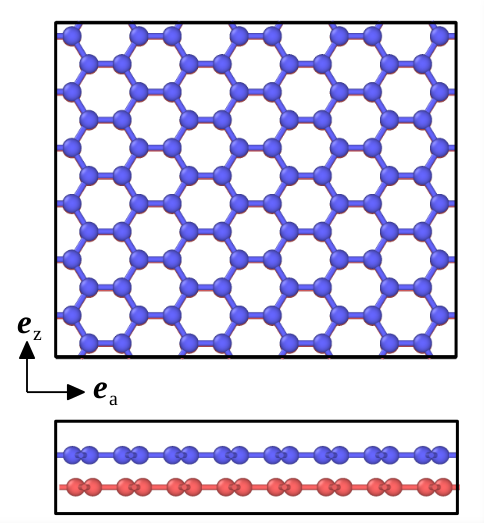}}
\put(-2,-0.3){\includegraphics[width=45mm, trim={0 0 3mm 0},clip]{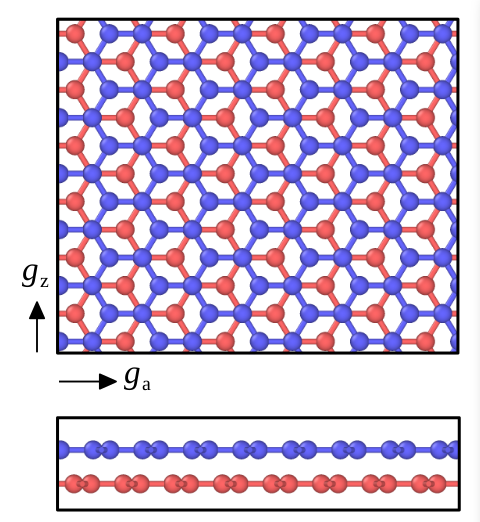}}
\put(3.5,-0.3){\includegraphics[width=45mm, trim={0 0 3mm 0},clip]{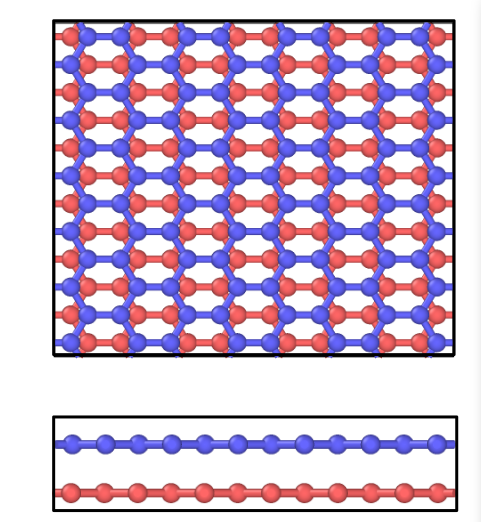}}
\put(-7.6,-0.5){(a)}
\put(-2.1,-0.5){(b)}
\put(3.5,-0.5){(c)}
\end{picture}
\vspace{5mm}
\caption{Different stackings of bi-layered graphene: (a)~AA, (b)~AB and (c)~SP stacking. $g_\mra$ and $g_\mrz$ specify the relative displacement between the two sheets along the armchair and zigzag directions (denoted $\be_\mra$ and $\be_\mrz$), respectively. } 
\label{stacking}
\end{center}
\end{figure}

We relax the system before applying tangential sliding between the layers.  The bilayer graphene system is brought to the minimum energy configuration using the Polak-Ribiere conjugate gradient method \citep{Polak1969}. Subsequently, the system is thermally equilibrated at 0.1 K employing the Nos\'e-Hoover thermostat \citep{Evans1985} with three Nos\'e-Hoover chains. Once the relaxed state is achieved, the lower layer is kept fixed by constraining all the degrees of freedom of the atoms.  The top sheet is then pulled along the armchair or zigzag direction (${ \boldsymbol{e}}_\mathrm{a}$ or ${ \boldsymbol{e}}_\mathrm{z}$ in Fig.~\ref{stacking}) parallel to the bottom layer, by providing a constant velocity of 0.01 \AA$/\text{ps}$ to all atoms lying on the four edges. While sliding, the lateral movement of the edge atoms of the top sheet is constrained. The resistance of the top sheet to sliding, i.e.~the tangential traction, is determined from 
\eqb{lll}
\ds {{\bt}_\mathrm{t}} = \ds \sum_{I=1}^N {{\bF}}_I/A\,,
\eqe
where ${{\bF}}_I$ is the tangential component of the van der Waals (vdW) force acting on atom $I$ due to the bottom layer, $A$ is the surface area of a relaxed sheet, and $N=4032$ is the total number of atoms of the top layer.

A timestep of 1 fs, suitable for the considered potentials \citep{Mokhalingam2020}, is employed to integrate the equations of motion by the velocity Verlet algorithm \citep{Swope1982}. Periodic boundary conditions are employed along the ${ \boldsymbol{e}}_\mathrm{a}$ and ${ \boldsymbol{e}}_\mathrm{z}$ directions. All the MD simulations are performed using the Large-scale Atomic/Molecular Massively Parallel Simulator (LAMMPS) \citep{lammps}.
\section{Continuum interaction model} \label{continuum_model}
This section presents the proposed continuum contact interaction model for flat and curved graphene sheets following classical nonlinear contact formulations. 
\subsection{Interaction potential for flat bilayer graphene}\label{s:flat}
For moderate contact pressures, mechanical dissipation is negligible and the surface interaction can be modeled using a surface potential. The interaction potential for two flat graphene sheets is commonly written in the form \citep{Ver2004, Lebedeva2010, Lebedeva2011}
\eqb{l}
\ds \Psi_\mathrm{flat}(\bg) = \ds \Psi_1(\gn) +  \Psi_2(\gn)\,\bar\Psi_\mrt(\ga,\gz) \,.
\label{e:Psif}\eqe
It gives the energy per undeformed area of a graphene unit cell interacting with an underlying graphene sheet.
\begin{figure}[!htbp]
\begin{center} \unitlength1cm
\begin{picture}(0,5.4)
\put(-3.6,-.1){\includegraphics[height=54mm]{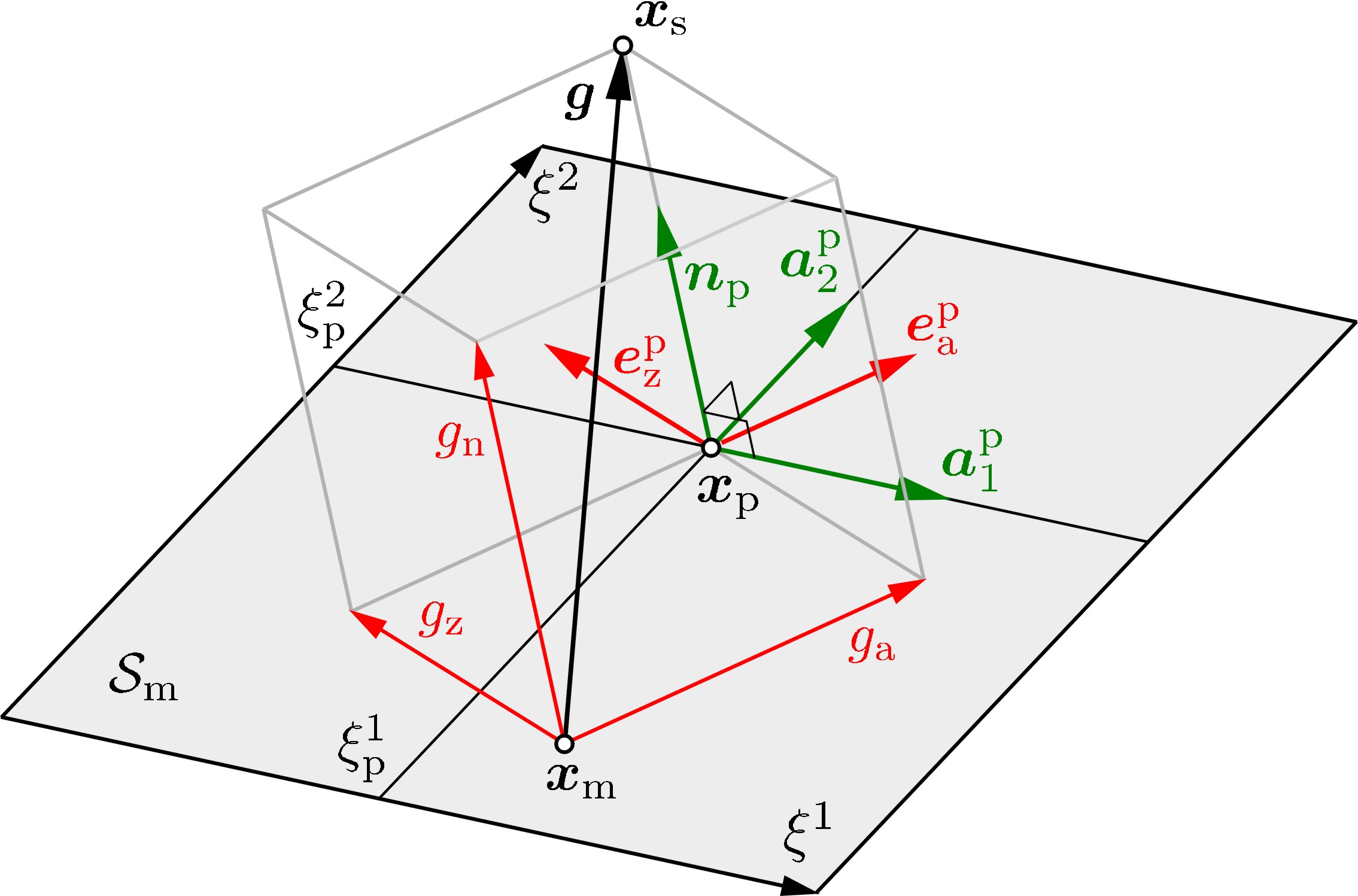}}
\end{picture}
\caption{Gap vector $\bg$ between slave point  $\bx_\mrs$ and master point $\bx_\mrm$ on $\sS_\mrm$, and its components $g_\mrn$, $g_\mra$ and $g_\mrz$. These are generally not equal to the surface coordinates $\xi^1$ and $\xi^2$ that are usually aligned with the surface geometry.} 
\label{fig:n}
\end{center}
\end{figure}
The gap vector $\bg$ with components $\gn$, $\ga$ and $\gz$ (see Fig.~\ref{fig:n}) admits any value, but the unit cell has to be aligned with the underlying sheet.
An example for functions $ \Psi_1(\gn)$, $\Psi_2(\gn)$ and $\bar\Psi_\mrt(\ga,\gz)$ is given in Sec.~\ref{s:CCex}.
There, $\ga = \gz = 0$ corresponds to the AA stack.  
Integrating \eqref{e:Psif} over the undeformed surface gives the total interaction energy
\eqb{l}
\Pi = \ds\int_{\sS} \Psi_\mathrm{flat}\,\dif A\,.
\label{e:Pi}\eqe
This has to be equal for integration over top and bottom layer.

\begin{remark}\label{r:Psi}
$\Psi_\mathrm{flat}$ contains the atomic densities of the two sheets. 
By choice the density of the unit cell is taken as the initial density, such that $\Psi_\mathrm{flat}$ is the energy per initial area.
The density of the lower sheet, however, should be taken as its current density, to account for the change in energy (and forces) due to deformation.
Thus $\Psi_\mathrm{flat}$ depends on the area change of the neighboring sheet, $J_\ell$, as described in Sauer and Wriggers \citep{sauer09b}.
For commensurate sheets the deformation in both sheets is equal, such that $J_2=J_1$ and the integration equivalence of Eq.~\eqref{e:Pi} is ensured.
For incommensurate sheets potential \eqref{e:Psif} can be modified, see Remark~\ref{r:strain}.
For small deformations, the dependency of $\Psi_\mathrm{flat}$ on $J_\ell$ and a resulting incommensurability can be neglected.
\end{remark}
\subsection{Interaction potential for curved bilayer graphene}\label{s:curv}
Two curved graphene sheets, like the walls of two nested CNTs, have different surface area.
Hence integral \eqref{e:Pi} will not be identical for both the walls unless the potential is modified.
Integrating Eq.~\eqref{e:Psif} over a common reference surface $\sS_0$ yields
\eqb{l}
\Pi = \ds\int_{\sS_0} \Psi_\mathrm{flat}\,\dif A_0\,.
\eqe 
The curved area element $\dif A_0$ of the reference surface can be related to an aligned curved area element $\dif A$ located at distance $\xi_0$ by 
\eqb{l}
\dif A_0 = S(\xi_0)\,\dif A\,,\quad S(\xi_0) = 1-2H_0\,\xi_0 + \kappa_0\,\xi_0^2\,,
\label{e:dA0}\eqe
where $H_0$ and $\kappa_0$ are the mean and Gaussian curvature of $\dif A$, respectively. 
Their sign is defined with respect to the direction of positive $\xi_0$.
Eq.~\eqref{e:dA0} is a well-known result from shell theory, e.g.~see  Ba{\c{s}}ar and Ding \citep{basar96} and Arciniega and Reddy \citep{arciniega05}.
Choosing the imaginary mid-surface $\bar\sS$ of the bilayer as the reference surface, which has the initial distance $G_\mrn/2$ from either graphene layer, then gives
\eqb{l}
\Pi = \ds\int_{\sS} \Psi_\mrc\,\dif A\,,
\label{e:Pic}\eqe
where
\eqb{l}
\Psi_\mrc = \Sm\,\Psi_\mathrm{flat}\,,\quad \Sm := S\big(\frac{G_\mrn}{2}\big) = 1-H_0\,G_\mrn + \kappa_0\ds\frac{G_\mrn^2}{4}\,,
\label{e:Psic}\eqe
is the potential for a curved graphene unit cell above a graphene sheet.

As an example, consider two CNTs with radii 
$R_\mathrm{in} = \bar R - G_\mrn/2$ and $R_\mathrm{out} = \bar R + G_\mrn/2$, where $\bar R$ is the radius of the mid-surface.
In this case $H_\mathrm{in} = -1/(2R_\mathrm{in})$ and $H_\mathrm{out} = +1/(2R_\mathrm{out})$, while $\kappa_\mathrm{in}=\kappa_\mathrm{out}=0$. 
Therefore, 
the value of $\Sm$ with respect to the outer and inner surfaces becomes $\bar S_\mathrm{in} = \bar R/R_\mathrm{in}$ and $\bar S_\mathrm{out} = \bar R/R_\mathrm{out}$, respectively, and the integration correctly yields
\eqb{l}
\Pi = \ds\int_{\sS_\mathrm{in}} \Psi_\mrc^\mathrm{in}\,\dif A_\mathrm{in} = \ds\int_{\sS_\mathrm{out}} \Psi_\mrc^\mathrm{out}\,\dif A_\mathrm{out}
= \ds\int_{\sS_\mrm} \Psi_\mathrm{flat}\,\bar R\,\dif\theta\,\dif L\,, 
\eqe
for $\Psi_\mrc^\mathrm{in} = \bar S_\mathrm{in}\,\Psi_\mathrm{flat}$ and $\Psi_\mrc^\mathrm{out} = \bar S_\mathrm{out}\,\Psi_\mathrm{flat}$,
since one can write $\dif A_\mathrm{in} = R_\mathrm{in}\,\dif \theta\,\dif L$ and $\dif A_\mathrm{out} = R_\mathrm{out}\,\dif \theta\,\dif L$, where $L$ is the length of the CNT.
\subsection{Contact kinematics}
In frictional contact, the interaction of two surfaces depends on their relative normal and tangential displacement.
This leads to the notion of a gap vector $\bg$ -- with normal and tangential components -- 
defined at every surface point.
Following classical contact notation \citep{laursen,wriggers-contact}, the two interacting surfaces are distinguished into slave and master surface (see Fig.~\ref{fig:n}).
Given the surface point $\bx_\mrs$ on the slave surface $\sS_\mrs$, its counterpart $\bx_\mrm$ on the neighboring master surface $\sS_\mrm$ is determined, as described below.
The current contact gap then is
\eqb{l}
\bg := \bx_\mrs - \bx_\mrm\,.
\eqe
%
Given a parameterization of the master surface in the form
\eqb{l}
\bx_\mrm = \bx_\mrm(\xi^1,\xi^2)\,,
\eqe
one can determine the closest projection point $\bx_\mrp := \bx_\mrm(\xi^1_\mrp,\xi^2_\mrp)$ and its corresponding gap vector $\bg_\mrp := \bx_\mrs - \bx_\mrp$ by solving the two equations 
($\alpha=1,2$)
\eqb{l}
\bg_\mrp\cdot\ba_\alpha^\mrp = 0 \,, 
\eqe
for the local coordinates $\xi^1_\mrp$ and $\xi^2_\mrp$.
Here
\eqb{l}
\ba^\mrp_\alpha := \ds\pa{\bx_\mrp}{\xi^\alpha_\mrp}\,, \quad \alpha=1,2\,,
\eqe
denote the two tangent vectors of master surface $\sS_\mrm$ at point $\bx_\mrp$ along coordinates $\xi^1$ and $\xi^2$.
Generally this is done by a local Newton-Raphson iteration for every $\bx_\mrs$ \citep{wriggers-contact}.
But for simple surfaces, such as cylinders, $\xi^\alpha_\mrp$ can be determined in closed form, as discussed below. 
Given $\bx_\mrp$ the normal gap can then be determined from
\eqb{l}
\gn = \bg_\mrp\cdot\bn_\mrp\,,
\label{e:gn}\eqe
where $\bn_\mrp$ is the surface normal of $\sS_\mrm$ at $\bx_\mrp$.
The tangential gap, on the other hand, follows directly from the coordinates $\xi^1_\mrp$ and $\xi^2_\mrp$.

\begin{figure}[h]
\begin{center} \unitlength1cm
\begin{picture}(0,6.2)
\put(-7.8,0){\includegraphics[height=60mm]{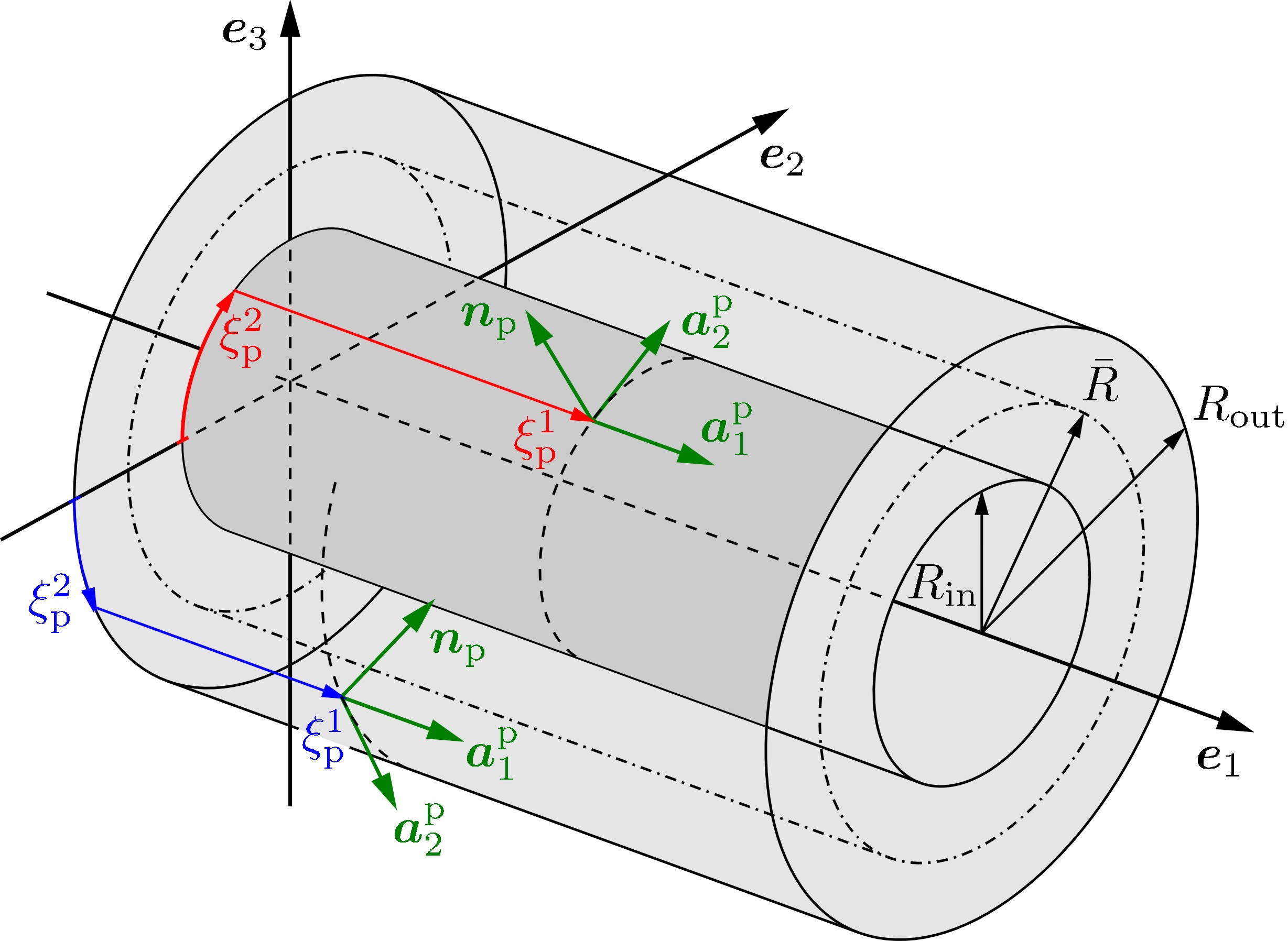}} 
\put(1.4,0){\includegraphics[height=60mm]{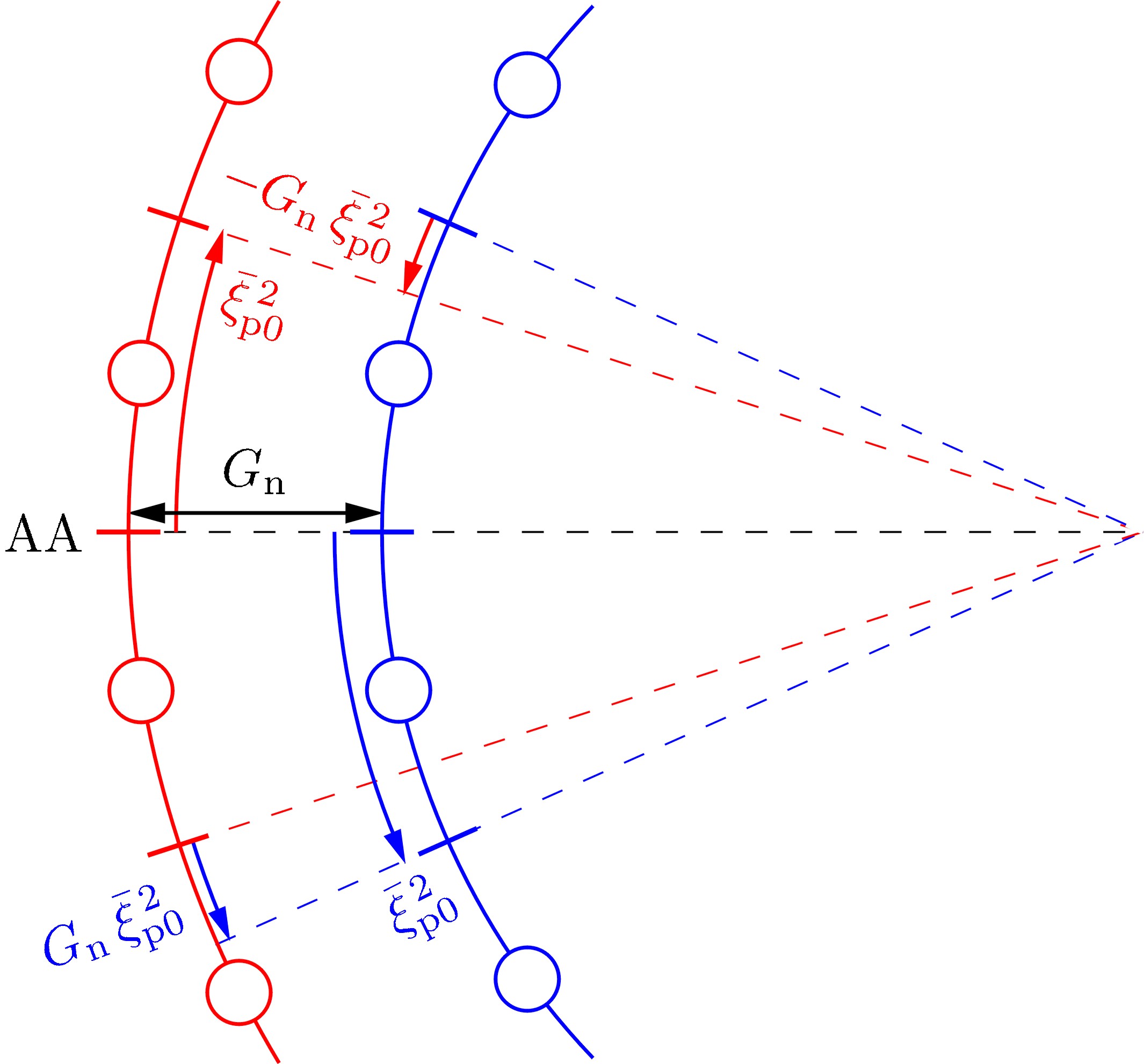}} 
\put(-7.8,0){\small (a)}
\put(1.0,0){\small (b)}
\end{picture}
\caption{Contact kinematics of interacting CNTs: (a) coordinates $\xi^\alpha_\mrp$ and master basis $\{\ba_1^\mrp,\,\ba_2^\mrp,\,\bn_\mrp\}$ on inner and outer CNT; normal vector $\bn_\mrp$ is chosen to point towards the neighboring slave CNT; $\xi^2_\mrp$ and $\ba_2^\mrp$ thus point in opposite direction on both surfaces to ensure right-handed bases; (b) initial interference between inner and outer CNT with respect to the AA stack: the location $\bar\xi^{\,2}_{\mrp0}$ on the inner CNT is ahead of the outer tube by the amount $G_\mrn\,\bar\xi^{\,2}_{\mrp0}$ (marked in blue), while the location $\bar\xi^{\,2}_{\mrp0}$ on the outer CNT lags behind by the amount $G_\mrn\,\bar\xi^{\,2}_{\mrp0}$ (marked in red).}
\label{f:cyl}
\end{center}
\end{figure}

In the following examples the interaction of two nested CNTs during pull-out and twisting is considered. 
As deformations are expected to be small, the master surface is taken to be rigid (but movable), which simplifies the contact description greatly.
The influence of this assumption on the sliding examples considered here is very small, as is seen later.
The master surface thus is a rigid cylinder.
The slave CNT and its surface point $\bx_\mrs$, on the other hand, are still considered general.
The master CNT axis is denoted by vector $\be_1$, and vectors $\be_2$ and $\be_3$ span the cylinder cross section, see Fig.~\ref{f:cyl}a. 
Vectors $\be_i$ ($i=1, 2, 3$) are taken as unit vectors and form a Cartesian basis. 
As Fig.~\ref{f:cyl}a shows, coordinates $\xi^1_\mrp$ and $\xi^2_\mrp$ are considered aligned with the axial and circumferential direction, respectively.
Either the inner CNT is the master surface and the outer CNT serves as slave (shown in red), or the outer CNT is the master surface and the inner CNT serves as slave (shown in blue).
The axial projection point coordinate can then be written as
\eqb{l}
\xi^1_\mrp = \ds \bx_\mrs \cdot \be_1 - u_{\mrm}\,, 
\label{e:xi1}\eqe
where 
$u_\mrm$ describes an axial rigid body displacement of the master surface.
The circumferential projection point coordinate can be written as
\eqb{l}
\xi^2_\mrp = R_\mrm\,\bar\xi^{\,2}_\mrp\,,\quad
\bar\xi^{\,2}_\mrp = \sign(\bn_\mrp\cdot\be_3)\, \arccos( \mp\bn_\mrp\cdot\be_2) \,, 
\label{e:xi2}\eqe
where $R_\mrm$ is the master cylinder radius and $\bar\xi^{\,2}_\mrp$ is the circumferential angle. 
The upper sign in Eq.~\eqref{e:xi2} corresponds to the case where the master cylinder is inside, while the lower sign is for the case where the master cylinder is outside, see Fig.~\ref{f:cyl}a. 
Axial rigid body rotations of the master cylinder are captured by a corresponding rotation of $\be_2$ and $\be_3$.
The surface normal $\bn_\mrp$, needed for Eq.~\eqref{e:xi2} can be determined from
\eqb{l}
\bn_\mrp = \pm\ds\frac{\bP(\bx_\mrs-\bx_0)}{\norm{\bP(\bx_\mrs-\bx_0)}} \,,
\label{e:np}\eqe
where $\bx_0$ is some point on the cylinder axis and $\bP := \bone-\be_1\otimes\be_1$ is a projection tensor.
The sign in Eq.~\eqref{e:np} follows the previous convention. 
Accordingly, $\bn_\mrp$ always points towards the other surface, see Fig.~\ref{f:cyl}a. 

\begin{figure}[h]
\begin{center} \unitlength1cm
\begin{picture}(0,4.3)
\put(-7.6,-.3){\includegraphics[height=44mm]{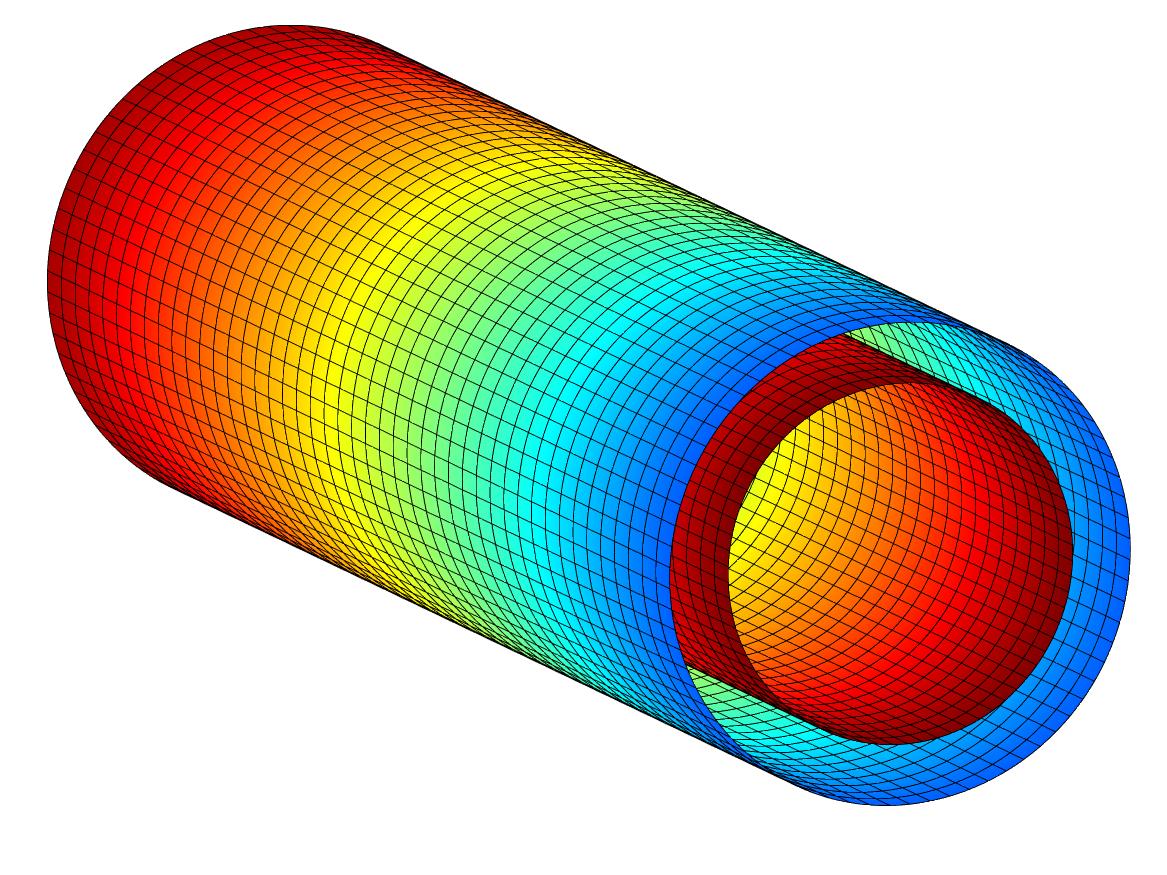}}
\put(-1.5,-.5){\includegraphics[height=48mm]{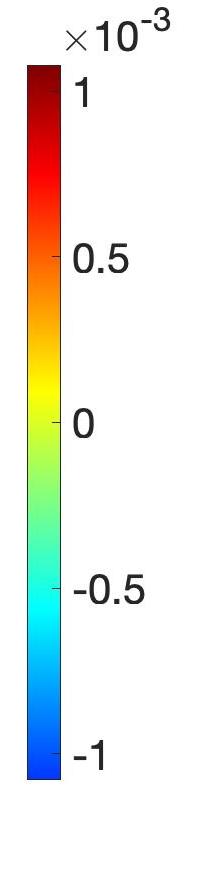}}
\put(0.4,-.3){\includegraphics[height=44mm]{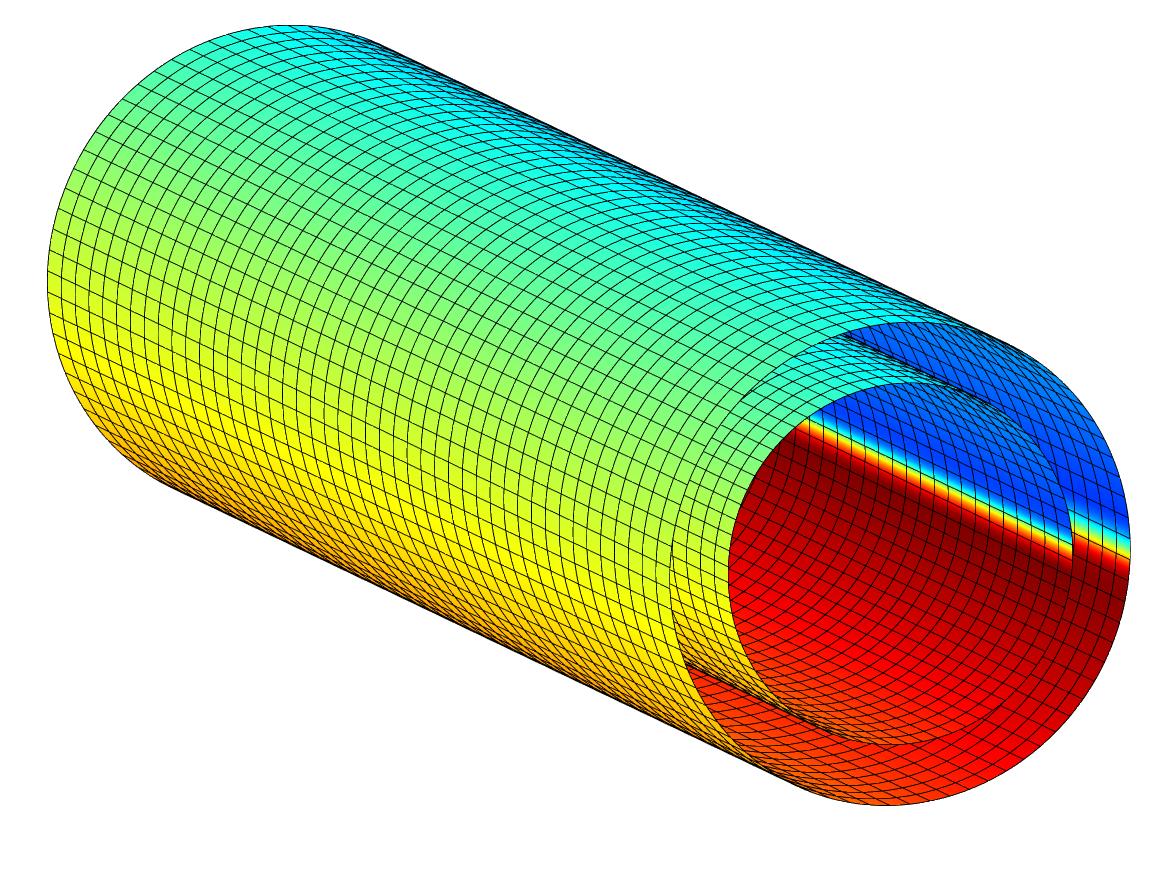}}
\put(6.5,-.5){\includegraphics[height=48mm]{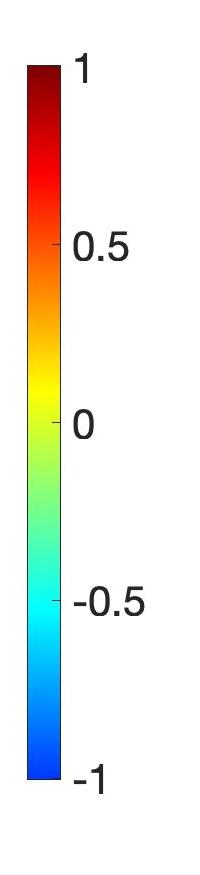}}
\put(-7.4,0){\small (a)}
\put(0.6,0){\small (b)}
\end{picture}
\caption{Contact kinematics of interacting CNTs: Color plot of initial tangential gaps $g^1$ (a) and $g^2$ (b) in [nm] for relaxed CNT(15,15) inside relaxed CNT(20,20) with respect to the AA stack of the central cross-section. The initial CNT length is $9.9207$ nm. $g^1$ has opposite sign on the two surfaces, as $\xi^1_\mrp$ runs in the same direction on the two surfaces, see Fig.~\ref{f:cyl}.
It is the other way around for $g^2$.}
\label{f:gap}
\end{center}
\end{figure}

Given the quantities and sign convention in Eqs.~\eqref{e:xi1}-\eqref{e:np},
the normal gap follows as
\eqb{l}
\gn = \ds \pm\big(\norm{\bP(\bx_\mrs-\bx_0)} - R_\mrm\big)\,,
\eqe
while the axial and circumferential gaps are
\eqb{l}
g^1 = \ds \xi^1_\mrp - \xi^1_{\mrp0} \,, 
\label{e:g1}\eqe
and
\eqb{l}
g^2 = \ds \xi^2_\mrp - \xi^2_{\mrp0} \mp G_\mrn\,\bar\xi^{\,2}_{\mrp0}\,,
\label{e:g2}\eqe
respectively.
Here $G_\mrn$ and $\xi^{\alpha}_{\mrp0}$ are the initial values of $g_\mrn$ and $\xi^{\alpha}_\mrp$ that follow from Eqs.~\eqref{e:xi1}-\eqref{e:np} for the initial (unrelaxed) location $\bX_\mrs$. 
The term $\xi^\alpha_\mrp - \xi^\alpha_{\mrp0}$ in Eqs.~\eqref{e:g1}-\eqref{e:g2} describes relative tangential motion with respect to the initial state.
The term $\mp G_\mrn\,\bar\xi^{\,2}_{\mrp0}$ in Eq.~\eqref{e:g2} is required in order to account for the circumferential lattice mismatch (i.e.~lattice interference) between the two CNTs, as is illustrated in Fig.~\ref{f:cyl}b.
Using the initial gap $G_\mrn = R_\mathrm{out}-R_\mathrm{in}$ in Eq.~\eqref{e:g2} ensures that the interference is an integer multiple of the unit cell size, which in turn ensures periodicity in $g^2$.
Fig.~\ref{f:gap} shows the initial tangential gaps $g^1$ and $g^2$ for CNT(15,15) inside CNT(20,20).
Here, $g^1$ arises due to the different stretching of the two CNTs -- the inner tube is stretched and the outer shortens due to contact pressure -- while $g^2$ is caused by the lattice mismatch term $\mp G_\mrn\,\bar\xi^{\,2}_{\mrp0}$. 
The former is negligible in comparison to the latter.

In the chosen parameterization given above, the tangent vector along $\xi^1_\mrp$ and $g^1$ becomes
\eqb{l}
\ba^\mrp_1 = \ds \be_1\,,
\label{e:ap1}\eqe
while the tangent vector along $\xi^2_\mrp$ and $g^2$ is
\eqb{l}
\ba^\mrp_2 = \sin\bar\xi^{\,2}_\mrp\,\be_2 \pm \cos\bar\xi^{\,2}_\mrp\,\be_3\,.
\label{e:ap2}\eqe
Out of these, only $\ba^\mrp_1$ is constant.
But both $\ba^\mrp_1$ and $\ba^\mrp_2$ are normalized and orthogonal to each other.
This implies that $g^1$ and $g^2$ measure the actual physical sliding distances.
It also implies that the surface metric $a^\mrp_{\alpha\beta}=\ba^\mrp_\alpha\cdot\ba^\mrp_\beta$ is equal to the Kronecker delta, 
i.e.~the $2\times2$ matrix $[a^\mrp_{\alpha\beta}]$ is the identity matrix.

In the above description, slave motions are captured through the motion of $\bx_\mrs$, leading to changed $\xi^\alpha_\mrp$ and hence updated $g^\alpha$.
Master motion, on the other hand, is captured by changing $\xi^\alpha_\mrp$ through changing $u_\mrm$, $\be_2$ and $\be_3$.

The axial and circumferential gaps $g^1$ and $g^2$ are only aligned with the graphene lattice for armchair and zigzag CNTs.
For general CNTs, described by the chirality parameters $n$ and $m$ and denoted CNT($n,m$), the armchair and zigzag gaps are given by 
\eqb{llrlr}
\ds g_\mra \is \ds g^1\cos\theta \plus g^2\sin\theta\,,  \\[3mm]
\ds g_\mrz \is \ds -g^1\sin\theta \plus g^2\cos\theta\,, 
\label{e:gaz}\eqe
with \citep{DRESSELHAUS1995}
\eqb{l}
\cos\theta := \ds\frac{2n+m}{2\sqrt{n^2 + nm + m^2}}\,,\quad
\sin\theta := \ds\frac{\sqrt{3}\,m}{2\sqrt{n^2 + nm + m^2}}\,.
\eqe
The special case $m=0$ gives zigzag CNTs ($g_\mra = g^1$ and $g_\mrz=g^2$), while $m = -2n$ gives armchair CNTs ($g_\mra=-g^2$ and $g_\mrz=g^1$). 
%
%

Introducing the matrices
\eqb{l}
[g^\alpha_{\mrc\mrc}] := \left[\begin{matrix}
g_\mra \\[3mm] g_\mrz
\end{matrix}\right],
\quad
[Q^\alpha_{\,\,\beta}] := \left[\begin{matrix}
\cos\theta & \sin\theta \\[3mm] -\sin\theta & \cos\theta
\end{matrix}\right],
\label{e:gQ}\eqe
and defining $\bar\xi^{\,1}_{\mrp0}:=0$ allows to simplify Eqs.~\eqref{e:g1}, \eqref{e:g2} and \eqref{e:gaz} into\footnote{Here and in the following, summation is implied on repeated Greek indices according to the rules of index notation.}
\eqb{l}
g_{\mrc\mrc}^\alpha = \ds Q^\alpha_{\,\,\beta}\, g^\beta \,,\quad
g^\beta = \ds \xi^\beta_\mrp - \xi^\beta_{\mrp0} \mp G_\mrn\,\bar\xi^{\,\beta}_{\mrp0}\,. 
\label{e:gcc}\eqe
Note that components $Q^\alpha_{\,\,\beta}$ need to be distinguished from the components $Q_{\!\beta}^{\,\,\alpha}$ of the transpose matrix $[Q_{\!\beta}^{\,\,\alpha}] = [Q^\alpha_{\,\,\beta}]^\mrT$ appearing in
\eqb{l}
g^\alpha = \ds  Q_{\!\beta}^{\,\,\alpha}\, g_{\mrc\mrc}^\beta \,.
\label{e:gccT}\eqe

For the derivation of the contact tractions (and their subsequent linearization) the kinematical quantities above need to be differentiated with respect to surface changes.
If only the slave surface is deformable, as is considered here, only the gradients with respect to slave point $\bx_\mrs$ are needed.
From Eqs.~\eqref{e:gn} and \eqref{e:gcc} follow
\eqb{l}
\ds\pa{g_\mrn}{\bx_\mrs} = \ds \bn_\mrp
\label{e:dgndx}\eqe
and
\eqb{l}
\ds\pa{g^\alpha_{\mrc\mrc}}{\bx_\mrs} = \ds Q^\alpha_{\,\,\beta}\,\ds\pa{\xi^\beta_\mrp}{\bx_\mrs}\,,
\label{e:dgccdxs}\eqe
since $\theta,\, G_\mrn$ and $\bar\xi^{\,\beta}_{\mrp0}$ are constant.
From Eqs.~\eqref{e:xi1}, \eqref{e:ap1} and \eqref{e:xi2}-\eqref{e:np} follow after some steps
\eqb{l}
\ds\pa{\xi^1_\mrp}{\bx_\mrs} = \ds \ba_1^\mrp\,,\quad
\ds\pa{\xi^2_\mrp}{\bx_\mrs} = \ds \frac{R_\mrm}{R_\mrm\pm g_\mrn}\ba_2^\mrp\,,
\eqe
or
\eqb{l}
\ds\pa{\xi^\alpha_\mrp}{\bx_\mrs} = c^{\alpha\beta}\,\ba_\beta^\mrp\,,\quad
[c^{\alpha\beta}] := \left[\begin{matrix}
1 & 0 \\[3mm] 0 & \ds\frac{R_\mrm}{R_\mrm\pm g_\mrn}
\end{matrix}\right].
\label{e:dxidxs}\eqe
These expressions are consistent with standard contact formulae, e.g.~see Wriggers \citep{wriggers-contact} and Sauer and De Lorenzis \citep{spbc}.
Inserting \eqref{e:dxidxs} into \eqref{e:dgccdxs} gives
\eqb{l}
\ds\pa{g^\alpha_{\mrc\mrc}}{\bx_\mrs} 
= Q^{\alpha\gamma}_\mrc\,\ba^\mrp_\gamma \,,
\label{e:dgcdx}\eqe
with 
\eqb{l}
Q^{\alpha\gamma}_\mrc := Q^\alpha_{\,\,\beta}\,c^{\beta\gamma}\,.
\label{e:Qc}\eqe
\subsection{Contact tractions}\label{s_cont_tra}
The contact traction at slave surface point $\bx_\mrs$ is given by
\eqb{l}
\bt_\mrs := -\ds\pa{\Psic}{\xs} = - \pa{\Psic}{\gn}\pa{\gn}{\xs} - \pa{\Psic}{g^\alpha_{\mrc\mrc}}\pa{g^\alpha_{\mrc\mrc}}{\xs}\,.
\eqe
Inserting Eq.~\eqref{e:dgndx} and Eq.~\eqref{e:dgcdx}, leads to
\eqb{l}
\bt_\mrs = p\,\bn_\mrp + t^\gamma\, \ba^\mrp_\gamma\,, 
\label{e:ts}\eqe
with the contact pressure 
\eqb{l}
p := - \ds\pa{\Psic}{\gn}
\label{e:p}\eqe
and tangential contact traction
\eqb{l}
t^\gamma = t_\alpha^\mrc\,Q^{\alpha\gamma}_\mrc\,,
\label{e:t}\eqe
based on
\eqb{l}
t^\mrc_\alpha := - \ds\pa{\Psic}{g^\alpha_{\mrc\mrc}}\,. 
\eqe
Here $t^1$ and $t^2$ are the traction components in axial and circumferential direction, while $t^\mrc_1$ and $t^\mrc_2$ are the traction components in armchair and zigzag direction, respectively. According to Eq.~\eqref{e:ts}, all components of $\bt_s$ are expressed in the master basis $\{\ba_1^\mrp,\,\ba_2^\mrp,\,\bn_\mrp\}$.
From Eq.~\eqref{e:Psic} and Eq.~\eqref{e:Psif} then follows
\eqb{l}
p = \Sm\,\big(p_1 + p_2\Psit  \big)\,,
\label{e:p2}\eqe
for
\eqb{l}\label{eq:ps}
p_1 := - \ds\pa{\Psi_1}{\gn}\,,\quad 
p_2 := - \ds\pa{\Psi_2}{\gn}\,,
\eqe
and
\eqb{l}
t^\mrc_\alpha = \Sm\,\Psi_2\,\bar t^\mrc_\alpha\,,
\label{e:tc2}\eqe
for
\eqb{l}
\bar t^\mrc_\alpha := -\ds\pa{\bar\Psi_\mrt}{g^\alpha_{\mrc\mrc}}\,.
\eqe
Note that expression \eqref{e:t} can also be written as 
$t^\gamma = Q^{\gamma\alpha}_{\mrc\mrT}\,t_\alpha^\mrc$ with $\big[Q^{\alpha\beta}_{\mrc\mrT}\big] := \big[Q^{\alpha\beta}_\mrc\big]^\mrT$.
Analogous to Eq.~(\ref{e:gQ}.1), we will also use
\eqb{l}
[\bar t^\mrc_\alpha] =: \left[\begin{matrix}
\bar t_\mra \\[3mm] \bar t_\mrz
\end{matrix}\right].
\label{e:tc3}\eqe
\section{Continuum model calibration}\label{s:CCex}
The continuum description in Sec.~\ref{continuum_model} is for general $\Psi_1(\gn)$, $\Psi_2(\gn)$ and $\bar\Psi_\mrt(g_\mra,g_\mrz)$.
Now specific choices for these functions are considered and calibrated from MD data for moderate contact pressures, where dissipation is negligible. At large contact pressures, mechanical energy is dissipated, and the proposed model becomes insufficient, as is shown in Sec.~\ref{model_limit}.
\vspace{-3mm}
\subsection{Potential functions} 
Considering
\eqb{l}\label{e:psi1}
\Psi_1(\gn) = p_{01}\,g_{01}\ds\Bigg(\frac{1}{10}\bigg(\frac{g_{01}}{\gn}\bigg)^{\!\!10} \!\!- \frac{1}{4}\bigg(\frac{g_{01}}{\gn}\bigg)^{\!\!4}\Bigg)\,
\eqe
leads to
\eqb{l}\label{e:p1}
p_1 = p_{01}\ds\Bigg(\bigg(\frac{g_{01}}{\gn}\bigg)^{\!\!11} \!\!- \bigg(\frac{g_{01}}{\gn}\bigg)^{\!\!5}\Bigg)\,,
\eqe
according to Eq.~(\ref{eq:ps}.1), and
\eqb{l}
p'_1 := \ds\pa{p_1}{\gn} = -\frac{p_{01}}{g_{01}}\ds\Bigg(11\bigg(\frac{g_{01}}{\gn}\bigg)^{\!\!12} \!\!- 5\bigg(\frac{g_{01}}{\gn}\bigg)^{\!\!6}\Bigg)\,,
\eqe
while the ansatz
\eqb{l}\label{e:psi2}
\Psi_2(\gn) := 
p_{02}\,g_{02}\exp\bigg(\!\!-\ds\frac{\gn}{g_{02}}\bigg)
\eqe
gives
\eqb{l}\label{e:p2}
p_2 = 
p_{02}\exp\bigg(\!\!-\ds\frac{\gn}{g_{02}}\bigg)\,,
\eqe
according to Eq.~(\ref{eq:ps}.2), and
\eqb{l}
p'_2 := \ds\pa{p_2}{\gn} 
= -\frac{p_{02}}{g_{02}}\exp\bigg(\!\!-\frac{\gn}{g_{02}}\bigg)\,.
\eqe
Here $p_{01}$, $g_{01}$, $p_{02}$, and $g_{02}$ are constants that are calibrated from MD simulations, which is discussed subsequently. The potential $\Psi_1(g_\mrn)$ specifies the mean of the interaction energy. The widely used surface-integrated LJ potential is chosen for $\Psi_1(g_\mrn)$ \citep{Girifalco_2000, Xue2022, Morovati2022}. Our proposed potential $\Psi_2(g_\mrn)$, on the other hand, is solely motivated from the obtained MD data.\\
Further, the tangential potential \citep{Ver2004, Lebedeva2010, Lebedeva2011}
\eqb{l}
\bar\Psi_\mrt(\ga,\gz) = \ds\cos\frac{4\pi\ga}{\ell_\mra} + 2\cos\frac{2\pi\ga}{\ell_\mra} \cos\frac{2\pi\gz}{\ell_\mrz}
\label{e:Psit}\eqe
yields
\eqb{lllll}
\bar t_\mra \is -\ds\pa{\bar\Psi_\mrt}{\ga} \is 
\ds\frac{4\pi}{\ell_\mra}\bigg(\!\sin\frac{4\pi\ga}{\ell_\mra} + \sin\frac{2\pi\ga}{\ell_\mra} \cos\frac{2\pi\gz}{\ell_\mrz}\bigg)\,,  \\[7mm]
\bar t_\mrz \is -\ds\pa{\bar\Psi_\mrt}{\gz} \is 
\ds\frac{4\pi}{\ell_\mrz} \cos\frac{2\pi\ga}{\ell_\mra} \sin\frac{2\pi\gz}{\ell_\mrz}\,,
\label{e:btaz}\eqe
and
\eqb{lll}
\ds\pa{\bar t_\mra}{\ga} \is 
\ds\frac{8\pi^2}{\ell^2_\mra}\bigg(\!2\cos\frac{4\pi\ga}{\ell_\mra}  + \cos\frac{2\pi\ga}{\ell_\mra} \cos\frac{2\pi\gz}{\ell_\mrz}\bigg)\,,  \\[7mm]
\ds\pa{\bar t_\mra}{\gz} \is \ds\pa{\bar t_\mrz}{\ga} = - \ds\frac{8\pi^2}{\ell_\mra\,\ell_\mrz} \sin\frac{2\pi\ga}{\ell_\mra} \sin\frac{2\pi\gz}{\ell_\mrz}\,, \\[7mm]
\ds\pa{\bar t_\mrz}{\gz} \is \ds\frac{8\pi^2}{\ell^2_\mrz} \cos\frac{2\pi\ga}{\ell_\mra} \cos\frac{2\pi\gz}{\ell_\mrz}\,.
\eqe
Here $\ell_\mra$ and $\ell_\mrz$ are treated as constants that follow from the graphene lattice, see Tab.~\ref{potential_param}, and $g_\mra$ and $g_\mrz$ are the relative displacement components between two graphene layers along the armchair and zigzag directions, respectively. Thus, function $\bar\Psi_t$ is fully specified and only $\Psi_1$ and $\Psi_2$ remain to be calibrated.
\begin{table}[!htbp]
\centering
\begin{tabular}{|c|c|c|c|c|}
  \hline
  Parameters & Value  \\[0mm] \hline 
$a_{\text{cc}}$ & 0.1397 nm  \\[.5mm]
$\bar{l}_{\mathrm{a}}$  & 3   \\[.5mm]
$\bar{l}_{\mathrm{z}}$ & $\sqrt{3}$  \\[.5mm]
$l_{\mathrm{a}}$  & $\bar{l}_{\mathrm{a}}\,a_{\text{cc}}$   \\[.5mm]
$l_{\mathrm{z}}$ & $\bar{l}_{\mathrm{z}}\,a_{\text{cc}}$  \\[.5mm]
$p_{01}$ & 5.8646 nN/nm$^2$  \\[.5mm]
$g_{01}$ & 0.3376 nm  \\[.5mm]
$p_{02}$ & $4.404\cdot10^6 $ nN/nm$^2$  \\[.5mm]
$g_{02}$  & $1.875\cdot10^{-2}$ nm  \\[.5mm]
   \hline
\end{tabular}
\caption{Potential parameters.}
\label{potential_param}
\end{table}
\begin{remark}\label{r:strain}
If the neighboring graphene lattice deforms, $\ell_\mra$ and $\ell_\mrz$ are not constant anymore.
Stretches along armchair and zigzag direction can be accounted for in Eq.~\eqref{e:Psit} by writing
\eqb{l}
\ell_\mra = \lambda_\mra\,L_\mra\,,\quad
\ell_\mrz = \lambda_\mrz\,L_\mrz\,,
\eqe
where $L_\mra$ and $L_\mrz$ are the initial lattice periods.
In the small deformation regime, the stretches are related to the corresponding infinitesimal strains by
\eqb{l}
\lambda_\mra = 1 + \varepsilon_\mra\,,\quad
\lambda_\mrz = 1 + \varepsilon_\mrz\,.
\eqe
Shear strains $\varepsilon_{\mra\mrz}$ are not accounted for in these expressions. 
It can be expected that they require changes of the functional form in Eq.~\eqref{e:Psit}.
\end{remark}
\subsection{Potential calibration}
\begin{figure}[!htbp]
\begin{center} \unitlength1cm
\begin{picture}(0,5.2)
\put(-7.8,0){\includegraphics[height=52mm]{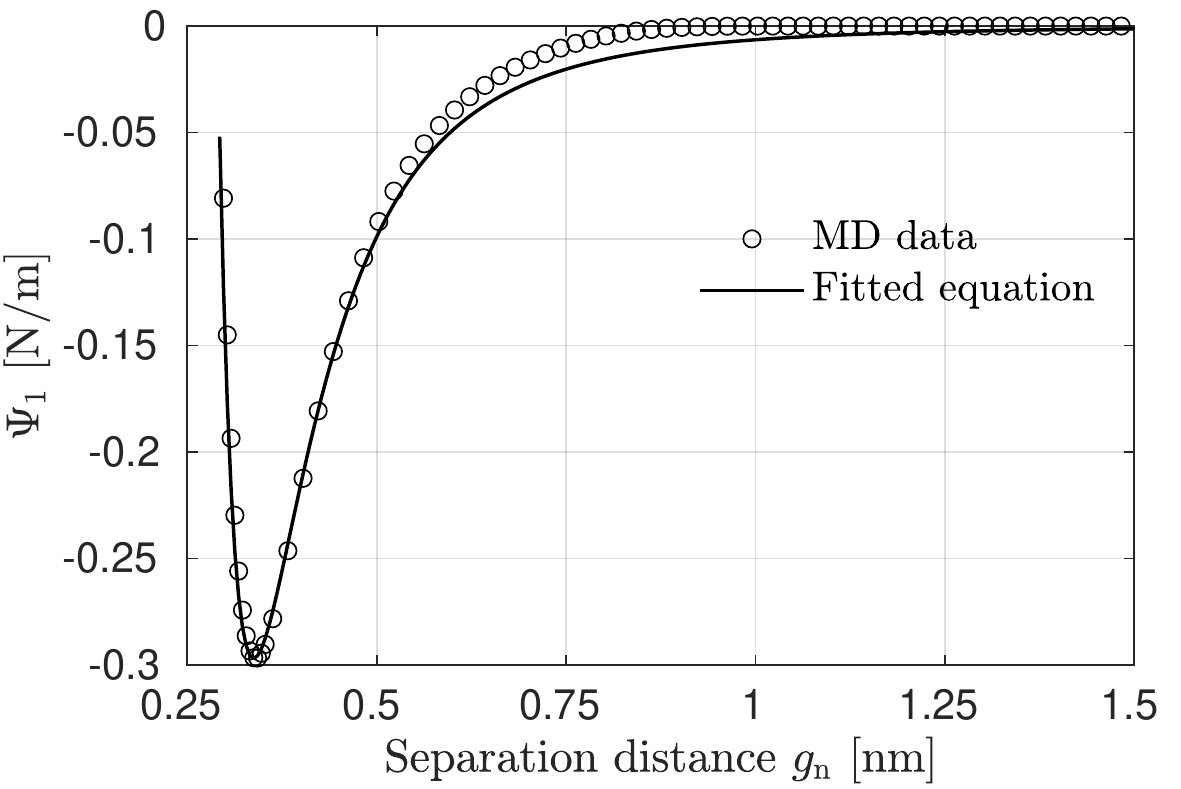}}
\put(0.2,0){\includegraphics[height=52mm]{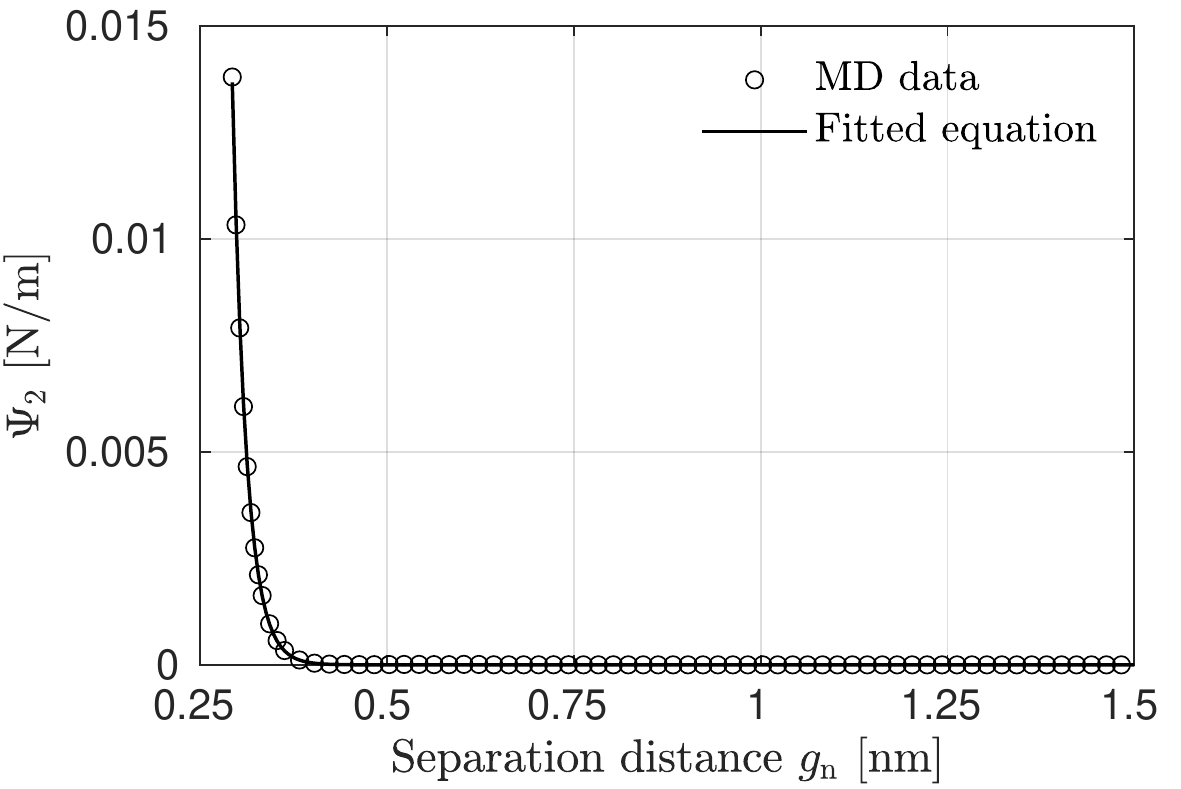}}
\put(-7.7,0){(a)}
\put(.3,0){(b)}
\end{picture}
\caption{Calibrated potential functions $\Psi_1(g_\mathrm{n})$ (a) and $\Psi_2(g_\mathrm{n})$ (b) using the shown MD data. The MD data for $\Psi_1$ is obtained at $g_\mathrm{a}=3 a_{\textnormal{cc}}/8$ and $g_\mathrm{z}=\sqrt{3}a_{\textnormal{cc}}/4$, while the MD data for $\Psi_2$ is averaged over several $g_\mathrm{a}$ and $g_\mathrm{z}$ values.  } 
\label{param_AB_AA_fit}
\end{center}
\end{figure}

To calibrate the potential parameters in Eq.~\eqref{e:psi1}, the top layer is displaced by $g_{\mathrm{a}} = 3 a_{\textnormal{cc}}/8$ and $g_{\mathrm{z}} = \sqrt{3}a_{\textnormal{cc}}/4$, and the interaction energy $\Psi_\mathrm{flat}$ between the two graphene sheets is recorded for varying $g_\mathrm{n}$ from MD simulations. For the selected values of $g_{\mathrm{a}}$ and $g_{\mathrm{z}}$, the tangential potential $\bar\Psi_\mrt$ is zero such that the interaction energy is equal to $\Psi_1$. The parameters of Eq.~\eqref{e:psi1} are then fitted to the obtained data using least square curve fitting, see Fig.~\ref{param_AB_AA_fit}a. The calibrated parameters thus become $g_{01}=0.3376$ nm and $p_{01}=5.8646$ nN/nm$^2$. \\
In order to determine the constants in Eq.~\eqref{e:psi2}, we choose different combinations of $g_{\mathrm{a}}$ and $g_{\mathrm{z}}$ and determine $\Psi_\mathrm{flat}$ from MD simulations for varying $g_\mathrm{n}$.
The values of $g_{\mathrm{a}}$ and $g_{\mathrm{z}}$ are varied over one period, respectively (see Fig.~\ref{stacking}). The potential $\Psi_2$ then follows as ($\Psi_\mathrm{flat}-\Psi_1)/\bar\Psi_\mrt$, according to Eq.~\eqref{e:Psif}. For each $g_\mathrm{n}$, the $\Psi_2$ data is averaged for all $g_{\mathrm{a}}$ and $g_{\mathrm{z}}$ values, and then used to calibrate the parameters of Eq.~\eqref{e:psi2} using least square curve fitting. This gives $p_{02}=4.404\cdot10^6 $ nN/nm$^2$ and $g_{02}=1.875\cdot10^{-2}$ nm. The two calibrated functions $\Psi_1$ and $\Psi_2$ are shown in Fig.~\ref{param_AB_AA_fit}. 

\begin{figure}[!htbp]
\begin{center} \unitlength1cm
\begin{picture}(0,5.2)
\put(-7.8,0){\includegraphics[height=52mm]{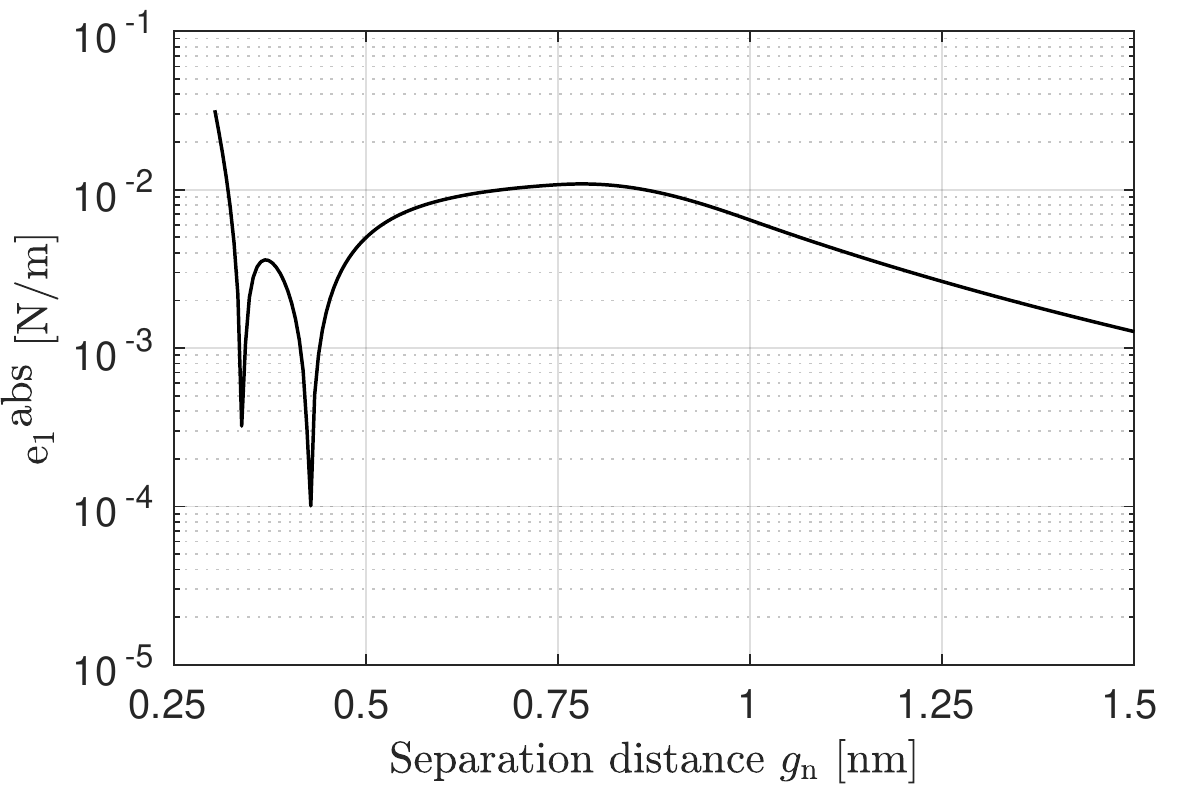}}
\put(0.2,0){\includegraphics[height=52mm]{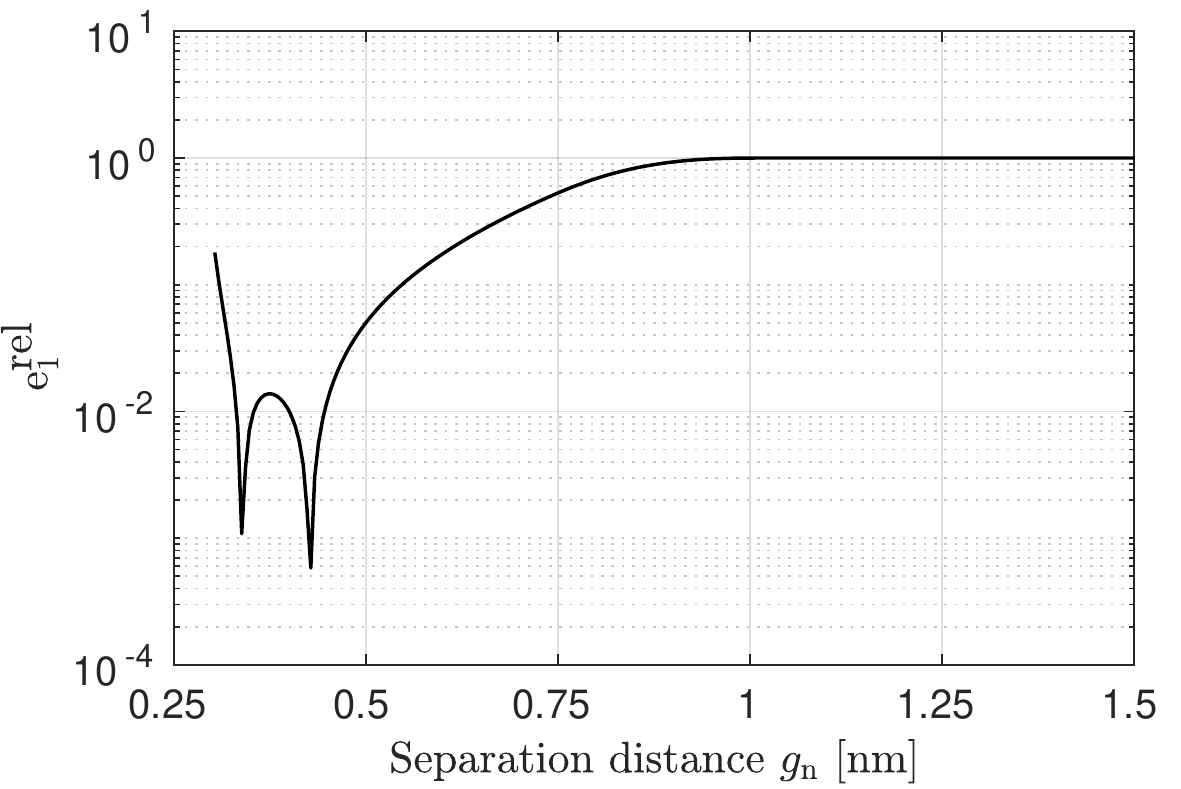}}
\put(-7.7,0){(a)}
\put(.3,0){(b)}
\end{picture}
\caption{(a)~Absolute and (b)~relative error of $\Psi_1$ as a function of separation distance $g_\mrn$.  } 
\label{U_1_errors}
\end{center}
\end{figure}

To check the accuracy of $\Psi_1$, the absolute and relative error{s} of function $\Psi_1(g_\mathrm{n})$ with respect to the MD data $\Psi_1^{\text{MD}}$ are calculated from $\mre_1^{\textnormal{abs}} = |\Psi_1^{\textnormal{MD}} - \Psi_1|$ and $\mre_1^{\textnormal{rel}} = \mre_1^{\textnormal{abs}}/|\Psi_1|$, respectively, and shown in Fig.~\ref{U_1_errors}. 
As seen, the absolute difference is less than $3\cdot10^{-2}$ N/m. The relative error approaches one as $g_\mathrm{n}$ increases, since $\Psi_1^{\text{MD}}$ reaches approximately zero for $g_\mathrm{n} > 1$ nm. All potential parameters are summarized in Tab.~\ref{potential_param}. 
\subsection{Resulting interaction behavior}\label{sec:int_beh}
The continuum model has been calibrated by fitting functions $\Psi_1$ and $\Psi_2$ for selected $g_\mathrm{a}$ and $g_\mathrm{z}$ values. We now show that this is sufficient to describe energy $\Psi_\mathrm{flat}$ and its resulting contact tractions $\bt_t$ and $p$ over a wide range of separation distances.
\subsubsection{Potential energy}
\begin{figure}[!htbp]
\begin{center} \unitlength1cm
\begin{picture}(0,4.7)
\put(-8.5,0.0){\includegraphics[height=47mm]{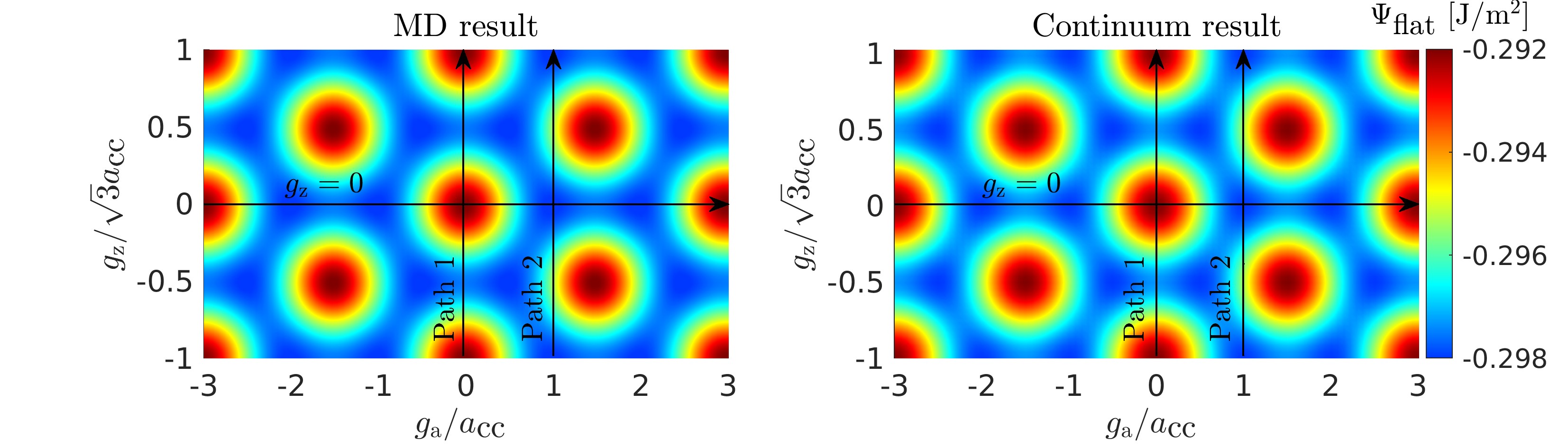}}
\end{picture}
    \caption{{Variation of interaction energy $\Psi_{\mathrm{flat}}$ obtained from MD simulations and continuum model. Three sliding paths are investigated: The armchair path at $g_\mrz = 0$, and two sliding paths at $g_\mra = 0$ and $g_\mra = a_{\text{cc}}$.} 
\label{fig:PES}}
\end{center}
\end{figure}

Fig.~\ref{fig:PES} shows that the interaction energies determined from the MD simulations and continuum model are in good agreement across the entire range of $g_\mra$ and $g_\mrz$, with an average absolute error of $\approx 3\%.$
Here, the MD results are based on the LJ potential for the interaction between the graphene layers. A comparison to other interaction potentials is discussed in Appendix~\ref{comp_pot}. {The three sliding paths shown in Fig.~\ref{fig:PES} are examined next. They are kept straight in order to sample all energy levels.}\footnote{{If lateral motions are allowed, the sliding trajectory will follow the minimum energy paths along the blue valleys \citep{Ouyang2018}.}} 

\begin{figure}[h]
\begin{center} \unitlength1cm
\begin{picture}(0,5.2)
\put(-7.8,0){\includegraphics[height=52mm]{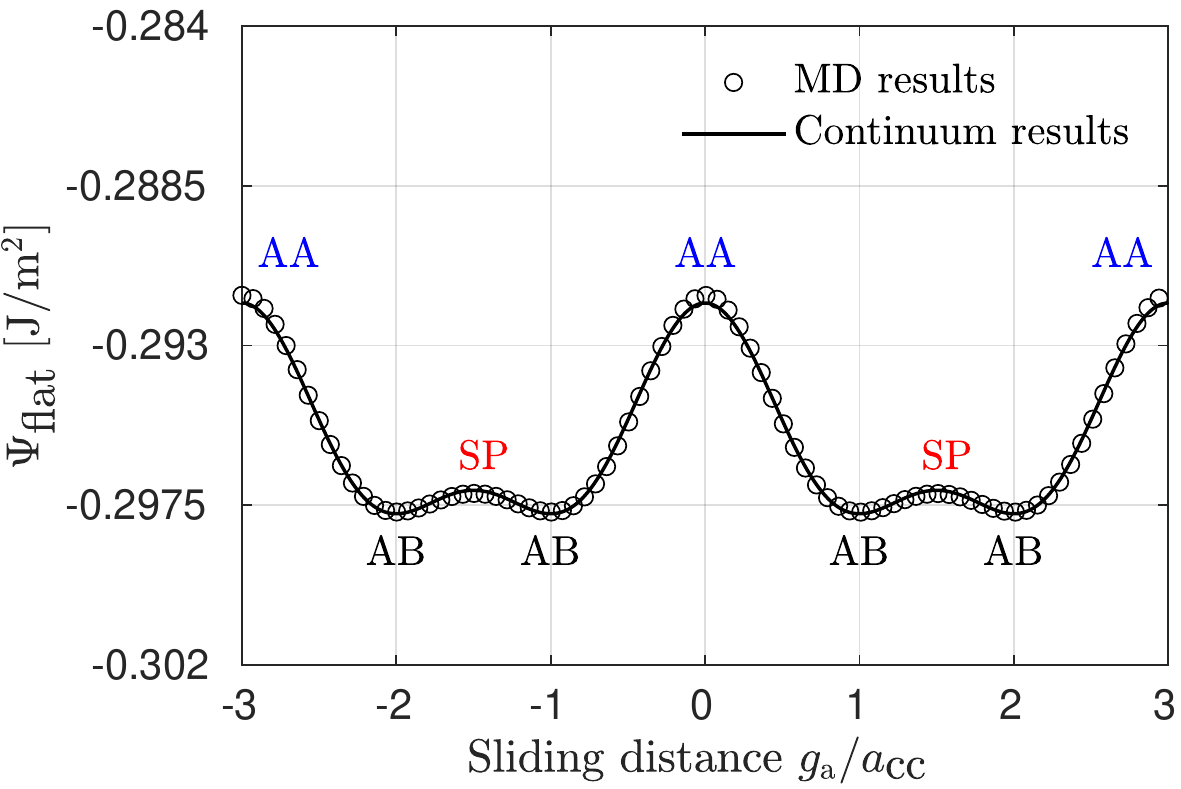}}
\put(0.2,0){\includegraphics[height=52mm]{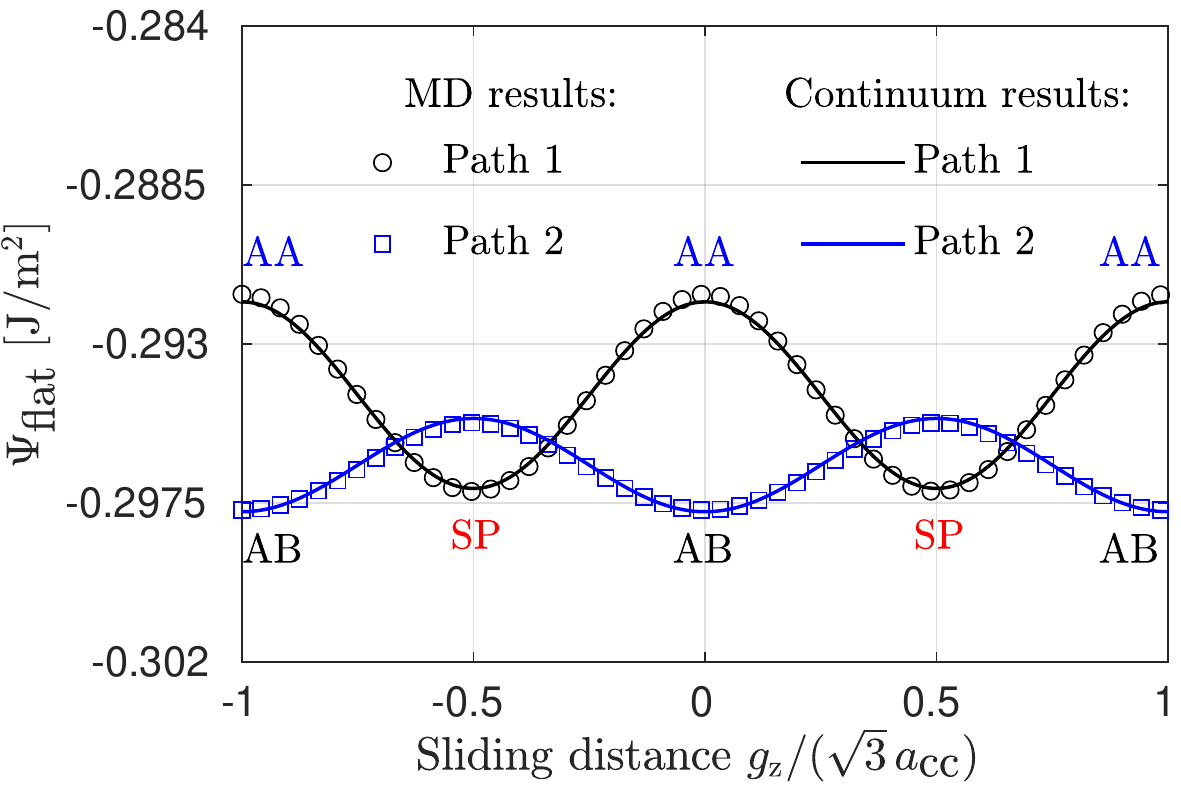}}
\put(-7.7,0){(a)}
\put(.3,0){(b)}
\end{picture}
\caption{Variation of interaction energy $\Psi_{\mathrm{flat}}$: Sliding along the (a)~armchair {path and (b)~two zigzag paths shown in Fig.~\ref{fig:PES}}, all at $g_\mathrm{n}=0.3366$ nm ($\sim$ equilibrium separation distance of the AB stacking). The difference between (a) and (b) illustrates the sliding anisotropy. } 
\label{fig:energy}
\end{center}
\end{figure}

Fig.~\ref{fig:energy}a shows the interaction energy $\Psi_{\mathrm{flat}}$ as a function of relative armchair displacement $g_\mathrm{a}$ for $g_\mathrm{z} = 0$. The separation distance between the two sheets is taken as $g_\mathrm{n}=0.3366$ nm, which corresponds to the equilibrium separation distance of the AB stacking.
The maxima of $\Psi_{\mathrm{flat}}~\text{vs.}~g_\mathrm{a}$ are located at $g_\mathrm{a}=0$  and multiples of $\ell_a$, which all correspond to the AA stacking. The global minima and local maxima are located at $g_\mathrm{a}=\pm a_{\textnormal{cc}},~g_\mathrm{a}=\pm 2a_{\textnormal{cc}}$ and $g_\mathrm{a}=\pm 1.5a_{\textnormal{cc}}$, which are the AB and SP stackings, respectively. This is in agreement with experimental studies \citep{Dienwiebel2005}. The binding energy of bilayer graphene at the equilibrium separation distance of the AB stacking is -46.9 meV/atom\footnote{The binding energy of bilayer graphene in meV/atom is calculated by normalizing the total interaction energy by the total number of atoms. 1 J/m$^2$ = 0.15758 eV.}, agreeing with the value -45.6 meV/atom from Lebedeva et al. \citep{Lebedeva2011_01}. The slight difference can be attributed to the elastic nature of the sheet, as well as the accuracy of the interatomic potential functions and the constants used. Further, the amplitude of $\Psi_{\mathrm{flat}}$, obtained as $\Delta \Psi_{\mathrm{flat}}^\textnormal{AA}=\Psi_{\mathrm{flat}}^\textnormal{AA}-\Psi_{\mathrm{flat}}^\textnormal{AB}$, is $6.129\cdot10^{-3}$ N/m, and the amplitude between the AB and SP stacking is $\Delta \Psi_{\mathrm{flat}}^\textnormal{SP}=\Psi_{\mathrm{flat}}^\textnormal{SP}-\Psi_{\mathrm{flat}}^\textnormal{AB}=6.770\cdot10^{-4}$ N/m.

{Fig.~\ref{fig:energy}b shows} the interaction energy $\Psi_{\mathrm{flat}}$ as a function of the relative zigzag displacement $g_\mathrm{z}$ for $g_\mathrm{a} = 0$ (Path 1) and $g_\mathrm{a} = a_\text{cc}$ (Path 2). The maxima of $\Psi_{\mathrm{flat}}~\text{vs.}~g_\mathrm{z}$ for Path 1 are located at $g_\mathrm{a} = 0$ and multiples of $\ell_z$, which all correspond to the AA stacking, while the minima are at $g_\mathrm{a} = \pm \ell_z /2$, which correspond to the SP stacking. For Path 2, the minima correspond to the AB stacking.
\subsubsection{Tangential traction}
\begin{figure}[!htbp]
\begin{center} \unitlength1cm
\begin{picture}(0,5.2)
\put(-7.8,0){\includegraphics[height=52mm]{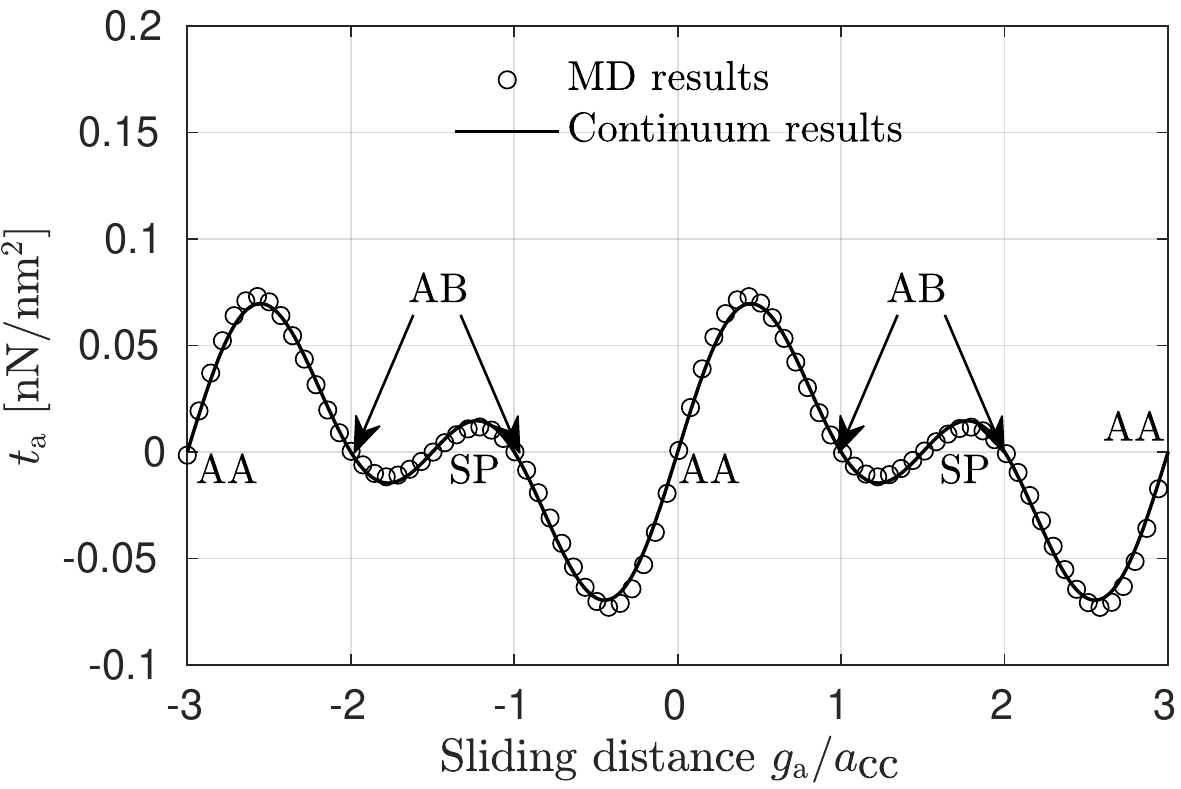}}
\put(0.2,0){\includegraphics[height=52mm]{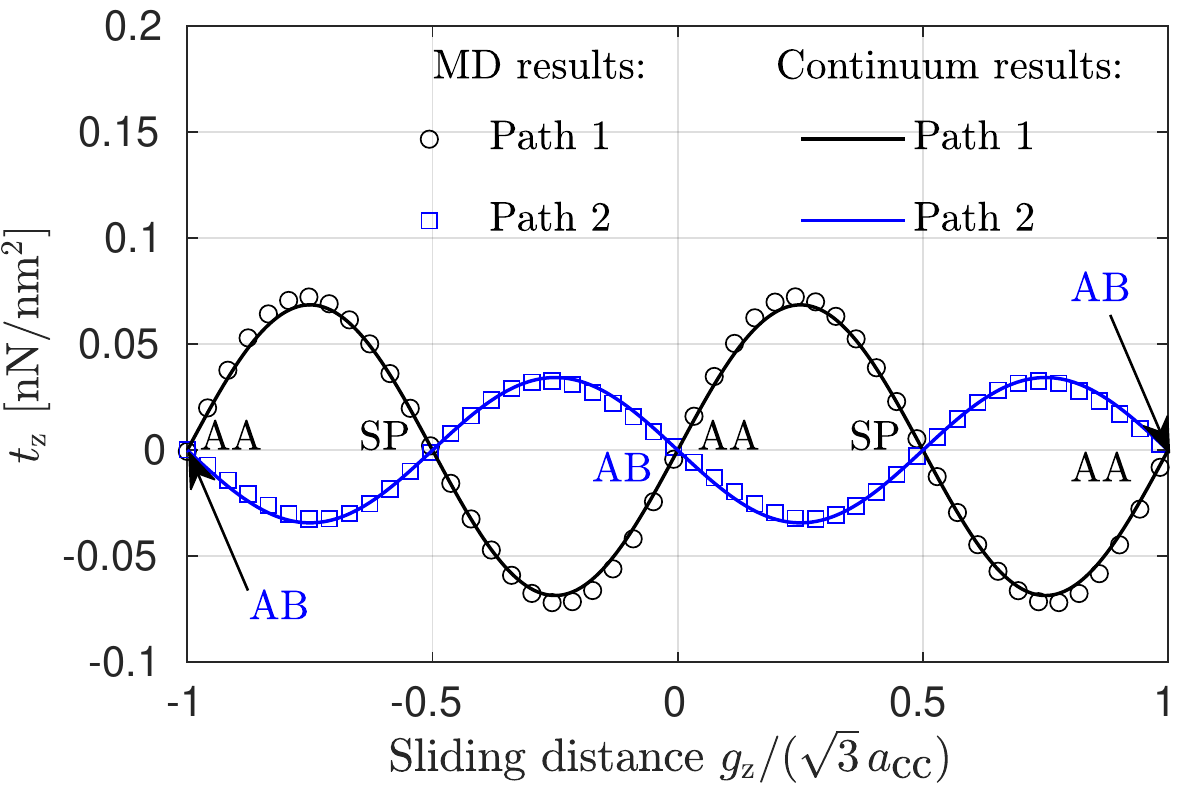}}
\put(-7.7,0){(a)}
\put(.3,0){(b)}
\end{picture}
\caption{Tangential traction: Sliding along the (a)~armchair direction for $g_\mathrm{z} = 0 $ and (b)~zigzag direction for $g_\mathrm{a}$ = 0 (Path 1) and $g_\mathrm{a} = a_{\textnormal{cc}}$ (Path 2), all at $g_\mathrm{n}=0.3366$ nm ($\sim$ equilibrium separation distance of AB stacking). The difference between (a) and (b) illustrates the sliding anisotropy.  } 
\label{F_AB1}
\end{center}
\end{figure}

For flat graphene sheets, the expressions in Sec.~\ref{s_cont_tra} simplify to the tangential traction
\eqb{lll}\label{eq:tan_trac}
\ds \bt_t \is \Psi_2\,\big(\bar t_\mra\,\be_\mra + \bar t_\mrz \, \be_\mrz \big)\,, 
\eqe
where $\bar t_\mra$ and $\bar t_\mrz$ are given by Eq.~\eqref{e:btaz} and $\Psi_2$ was calibrated above. Figs.~\ref{F_AB1}a and \ref{F_AB1}b show the comparison of these tractions with those determined from the MD simulations considering sliding along the armchair direction (for $g_\mathrm{z}$ = 0) and zigzag direction (for Path 1 and Path 2), respectively. 
As seen, the continuum tractions {agree well with} those obtained from the MD simulations across the entire range of sliding distances, with an average error of $\approx 3\%.$

\begin{figure}[!htbp]
\begin{center} \unitlength1cm
\begin{picture}(0,5.2)
\put(-7.8,0){\includegraphics[height=52mm]{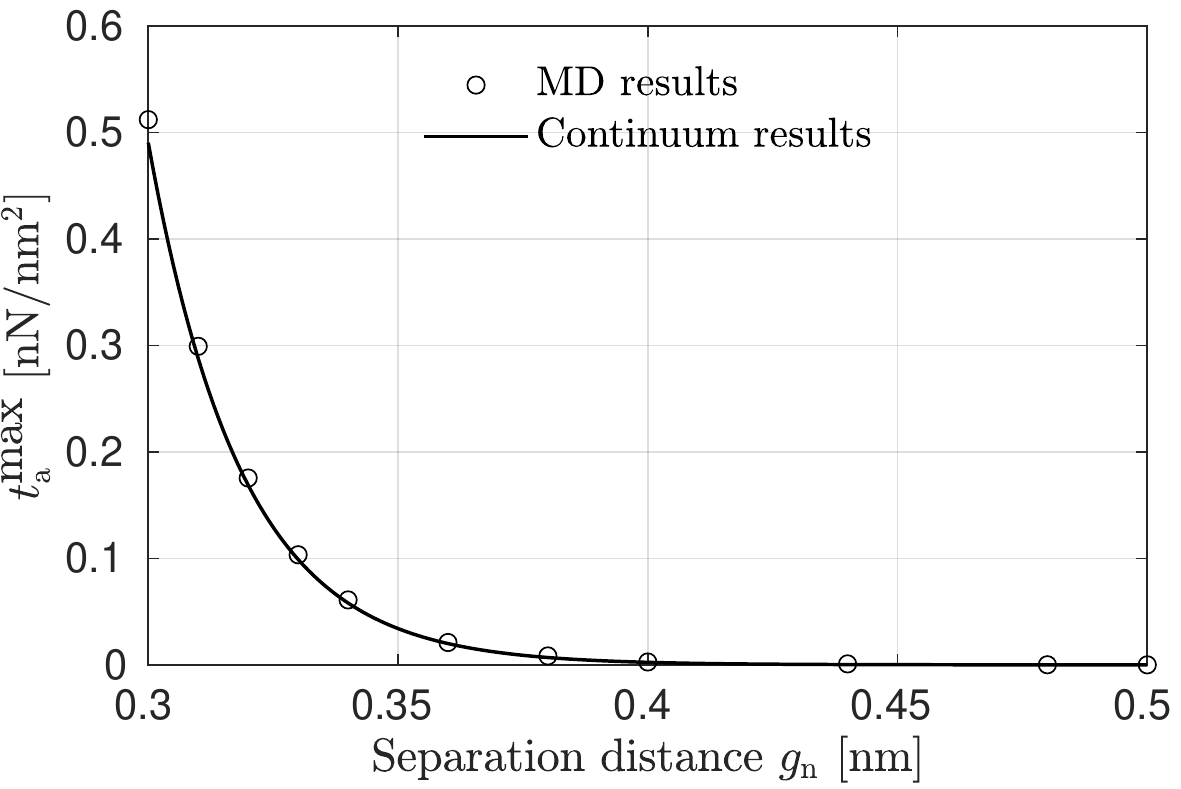}}
\put(0.2,0){\includegraphics[height=52mm]{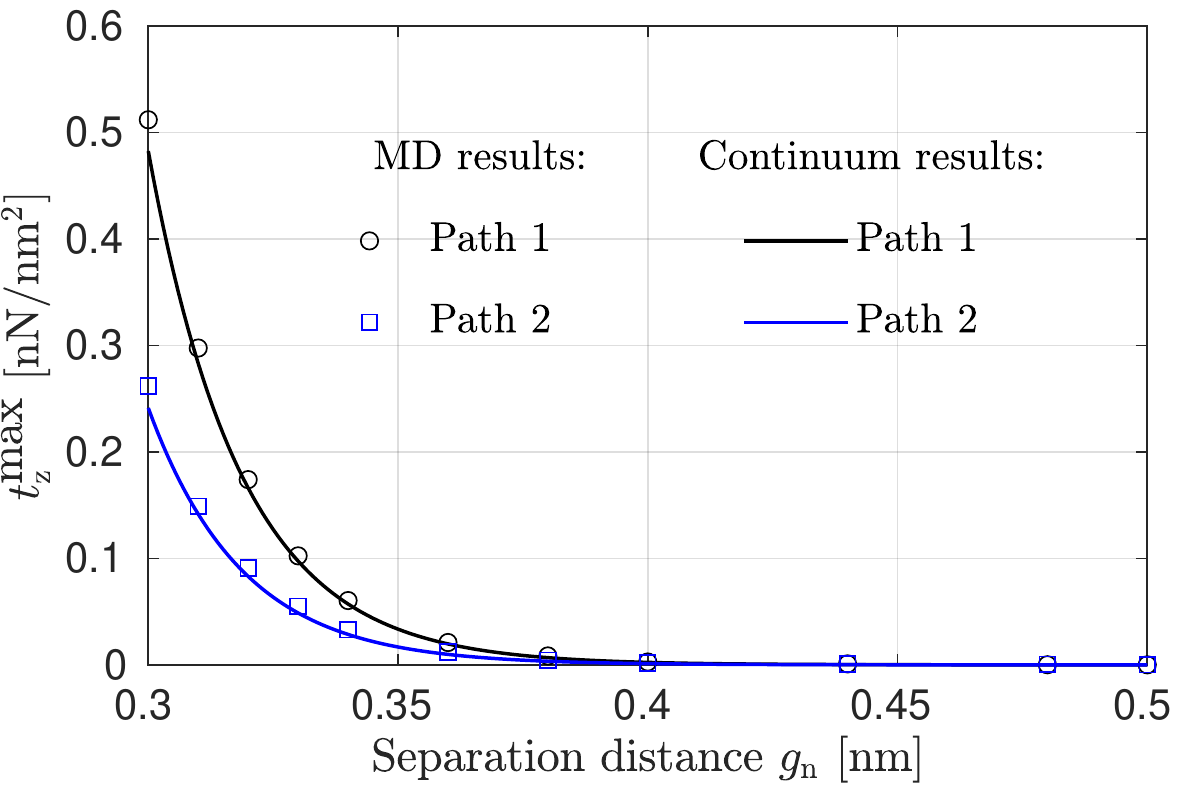}}
\put(-7.7,0){(a)}
\put(.3,0){(b)}
\end{picture}
\caption{Maximum tangential traction (= sticking limit) at different separation distances for sliding along the (a)~armchair direction for $g_\mathrm{z} = 0 $ and (b)~zigzag direction for $g_\mathrm{a}$ = 0 (Path 1) and $g_\mathrm{a} = a_{\textnormal{cc}}$ (Path 2).}     \label{F_max}
\end{center}
\end{figure}

The amplitude of the tangential traction, calculated as $t_{{s}}^{\text{max}}-t_{{s}}^{\text{min}}$, where subscript ${s}$ stands for the sliding direction, is 0.1460 N/m for the path in Fig.~\ref{F_AB1}a, which reduces by $\approx 1.2\%$ and $\approx 55\%$ for the two zigzag paths shown in Fig.~\ref{F_AB1}b. Further, the amplitude of the sticking limit i.e.~$t_{{s}}^{\text{max}}-t_{{s}}^{\text{min}}$ depends on the separation distance $g_\mathrm{n}$ (or normal pressure), as Fig.~\ref{F_max} shows. 
In all three cases, the sticking limit decays exponentially with increasing separation gap. 

\subsubsection{Normal traction}
\begin{figure}[h]
\begin{center} \unitlength1cm
\begin{picture}(0,5.2)
\put(-7.8,0){\includegraphics[height=52mm]{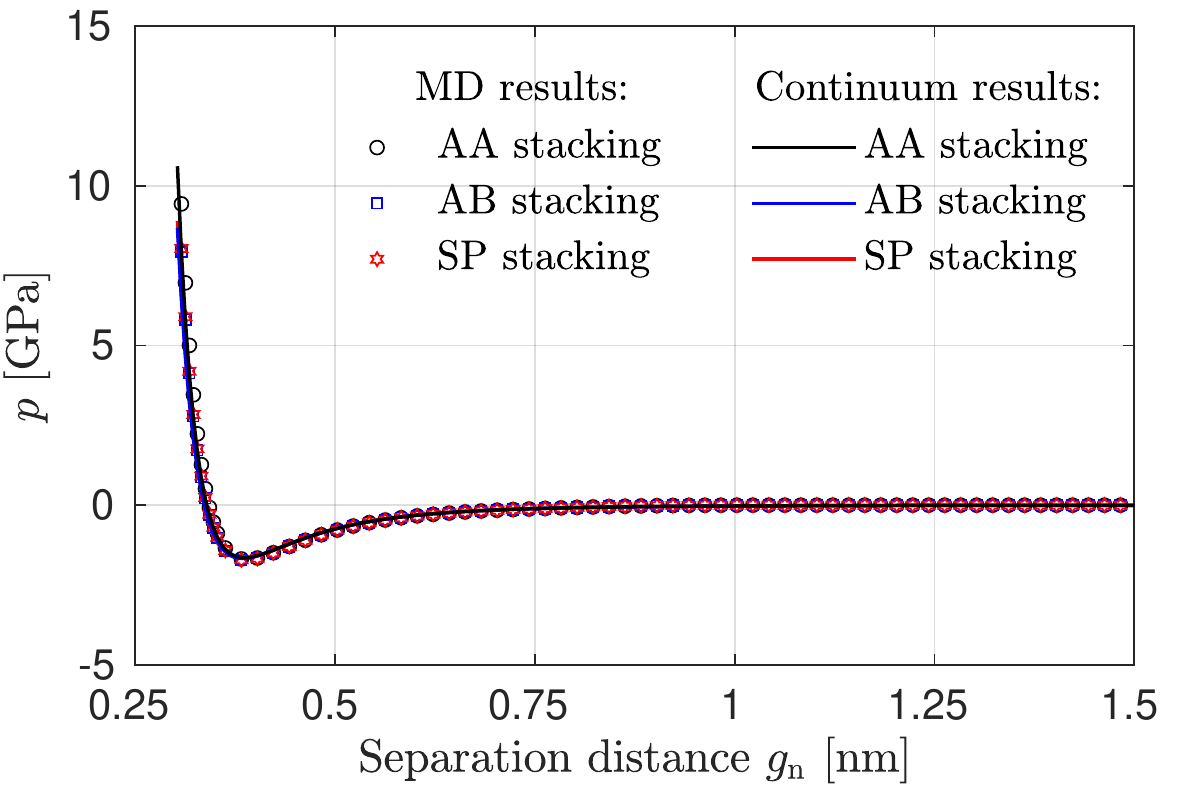}}
\put(0.2,0){\includegraphics[height=52mm]{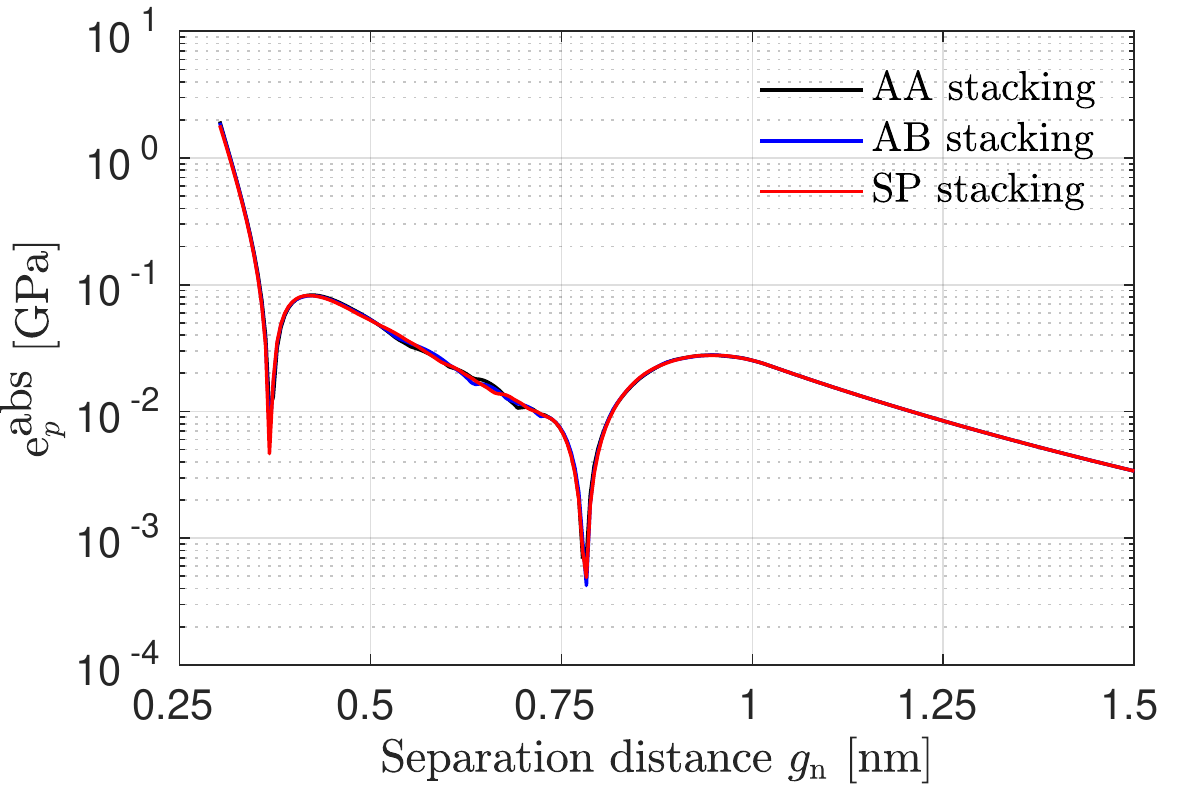}}
\put(-7.7,0){(a)}
\put(.3,0){(b)}
\end{picture}
\caption{(a)~Contact pressure or normal traction ($t_\mathrm{n}$) and (b)~absolute error of $t_\mathrm{n}$ as a function of separation distance for the three stackings. } 
\label{pressure_fit}
\end{center}
\end{figure}

For flat graphene sheets, the expression in Eq.~\eqref{e:p2} simplifies to the contact pressure (i.e. normal traction)
\eqb{lll} \label{eq:nor_tract}
\ds p \is \ds p_1 + \bar \Psi_\mrt \, p_2\,,
\eqe
which is specified through Eqs.~\eqref{e:p1}, \eqref{e:p2}, \eqref{e:Psit} and Tab.~\ref{potential_param}. The value of $\bar \Psi_\mrt(g_\mathrm{a}, g_\mathrm{z})$ in Eq.~\eqref{eq:nor_tract} for the three stackings is $\bar \Psi_\mrt^{\text{AA}}=3$, $\bar \Psi_\mrt^{\text{AB}}=-1.5$, and $\bar \Psi_\mrt^{\text{SP}}=-1$. The comparison of the continuum and MD results for $p(g_\mathrm{n})$ is shown in Fig.~\ref{pressure_fit}a, while Fig.~\ref{pressure_fit}b shows the absolute error defined as $\mre^{\text{abs}}_p=|p^{\textnormal{MD}}-p|$. 
The normal traction depends not only on the separation distance between the two layers but also on the type of stacking. At $g_\mathrm{n} = 0.3$ nm, $p$ equals $\approx 11.8$ nN/nm$^2$ for the AA stacking. At the same $g_\mathrm{n}$, these values are $\approx 15.8 \%$ and $\approx 10.9 \%$ less for the AB and SP stackings, respectively, compared to the AA stacking. The equilibrium separation distances are 0.3394, 0.3366, and 0.3370 nm for the AA, AB, and SP stackings, and Eq.~\eqref{eq:nor_tract} captures this behavior sufficiently well, as Fig.~\ref{pressure_fit} shows.
\subsection{High-pressure limitation}\label{model_limit}
\begin{figure}[h]
\begin{center} \unitlength1cm
\begin{picture}(0,5.2)
\put(-7.8,0){\includegraphics[height=52mm]{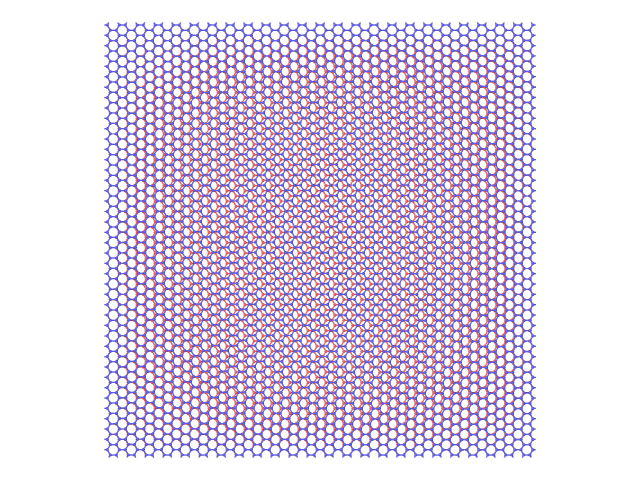}}
\put(-1,0){\includegraphics[height=52mm]{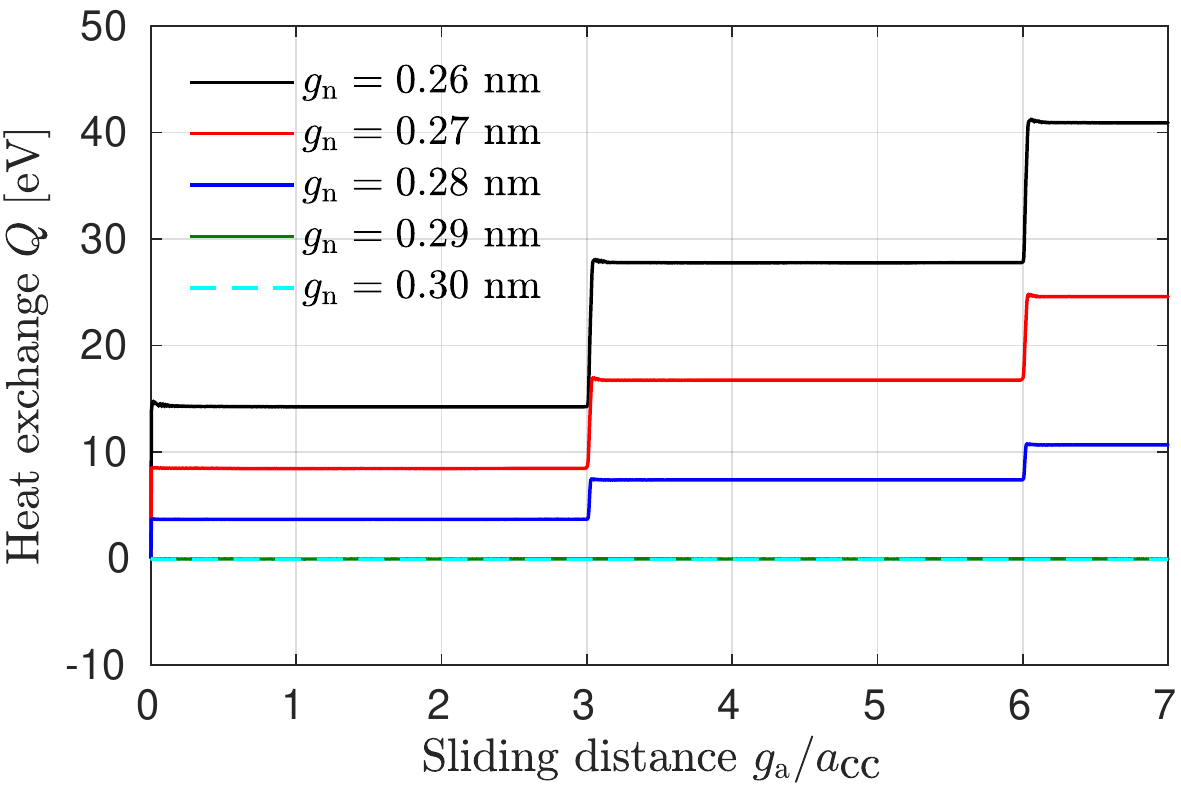}}
\put(-7.7,0){(a)}
\put(-0.9,0){(b)}
\end{picture}
\caption{Transition from dissipative to non-dissipative sliding friction: (a)~Bilayer configuration at $g_{\mathrm{a}}=0$, $g_{\mathrm{z}}=0$ and $g_\mathrm{n}=0.26$ nm.
(b)~Thermal energy exchange with the heat bath at different $g_\mathrm{n}$ for sliding along the armchair direction (for $g_\mathrm{z} = 0 $). The figure shows that sliding contact becomes non-dissipative at $g_\mrn \geq 0.29$ nm. } 
\label{f:heat_exchange}
\end{center}
\end{figure}

The calibration of the continuum model assumes small in-plane deformations of the graphene layers, which is accurate for separation distances larger than 0.29 nm. For $g_\mathrm{n} < 0.29$ nm, the contact pressure becomes very large leading to non-uniform tangential deformations in the graphene layer as observed in MD simulations\footnote{{At each applied pressure level, the system is relaxed using the Polak-Ribiere conjugate gradient method followed by thermal equilibration employing the Nos\'e-Hoover thermostat.} } (see Fig.~\ref{f:heat_exchange}a). 
As a result, the bilayer graphene system attains different stackings in different regions. As Fig.~\ref{f:heat_exchange}a shows, the center of the sheet is in the AB stacking while the edges remain in the AA stacking. This tangential deformation is solely a result of the large contact pressure, as no tangential displacements are applied to the boundary in this case. As a consequence of the tangential displacements  at the center, energy is dissipated when additional tangential displacements are applied to the boundary. Figure~\ref{f:heat_exchange}b therefore shows the amount of heat exchanged with the thermostat, $Q$, as a function of the sliding distance along the armchair direction (for $g_\mathrm{z}$ = 0 and various values for $g_\mathrm{n}$). As seen in the figure, for $g_\mrn \leq 0.28$ nm $Q$ increases sharply for every $3a_{\textnormal{cc}}$ of sliding distance. The increase is constant for fixed $g_\mrn$. Each increase corresponds to a sudden release of the strain energy in the two layers. Since the energy remains constant after each increase, it implies that energy is lost to the heat bath. This mechanical dissipation mechanism is not accounted for in the present continuum model. The present continuum model, which is conservative, is therefore only valid for $g_\mathrm{n} \geq $ 0.29 nm.  

\section{Finite element formulation} \label{s:fem_sec}
The calibrated continuum interaction model can be implemented straightforwardly within a nonlinear finite element contact code. The contact traction from Eq.~\eqref{e:ts} and its gradient {in} Eq.~\eqref{e:dtsdxs} enter the finite element contact force vector 
\eqb{l}
\mf^e_\mrc = - \ds\int_{\Omega^e_0}\mN^\mrT_e\,\bt_\mrs\,\dif A\,,
\eqe
and its associated stiffness matrix
\eqb{l}
\mk^e_\mrc = - \ds\int_{\Omega^e_0}\mN^\mrT_e\,\pa{\bt_\mrs}{\bx_\mrs}\,\mN_e\,\dif A
\eqe
\citep{sauer-phd,sauer09b}.
Vector $\mf^e_\mrc$ acts on the FE nodes of the slave surface and is integrated over the reference slave element domain $\Omega^e_0$ in accordance to the integration defined in Eq.~\eqref{e:Pic}.
Elemental array $\mN_e:=[\bone N_1,\, \bone N_2,\, ...,\, \bone N_{n_e}] $, where $\bone$ is the $3 \times 3$ identity tensor, contains the $n_e$ nodal shape functions $N_I$ that discretize the current and reference geometry according to
\eqb{lllll}
\bx \ais \ds\sum_I^{n_e} N_I\,\bx_I \is \mN_e\,\mx_e\,,\\[7.5mm]
\bX \ais \ds\sum_I^{n_e} N_I\,\bX_I \is \mN_e\,\mX_e\,,
\eqe
and the displacement field,
\eqb{l}
\bu := \bx - \bX \approx \ds\sum_I^{n_e} N_I\,\bu_I = \mN_e\,\muu_e\,,
\eqe
within each element $e$.
Here
\eqb{l}
\mx_e := \left[\begin{matrix}
\bx_1 \\[1.5mm] 
\bx_2 \\[1.5mm]
\vdots \\[1.5mm]
\bx_{n_e}
\end{matrix}\right],\quad
\mX_e := \left[\begin{matrix}
\bX_1 \\[1.5mm] 
\bX_2 \\[1.5mm]
\vdots \\[1.5mm]
\bX_{n_e}
\end{matrix}\right],\quad
\muu_e := \left[\begin{matrix}
\bu_1 \\[1.5mm] 
\bu_2 \\[1.5mm]
\vdots \\[1.5mm]
\bu_{n_e}
\end{matrix}\right],
\quad \bu_I = \bx_I-\bX_I\,,
\eqe
denote the arrays of all $n_e$ nodal positions and displacements of element $e$.
The elemental contributions $\mf^e_\mrc$ and $\mk^e_\mrc$ are assembled in the global arrays $\mf_\mrc$ and $\mk_\mrc$
that enter the discretized weak form and its linearization, which is required for a global Newton-Raphson {solution} procedure.
In the present formulation the master surface is treated rigidly (but movable). 
Hence only the deformation of the slave surface is discretized and computed by FE.
Alternating the designation of master and slave surface then allows to asses the error of treating the master surface rigidly. For the following examples, the preceding equations have been implemented in the isogeometric shell finite element formulation of Duong et al. \citep{solidshell} and Ghaffari et al. \citep{graphene} using the
contact interaction formulation of Sauer and De Lorenzis \citep{spbc,spbf}.

\section{Application examples: CNT pull-out \& twisting} \label{FEM}
We now turn to a set of application examples for validating the proposed continuum model. Considered is a CNT that is either pulled-out from or twisted within a second CNT.
\subsection{CNT geometry and loading}
\subsubsection{Initial CNT geometry}
\begin{figure}[!htbp]
\begin{center} \unitlength1cm
\begin{picture}(0,5.8)
\put(-4.0,-.1){\includegraphics[trim={1cm 13.0cm 0.8cm 4cm},clip,height=5.5cm]{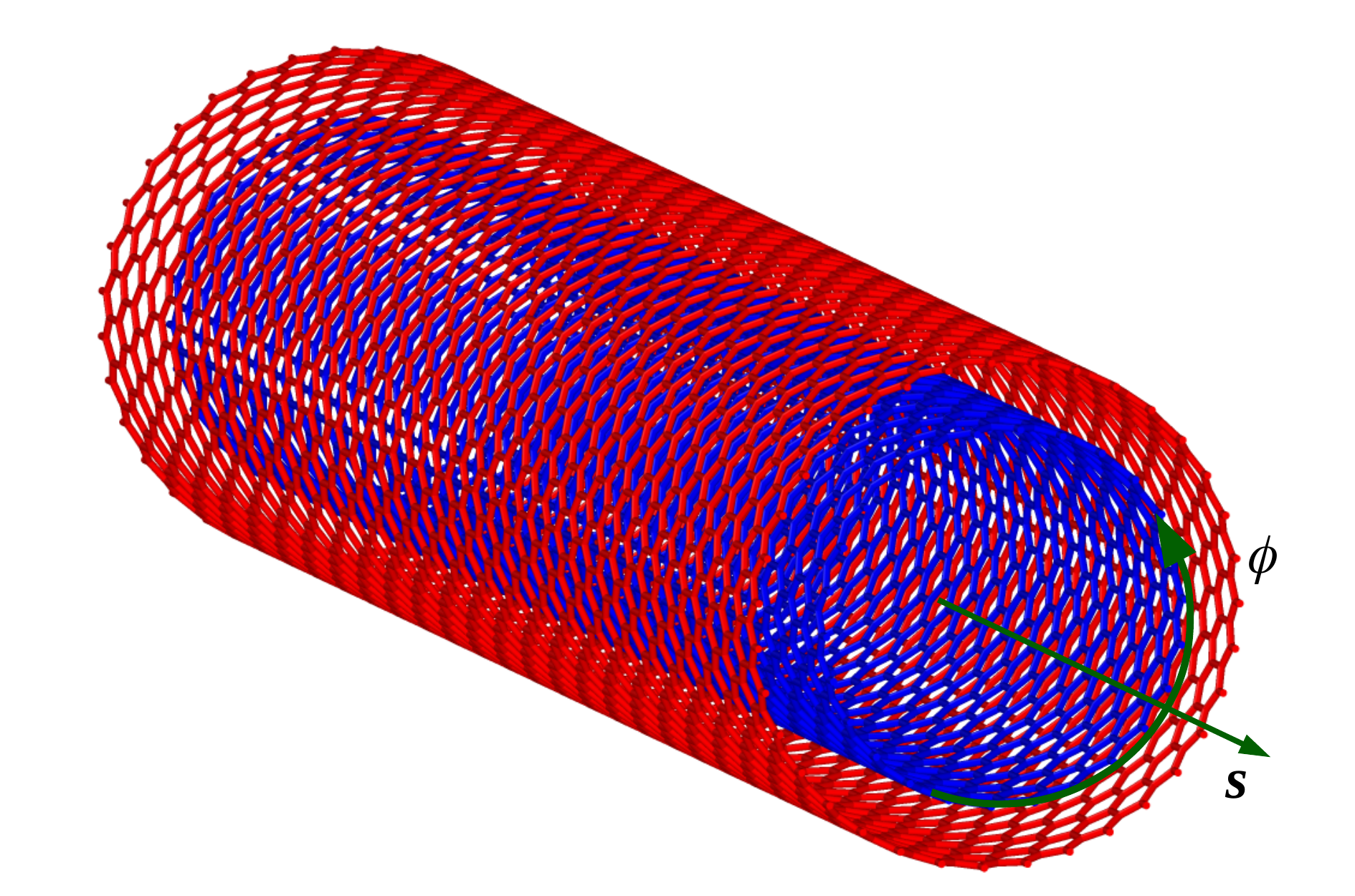}}
\end{picture}
\caption{Schematic representation of the pull-out and twisting of an inner CNT from/within an outer CNT. Shown here are CNT(26,0) and CNT(35,0).} 
\label{cnt_pull_img}
\end{center}
\end{figure}

The following three cases of DWCNTs are considered for pull-out and twisting: \\[-5mm]
\begin{enumerate}
\item CNT (26,0) inside CNT(35,0) \\[-5mm]
\item CNT (15,15) inside CNT(20,20) \\[-5mm]
\item CNT (21,9) inside CNT(28,12) \\[-5mm]
\end{enumerate}
The {initial geometry} parameters of these three cases, such as inner and outer CNT radii ($R_{\mathrm{i}}$ and $R_{\mathrm{o}}$), radial separation gap $G_{\mathrm{n}}$, and length $L$ are listed in Tab.~\ref{cnt_geo}. The initial radius \citep{DRESSELHAUS1995} of an undeformed CNT is given by
\eqb{lll}
\ds R \is ({\sqrt3a_{\text{cc}}}/{2\pi})\sqrt{n^2+m^2+nm}\,,
\eqe
where $n$ and $m$ are the chirality indices \citep{DRESSELHAUS1995}. A schematic representation of the setup is shown in Fig.~\ref{cnt_pull_img}.  
\begin{table}[ht]
\centering
\begin{tabular}{|c|c|c|c|c|c|c|c|}
  \hline
  \multirow{2}{*}{Case} &  \multicolumn{2}{c|}{Inner CNT} & \multicolumn{2}{c|}{Outer CNT} & Radial  gap  & \multicolumn{2}{c|}{Length ($L$)} \\[0mm] \cline{2-5} \cline{7-8}
& Chirality & $R_{\mathrm{i}}$ [nm]  & Chirality & $R_{\mathrm{o}}$ [nm] & ($G_\mrn$) [nm] & Unit cells & [nm]  \\[0mm]   \hline 
1 & (26,0) & 1.0013  & (35,0) & 1.3479  & 0.3466 & 24 & 10.0584 \\[.5mm]
2   &  (15,15) & 1.0005 &  (20,20)& 1.3340 & 0.3335 & 42 & 10.1626 \\[.5mm]
3   & (21,9) & 1.0269 & (28,12) & 1.3691  & 0.3423 & 3 & 11.1751 \\[.5mm]
   \hline
\end{tabular}
\caption{Geometric parameters of inner and outer CNTs of DWCNTs before relaxation.}
\label{cnt_geo}
\end{table}
{Initially, the CNTs are relaxed following the approach described in Sec.~\ref{atomic_simulations}}. The {geometry} parameters of three cases of CNTs after individual and combined relaxation are given in Tab.~\ref{cnt_geo_after_relax}. 
\begin{table}[ht]
\centering
\begin{tabular}{|c|c|c|c|c|c|c|c|c|}
  \hline
  \multirow{2}{*}{Case} &  \multicolumn{4}{c|}{individual CNT relaxation} & \multicolumn{4}{c|}{Combined DWCNT relaxation} \\[0mm] \cline{2-9}
& $r_\mri$ [nm] & $r_\mro$ [nm] & $g_{\mathrm{n}}$ [nm] & $\frac{(\ell_\mri+\ell_\mro)}{2}$ [nm] & $r_\mri$ [nm] & $r_\mro$ [nm] & $g_{\mathrm{n}}$ [nm] & $\ell$ [nm]  \\[0mm]   \hline 
1 & 1.0026 & 1.3487  & 0.3461 & 10.0611 & 1.0036 & 1.3466  & 0.3430 & 10.0573 \\[.5mm]
2   &  1.0014 & 1.3345 & 0.3331 & 10.1694 &  0.9997 & 1.3362 & 0.3366 & 10.1638 \\[.5mm]
3   & 1.0279 & 1.3698 & 0.3419 & 11.1803 & 1.0281 & 1.3689 & 0.3410 & 11.1766 \\[.5mm]
   \hline
\end{tabular}
\caption{Geometric parameters of inner and outer CNTs of DWCNTs after individual and combined relaxation. The {geometry} parameters after relaxation are denoted with lowercase letters.}
\label{cnt_geo_after_relax}
\end{table} 
As discussed earlier, the equilibrium separation distance between two graphene layers depends on their stacking. Due to the curvature, DWCNTs possess various stackings, which implies that the contact pressure between the CNTs is not constant, and the CNT radii thus vary across the surface. Tab.~\ref{cnt_geo_after_relax} reports the average radii. Tab.~\ref{cnt_geo_after_relax} also shows that the average gap $g_\mrn = r_\mro - r_\mri$ is either in a state of attraction (for Case 1 and 3) or in a state of repulsion (for Case 2). The three cases thus cover negative and positive contact pressures between the walls. Comparison between Tabs.~\ref{cnt_geo} and~\ref{cnt_geo_after_relax} shows that the inner radius increases and the outer radius decreases for Case 1 and 3, while it is the other way around for Case 2. This is due to the positive Poisson ratio of graphene (see Appendix~\ref{ele_prop}): Positive contact pressure leads to circumferential strains that are positive in the outer tube and negative in the inner tube. These circumferential strains lead to axial strains that are negative for the outer tube and positive for the inner tube. For negative contact stresses, the effect is reversed. 

\subsubsection{Load application in MD}
After obtaining the relaxed DWCNTs, the inner CNT is pulled quasi-statically by assigning {the} velocity 0.001 \AA/ps to the right edge atoms ({see Fig.~\ref{cnt_pull_img}).} For twisting, the right edge atoms of the inner CNT are rotated with an angular velocity of $6.28 \times 10^{-4}$ rad/ps. During pull-out or twisting, atoms on the left edge of the outer CNT and those on the right edge of the inner CNT are constrained in the tangential direction employing a torsional spring with stiffness 16.02 nN-nm/rad. Thus, a radial expansion of the CNTs is allowed. The resisting pull-out force is then calculated as the total vdW force in axial direction acting on the inner CNT due to the outer CNT. The torque is calculated as $\sum_{I=1}^N (F_I^y\cos(\phi)-F_I^x\sin(\phi))r$, where $F_I^x$ and $F_I^y$ are the inter-CNT vdW force {components} acting on atom $I$ along the $x$ and $y$ directions, respectively. Here, $r$ is the radius, $\phi$ is its angle of twist, and $N$ is its total number of atoms of the CNT.
\subsubsection{Load application in FE}
For pull-out, a constant displacement is prescribed to all FE nodes of the central cross-section of the CNT, while the rotation is constrained, but lateral motions are allowed, such that the lateral forces on the tubes remain zero. The tubes therefore do not remain exactly concentric during pull-out. 
For twisting, a constant rotation is applied by moving all atoms of the central cross-section in circumferential direction, while keeping the radial direction free. The longitudinal direction remains fixed during the rotation. 

The FE simulations show that enforcing concentricity of the inner and outer tubes during CNT pull-out leads to a small horizontal force (in $\be_2$-direction) and a twisting moment. Those are absent if the concentricity is not enforced. The FE simulations further show that it makes a difference where the displacements are prescribed. For the subsequent results, the displacements are applied at the center, and the forces are measured there. If the ends are used instead, the forces and moments are offset slightly on the horizontal axis in Figs.~\ref{cnt_pull_force} and \ref{cnt_twist_force}. 

\subsection{CNT pull-out} 
Next, the pull-out results are presented and discussed. The FE results are compared to analytical results {(derived in Appendix \ref{s_analytical_exp})} and MD results.
\subsubsection{General observations} \label{s:go}
The length of the axial unit cell of each CNT is defined by its two chiral indices $(n, m)$, and is given by $\ell = \sqrt{3}{L_c}/d_R$, where ${L_c}$ is the circumference of the CNT, and $d_R$ is the greatest common divisor (GCD) of $2n+m$ and $2m+n$ \citep{DRESSELHAUS1995}. This implies that the pull-out force has a periodicity equal to its unit cell length along the axial direction. For the three cases of CNTs considered here, the unit cell lengths are, respectively, $\ell_1=3a_{\text{cc}}=0.4191$ nm, $\ell_2=\sqrt{3}a_{\text{cc}}=0.2420$ nm, and $\ell_3=3\sqrt{79}a_{\text{cc}}=3.7250$ nm.
\begin{figure}[!htbp]
\begin{center} \unitlength1cm
\begin{picture}(0,10)
\put(-7.8,5){\includegraphics[height=48mm]{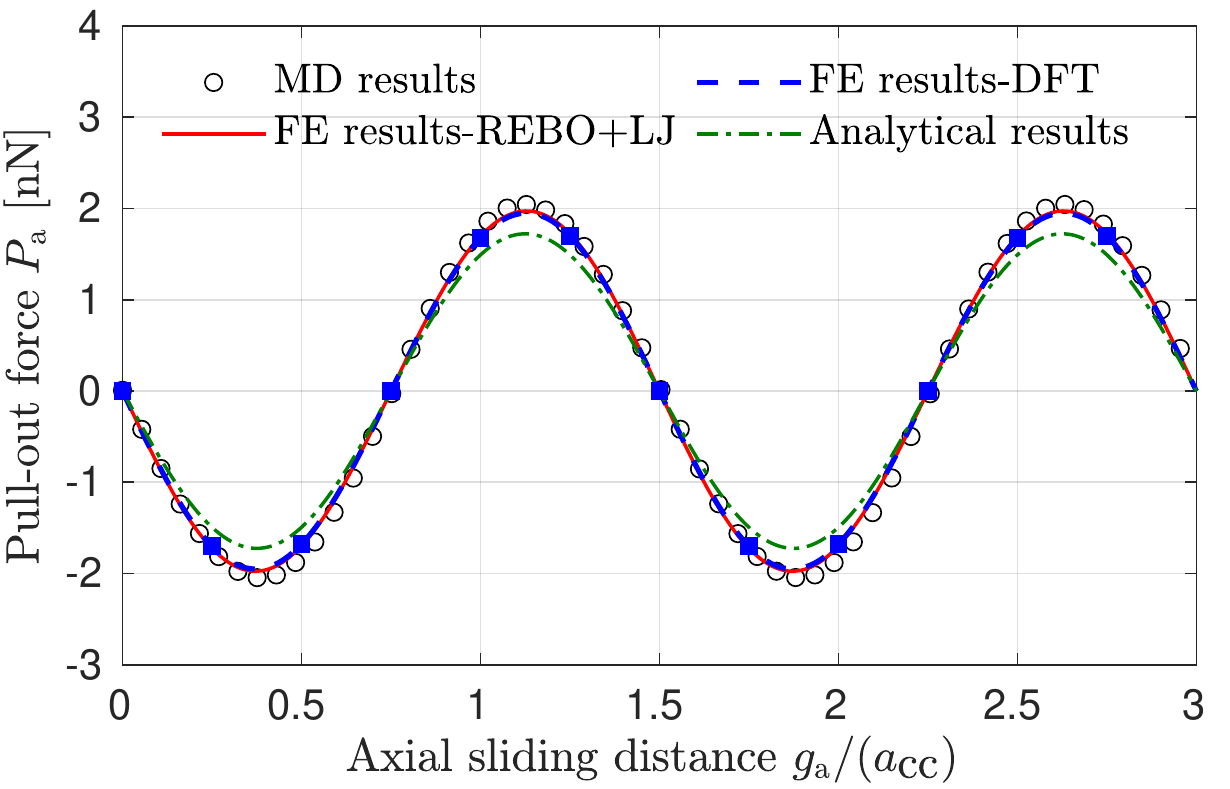}}
\put(0.2,5){\includegraphics[height=48mm]{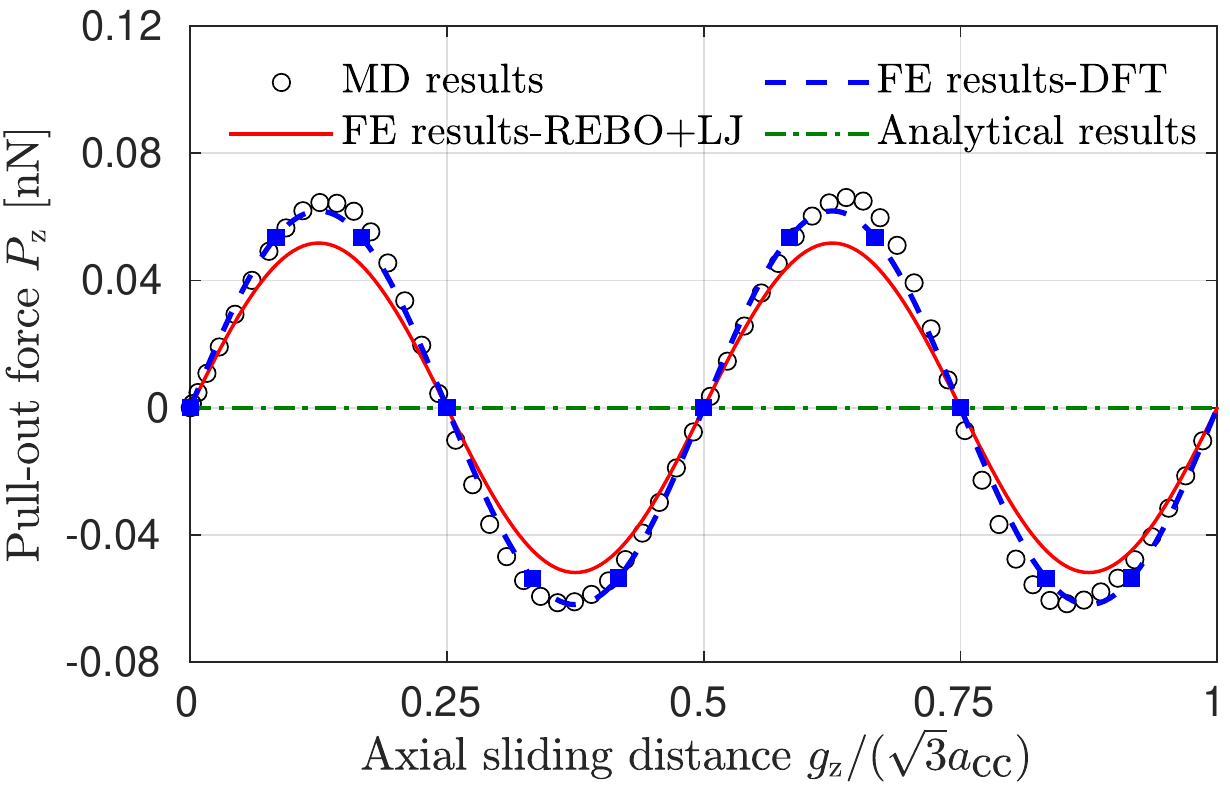}}
\put(-4.5,0){\includegraphics[height=48mm]{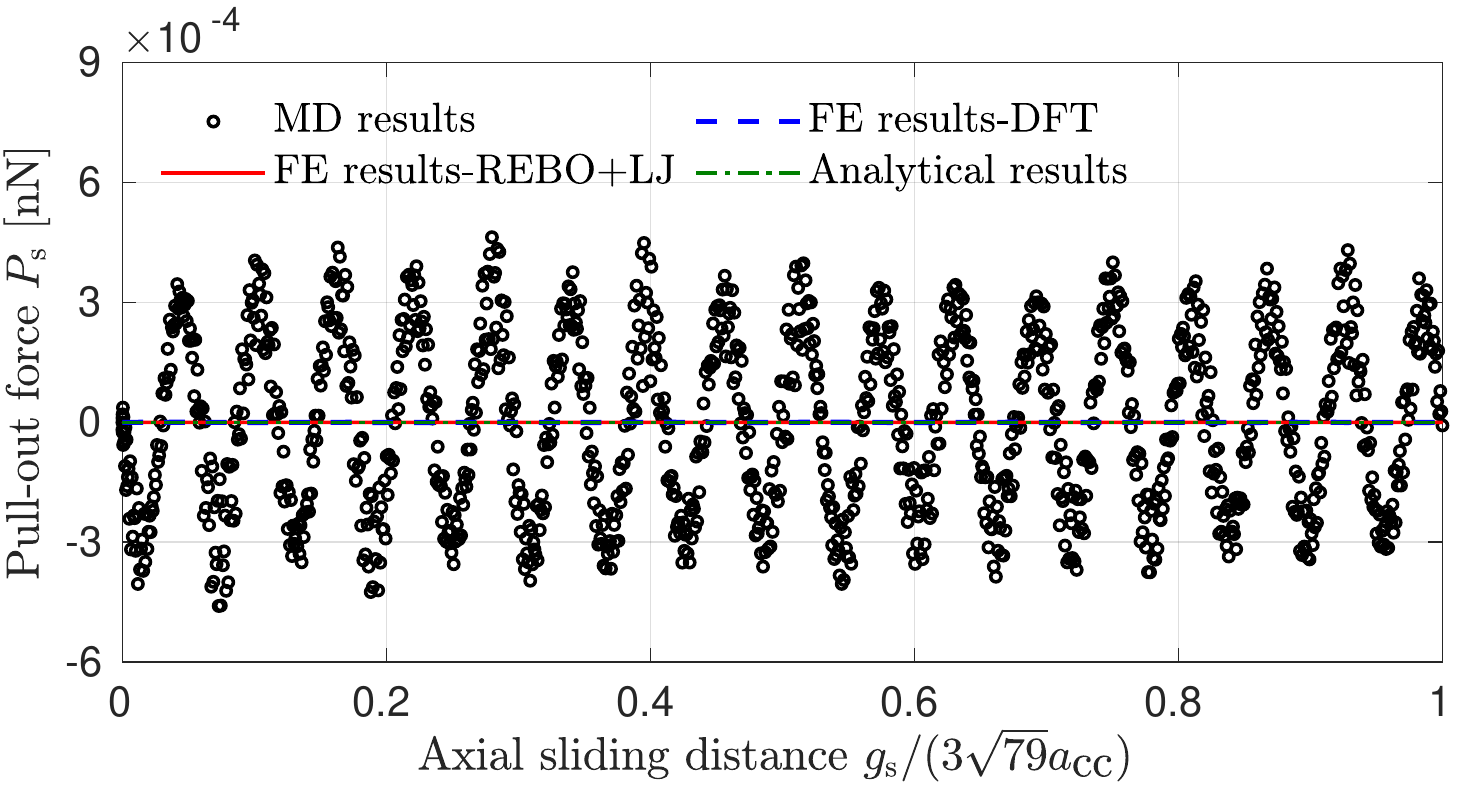}}
\put(-7.7,4.8){(a)}
\put(0.3,4.8){(b)}
\put(-3.5,0){(c)}
\end{picture}
\caption{Pull-out force for (a)~Case 1, (b)~Case 2, and (c)~Case 3 in dependence of the axial sliding distance. The corresponding contact tractions for the 12 locations marked by blue squares in (a) and (b) are illustrated in Figs.~\ref{f:pullout2p} -- \ref{f:pullout1t}.  }
\label{cnt_pull_force}
\end{center}
\end{figure}
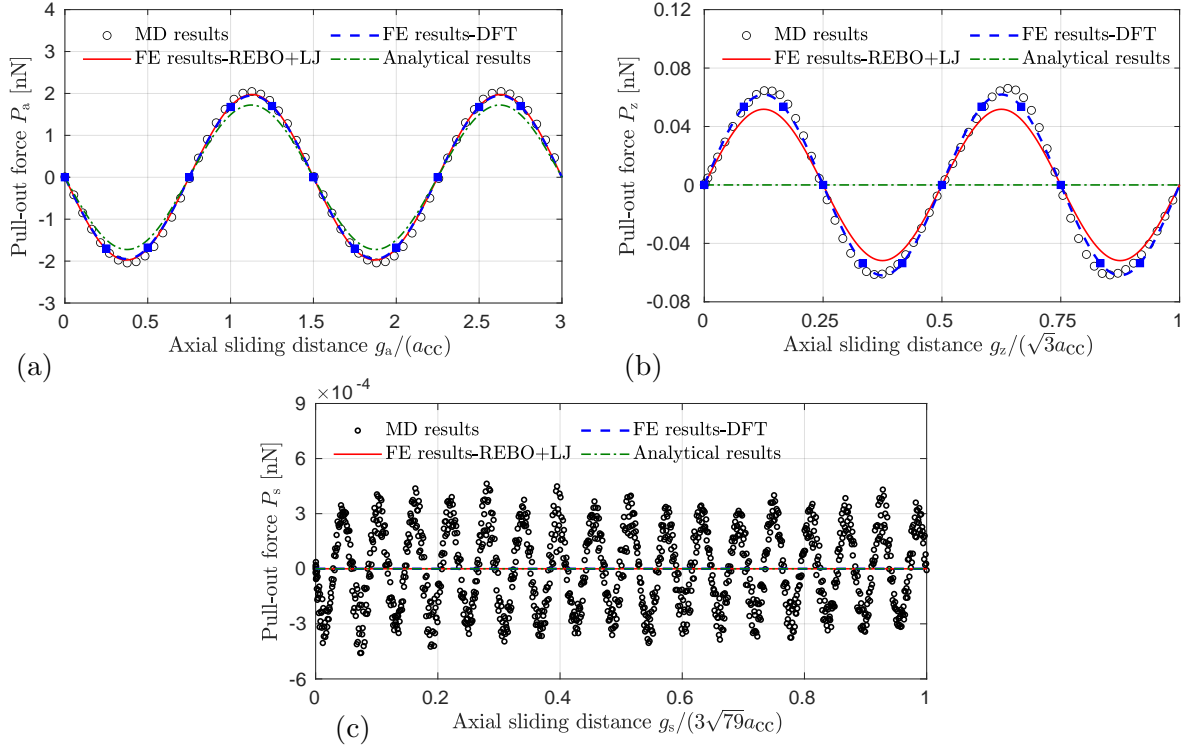

The FE simulations are carried out using two sets of material properties: Determined from 1) DFT \citep{Shirazian2018_01} and 2) MD using REBO+LJ potential (see Appendix~\ref{ele_prop} and Tab.~\ref{coil_param} for details). Further, the FE simulations based on the contact formulation of Sec.~\ref{continuum_model} assume that the neighboring CNT (outer or inner, respectively) is treated rigidly, while deformations are accounted for in the considered CNT itself. Still, the initial deformation of the neighboring CNT is accounted for by using the relaxed radii from Tab.~\ref{cnt_geo_after_relax}. Comparison plots of the pull-out forces determined from FE, MD, and analytical results are shown in Fig.~\ref{cnt_pull_force}. The periodicity obtained from these approaches agrees well with the theoretical predictions. Also, the amplitude agrees well for the FE and MD results. This is also seen in Tab.~\ref{cnt_pull_out_d}, which compares the absolute difference between the maximum pull-out force. 

\begin{table}[ht]
\centering
\begin{tabular}{|c|c|c|}
  \hline
Case &  $P_s^{\text{max}}(\text{MD})-P_s^{\text{max}}$(FE-DFT) [nN] & $P_s^{\text{max}}(\text{MD})-P_s^{\text{max}}$(FE-REBO+LJ) [nN] \\[0mm]
 \hline 
1 & $95.0626\cdot 10^{-3}$ & $70.9569\cdot 10^{-3}$   \\[.5mm]
2   &  $4.6490\cdot 10^{-3}$ & $14.7392\cdot 10^{-3}$  \\[.5mm]
3   &  $0.4739\cdot 10^{-3}$ & $0.4740\cdot 10^{-3}$  \\[.5mm]
   \hline
\end{tabular}
\caption{Absolute differences in the maximum pull-out force between MD and FE.}
\label{cnt_pull_out_d}
\end{table}
The amplitude of the pull-out force, defined as $P_s^{\text{max}}-P_s^\text{min}$ during the sliding for Case 1 obtained from the MD simulations, is 4.091 nN. These amplitudes are 0.132 and $9.6486\cdot10^{-4}$ nN, respectively, for Cases 2 and 3, which are $\approx 96.77\%$ and $\approx 99.98\%$ less than that for Case 1. Thus, the pull-out force amplitudes are sensitive to the chirality of the CNTs, with the maximum for zigzag and the minimum for chiral CNTs. While the relative differences between MD and FE results increase from Case 1 to 3, their absolute difference decreases, as seen in Tab.~\ref{cnt_pull_out_d}. 
\subsubsection{Case 1: Pull-out of CNT(26,0) from within CNT(35,0)}
\begin{figure}[!htbp]
\begin{center} \unitlength1cm
\begin{picture}(0,4.2)
\put(-7.95,1.7){\includegraphics[height=25mm]{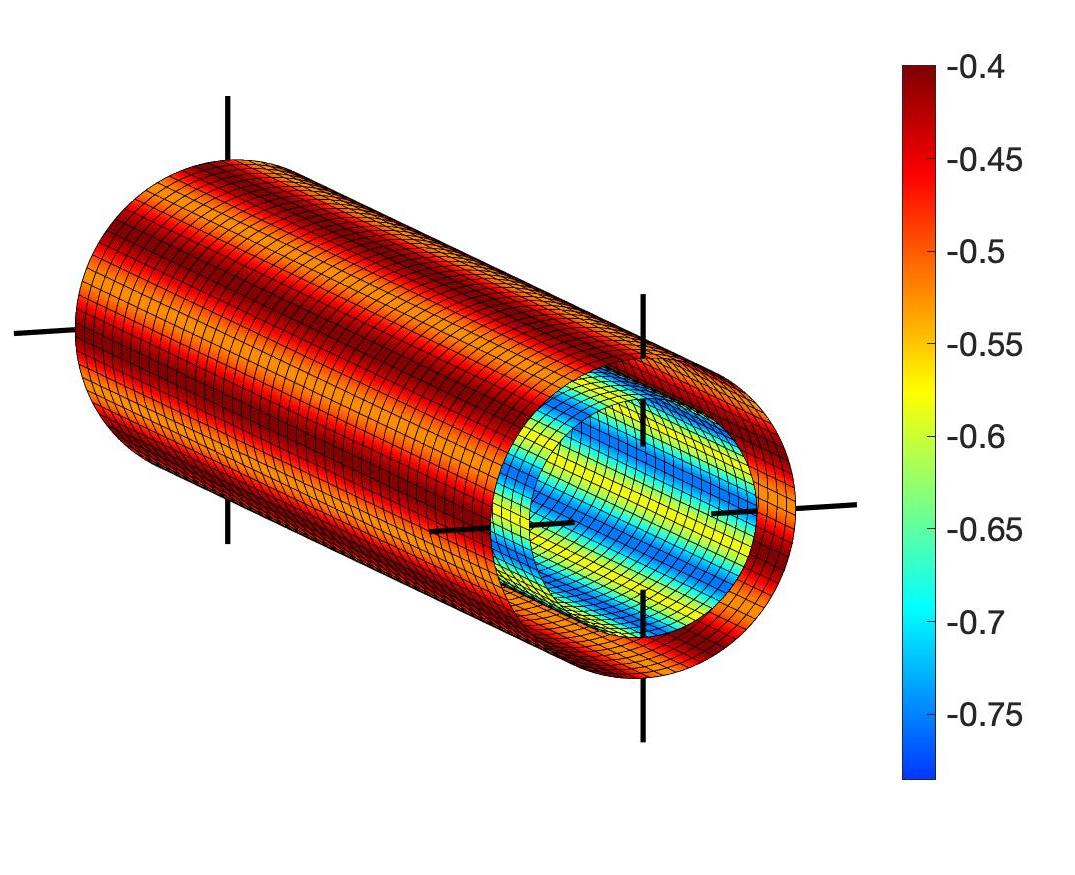}}
\put(-5.45,1.7){\includegraphics[height=25mm]{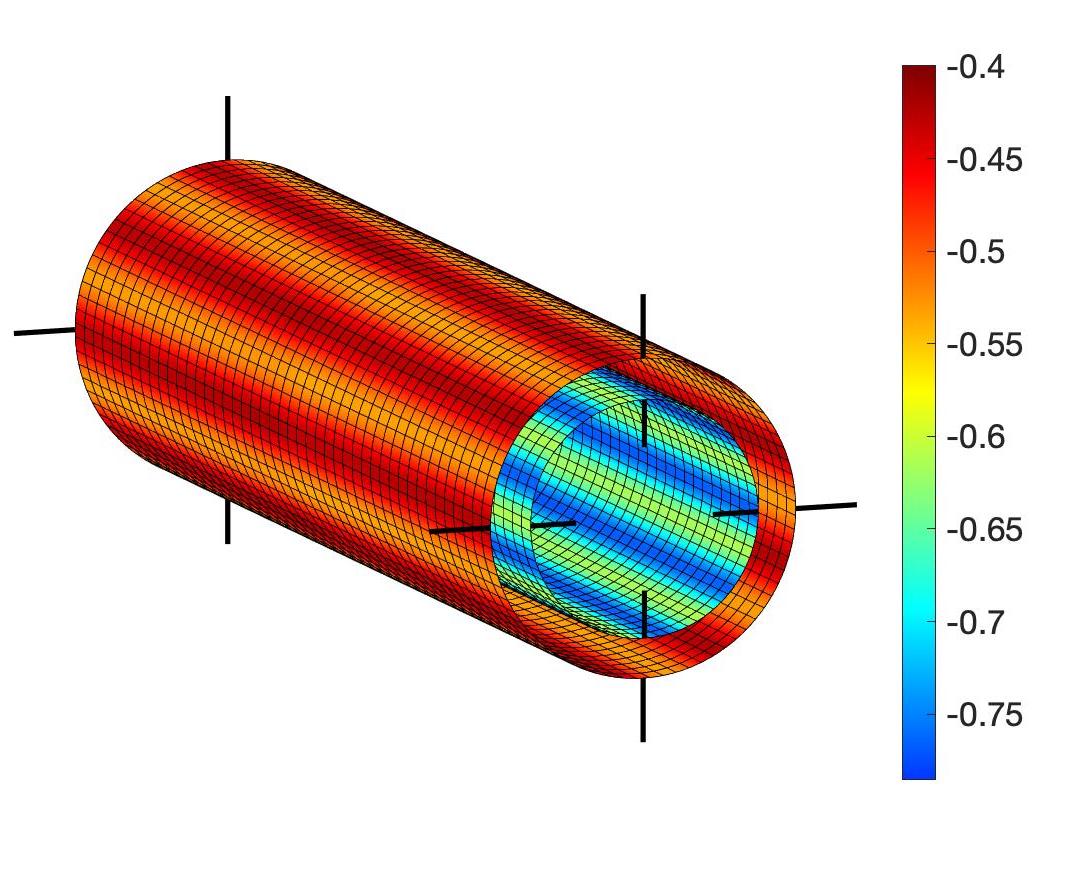}}
\put(-2.95,1.7){\includegraphics[height=25mm]{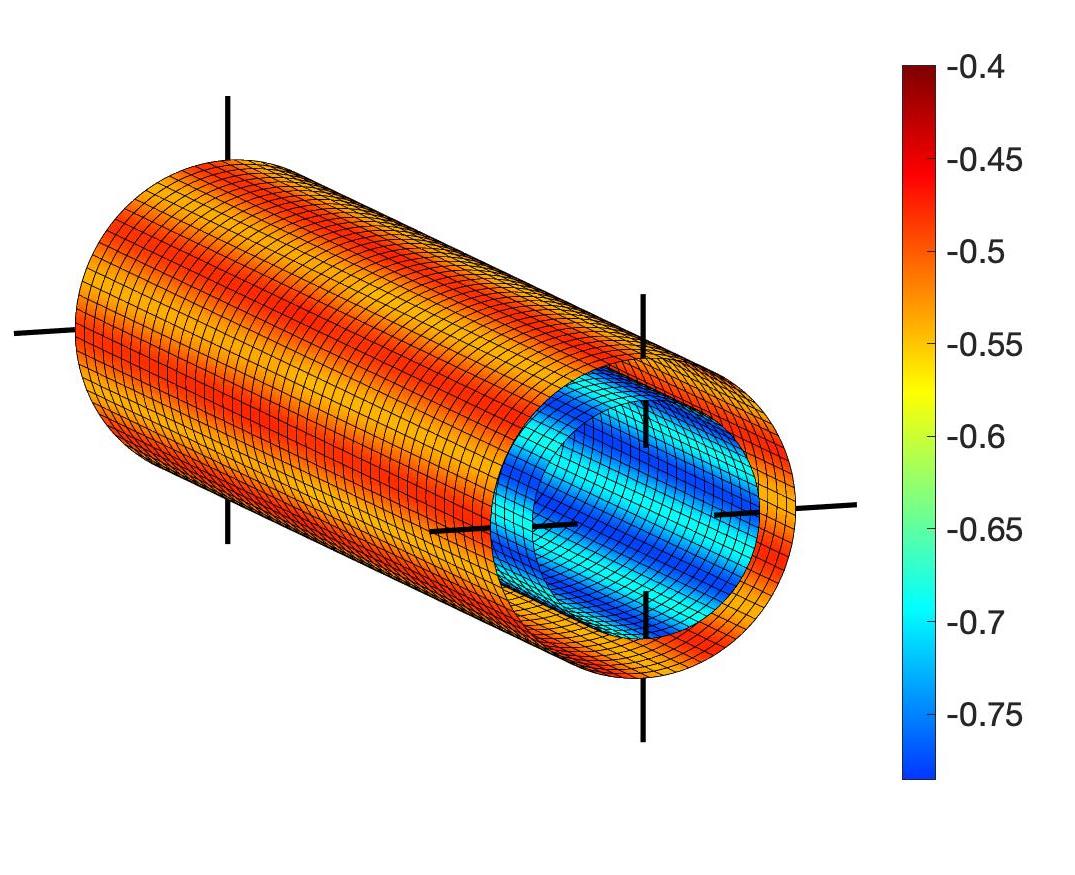}}
\put(-0.45,1.7){\includegraphics[height=25mm]{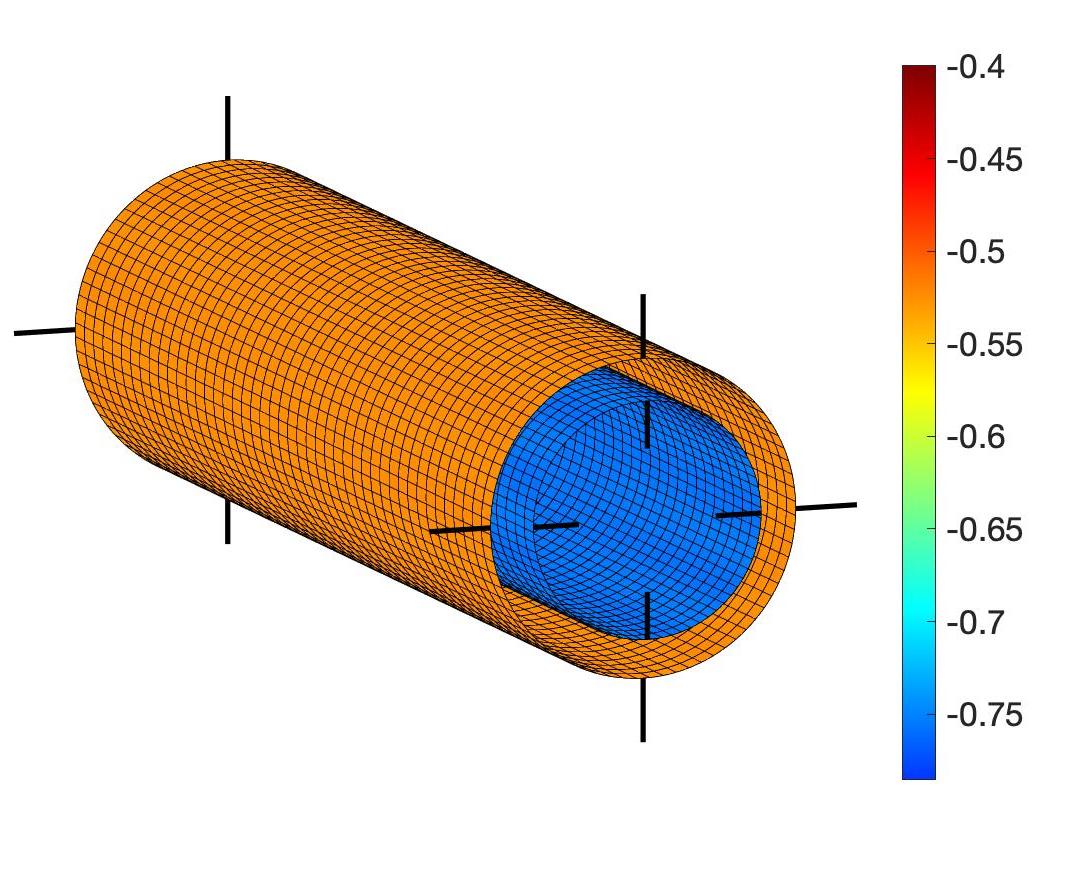}}
\put(2.05,1.7){\includegraphics[height=25mm]{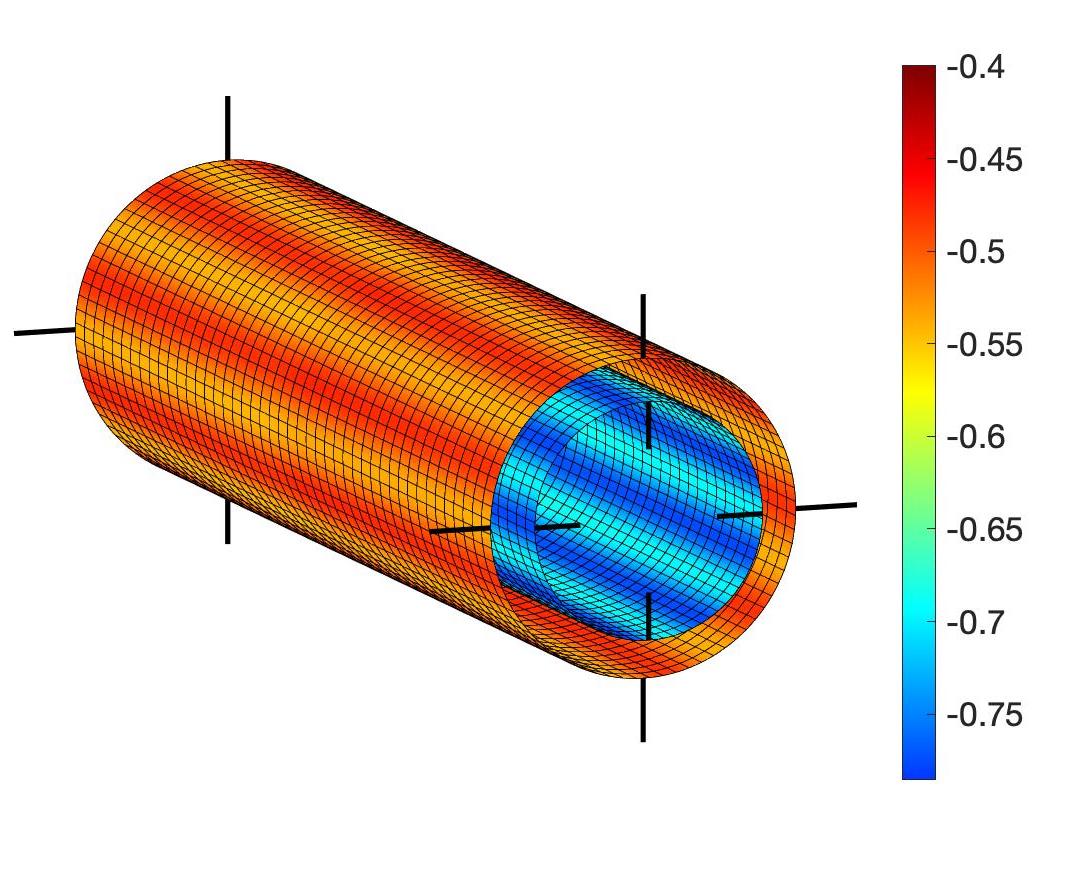}}
\put(4.55,1.7){\includegraphics[height=25mm]{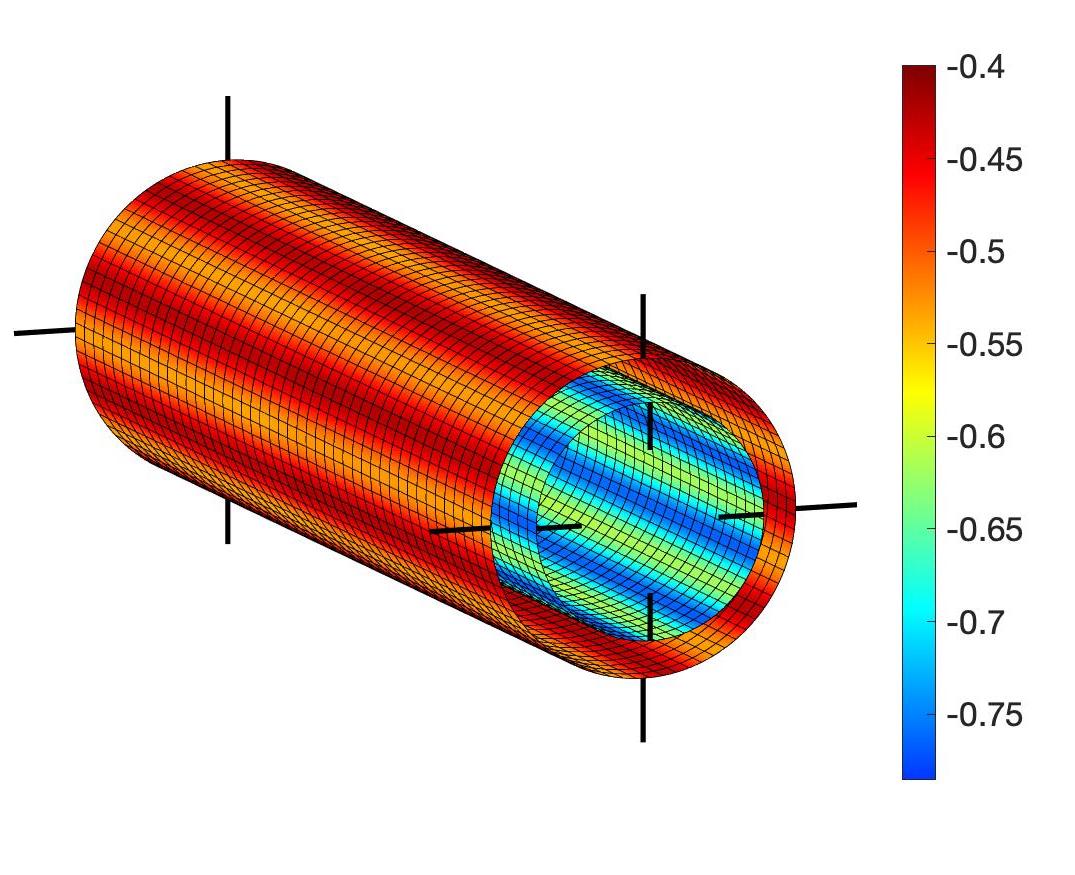}}
\put(-7.95,-.4){\includegraphics[height=25mm]{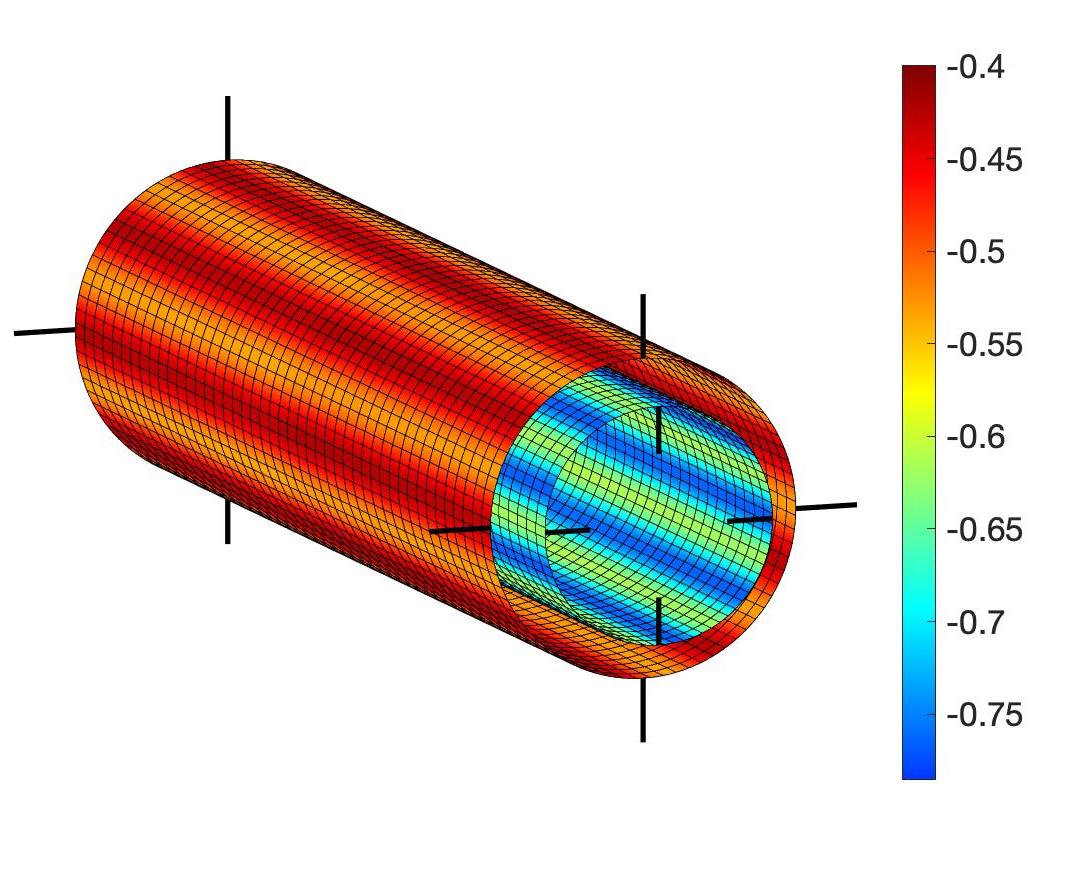}}
\put(-5.45,-.4){\includegraphics[height=25mm]{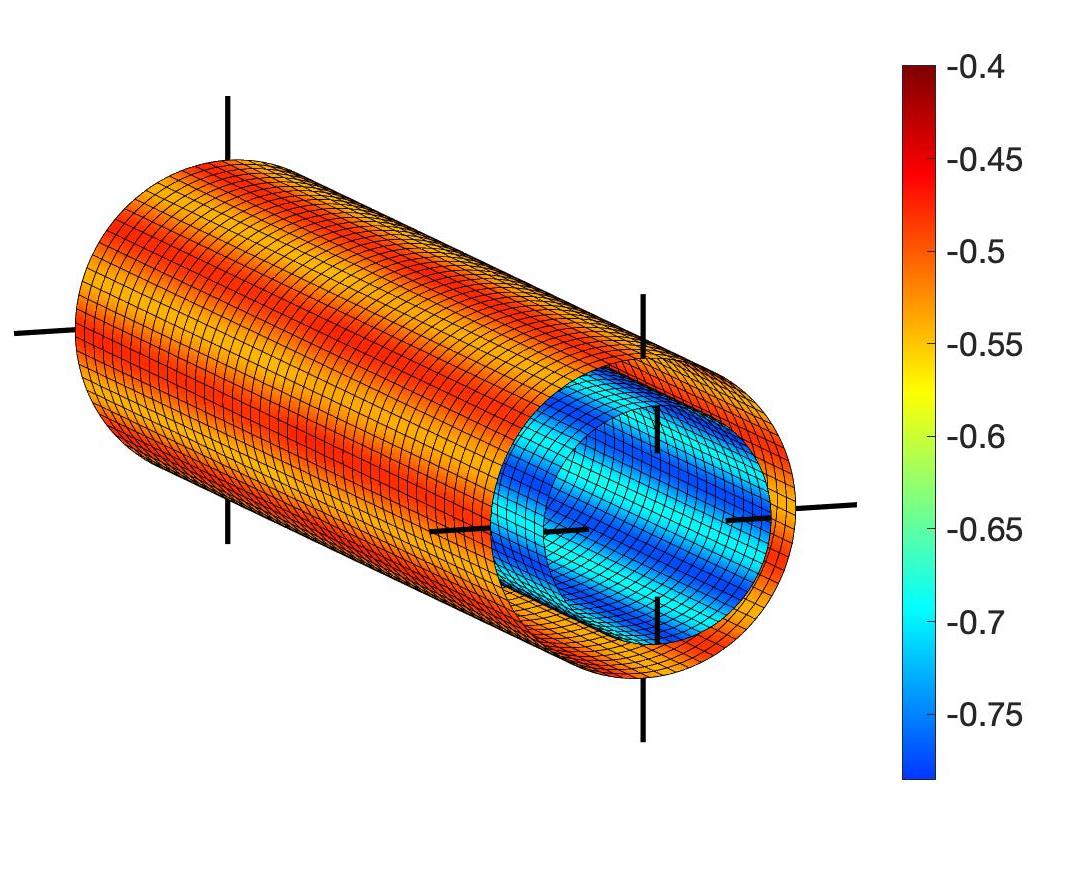}}
\put(-2.95,-.4){\includegraphics[height=25mm]{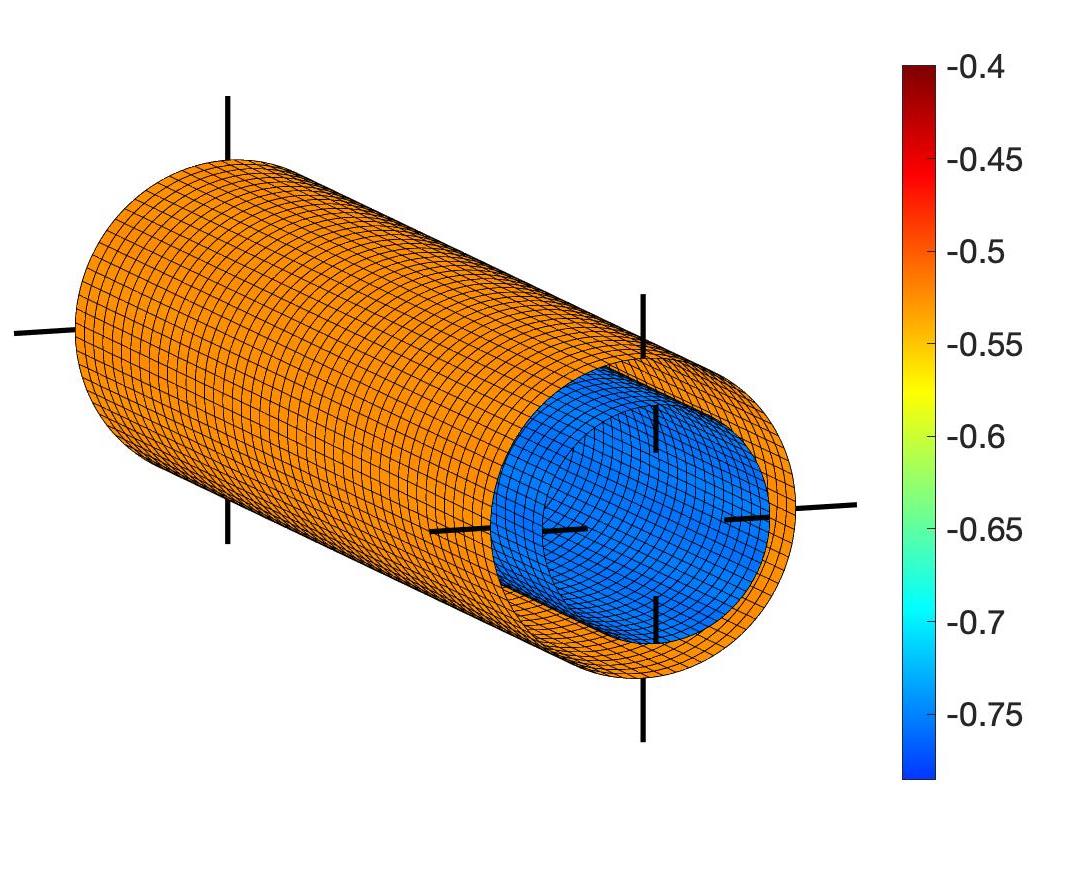}}
\put(-0.45,-.4){\includegraphics[height=25mm]{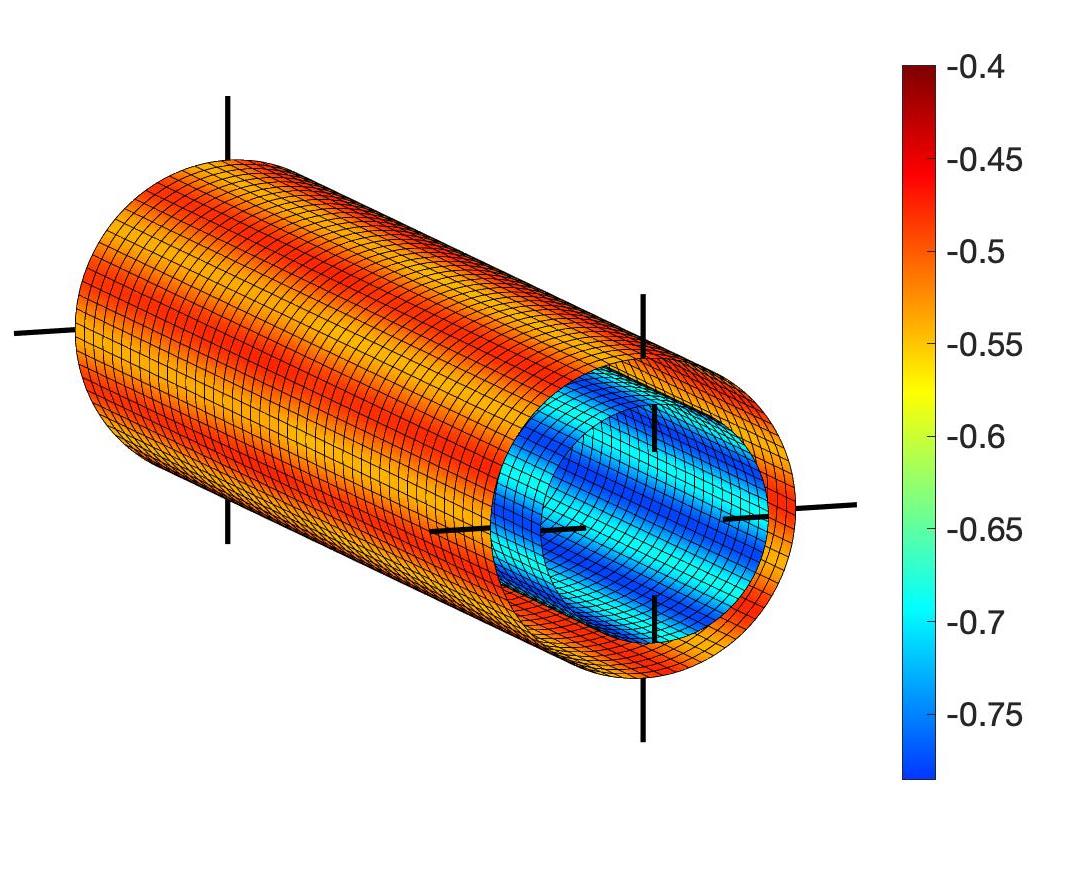}}
\put(2.05,-.4){\includegraphics[height=25mm]{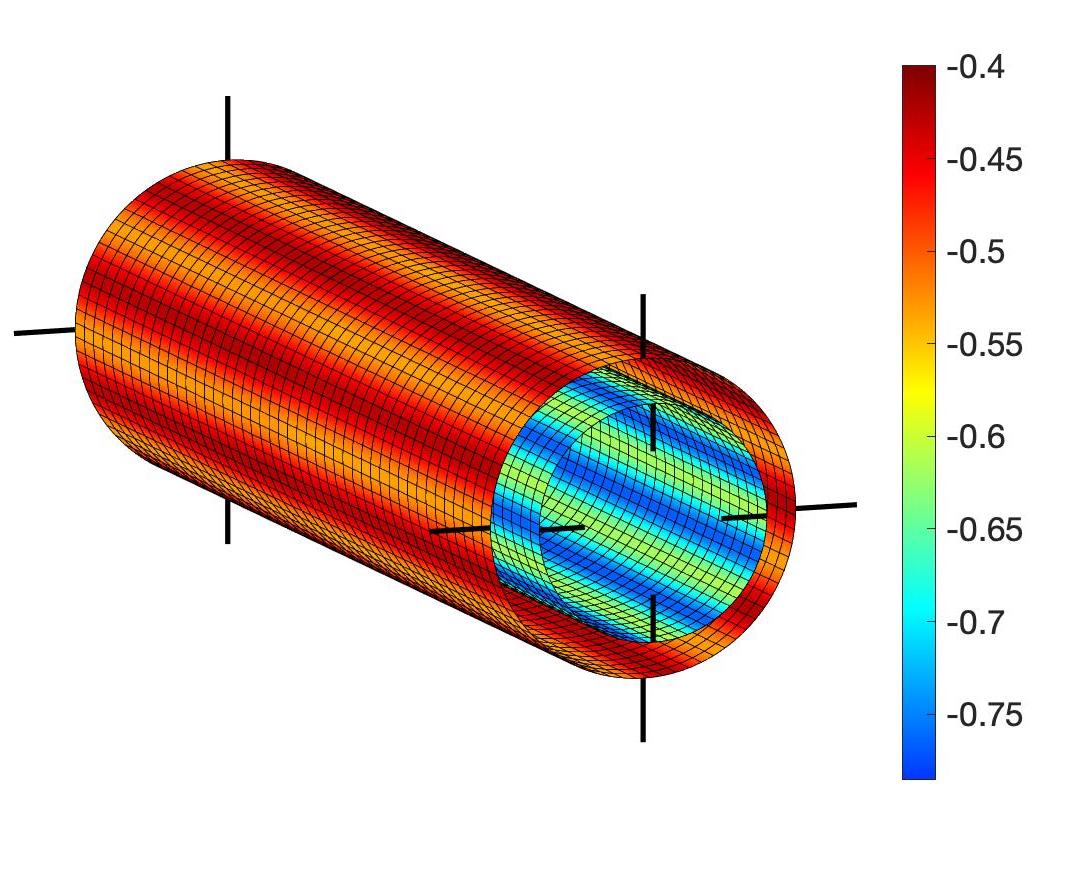}}
\put(4.55,-.4){\includegraphics[height=25mm]{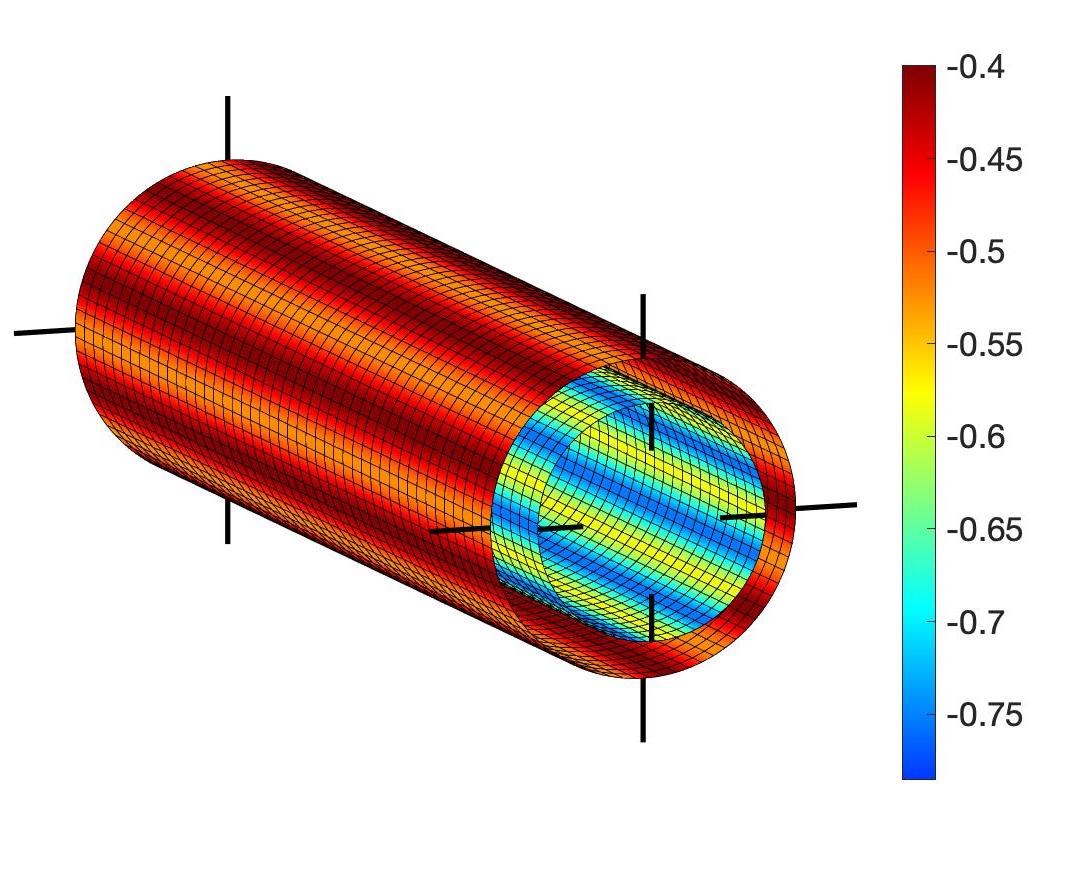}}
\put(7.1,-.55){\includegraphics[height=48mm]{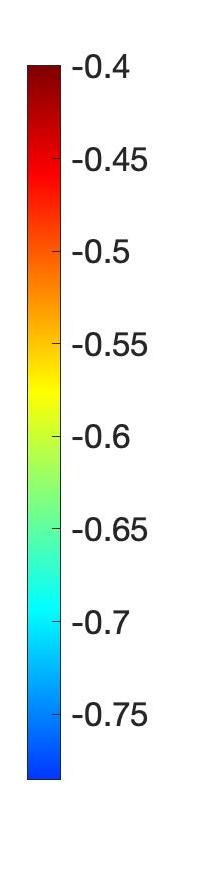}}
\put(-7.85,2.1){\scriptsize{0}}
\put(-5.35,2.1){\scriptsize{1}}
\put(-2.85,2.1){\scriptsize{2}}
\put(-0.35,2.1){\scriptsize{3}}
\put(2.15,2.1){\scriptsize{4}}
\put(4.65,2.1){\scriptsize{5}}
\put(4.65,0.){\scriptsize{6}}
\put(2.15,0.){\scriptsize{7}}
\put(-0.35,0.){\scriptsize{8}}
\put(-2.85,0.){\scriptsize{9}}
\put(-5.45,0.){\scriptsize{10}}
\put(-7.95,0.){\scriptsize{11}}
\end{picture}
\caption{Pull-out of CNT(26,0) from within CNT(35,0) (Case 1): Color plot of contact pressure $p$ in [GPa] at $g_\mra \in [0,\,1,\,2,\,...,\,11]\cdot\ell_\mra/12$ (clockwise, starting top left). 
}
\label{f:pullout2p}
\end{center}
\end{figure}

\begin{figure}[!htbp]
\begin{center} \unitlength1cm
\begin{picture}(0,4.2)
\put(-7.95,1.7){\includegraphics[height=25mm]{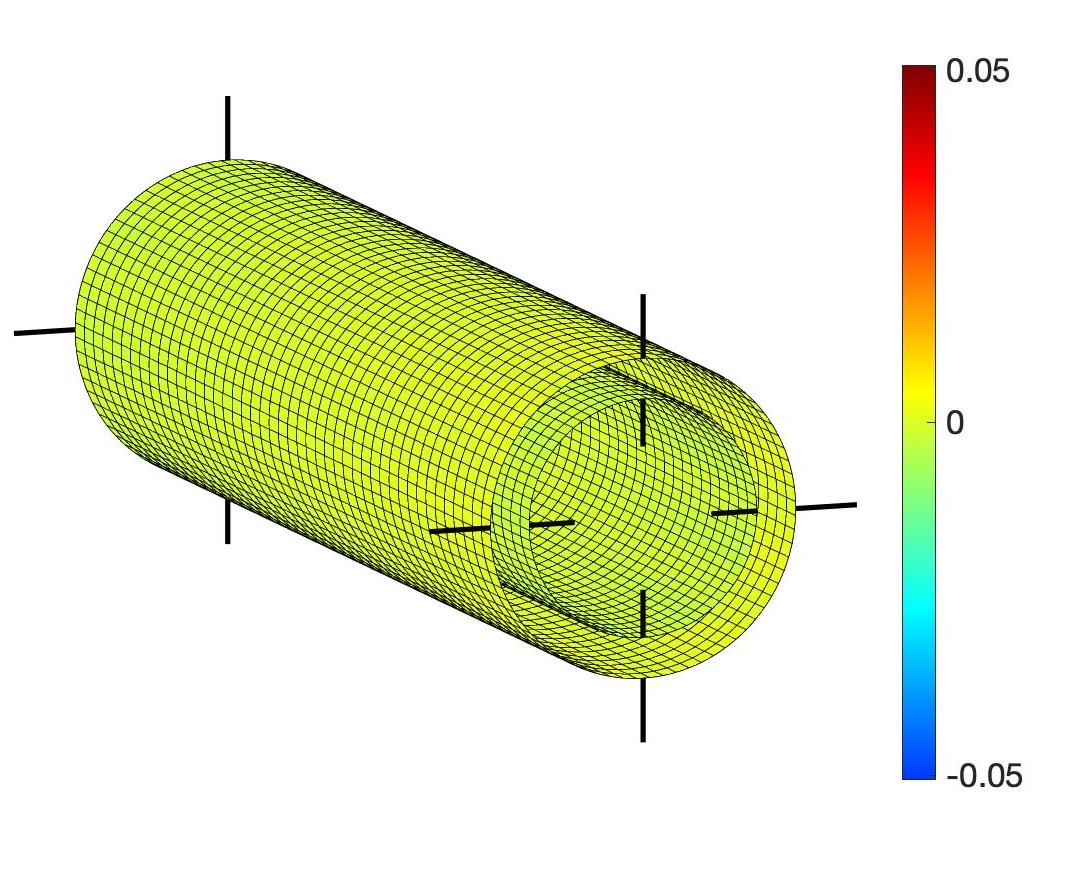}}
\put(-5.45,1.7){\includegraphics[height=25mm]{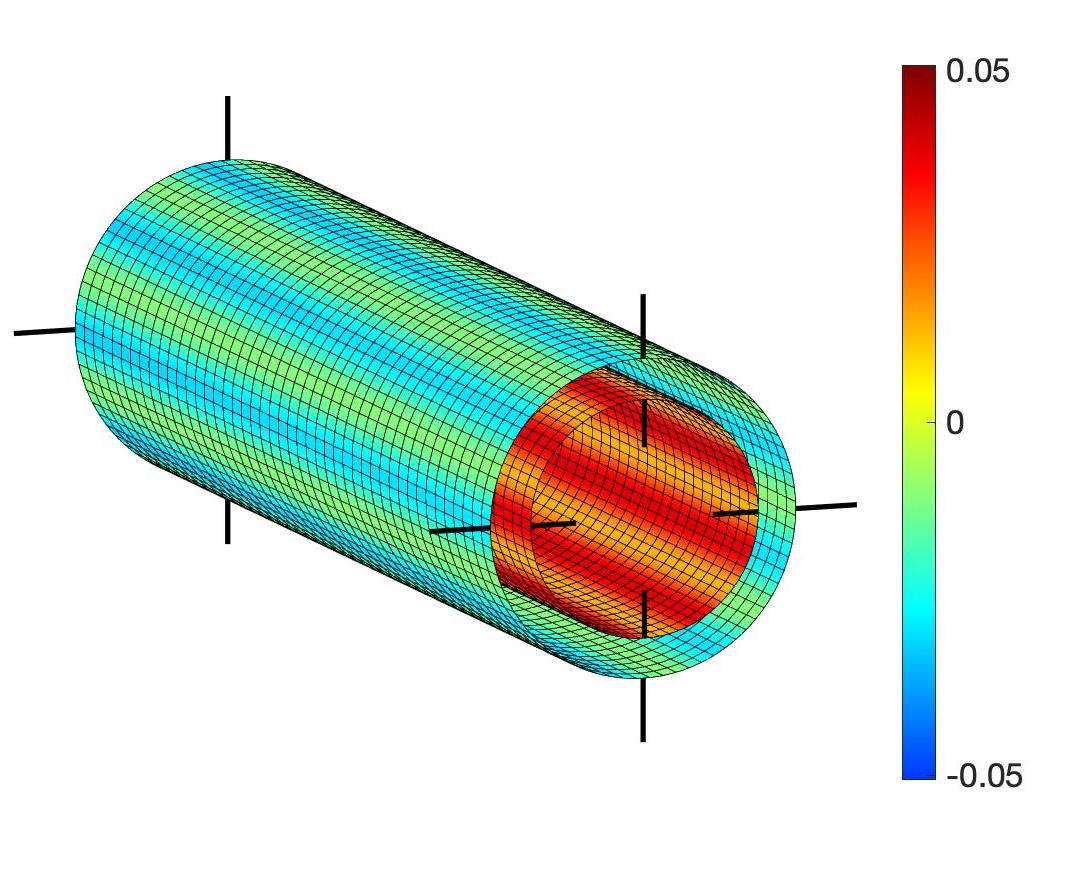}}
\put(-2.95,1.7){\includegraphics[height=25mm]{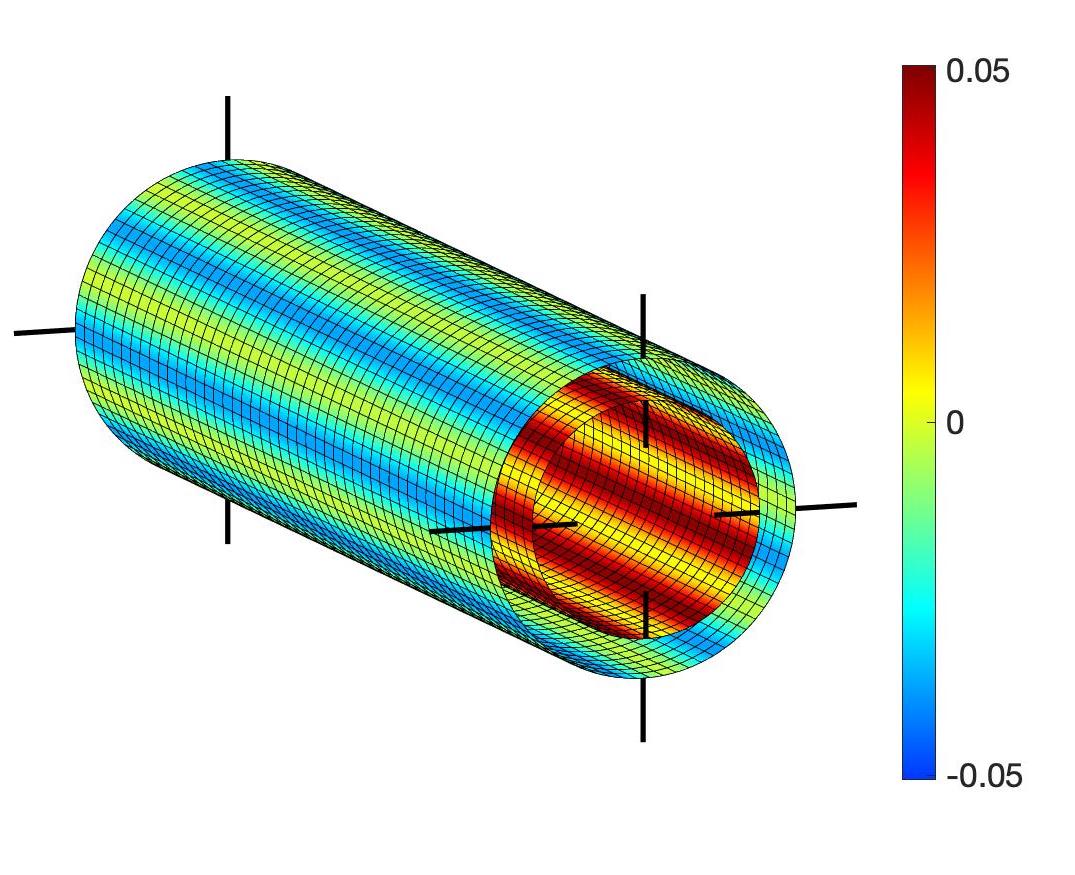}}
\put(-0.45,1.7){\includegraphics[height=25mm]{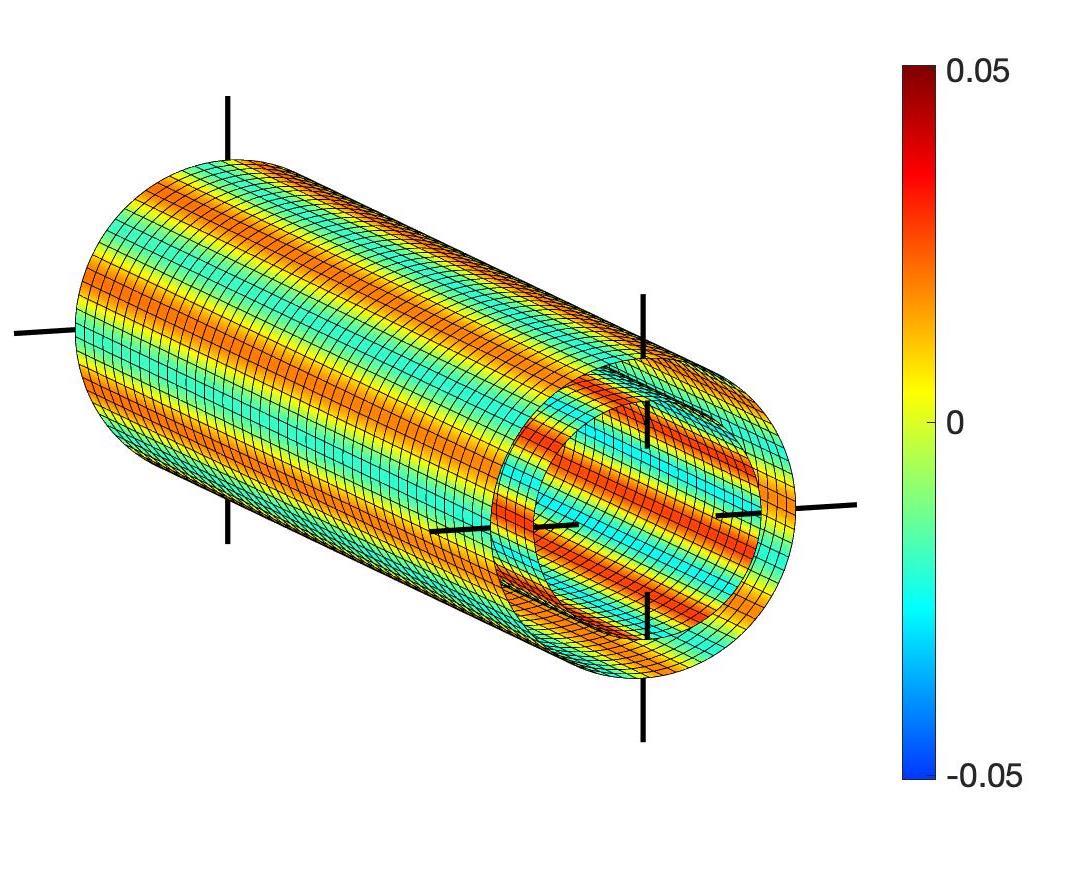}}
\put(2.05,1.7){\includegraphics[height=25mm]{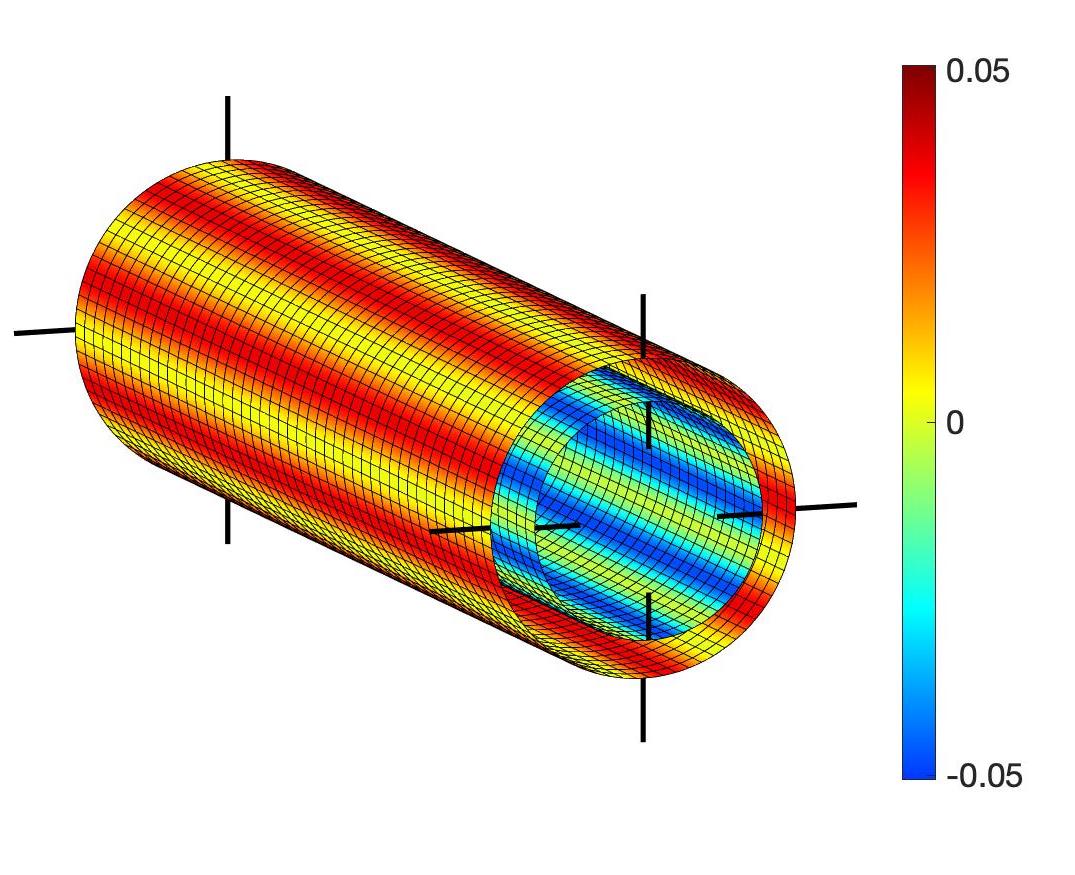}}
\put(4.55,1.7){\includegraphics[height=25mm]{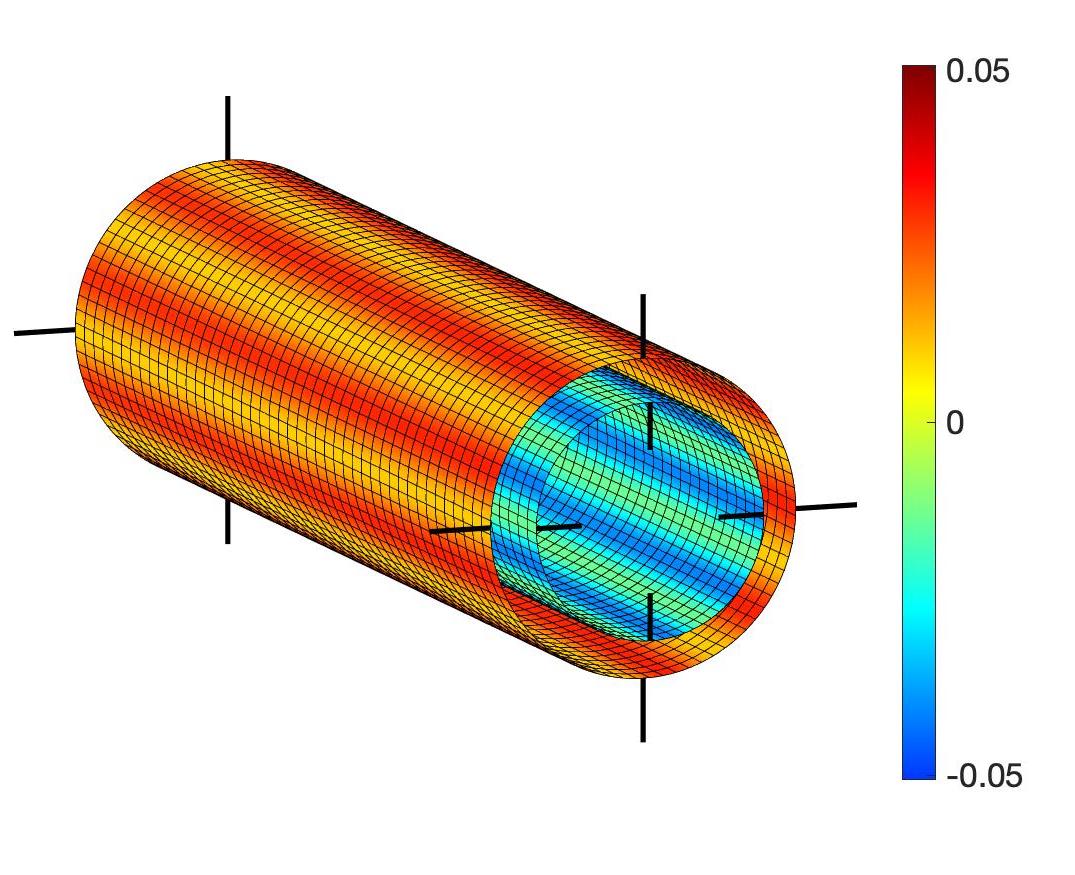}}
\put(-7.95,-.4){\includegraphics[height=25mm]{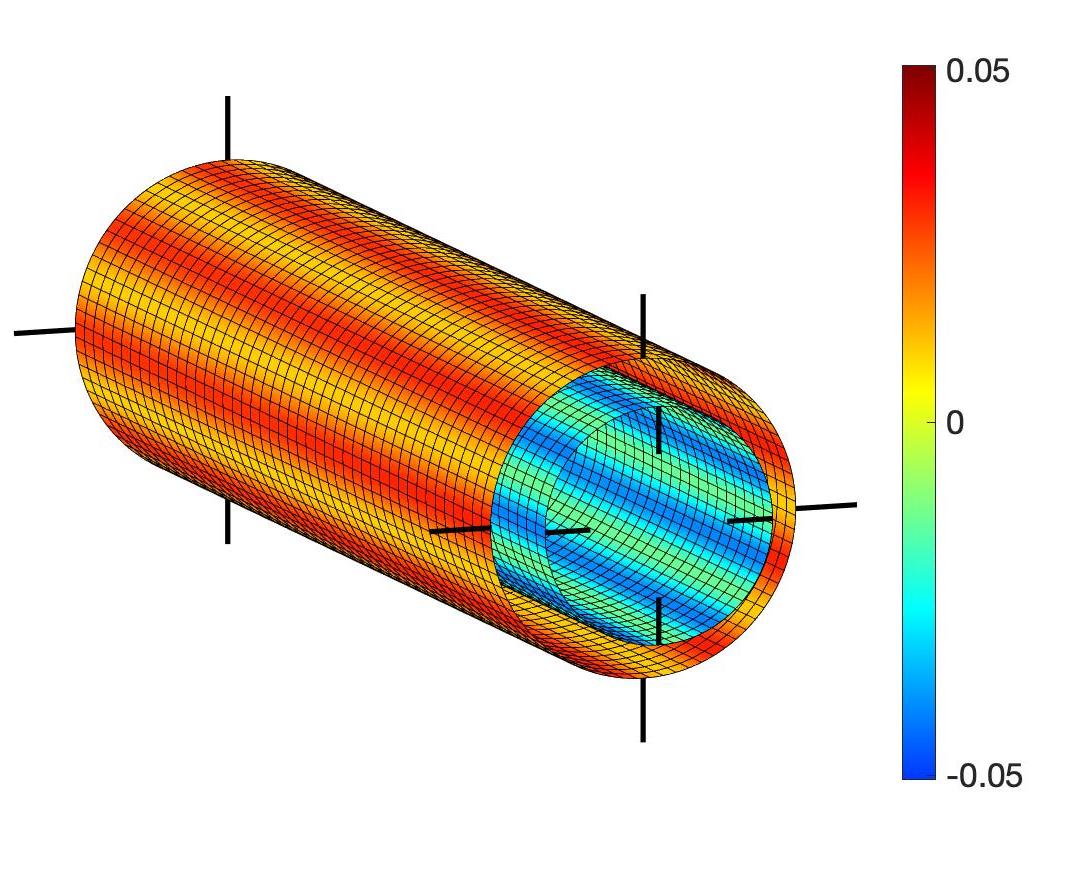}}
\put(-5.45,-.4){\includegraphics[height=25mm]{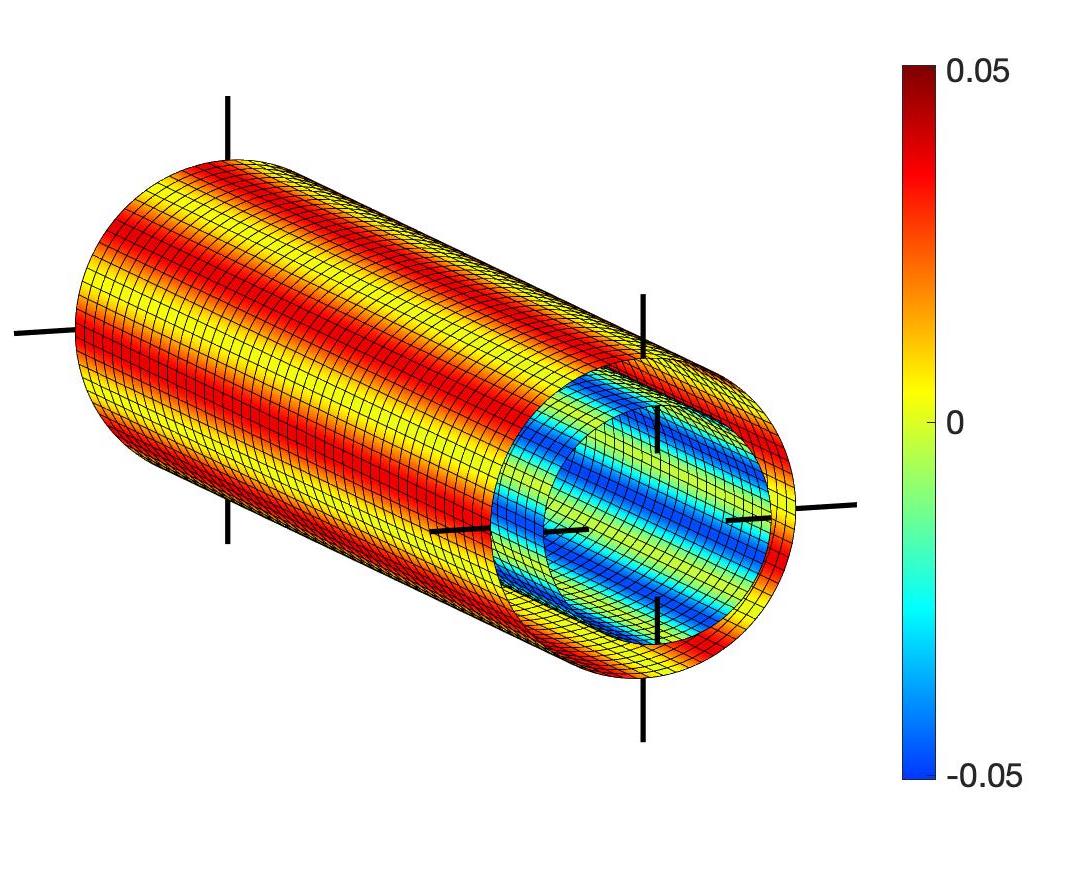}}
\put(-2.95,-.4){\includegraphics[height=25mm]{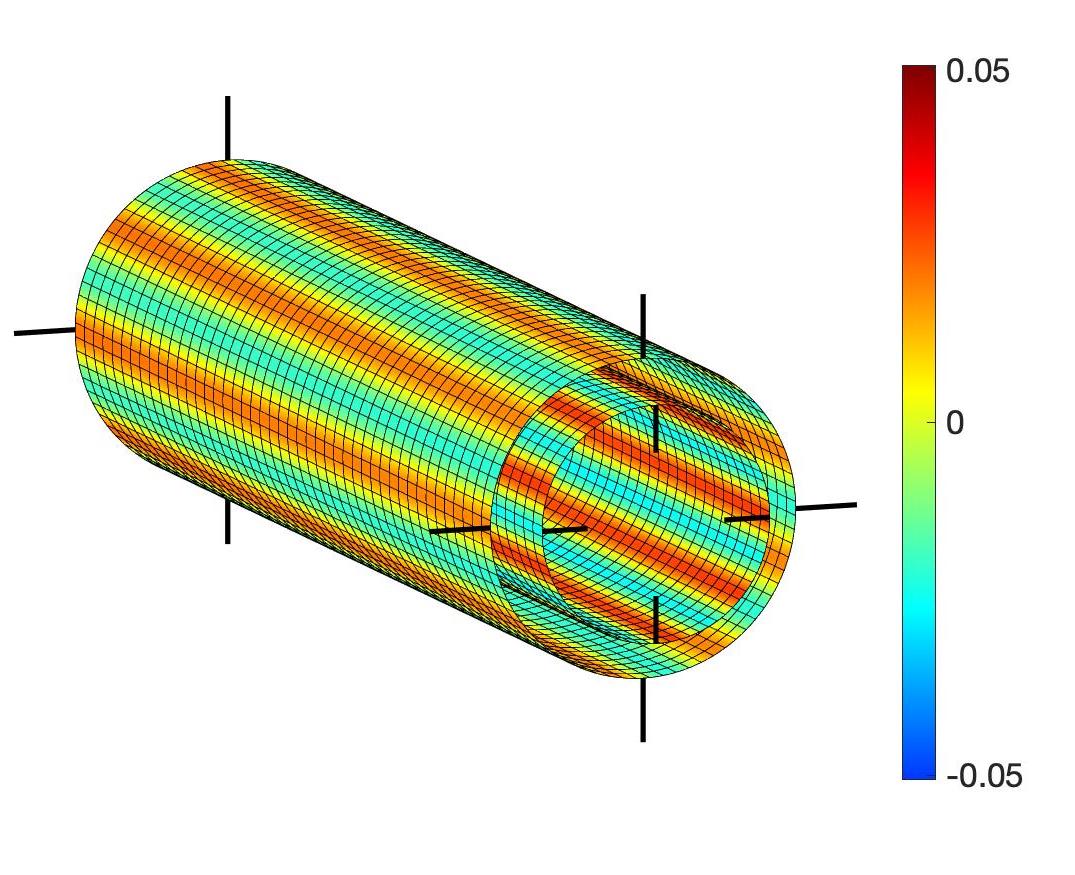}}
\put(-0.45,-.4){\includegraphics[height=25mm]{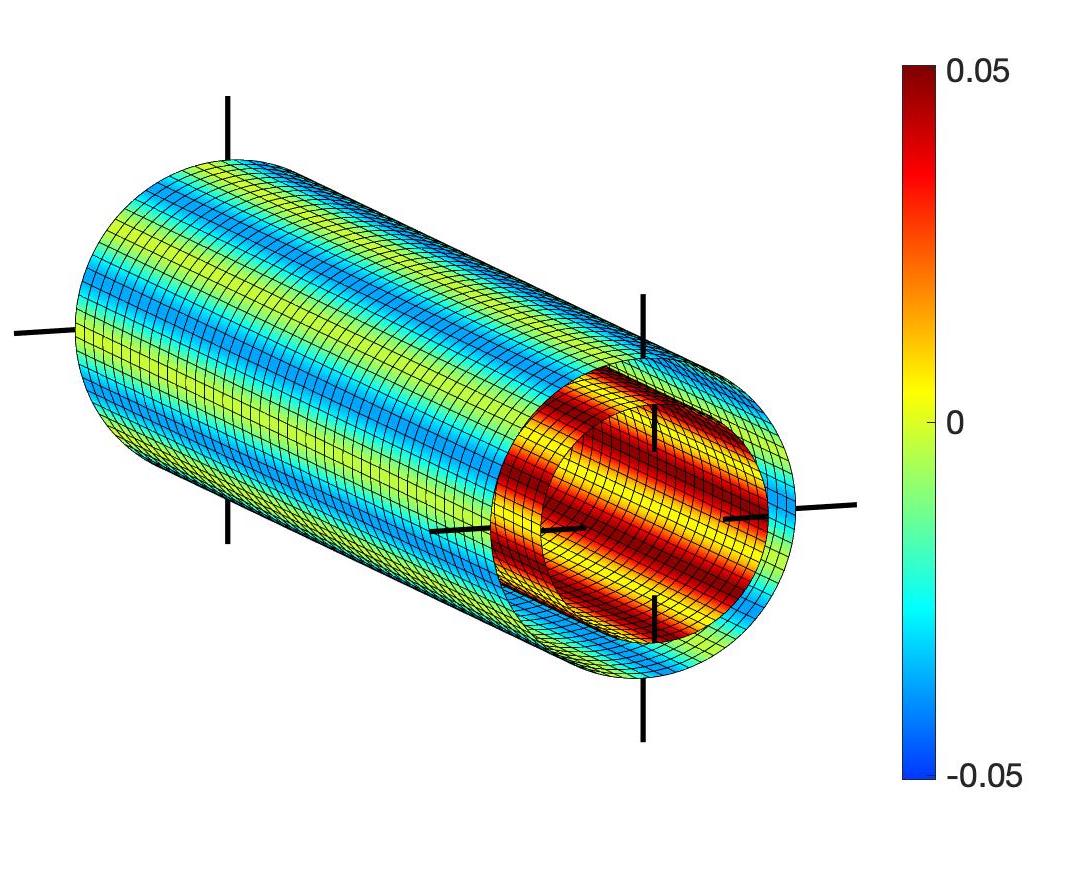}}
\put(2.05,-.4){\includegraphics[height=25mm]{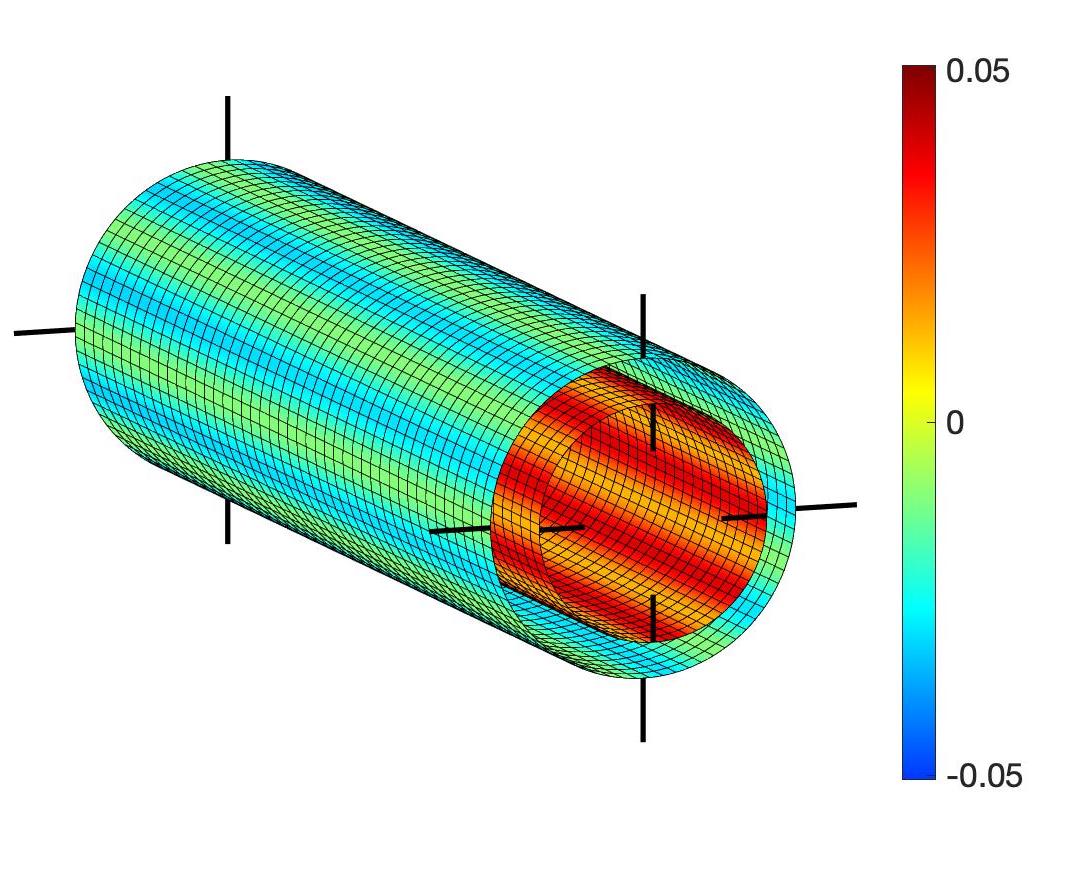}}
\put(4.55,-.4){\includegraphics[height=25mm]{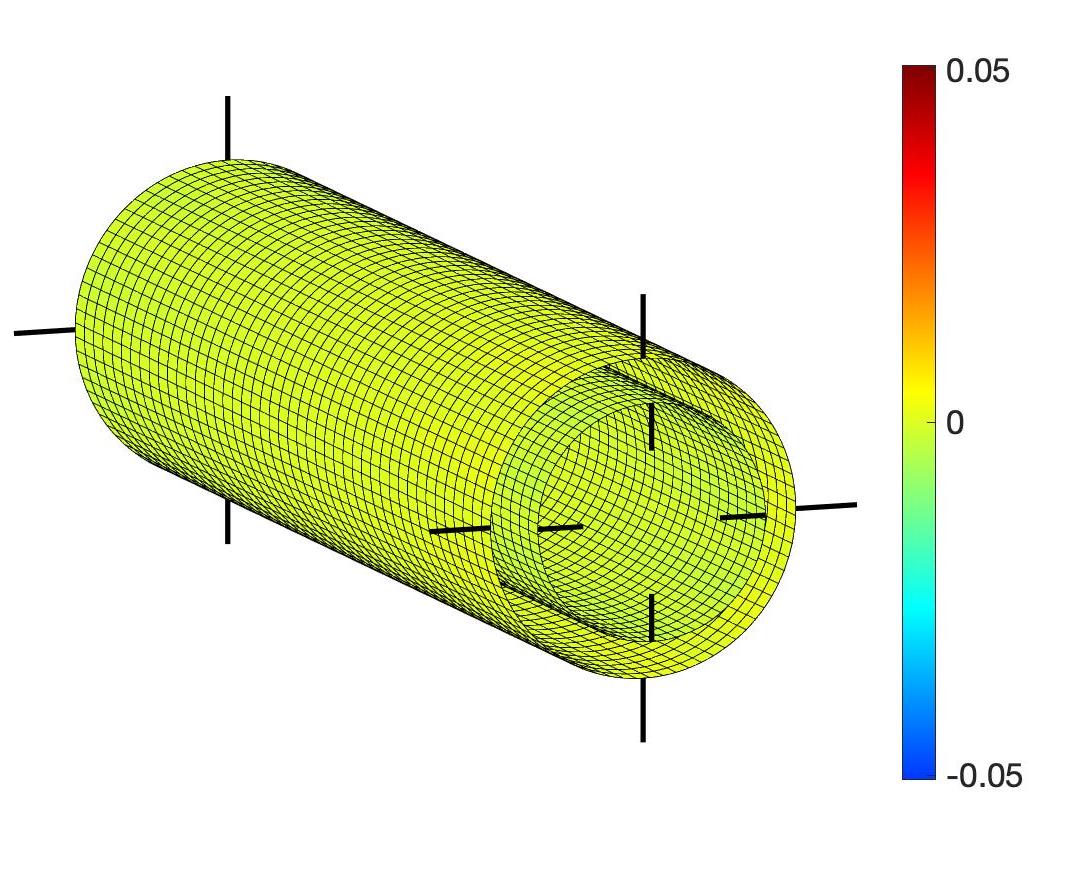}}
\put(7.1,-.55){\includegraphics[height=48mm]{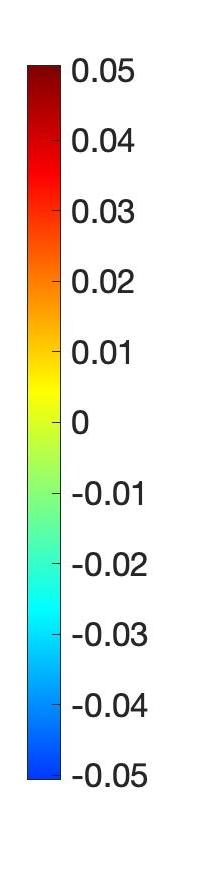}}
\put(-7.85,2.1){\scriptsize{0}}
\put(-5.35,2.1){\scriptsize{1}}
\put(-2.85,2.1){\scriptsize{2}}
\put(-0.35,2.1){\scriptsize{3}}
\put(2.15,2.1){\scriptsize{4}}
\put(4.65,2.1){\scriptsize{5}}
\put(4.65,0.){\scriptsize{6}}
\put(2.15,0.){\scriptsize{7}}
\put(-0.35,0.){\scriptsize{8}}
\put(-2.85,0.){\scriptsize{9}}
\put(-5.45,0.){\scriptsize{10}}
\put(-7.95,0.){\scriptsize{11}}
\end{picture}
\caption{Pull-out of CNT(26,0) from within CNT(35,0) (Case 1): Color plot of axial contact traction $t^1$ in [GPa] at $g_\mra \in [0,\,1,\,2,\,...,\,11]\cdot\ell_\mra/12$ (clockwise, starting top left). The axial tractions in these 12 configurations sum up to the pull-out forces marked in Fig.~\ref{cnt_pull_force}a.}
\label{f:pullout2t}
\end{center}
\end{figure}
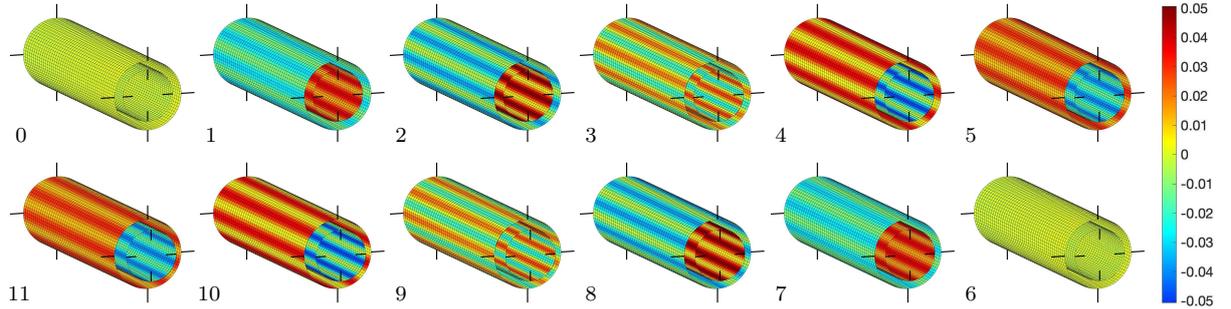

Figs.~\ref{f:pullout2p} and \ref{f:pullout2t} show the FE contact pressure and axial contact traction during pull-out for Case 1 using the DFT parameters. The contact forces vary in circumferential direction as a consequence of the circumferential interference noted in Sec.~\ref{continuum_model}.
Only at $g_\mra = \ell_a/4$ and $g_\mra  = 3\ell_a/4$ is the pressure uniform, while the axial traction is zero at $g_\mra = 0$ and $g_\mra = \ell_a/2$. 
As noted in Tab.~\ref{cnt_geo_after_relax}, the CNTs in Case 1 are in a state of attraction. As a consequence, there are circumferential stresses in the tubes, leading to the axial strains $\eps_\mathrm{out} = 3.616 \cdot 10^{-4}$ and $\eps_\mathrm{in} = -3.887 \cdot 10^{-4}$ in the initially relaxed configuration of the outer and inner tube, respectively. As seen in Fig.~\ref{cnt_pull_force}a, the periodicity and amplitude of the pull-out force determined from the FE simulations using the DFT parameters agree well with those obtained from the MD simulations. There is no error in the period, while the amplitudes differ by $\approx 5\%$ in relative and $\approx 95.0626\cdot10^{-3}$ nN in absolute terms. The FE results with the REBO+LJ parameters are even more accurate as Fig.~\ref{cnt_pull_force}a and Tab.~\ref{cnt_pull_out_d} show. The accuracy of the FE results can be further assessed by examining the difference of the pull-out forces acting on the inner and outer CNT. In theory, these forces should be in exact equilibrium. But due to the rigid master assumption made in Sec~\ref{continuum_model}, a slight difference can appear. Here this difference is below 2.82\% compared to the average pull-out force shown in Fig.~\ref{cnt_pull_force}a. This difference is very small and thus justifies the rigid master CNT assumption. 

\subsubsection{Case 2: Pull-out of CNT(15,15) from within CNT(20,20)}
\begin{figure}[!htbp]
\begin{center} \unitlength1cm
\begin{picture}(0,4.2)
\put(-7.95,1.7){\includegraphics[height=25mm]{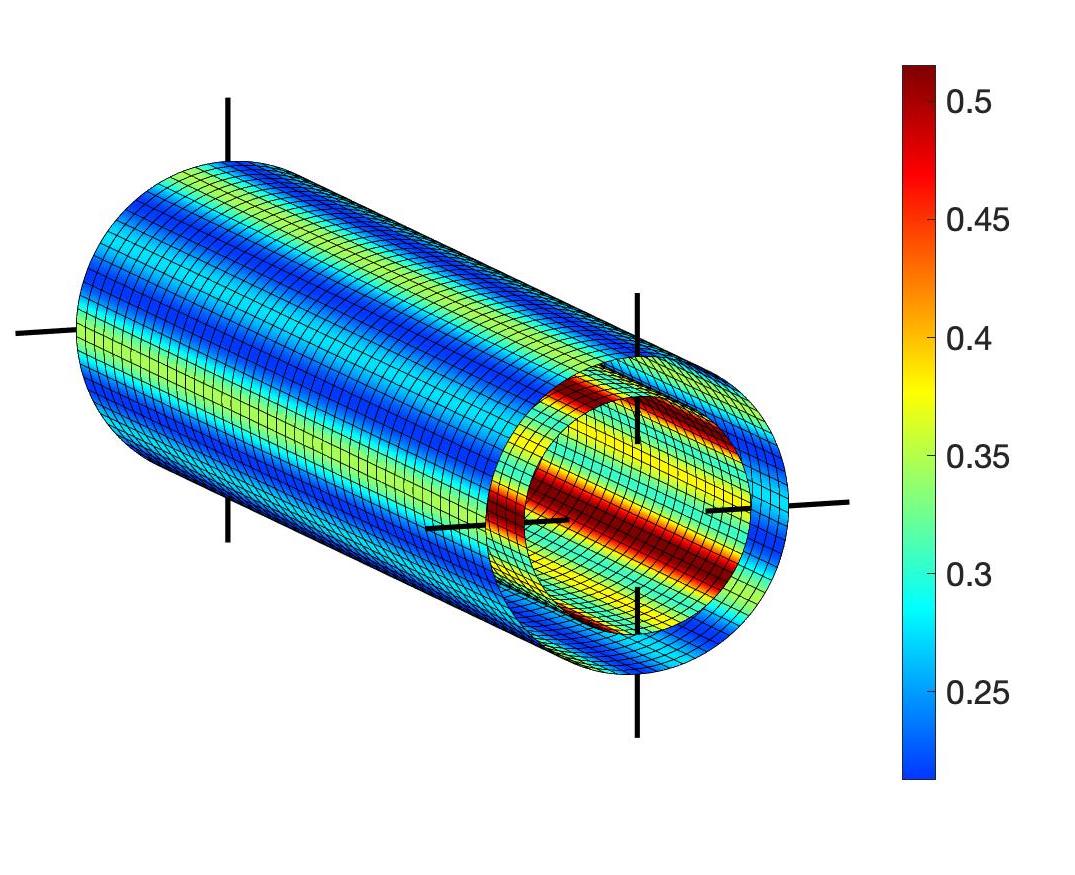}}
\put(-5.45,1.7){\includegraphics[height=25mm]{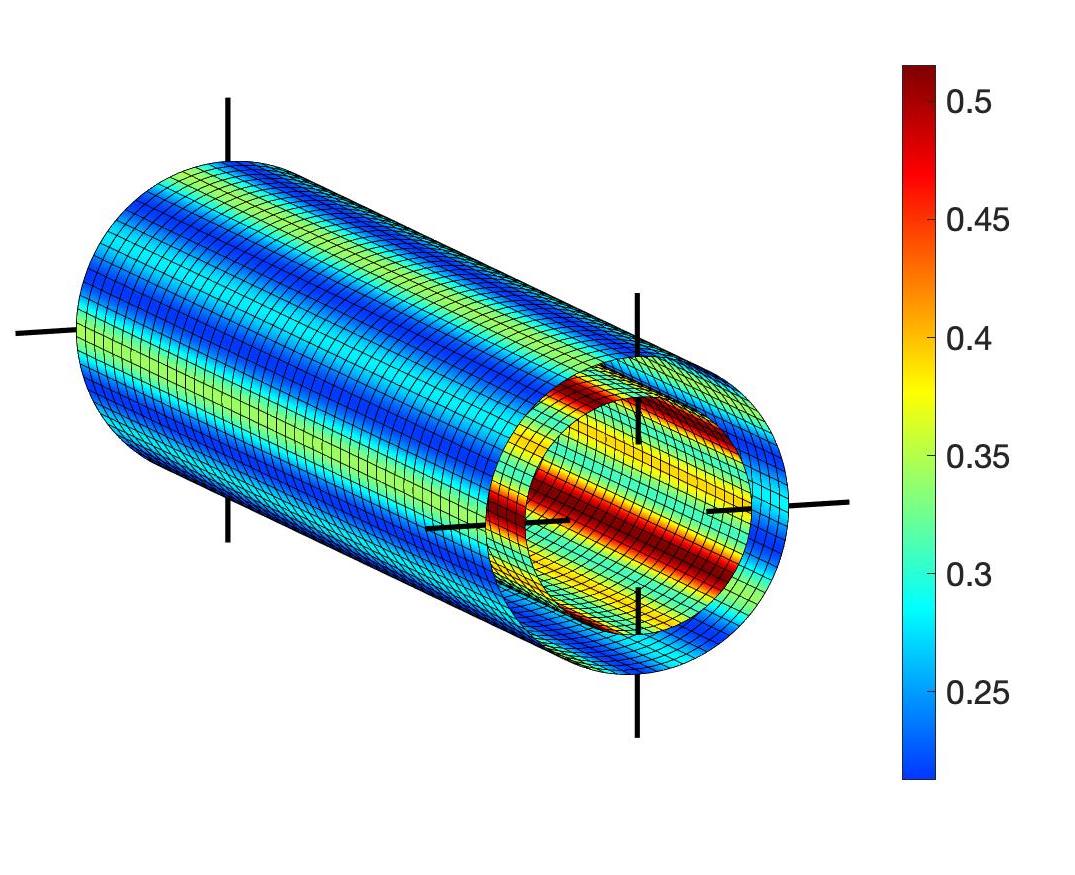}}
\put(-2.95,1.7){\includegraphics[height=25mm]{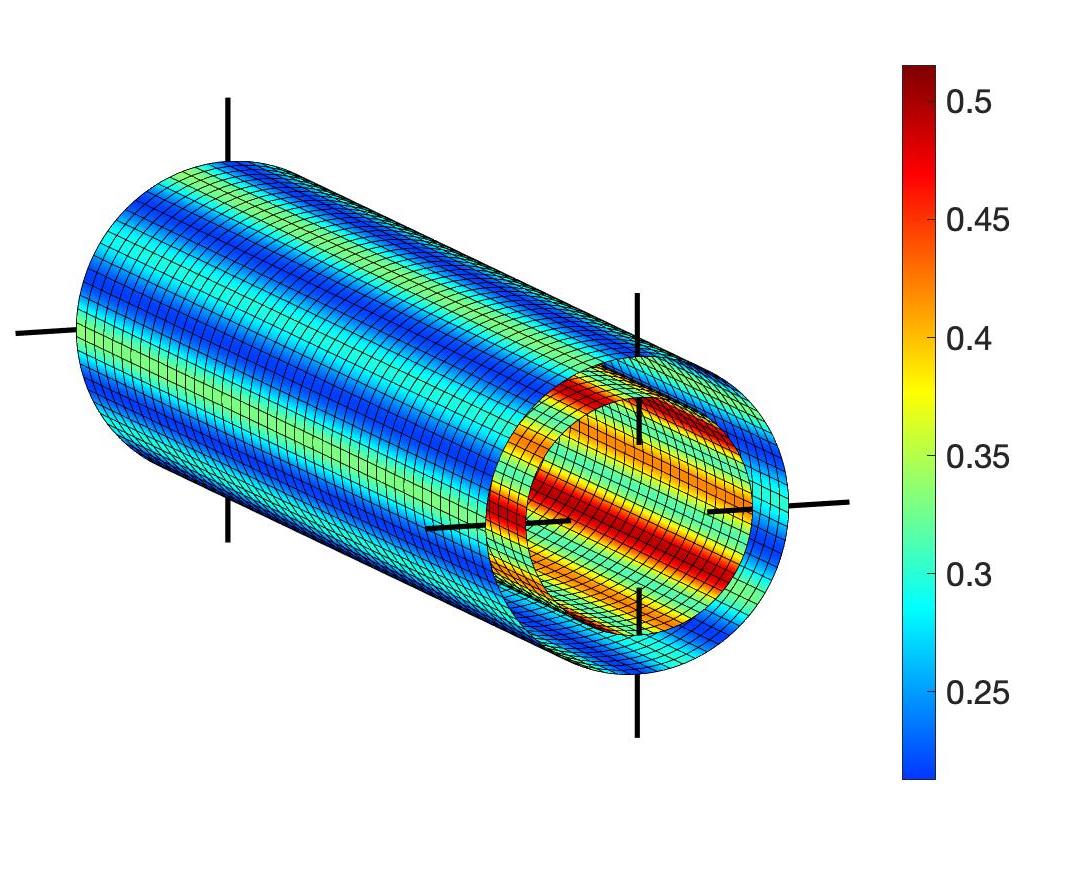}}
\put(-0.45,1.7){\includegraphics[height=25mm]{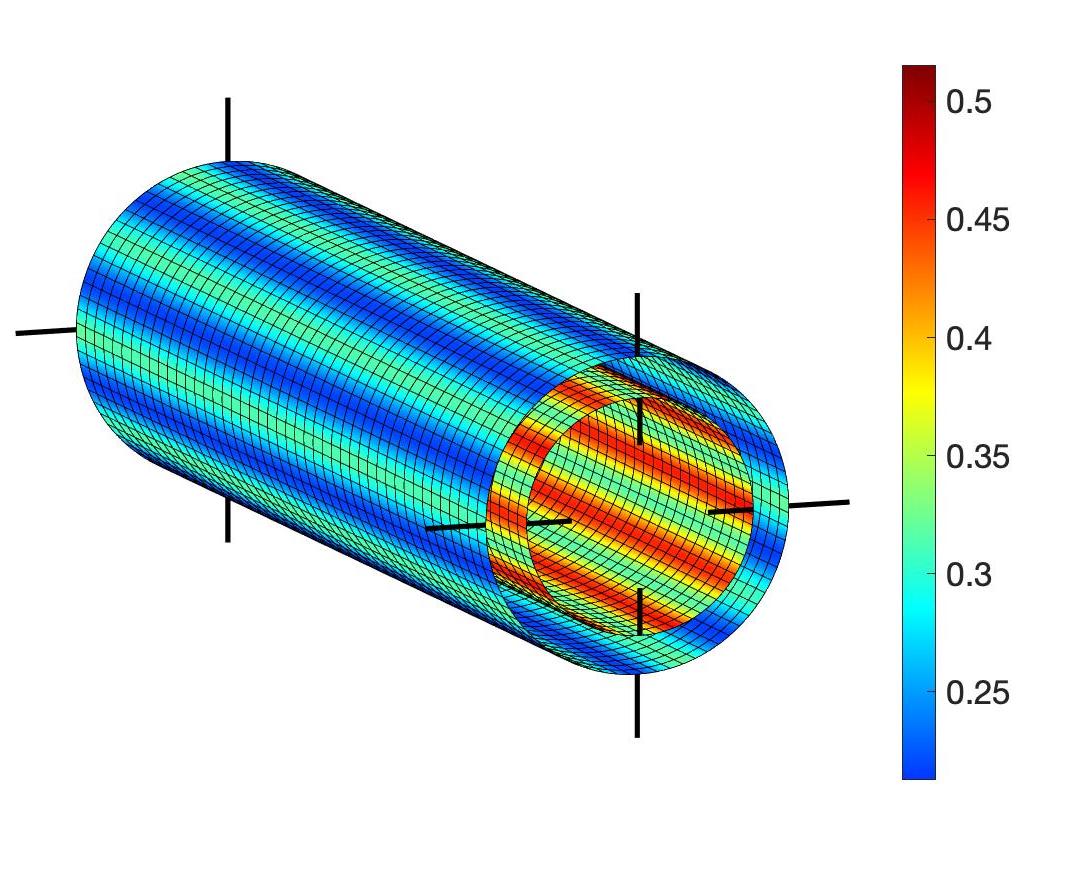}}
\put(2.05,1.7){\includegraphics[height=25mm]{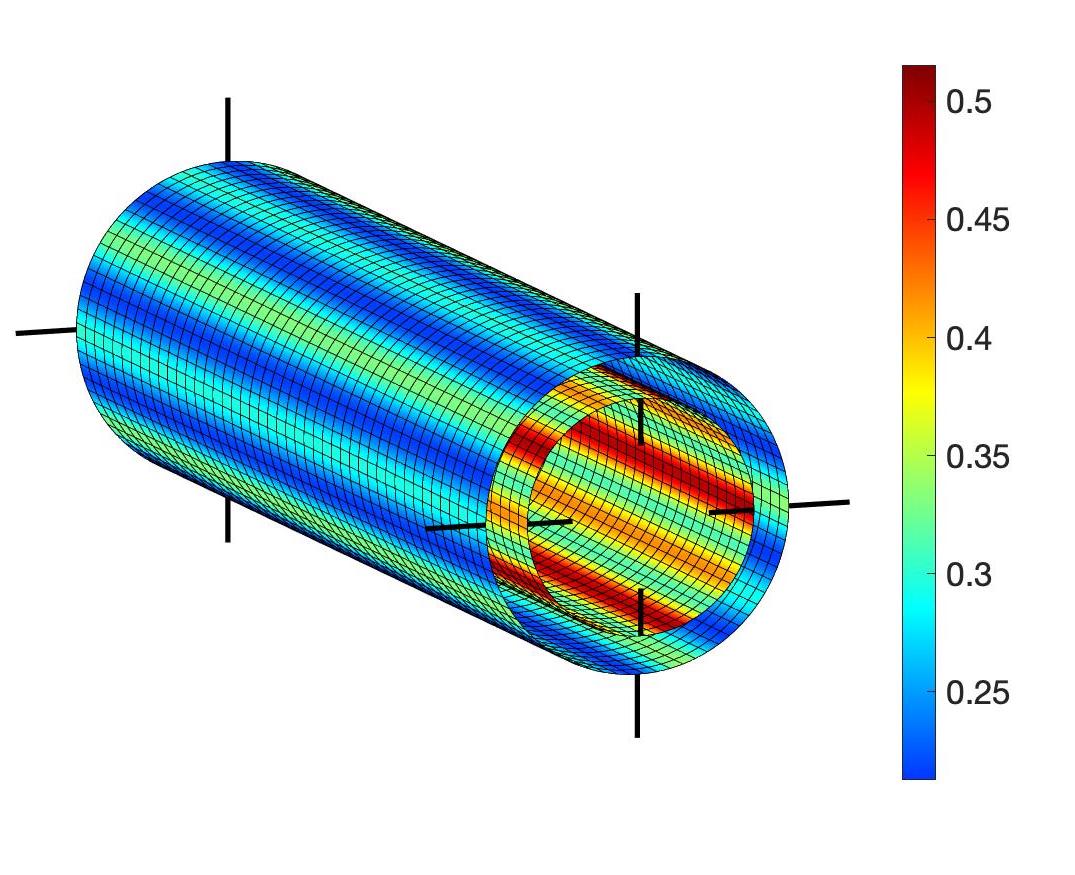}}
\put(4.55,1.7){\includegraphics[height=25mm]{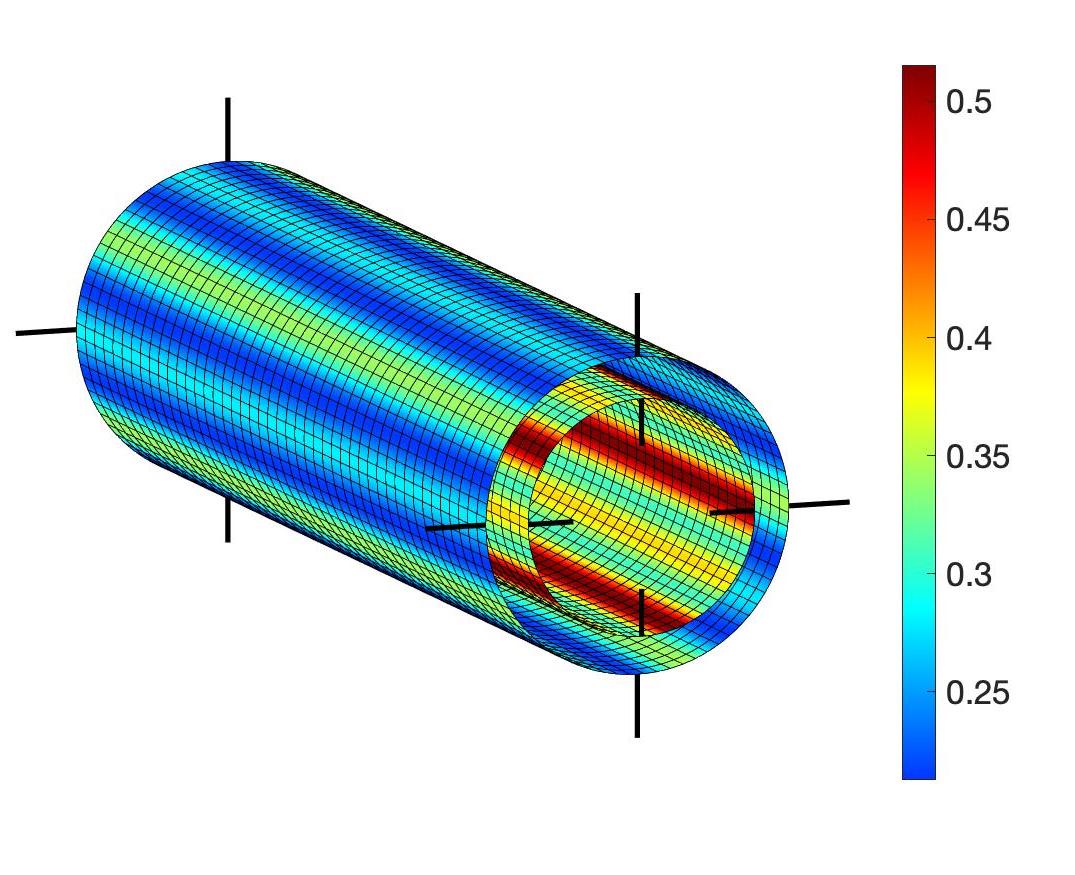}}
\put(-7.95,-.4){\includegraphics[height=25mm]{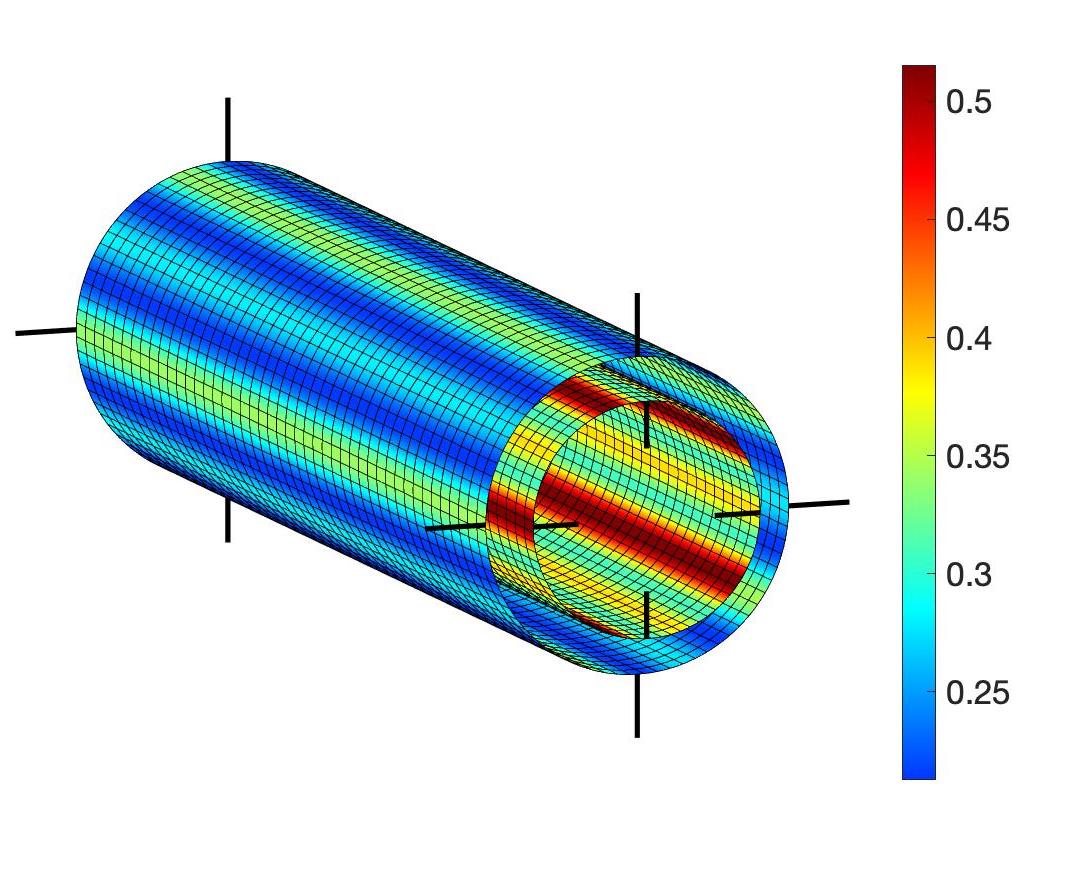}}
\put(-5.45,-.4){\includegraphics[height=25mm]{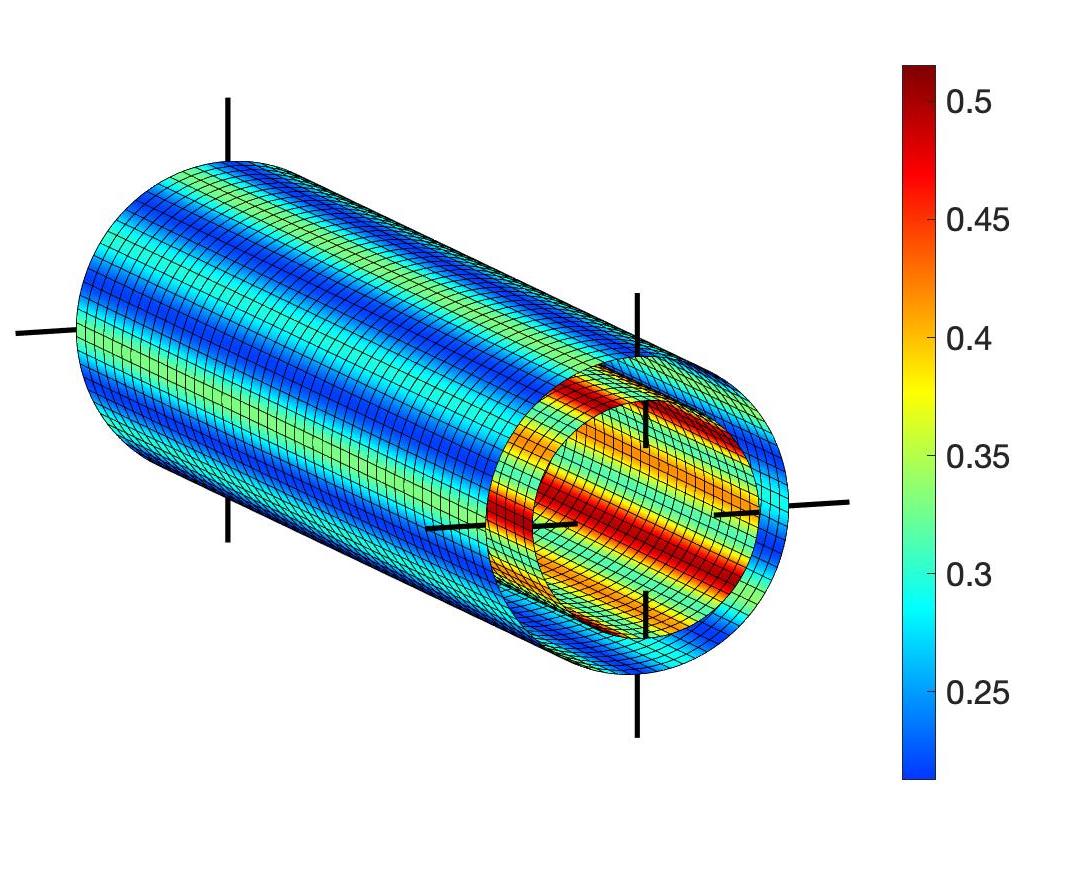}}
\put(-2.95,-.4){\includegraphics[height=25mm]{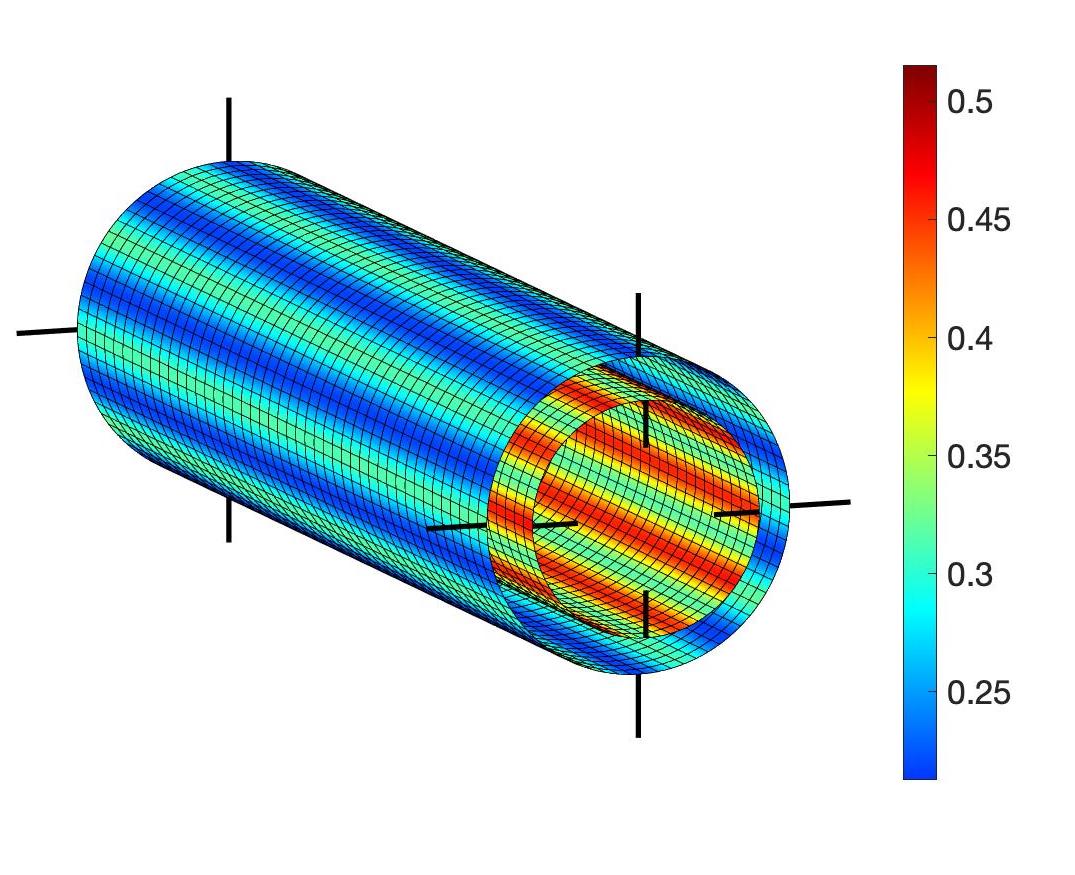}}
\put(-0.45,-.4){\includegraphics[height=25mm]{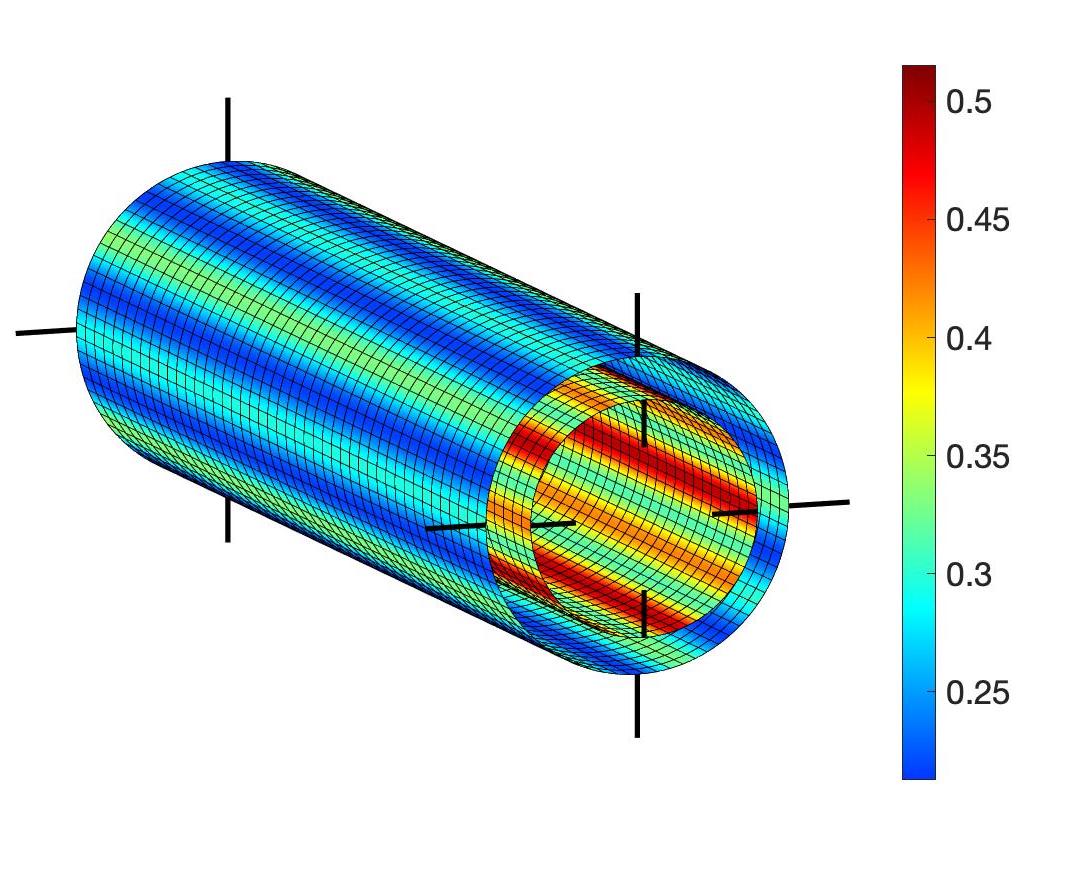}}
\put(2.05,-.4){\includegraphics[height=25mm]{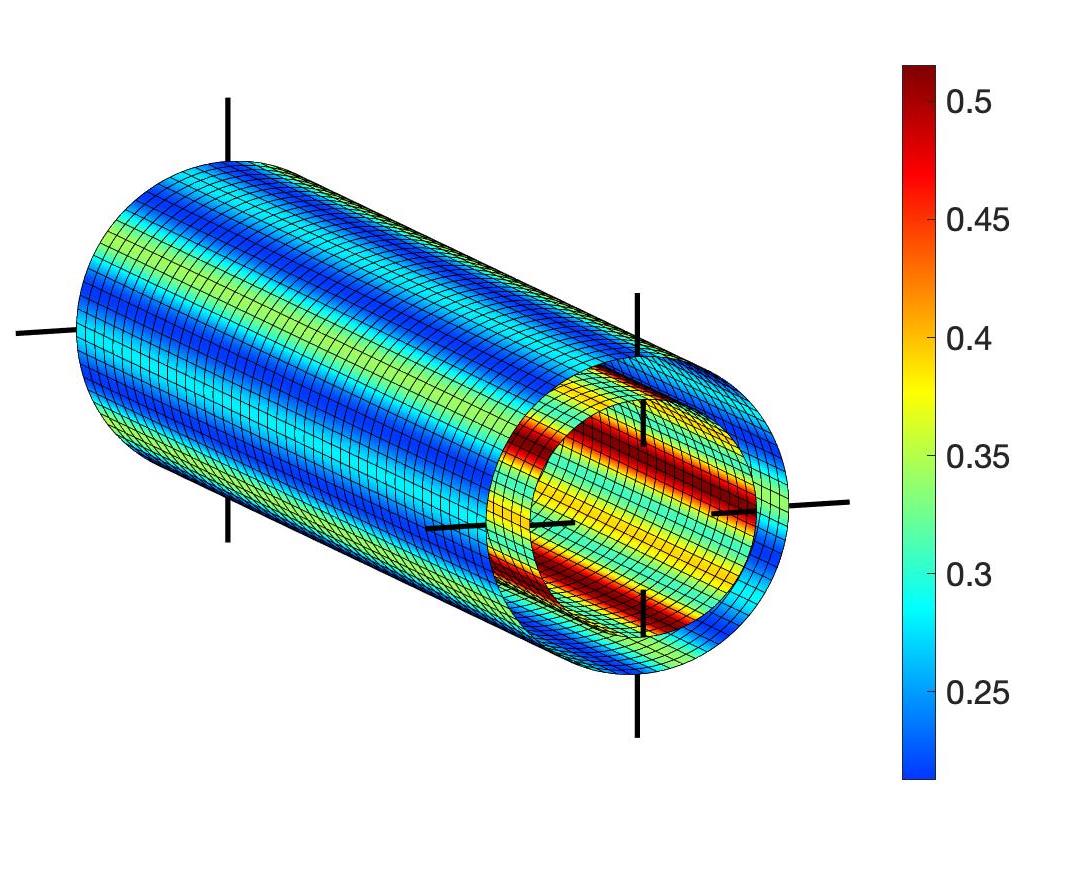}}
\put(4.55,-.4){\includegraphics[height=25mm]{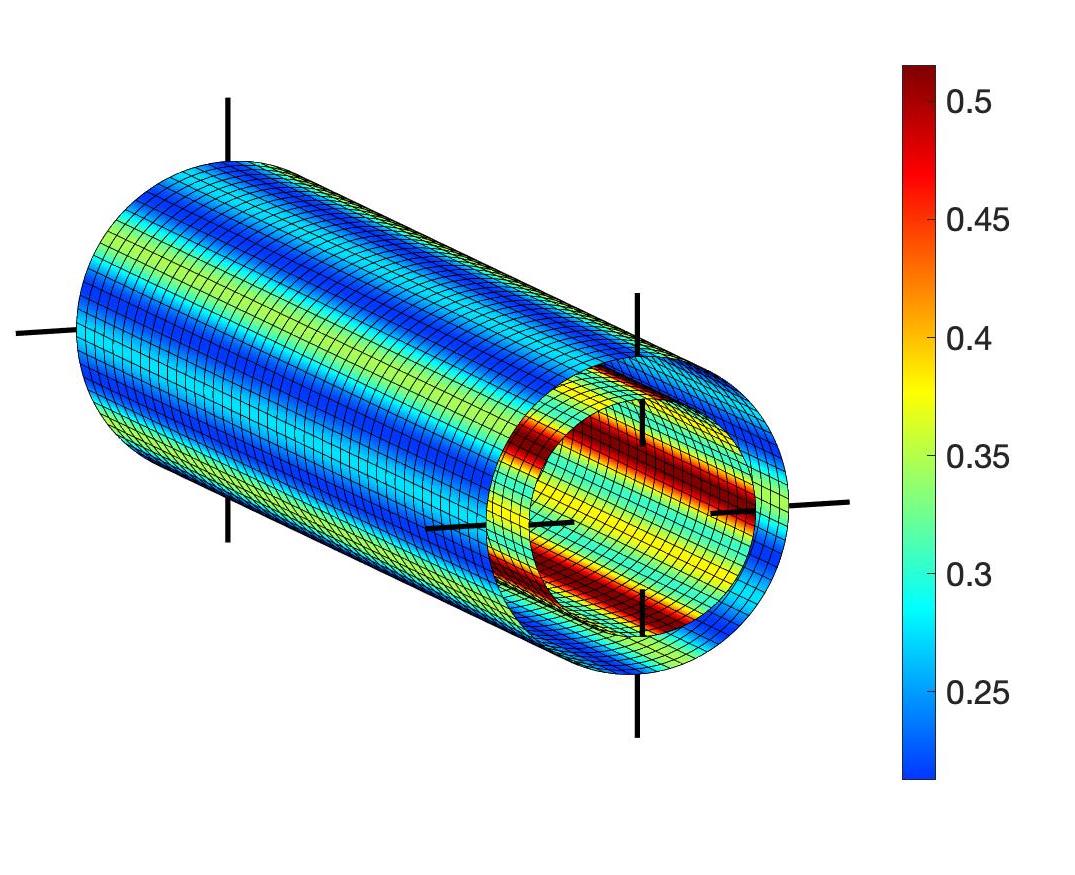}}
\put(7.1,-.55){\includegraphics[height=48mm]{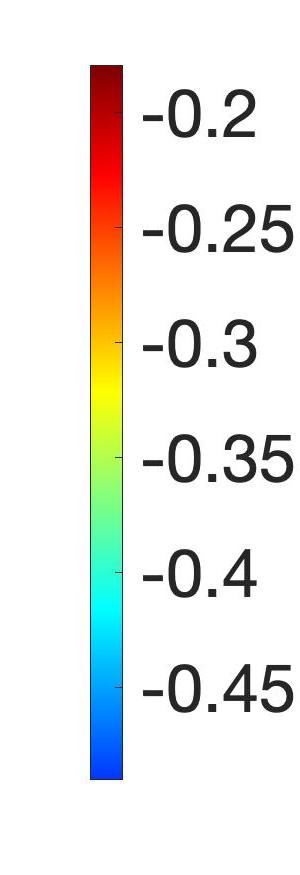}}
\put(-7.85,2.1){\scriptsize{0}}
\put(-5.35,2.1){\scriptsize{1}}
\put(-2.85,2.1){\scriptsize{2}}
\put(-0.35,2.1){\scriptsize{3}}
\put(2.15,2.1){\scriptsize{4}}
\put(4.65,2.1){\scriptsize{5}}
\put(4.65,0.){\scriptsize{6}}
\put(2.15,0.){\scriptsize{7}}
\put(-0.35,0.){\scriptsize{8}}
\put(-2.85,0.){\scriptsize{9}}
\put(-5.45,0.){\scriptsize{10}}
\put(-7.95,0.){\scriptsize{11}}
\end{picture}
\caption{Pull-out of CNT(15,15) from within CNT(20,20) (Case 2): Color plot of contact pressure $p$ in 
[GPa] at $g_\mrz \in [0,\,1,\,2,\,...,\,11]\cdot\ell_\mrz/12$ (clockwise, starting top left). 
}
\label{f:pullout1p}
\end{center}
\end{figure}
\begin{figure}[!htbp]
\begin{center} \unitlength1cm
\begin{picture}(0,4.2)
\put(-7.95,1.7){\includegraphics[height=25mm]{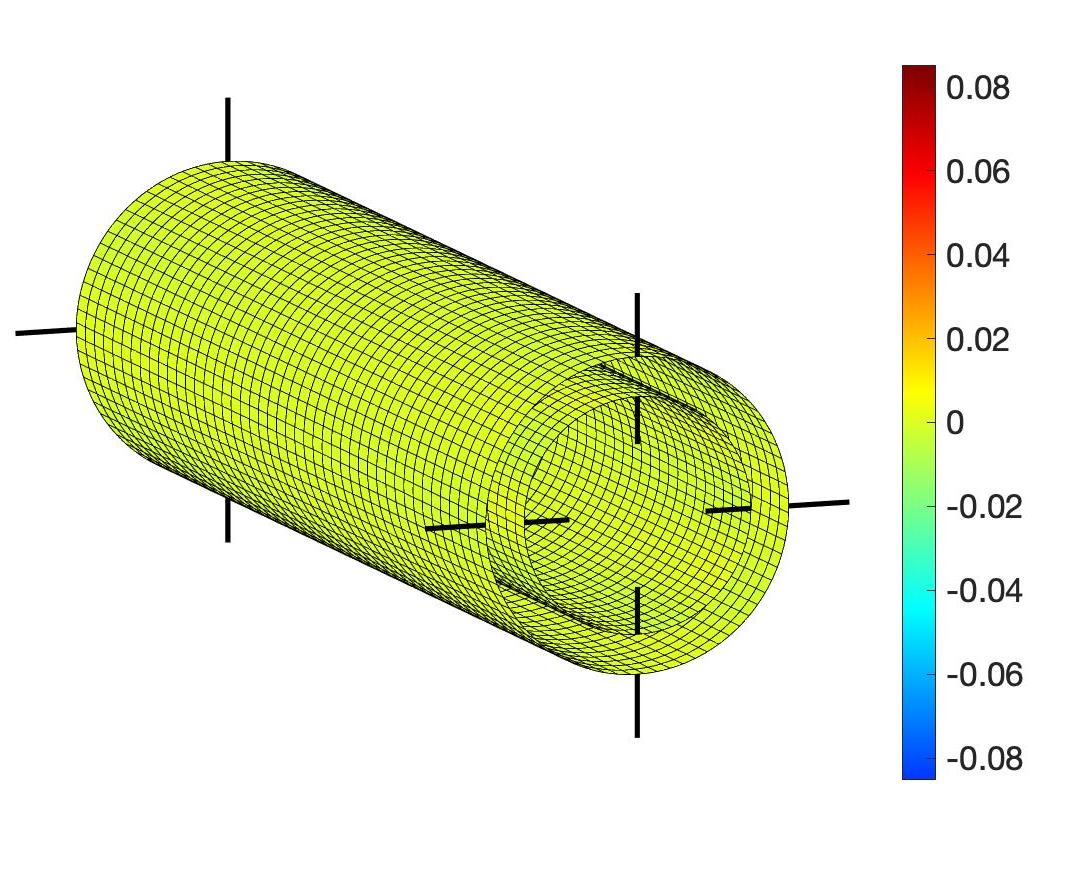}}
\put(-5.45,1.7){\includegraphics[height=25mm]{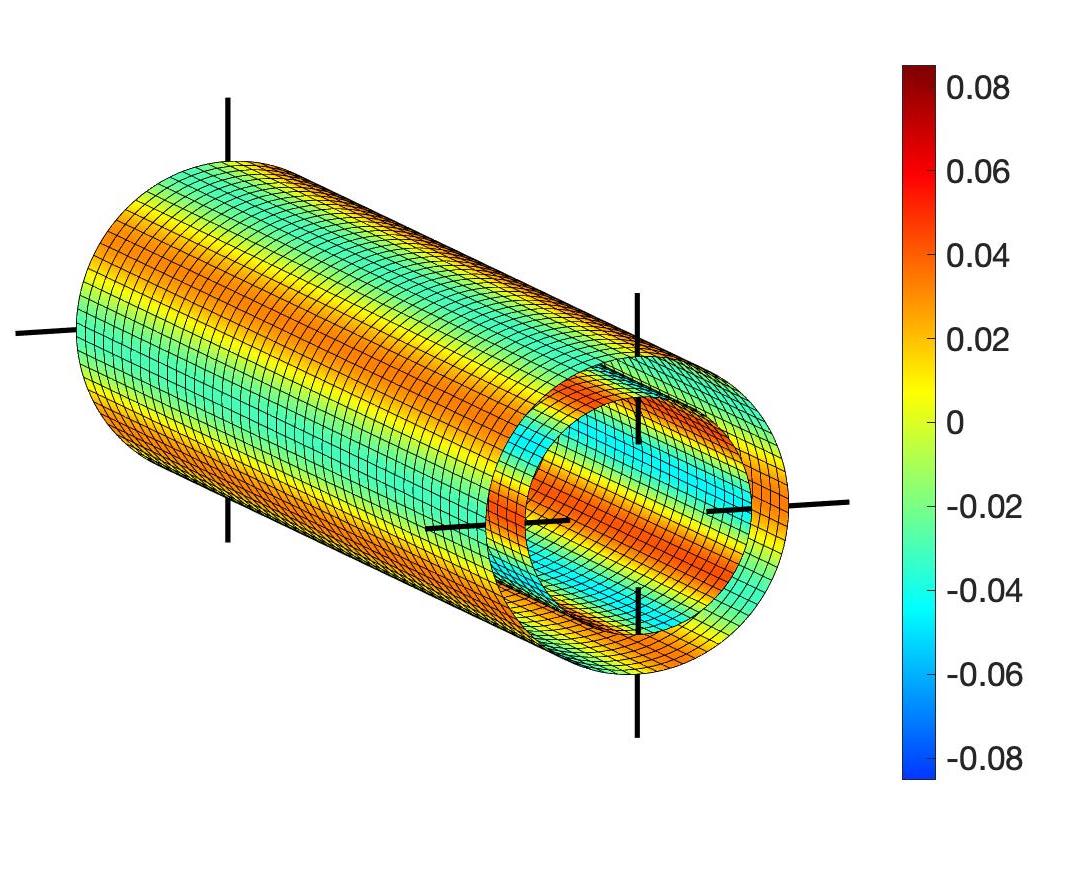}}
\put(-2.95,1.7){\includegraphics[height=25mm]{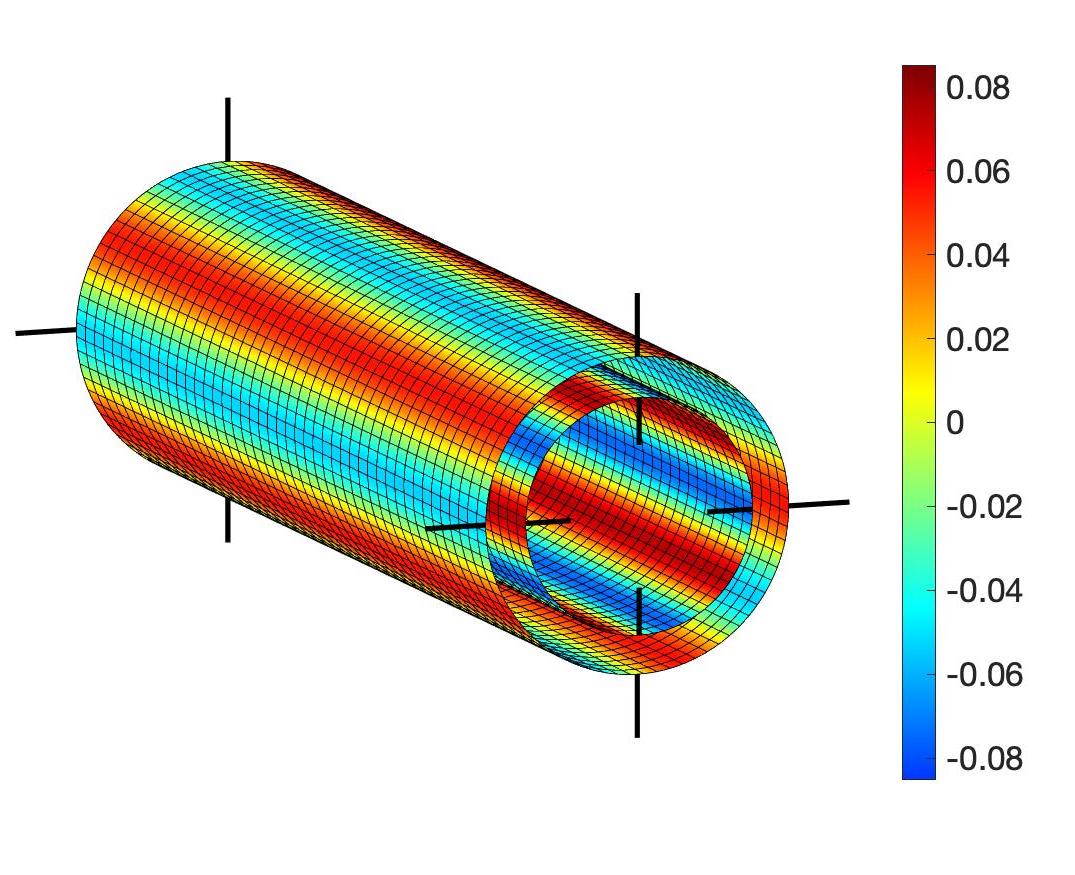}}
\put(-0.45,1.7){\includegraphics[height=25mm]{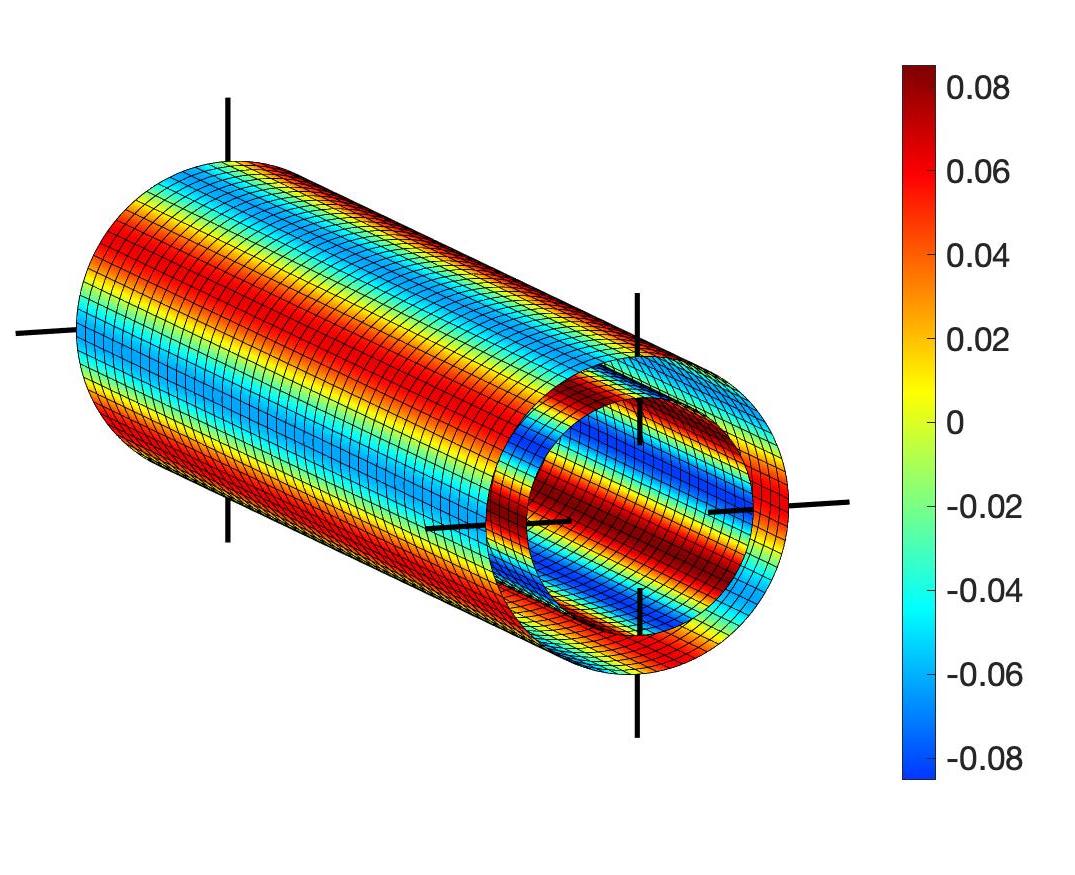}}
\put(2.05,1.7){\includegraphics[height=25mm]{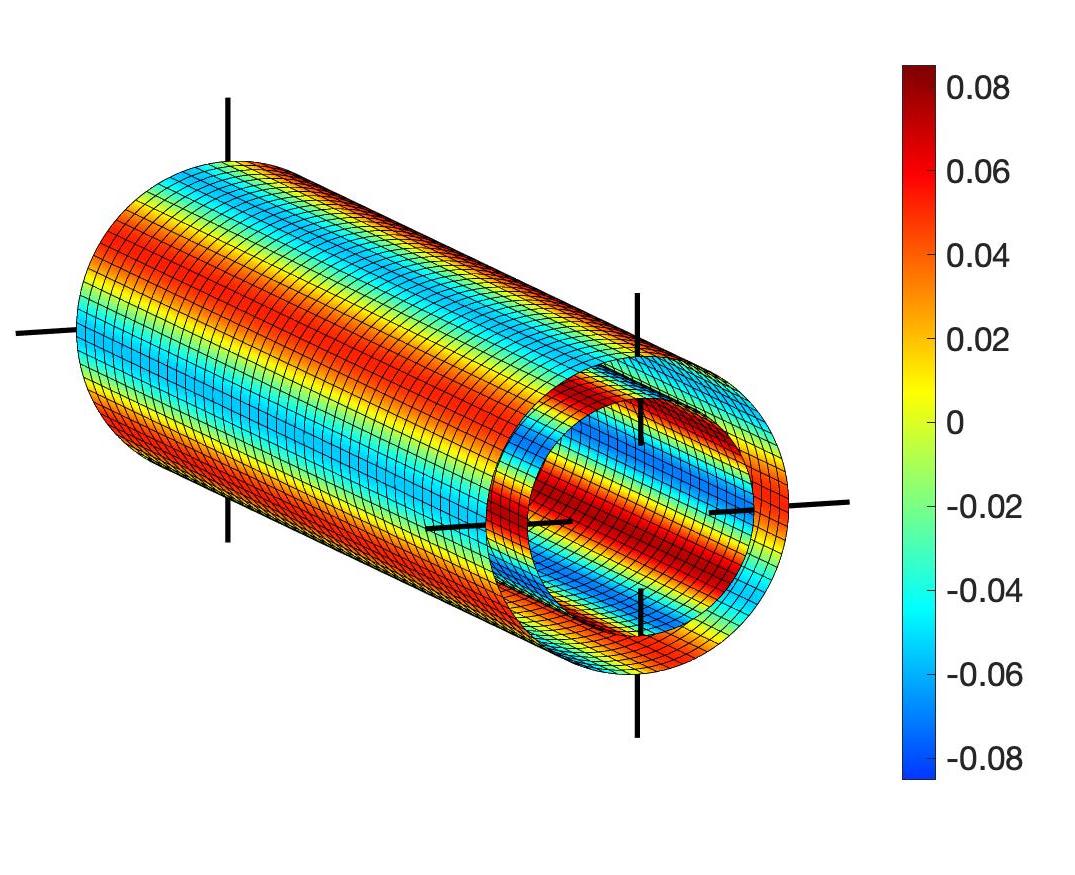}}
\put(4.55,1.7){\includegraphics[height=25mm]{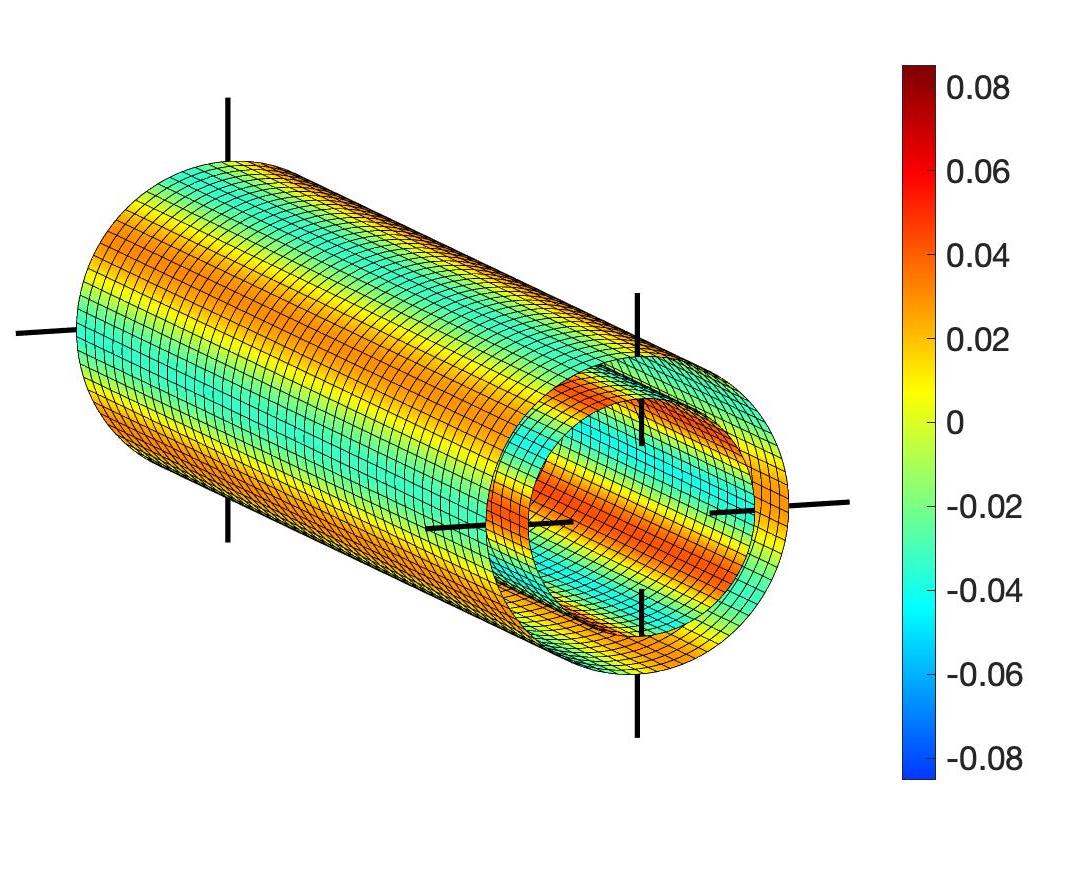}}
\put(-7.95,-.4){\includegraphics[height=25mm]{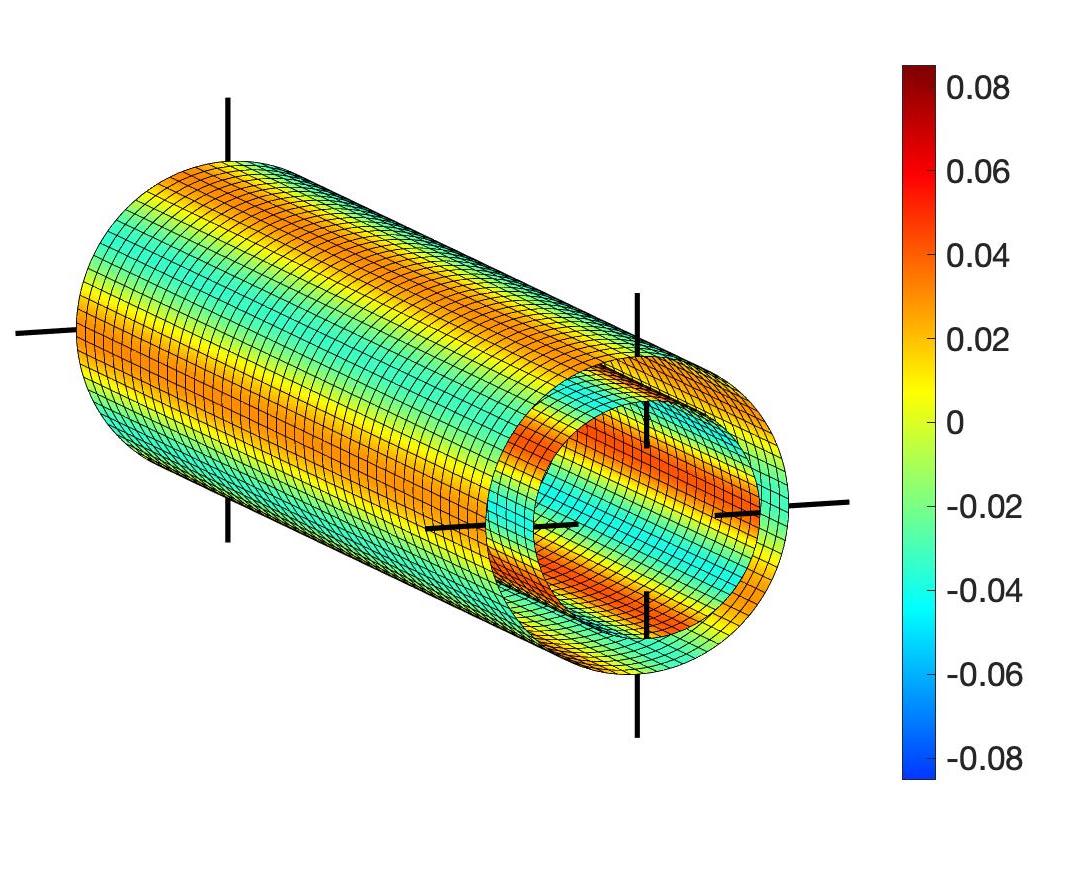}}
\put(-5.45,-.4){\includegraphics[height=25mm]{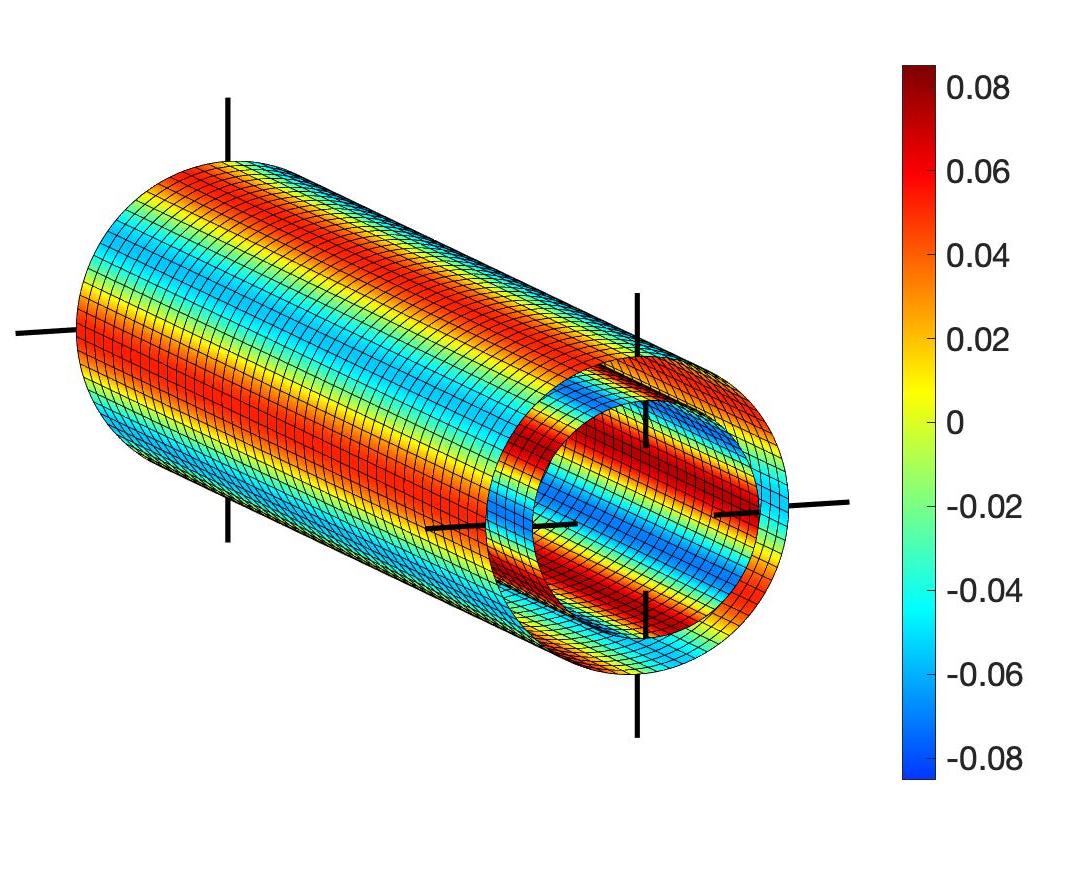}}
\put(-2.95,-.4){\includegraphics[height=25mm]{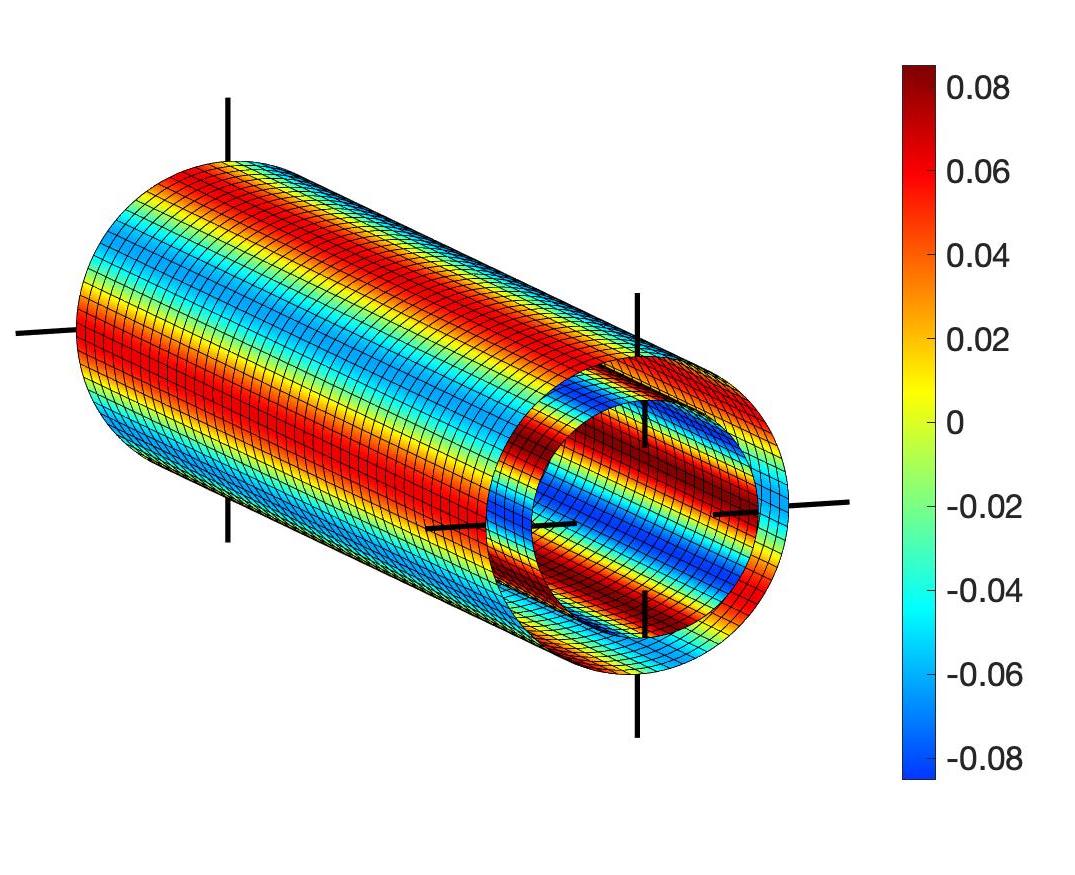}}
\put(-0.45,-.4){\includegraphics[height=25mm]{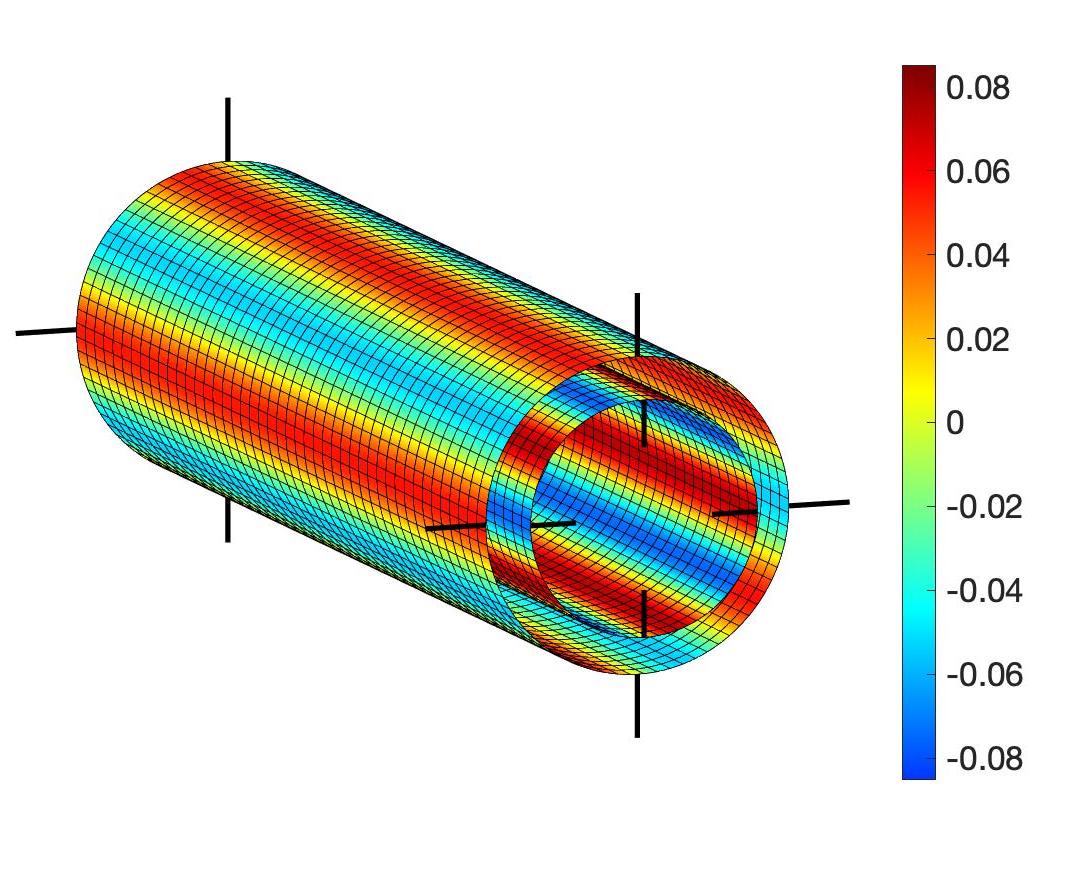}}
\put(2.05,-.4){\includegraphics[height=25mm]{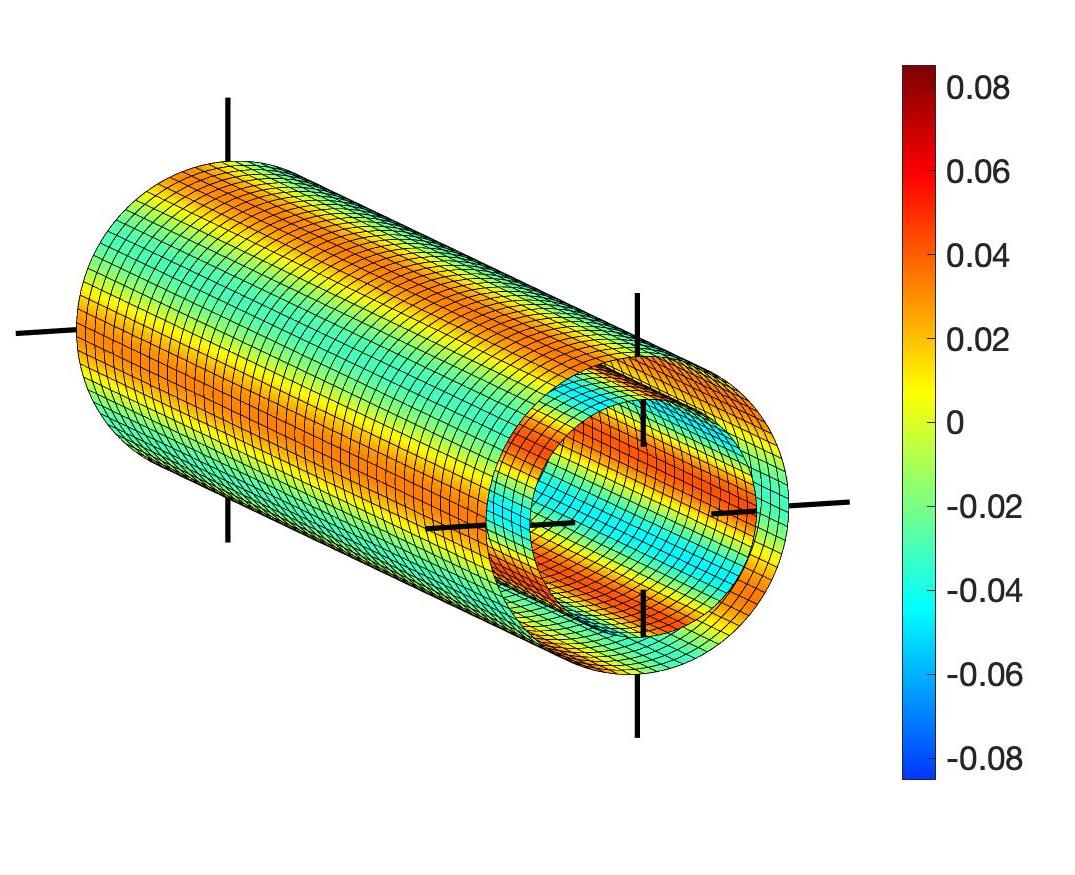}}
\put(4.55,-.4){\includegraphics[height=25mm]{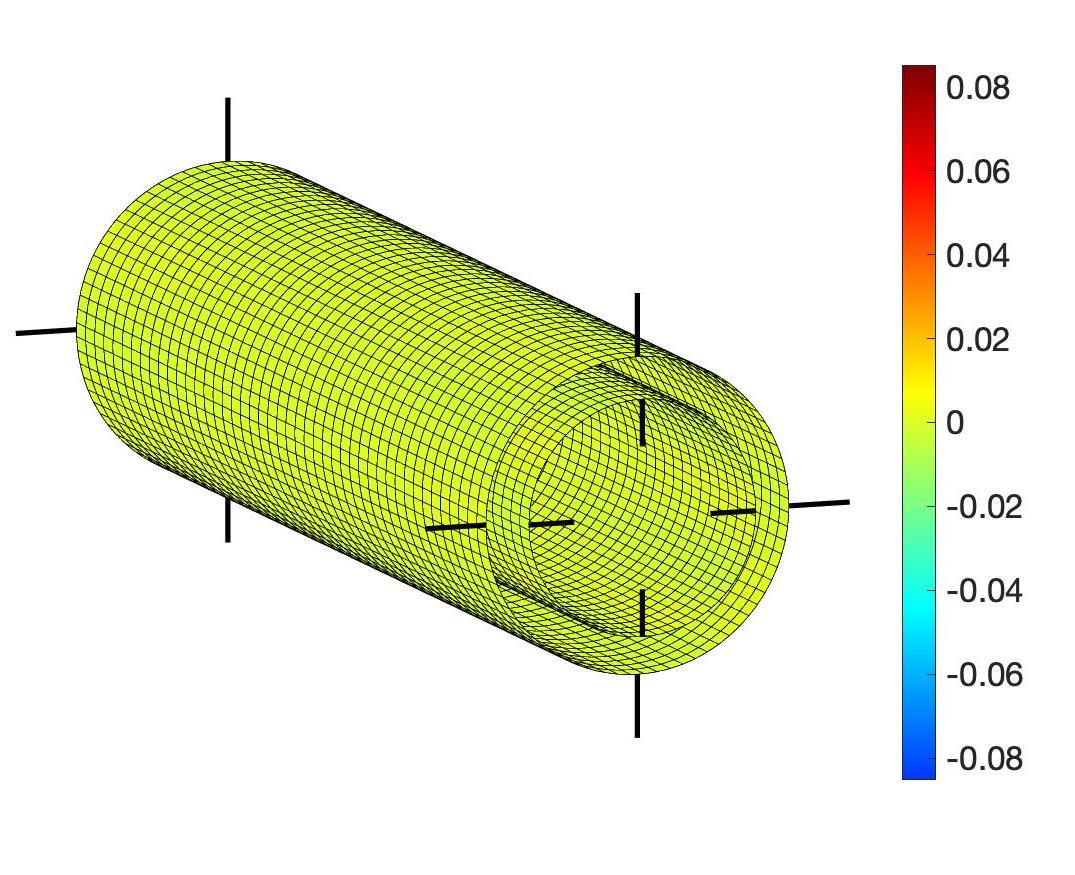}}
\put(7.1,-.55){\includegraphics[height=48mm]{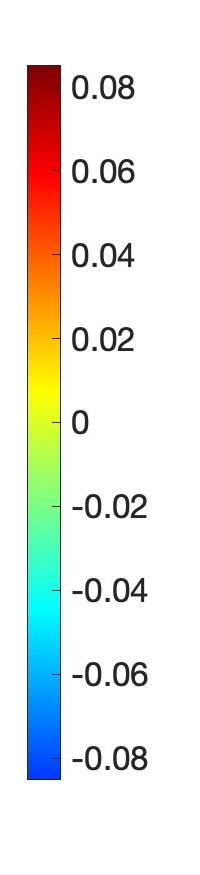}}
\put(-7.85,2.1){\scriptsize{0}}
\put(-5.35,2.1){\scriptsize{1}}
\put(-2.85,2.1){\scriptsize{2}}
\put(-0.35,2.1){\scriptsize{3}}
\put(2.15,2.1){\scriptsize{4}}
\put(4.65,2.1){\scriptsize{5}}
\put(4.65,0.){\scriptsize{6}}
\put(2.15,0.){\scriptsize{7}}
\put(-0.35,0.){\scriptsize{8}}
\put(-2.85,0.){\scriptsize{9}}
\put(-5.45,0.){\scriptsize{10}}
\put(-7.95,0.){\scriptsize{11}}
\end{picture}
\caption{Pull-out of CNT(15,15) from within CNT(20,20) (Case 2): Color plot of axial contact traction $t_1$ in [GPa] at $g_\mrz \in [0,\,1,\,2,\,...,\,11]\cdot\ell_\mrz/12$ (clockwise, starting top left).}
\label{f:pullout1t}
\end{center}
\end{figure}
Figs.~\ref{f:pullout1p} and \ref{f:pullout1t} show the contact pressures and axial contact tractions for Case 2 determined from the FE simulations using the DFT parameters. Again, the contact tractions vary in circumferential direction during sliding. The axial traction is zero at $g_\mrz = 0$ and $g_\mrz = \ell_z/2$.
The CNTs for this case are in {a} state of repulsion, {which} leads to positive contact pressures. As a consequence, the axial strain of the outer and inner CNT are $\eps_\mathrm{out} = -1.996 \cdot 10^{-4}$ and $\eps_\mathrm{in} = 2.189 \cdot 10^{-4}$, respectively, in the initially relaxed configuration. The maximum pull-out force from FE simulations compares well with that determined from MD simulations, with an absolute error of $4.6490\cdot 10^{-3}$ nN (see Tab.~\ref{cnt_pull_out_d}), which is much smaller than for Case 1. Relative errors, however, have increased as Fig.~\ref{cnt_pull_force}b shows. Also, the relative difference of the FE pull-out force on inner and outer CNT has increased, and is now below $\approx 18.42 \%$ compared to the average pull-out force shown in Fig.~\ref{cnt_pull_force}b. The rigid master assumption thus introduces a significant inaccuracy in Case 2. Furthermore, the FE pull-out force based on the elasticity parameters computed from the REBO+LJ potential is significantly smaller than that based on the elasticity parameters computed from DFT.

\subsubsection{Case 3: Pull-out of CNT(21,9) from within CNT(28,12)}
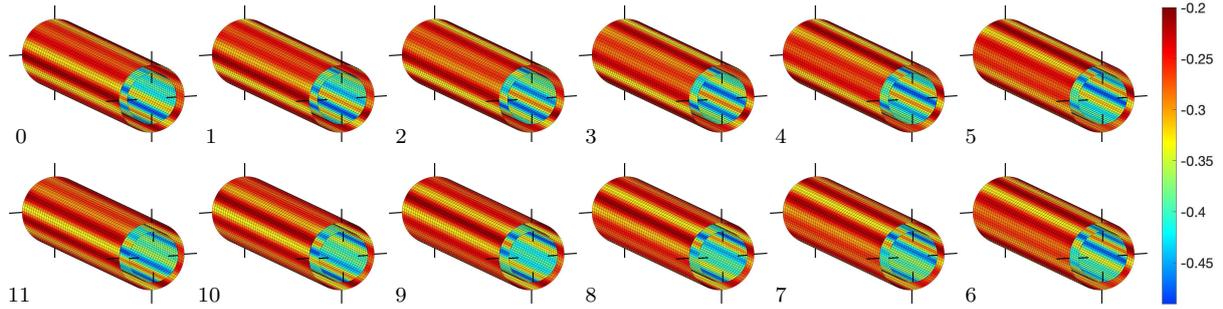
\begin{figure}[!htbp]
\begin{center} \unitlength1cm
\begin{picture}(0,4.2)
\put(-7.95,1.7){\includegraphics[height=25mm]{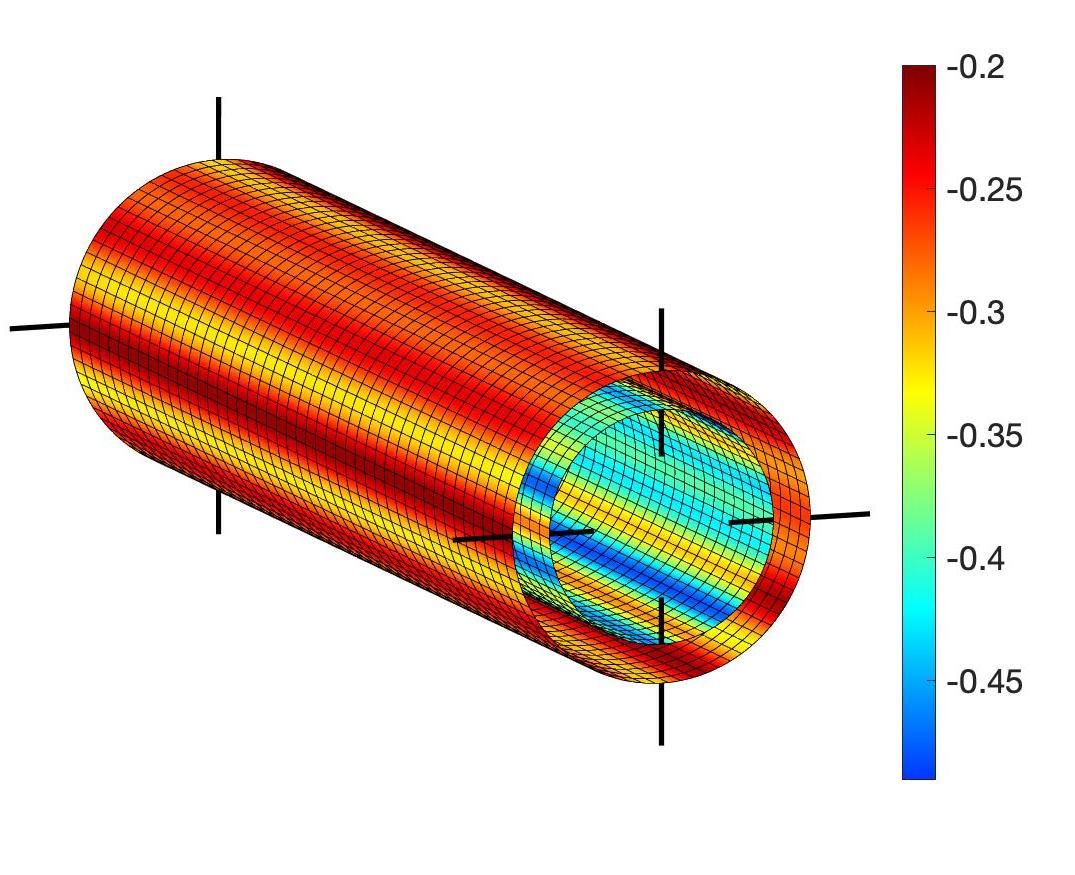}}
\put(-5.45,1.7){\includegraphics[height=25mm]{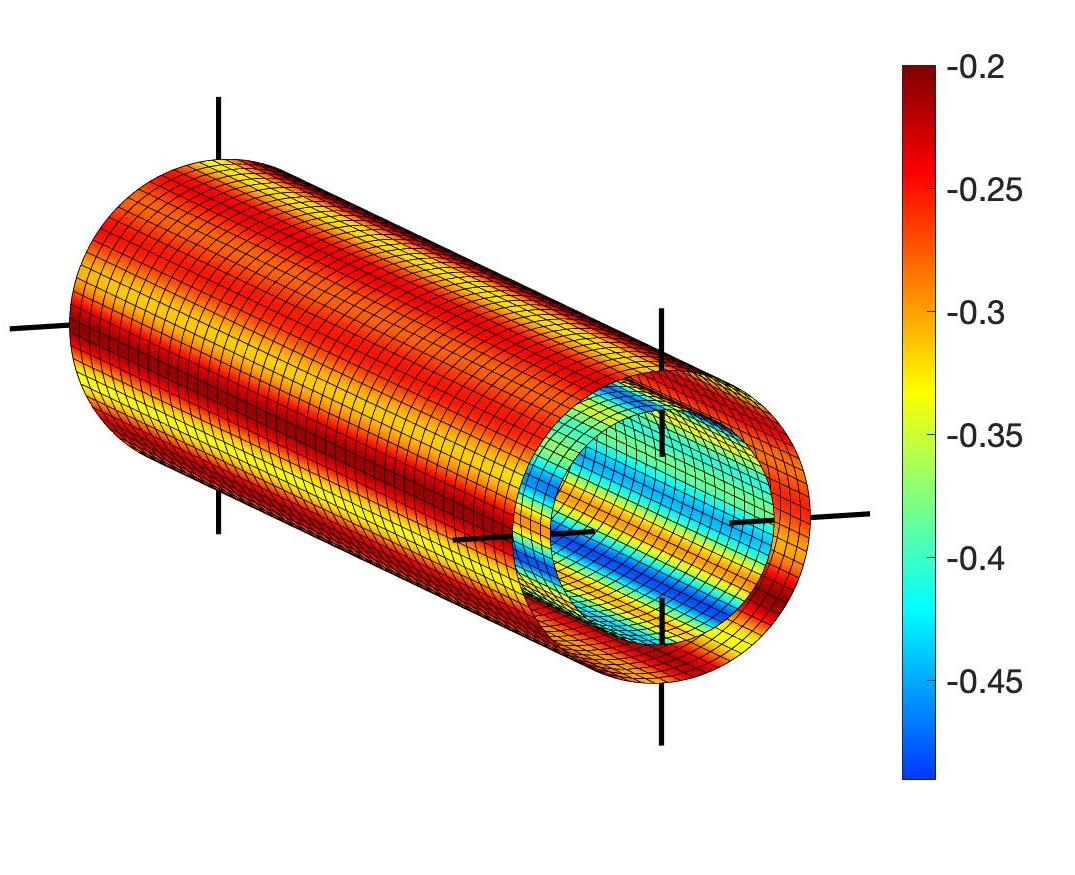}}
\put(-2.95,1.7){\includegraphics[height=25mm]{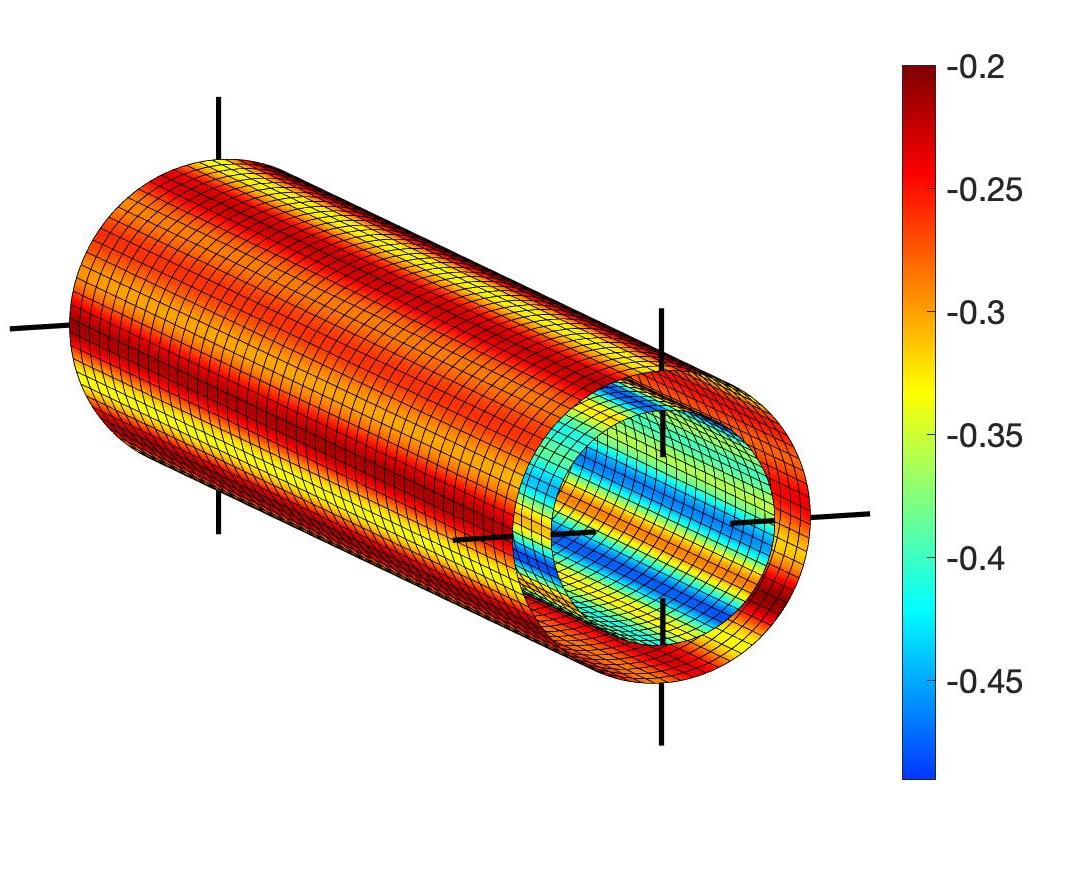}}
\put(-0.45,1.7){\includegraphics[height=25mm]{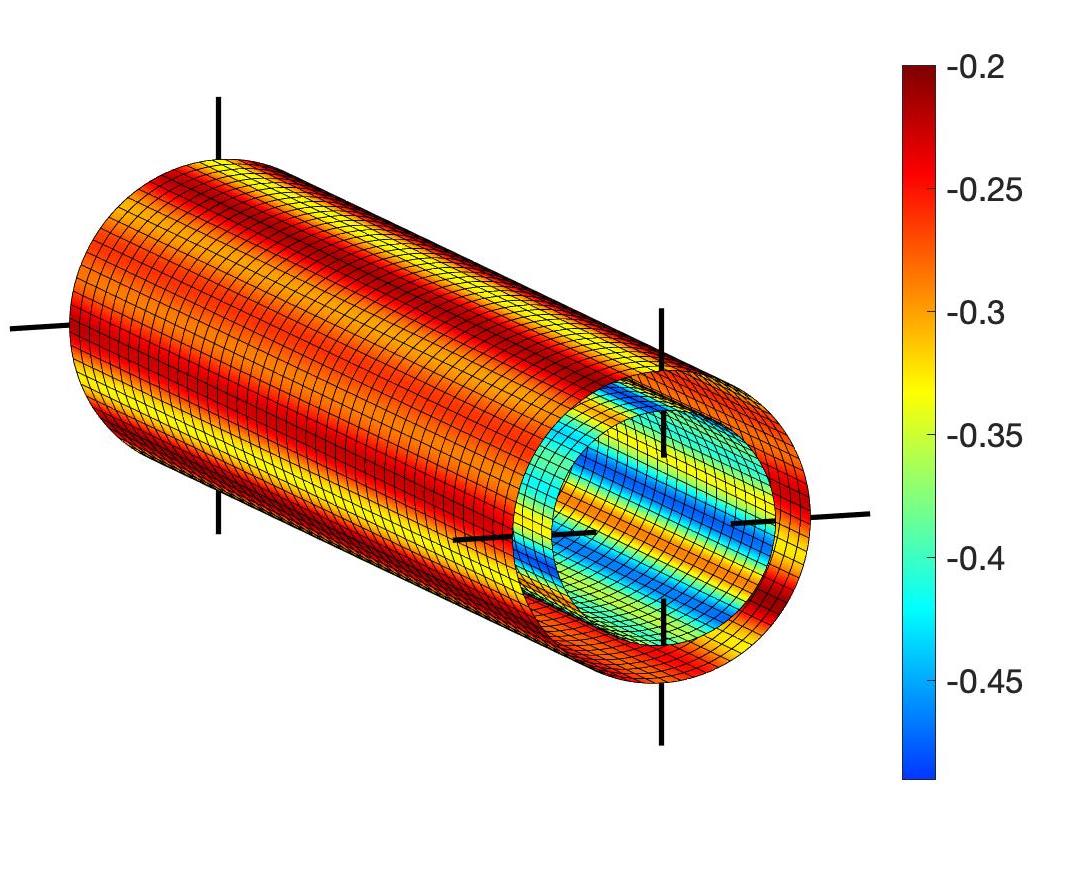}}
\put(2.05,1.7){\includegraphics[height=25mm]{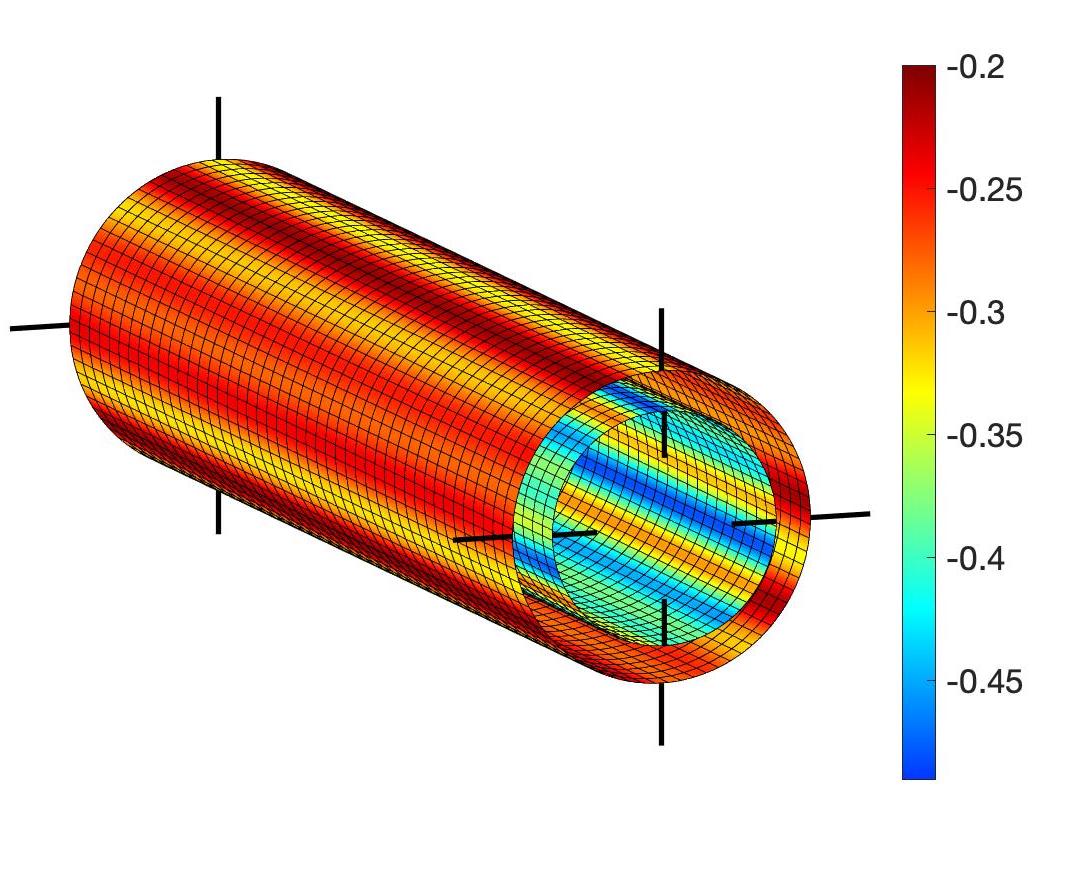}}
\put(4.55,1.7){\includegraphics[height=25mm]{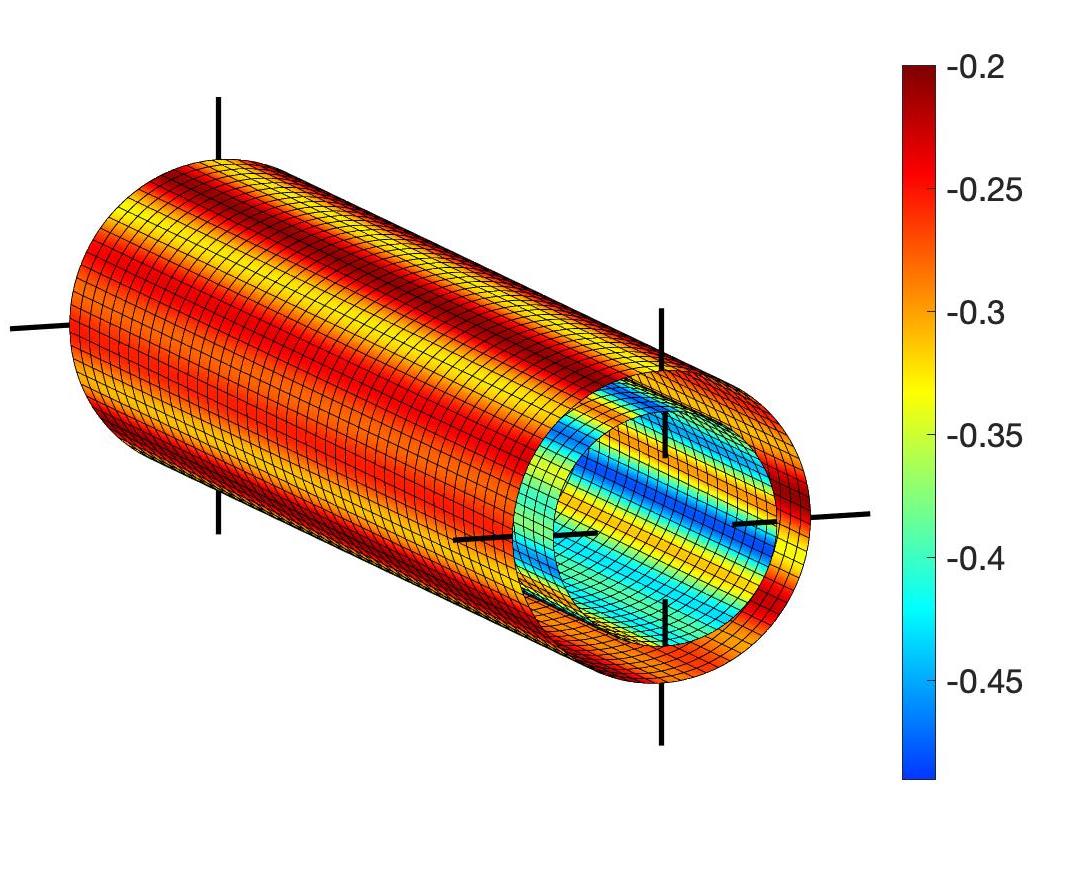}}
\put(-7.95,-.4){\includegraphics[height=25mm]{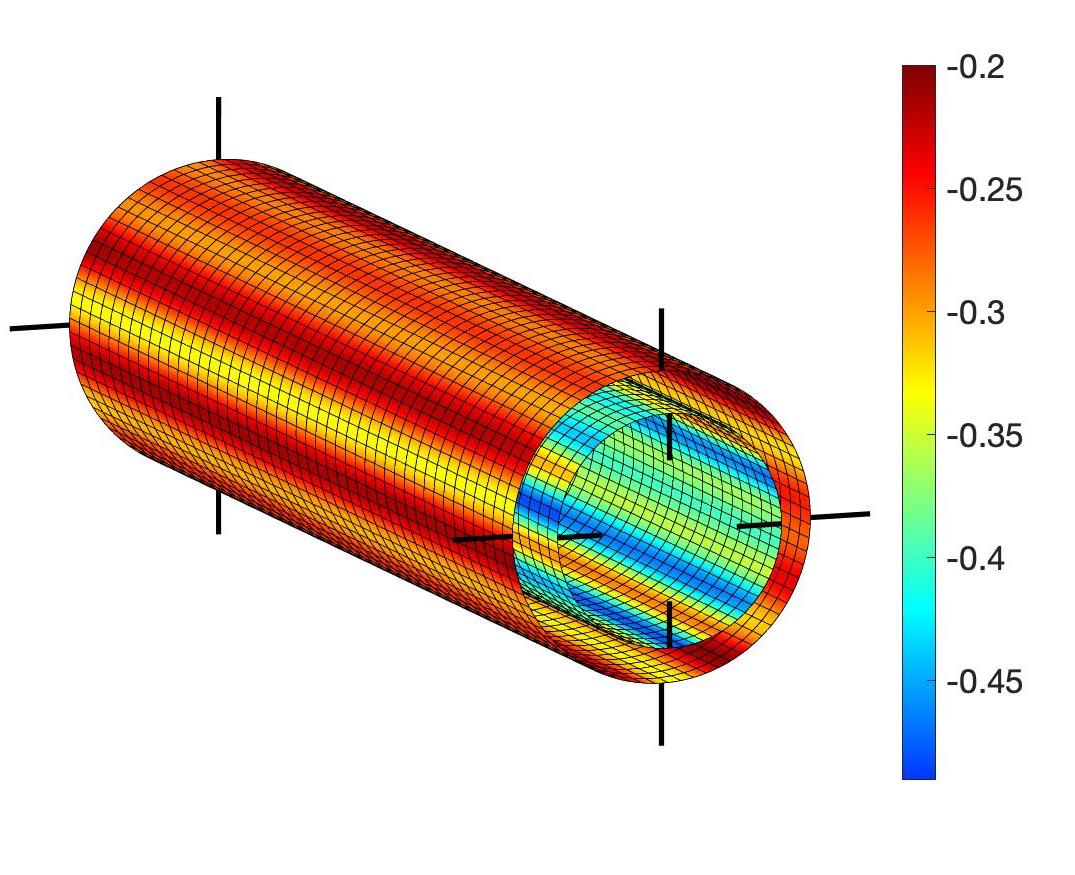}}
\put(-5.45,-.4){\includegraphics[height=25mm]{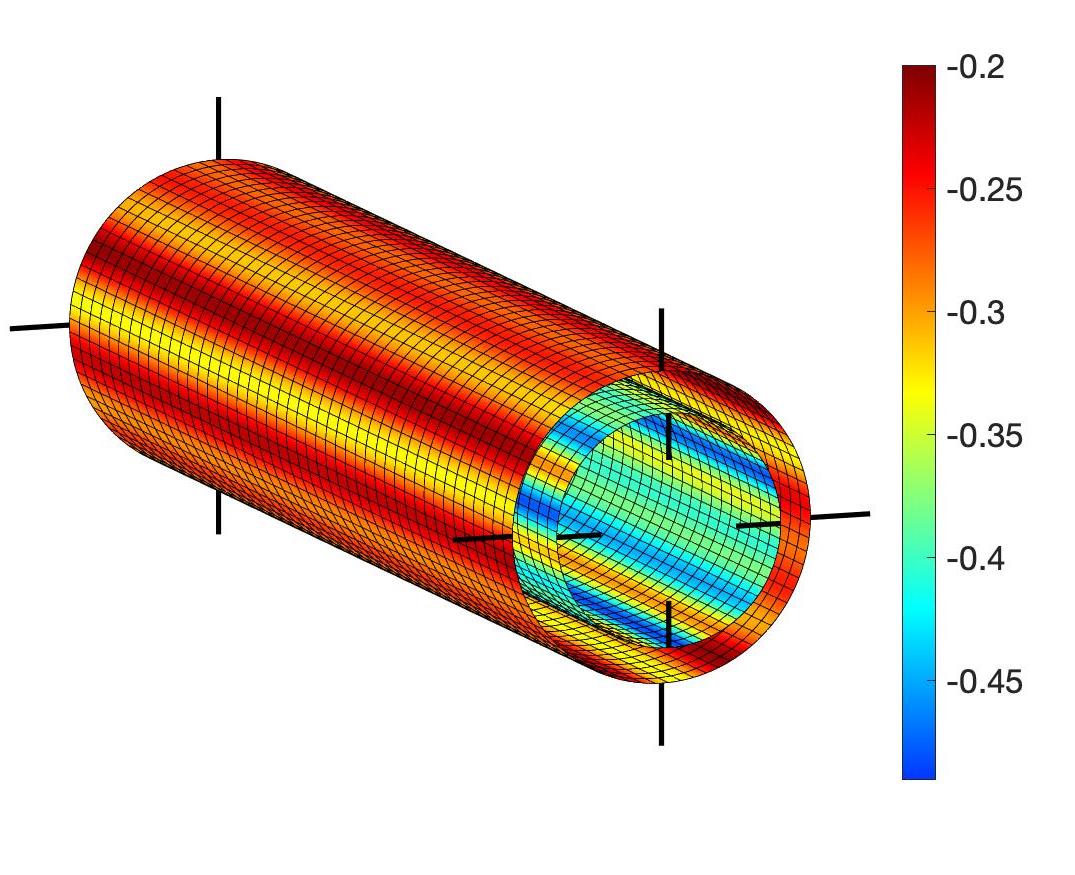}}
\put(-2.95,-.4){\includegraphics[height=25mm]{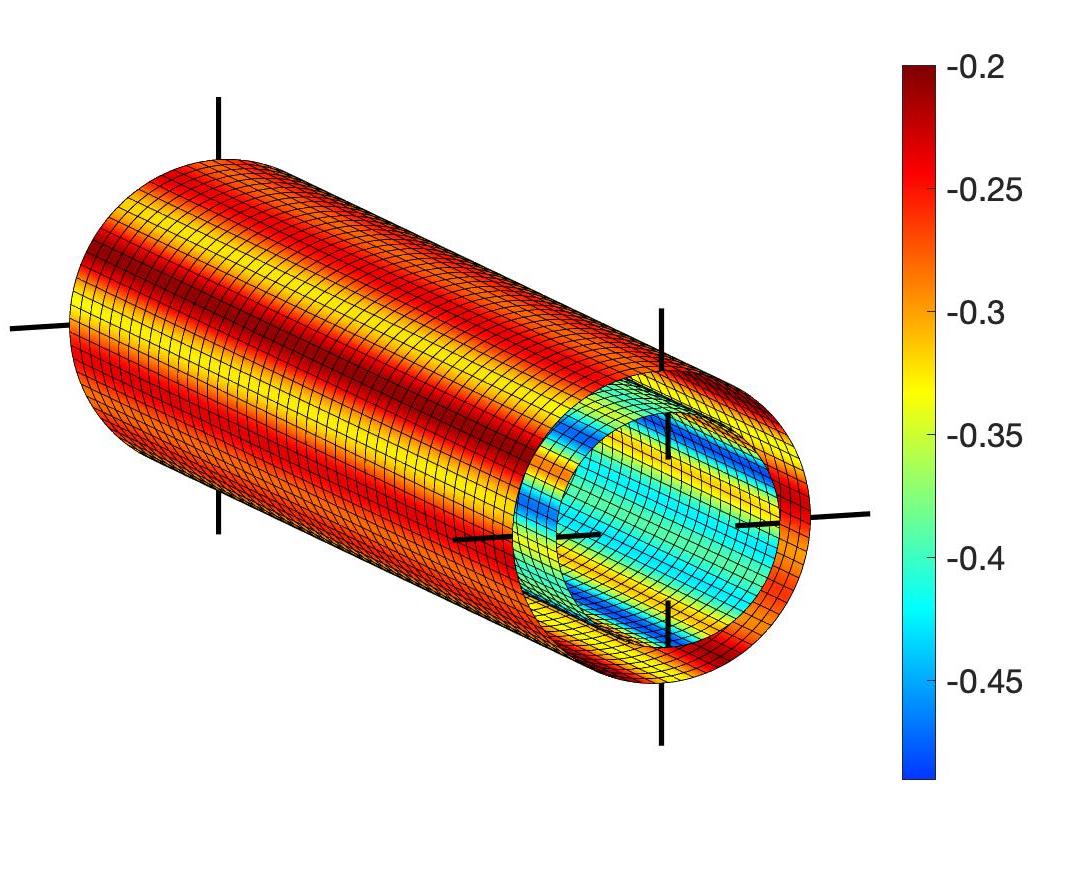}}
\put(-0.45,-.4){\includegraphics[height=25mm]{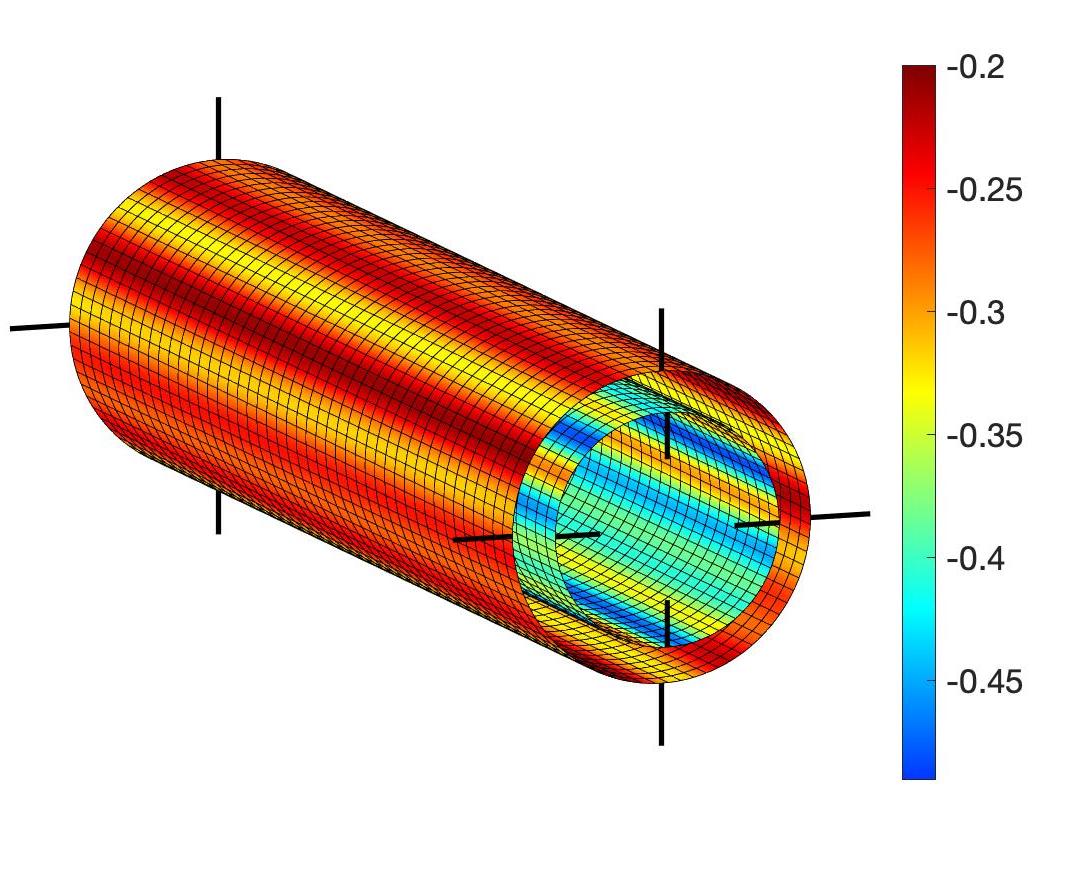}}
\put(2.05,-.4){\includegraphics[height=25mm]{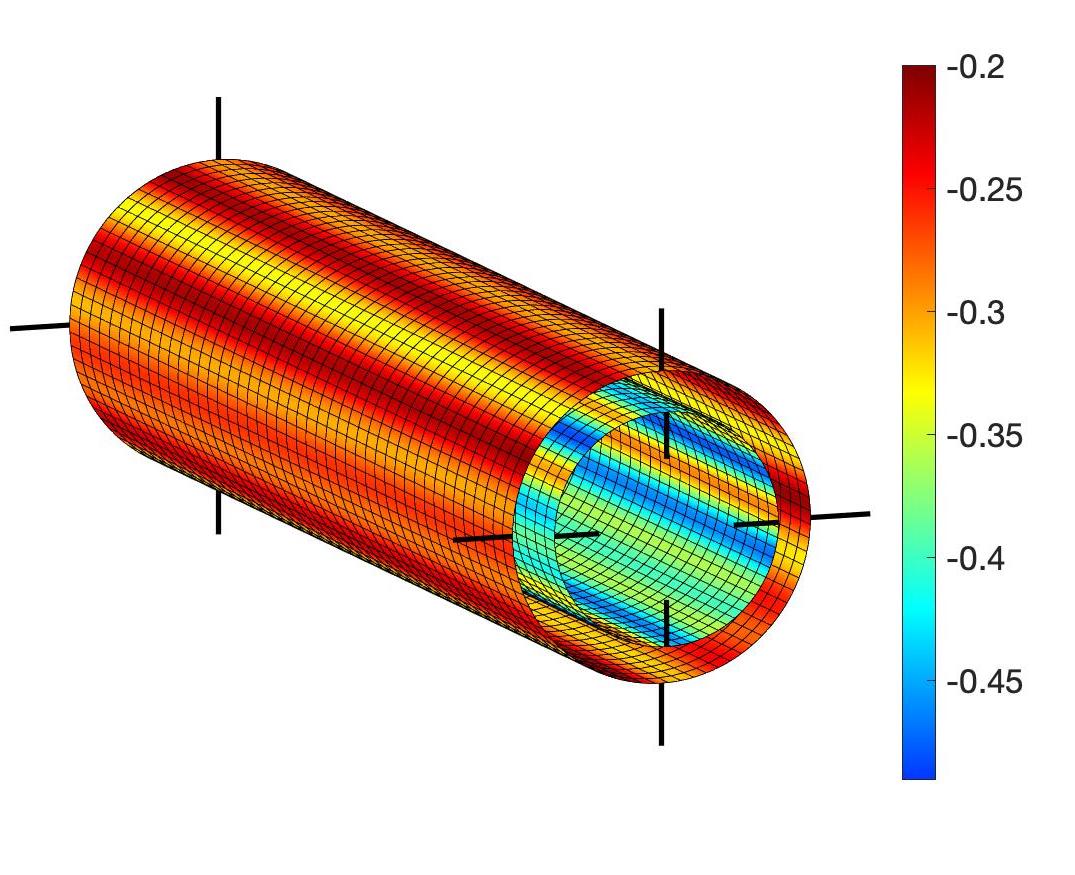}}
\put(4.55,-.4){\includegraphics[height=25mm]{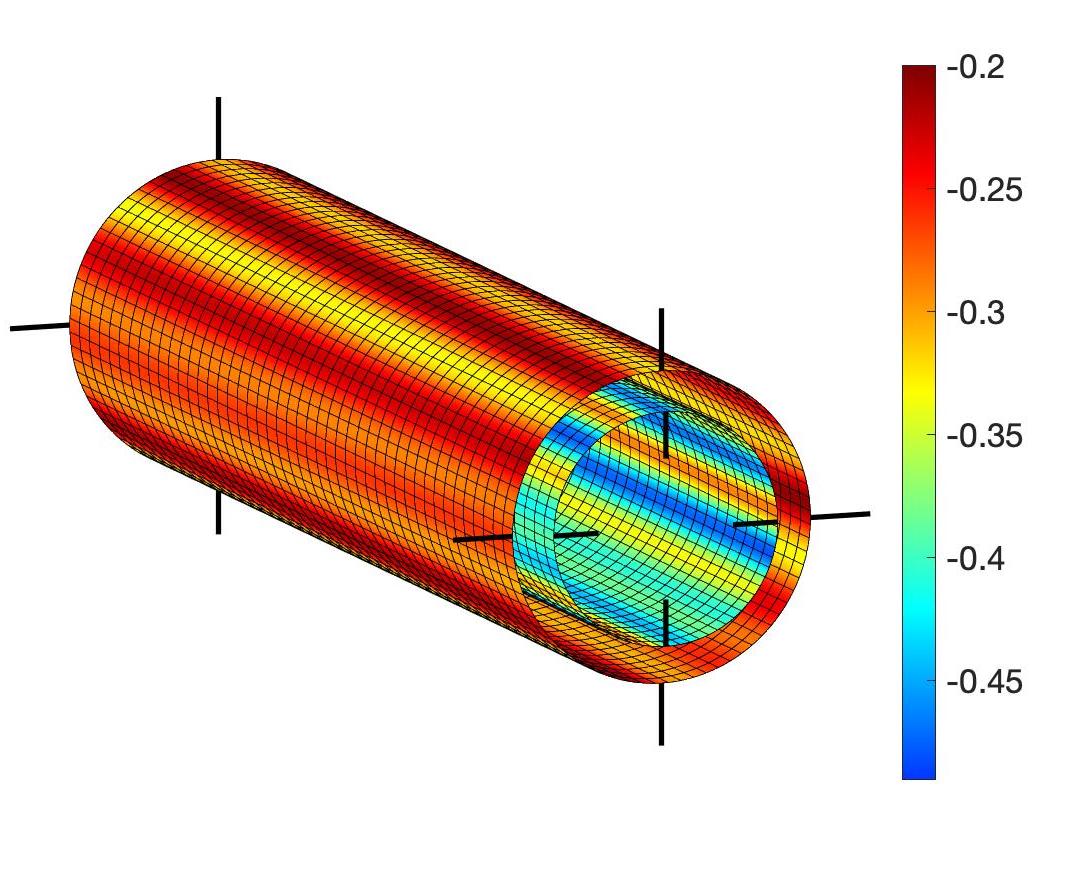}}
\put(7.1,-.55){\includegraphics[height=48mm]{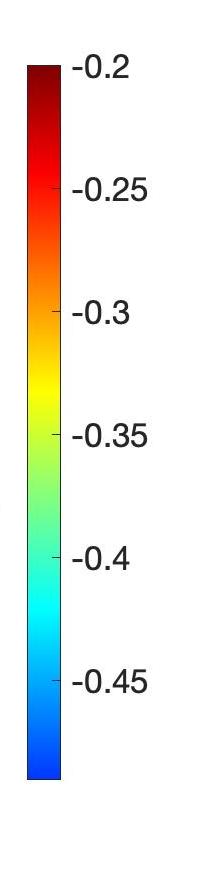}}
\put(-7.85,2.1){\scriptsize{0}}
\put(-5.35,2.1){\scriptsize{1}}
\put(-2.85,2.1){\scriptsize{2}}
\put(-0.35,2.1){\scriptsize{3}}
\put(2.15,2.1){\scriptsize{4}}
\put(4.65,2.1){\scriptsize{5}}
\put(4.65,0.){\scriptsize{6}}
\put(2.15,0.){\scriptsize{7}}
\put(-0.35,0.){\scriptsize{8}}
\put(-2.85,0.){\scriptsize{9}}
\put(-5.45,0.){\scriptsize{10}}
\put(-7.95,0.){\scriptsize{11}}
\end{picture}
\caption{Pull-out of CNT(21,9) from within CNT(28,12) (Case 3): Color plot of contact pressure $p$ in [GPa] at $u = [0,\,1,\,2,\, ...,\,11]\cdot\ell_3/200$ (clockwise, starting top left). 
}
\label{f:pullout3p}
\end{center}
\end{figure}

\begin{figure}[!htbp]
\begin{center} \unitlength1cm
\begin{picture}(0,4.2)
\put(-7.95,1.7){\includegraphics[height=25mm]{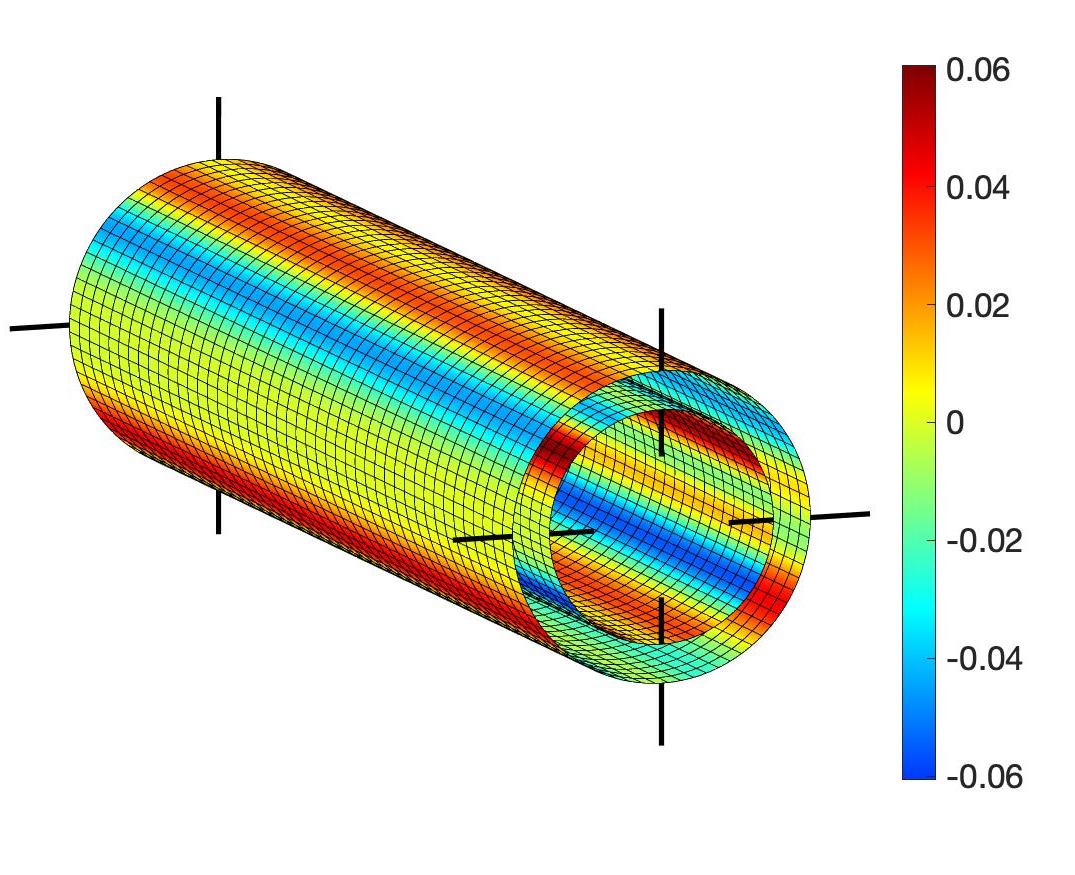}}
\put(-5.45,1.7){\includegraphics[height=25mm]{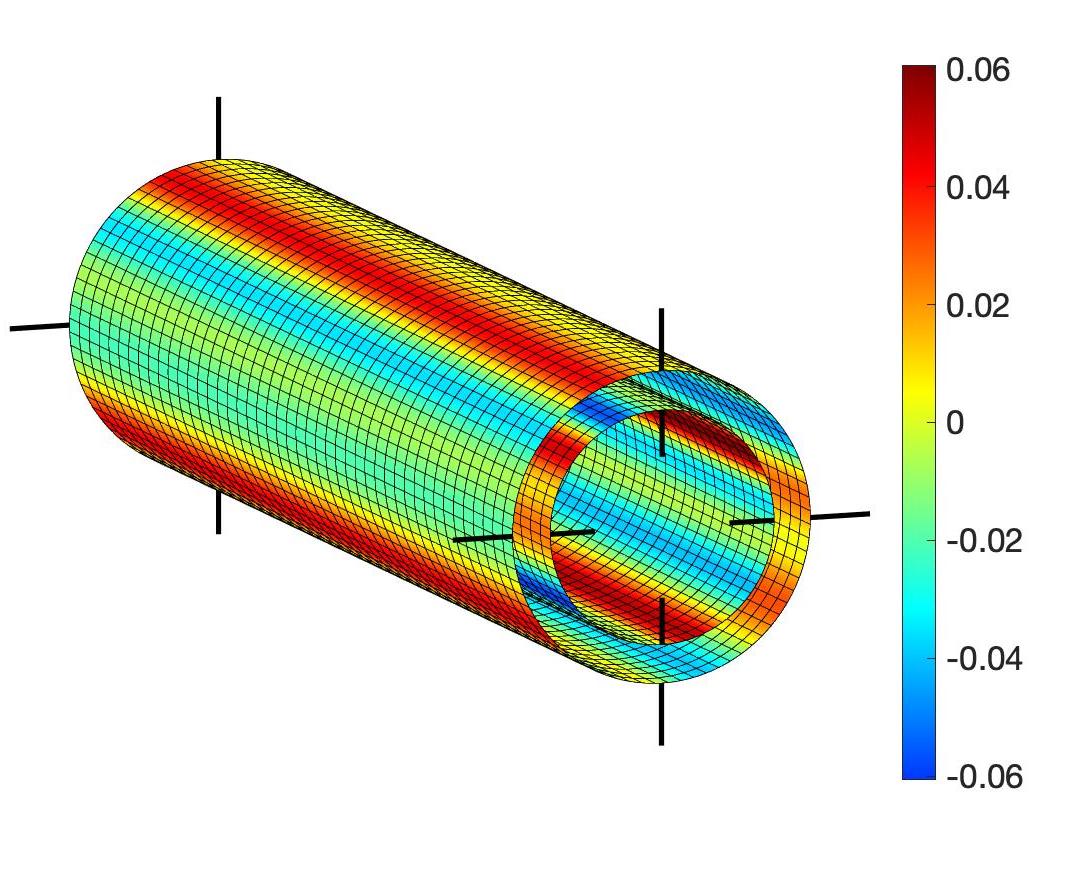}}
\put(-2.95,1.7){\includegraphics[height=25mm]{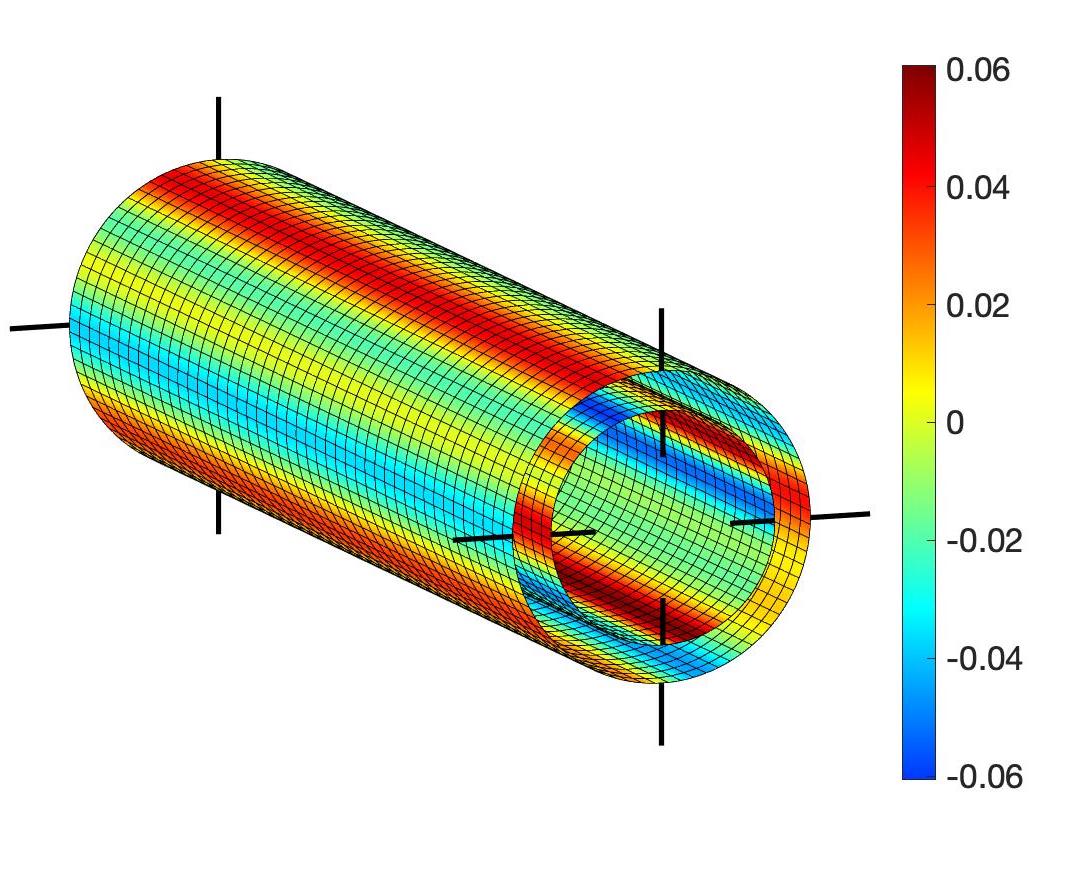}}
\put(-0.45,1.7){\includegraphics[height=25mm]{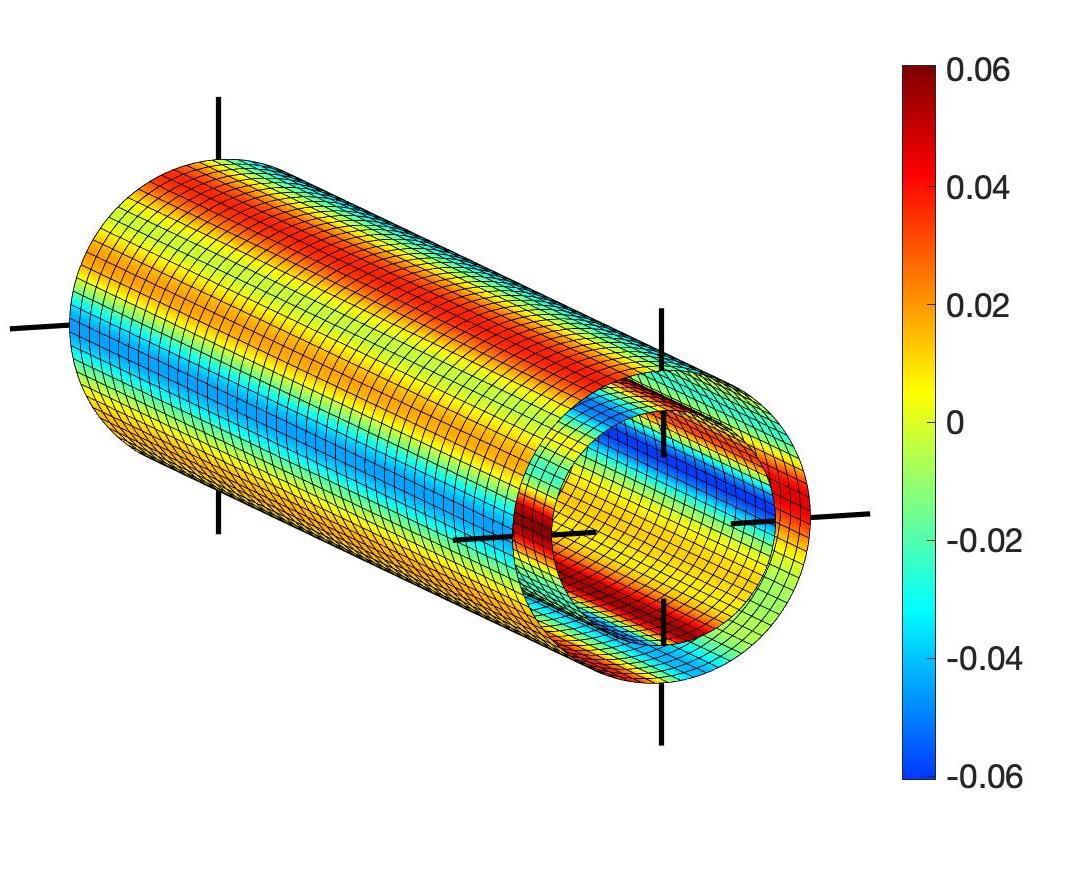}}
\put(2.05,1.7){\includegraphics[height=25mm]{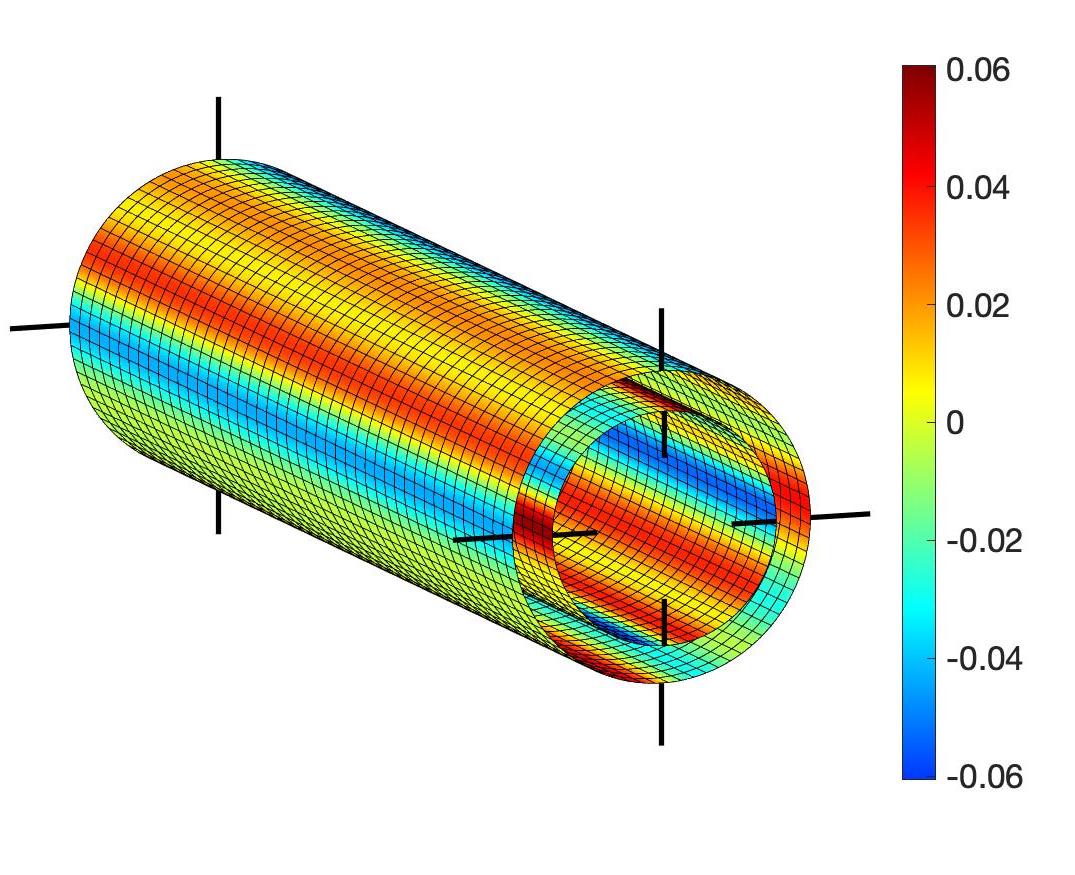}}
\put(4.55,1.7){\includegraphics[height=25mm]{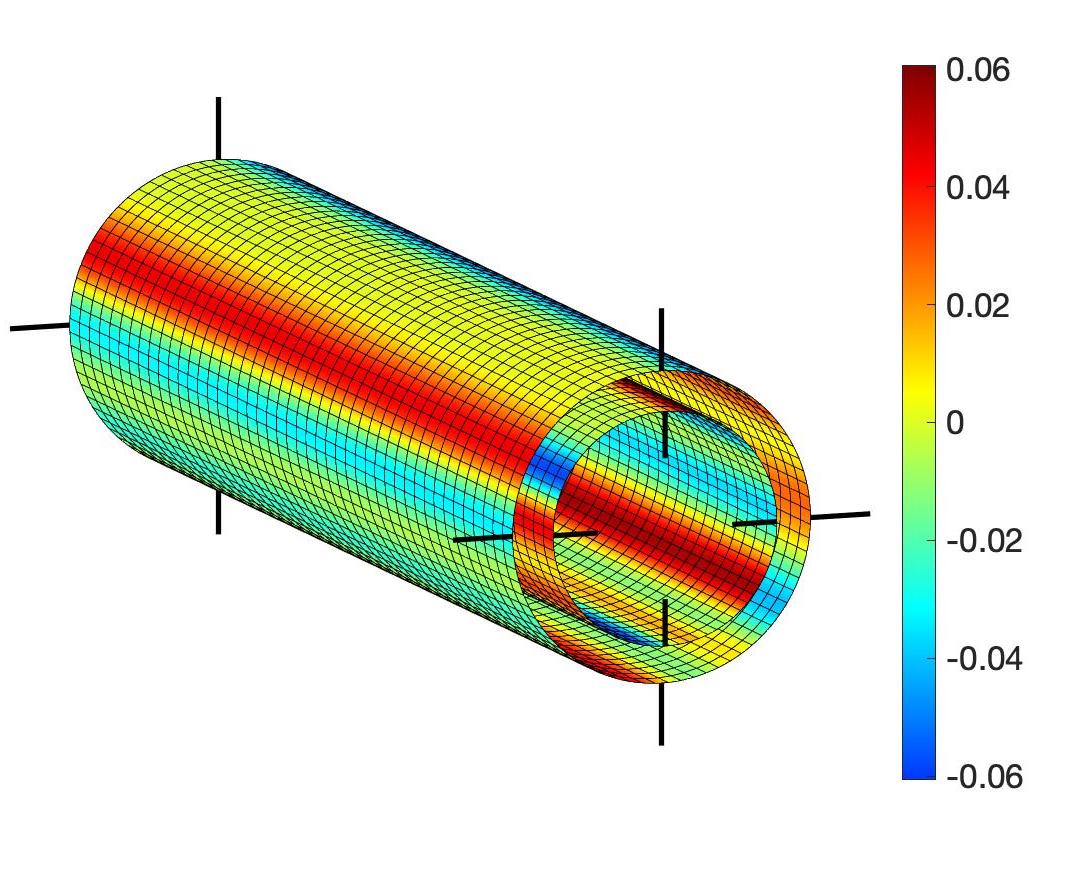}}
\put(-7.95,-.4){\includegraphics[height=25mm]{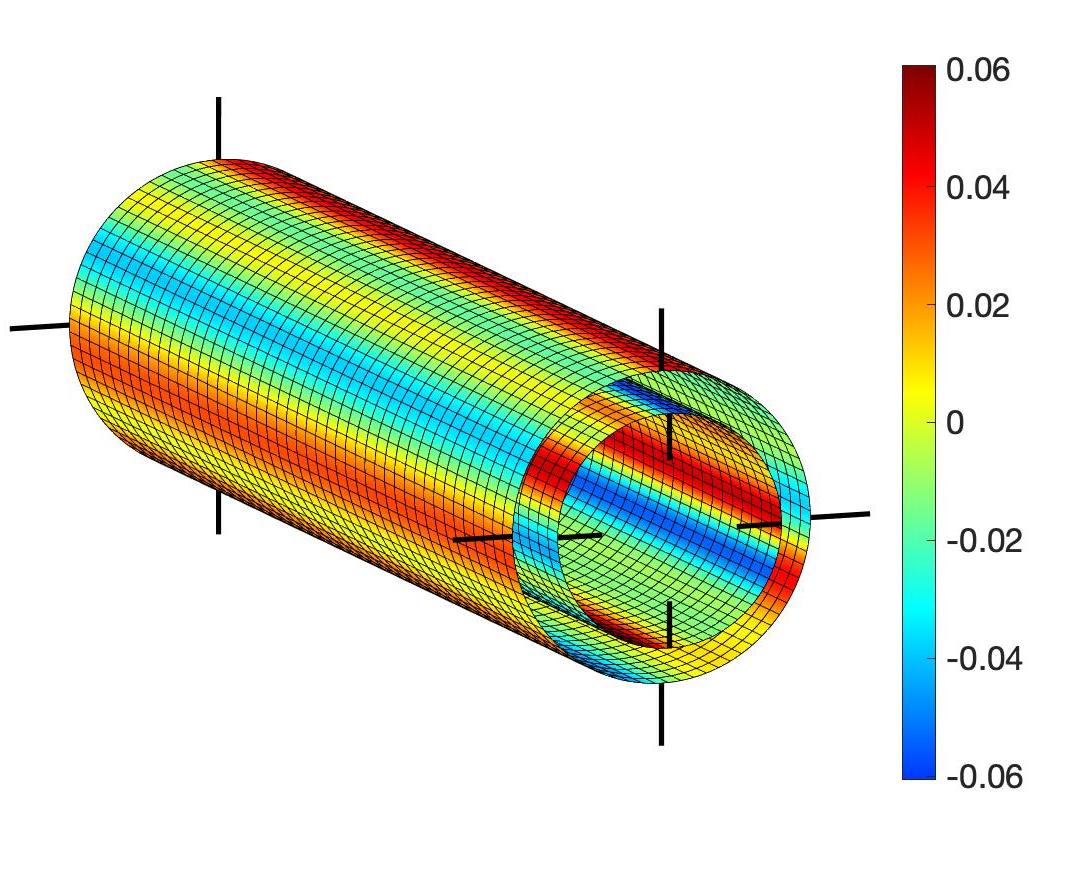}}
\put(-5.45,-.4){\includegraphics[height=25mm]{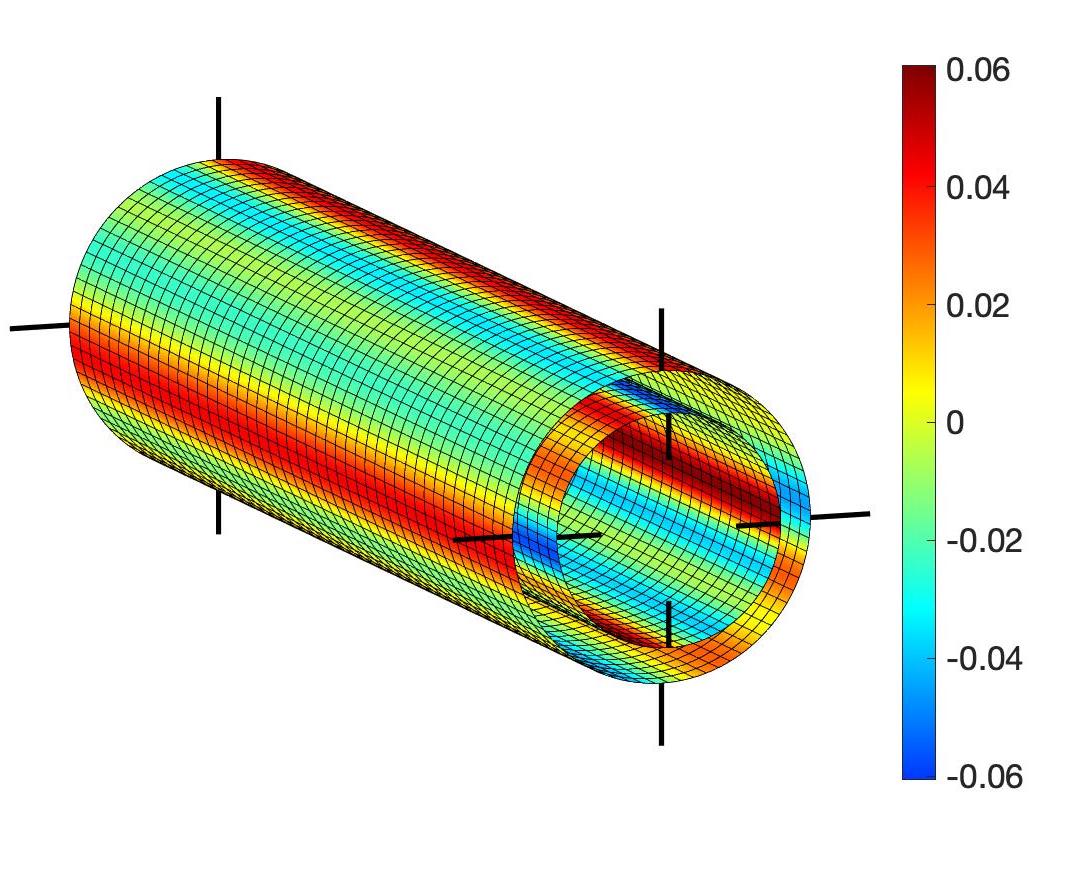}}
\put(-2.95,-.4){\includegraphics[height=25mm]{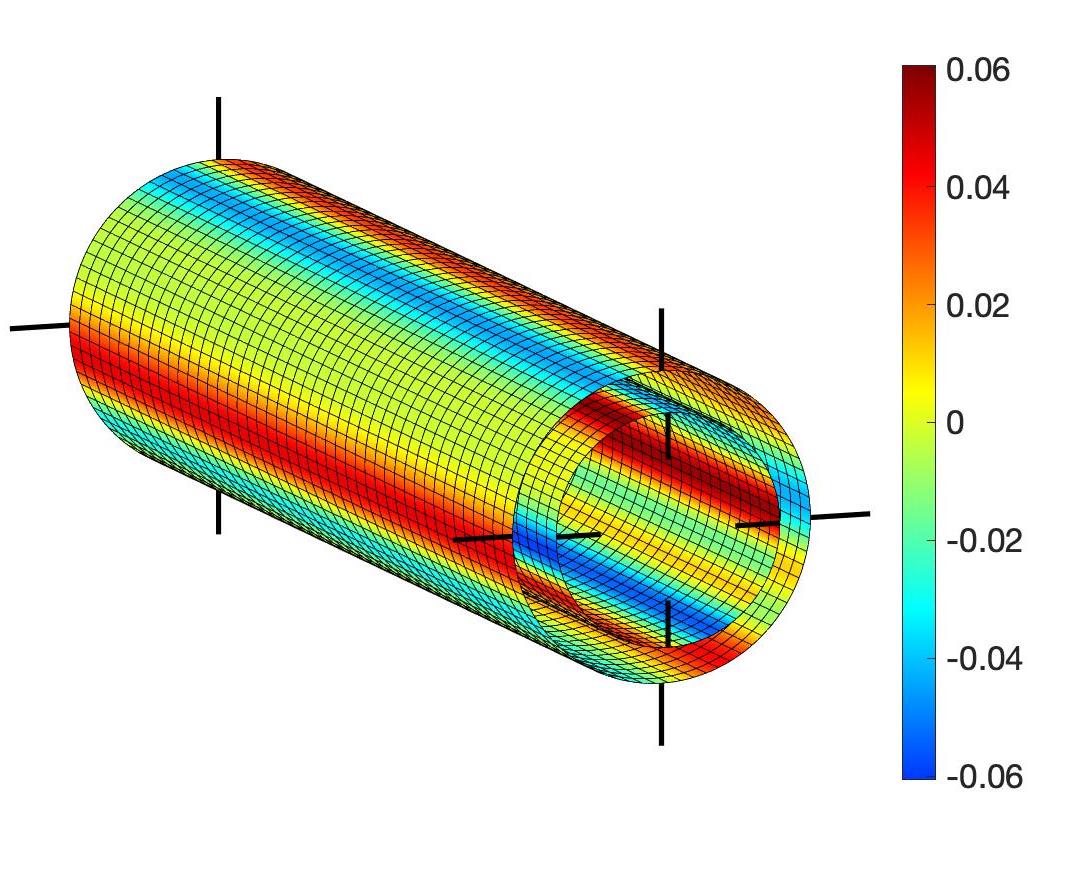}}
\put(-0.45,-.4){\includegraphics[height=25mm]{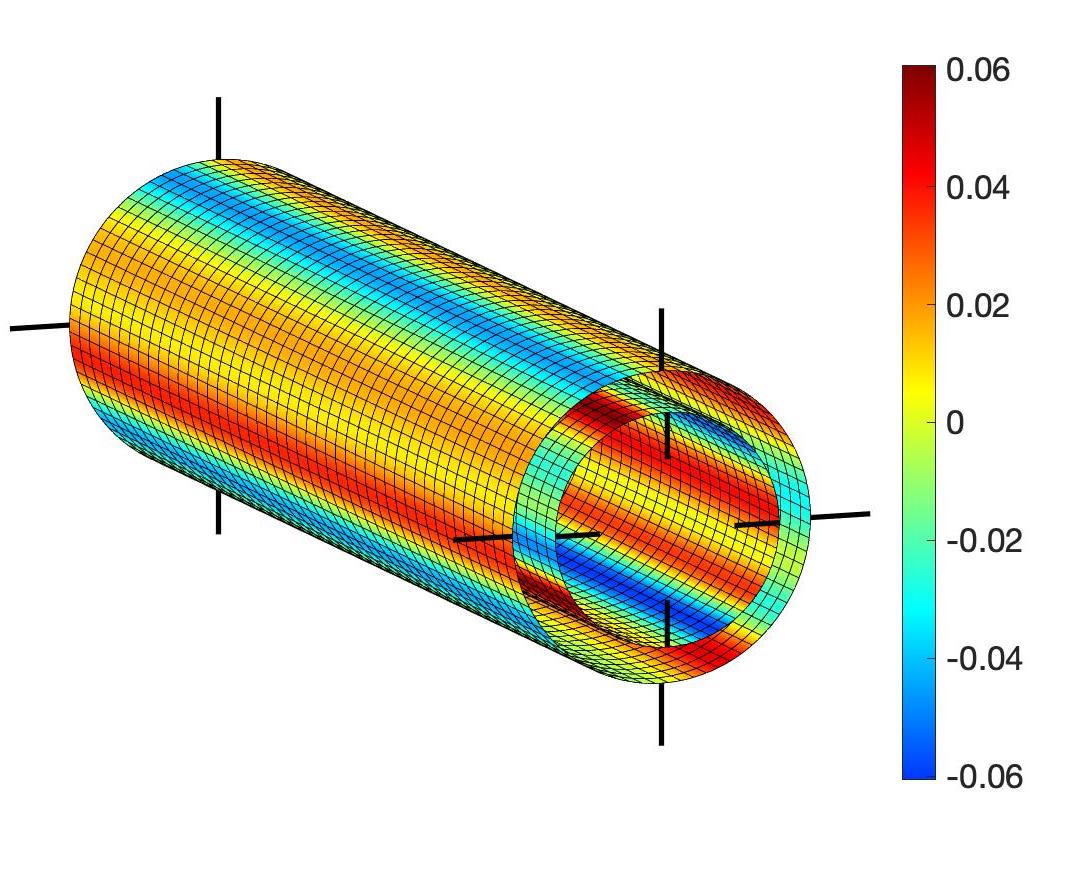}}
\put(2.05,-.4){\includegraphics[height=25mm]{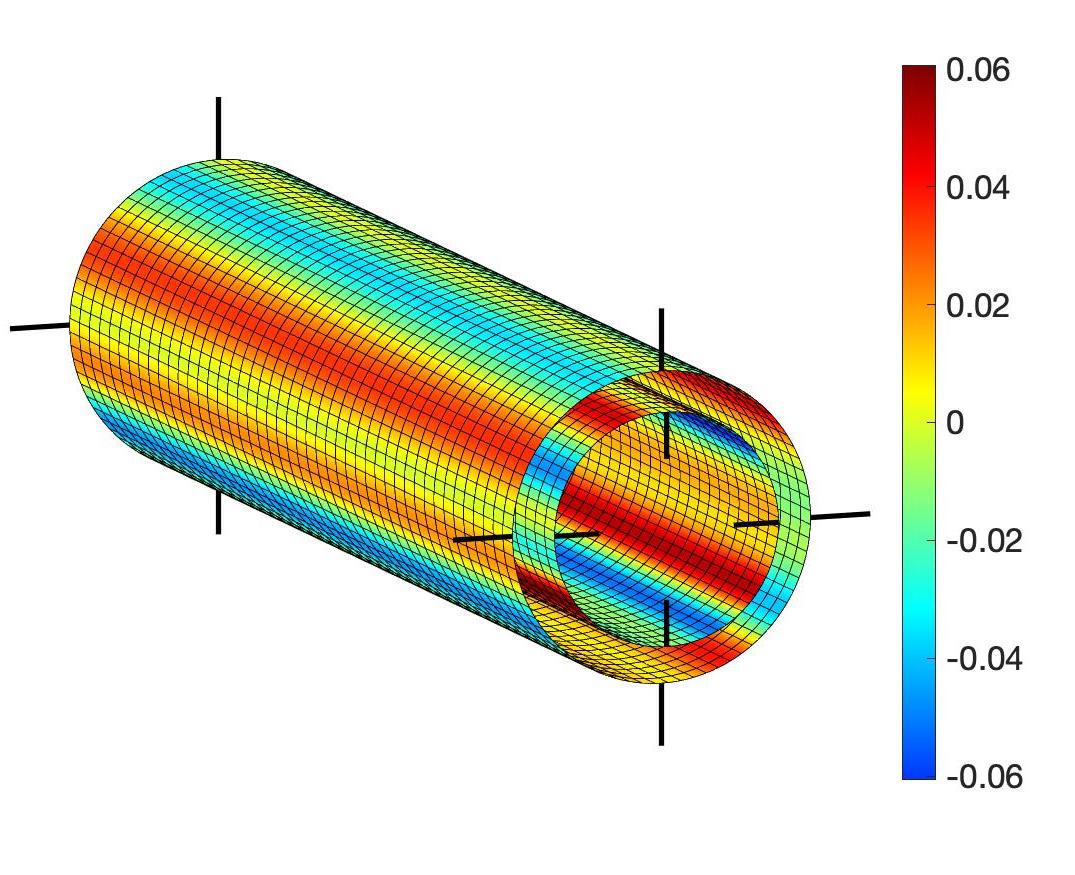}}
\put(4.55,-.4){\includegraphics[height=25mm]{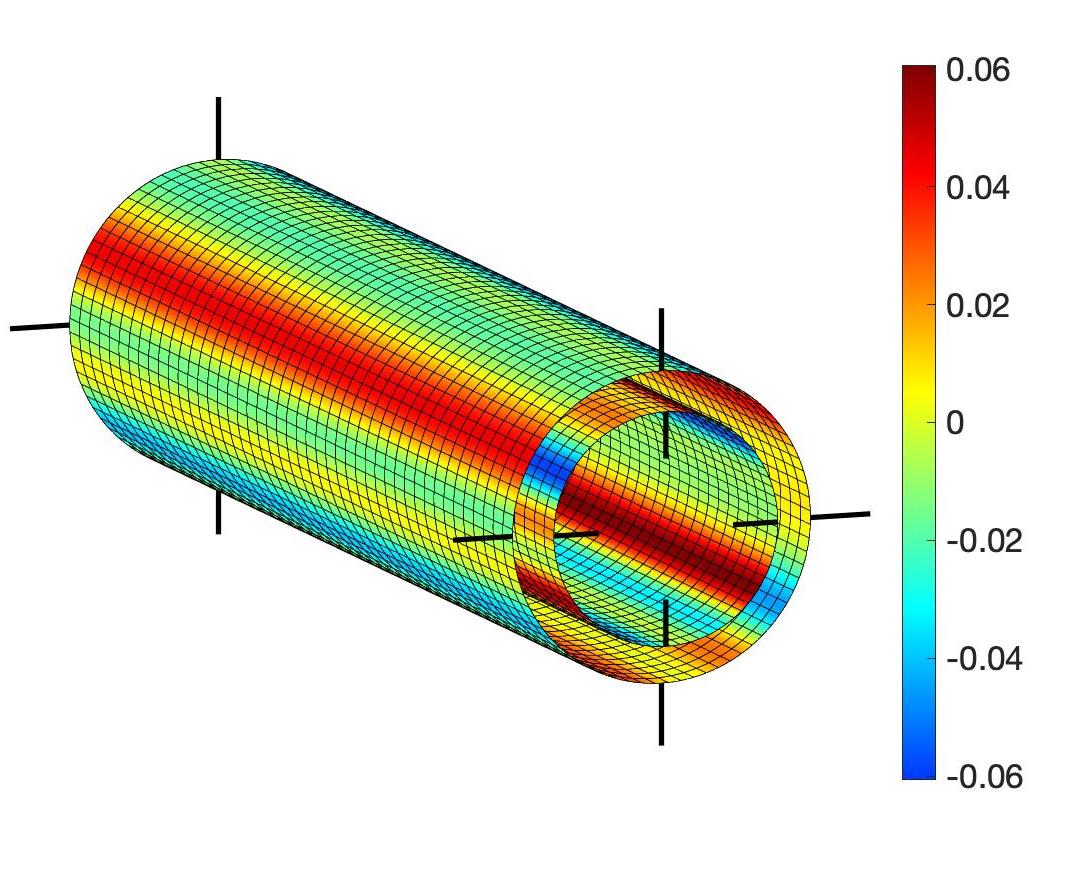}}
\put(7.1,-.55){\includegraphics[height=48mm]{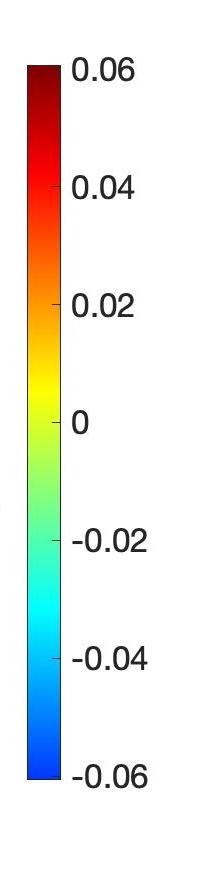}}
\put(-7.85,2.1){\scriptsize{0}}
\put(-5.35,2.1){\scriptsize{1}}
\put(-2.85,2.1){\scriptsize{2}}
\put(-0.35,2.1){\scriptsize{3}}
\put(2.15,2.1){\scriptsize{4}}
\put(4.65,2.1){\scriptsize{5}}
\put(4.65,0.){\scriptsize{6}}
\put(2.15,0.){\scriptsize{7}}
\put(-0.35,0.){\scriptsize{8}}
\put(-2.85,0.){\scriptsize{9}}
\put(-5.45,0.){\scriptsize{10}}
\put(-7.95,0.){\scriptsize{11}}
\end{picture}
\caption{Pull-out of CNT(21,9) from within CNT(28,12) (Case 3): Color plot of axial contact traction $t^1$ in [GPa] at 
$u = [0,\,1,\,2,\, ...,\,11]\cdot\ell_3/200$
(clockwise, starting top left).}
\label{f:pullout3t}
\end{center}
\end{figure}
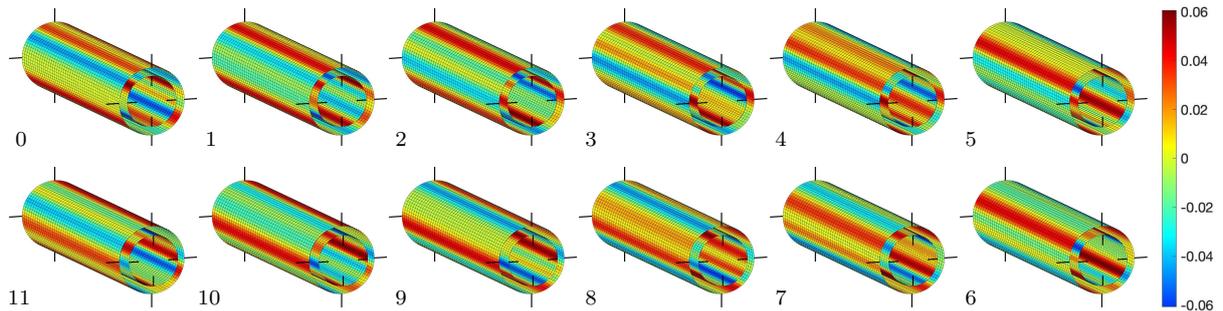

Figs.~\ref{f:pullout3p} and \ref{f:pullout3t} show the contact pressures and axial tractions for Case 3 determined from FE using the DFT parameters. Again, the contact pressure and axial tractions vary circumferentially during axial sliding. 
Case 3 is in a state of attraction; therefore, there is adhesion between the tubes, leading to the axial strains $\eps_\mathrm{out} = 2.107 \cdot 10^{-4}$ and $\eps_\mathrm{in} = -1.982 \cdot 10^{-4}$ in the initially relaxed configuration for the outer and inner tube, respectively. 
The periodicity of the pull-out force determined from FE is $\ell_3 = 3\sqrt{79}a_{\text{cc}}$, which is exactly the same as that obtained from the MD and theoretical calculations (see Fig.~\ref{cnt_pull_force}c and Sec.~\ref{s:go}).

The maximum amplitude of the pull-out forces determined from MD and FE using DFT and REBO+LJ parameters are, respectively, $4.7\cdot10^{-4}$, $1.4\cdot10^{-7}$, and $8.4\cdot10^{-8}$ nN. 
The maximum pull-out forces obtained from FE differ with respect to MD by $0.4739\cdot 10^{-3}$ and $0.4740\cdot 10^{-3}$ nN for the two material parameters considered, see Tab.~\ref{cnt_pull_out_d}. These small differences can be attributed to several small discrepancies between MD and FE. One is the rigidity assumption of the contact master surface, which introduces the relative difference of $\approx 3.35\cdot 10^{-4}$ compared to the maximum MD force, which seems insignificant. A second are the differences between the chosen continuum ansatz \eqref{e:Psif} and the MD data. But also those are very small as seen in Sec.~\ref{sec:int_beh}. A third are the boundary conditions: In the FE simulations, the rotation of the CNT about its axis is constrained, which is a displacement boundary condition. On the other hand, due to the limitation in LAMMPS, this boundary condition cannot be applied directly, and is mimicked by applying torsional springs to the circumferential atoms, which corresponds to a force boundary condition. As a result, the reaction torques during pull-out cause small rotational oscillations leading to fluctuations in the pull-out forces. The MD data plotted in Fig.~\ref{cnt_pull_force} is filtered using a moving average approach with window/sample length 5, 20, and 100 for cases 1, 2 and 3, respectively. These fluctuations are minute for Case 1. However, at smaller pull-out forces, these fluctuations become much more significant, particularly for Case 3. Thus, the differences in the pull-out forces determined from FE and MD may be due to the rotational oscillations in the MD simulations. 

\subsubsection{Pull-out summary}
For all three cases, the FE simulations predict pull-out behavior very similar to that of MD. The amplitudes are found to be sensitive to the continuum material properties used, as seen in Fig.~\ref{cnt_pull_force} and Tab.~\ref{cnt_pull_out_d}. Those material parameters are based on decoupled membrane and bending models. In MD simulations, on the other hand, the bending stiffness of graphene is calculated by computing the potential energies of relaxed CNTs with respect to the ground-state potential energy of graphene. In such cases, the energy of relaxed CNTs is not just associated with the curvature of the CNTs but also the membrane strain energy.

In contrast to MD and FE simulations, the analytical results shown in Fig.~\ref{cnt_pull_force} and derived in Appendix~\ref{s_analytical_exp} are for rigid CNTs. Therefore, the differences in the pull-out forces determined from the FE/MD and analytical expression show the contribution of the elastic nature of CNTs. Case 1 is the only case where a pull-out force is generated independent of the CNT deformation (i.e.~corresponding to the analytical result). In all other cases the pull-out force is a higher-order effect coming solely from the CNT deformation. This is seen through the analytical solution as it does not capture this higher-order effect. 
\subsection{CNT within CNT twisting} 
Finally, the twisting results are presented and discussed. As before, the FE results are compared to analytical results {(derived in Appendix \ref{s_analytical_exp})} and MD results.

\subsubsection{Rotational symmetry}
Similar to the axial symmetry, the rotational symmetry of DWCNTs also depends on the chiral indices of each CNT. According to the elementary number theory, the inter-wall interaction energy due to inner CNT rotation and thus their resisting torque has periodicity of $\text{GCD}\,(n_1,n_2)/(n_1n_2) \times 360^{\circ}$, where the inner and outer CNTs have $n_1$-fold and $n_2$-fold symmetries, respectively \citep{Merkle_1993}.  For Case 1, the inner CNT thus has 26-fold rotational symmetry, while the outer one has 35. Therefore, the resisting torque must have a $0.3956^{\circ}$ periodicity. Similarly, for Cases 2 and 3 the inner CNTs have 15- and 3-fold rotational symmetries, while outer ones have 20- and 4-fold, respectively. Therefore, the resisting torque in Cases 2 and 3 must have a rotational periodicity of $6^{\circ}$ and $30^{\circ}$, respectively.  

\subsubsection{Twisting results}   
\begin{figure}[!htbp]
\begin{center} \unitlength1cm
\begin{picture}(0,9.5)
\put(-7.8,5){\includegraphics[height=48mm]{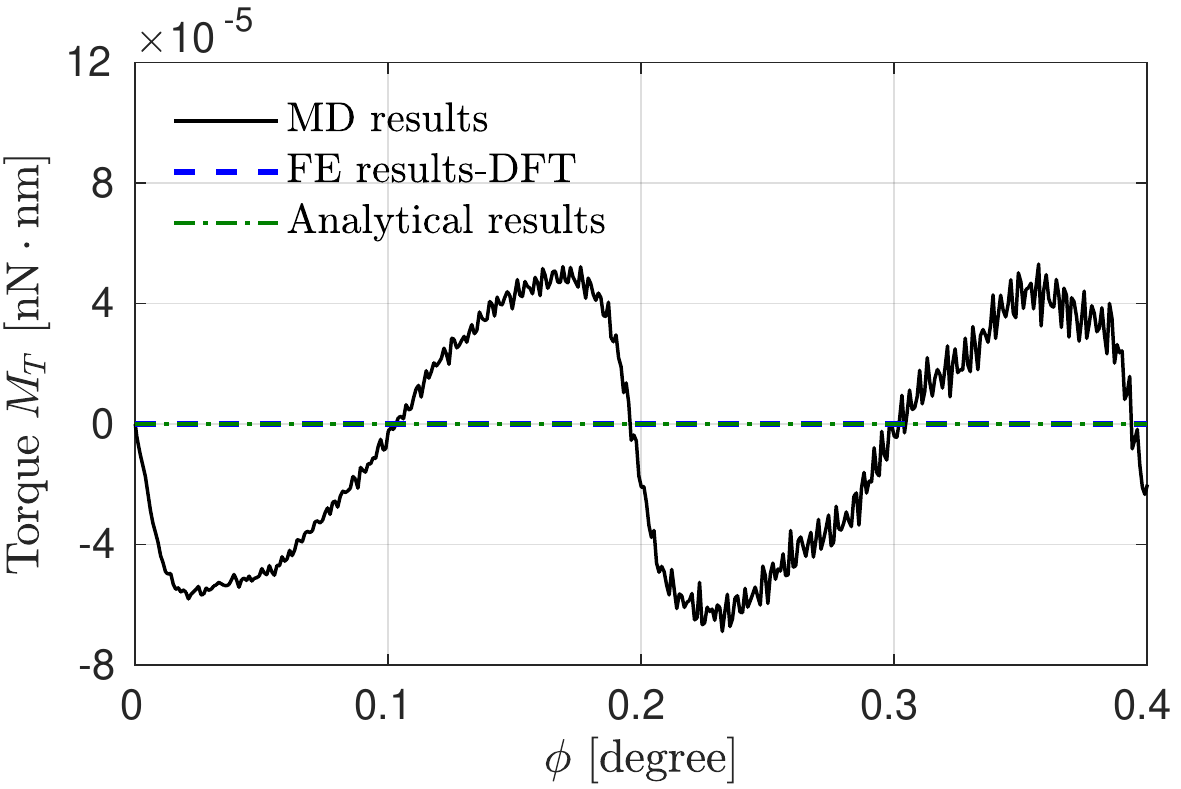}}
\put(0.2,5){\includegraphics[height=48mm]{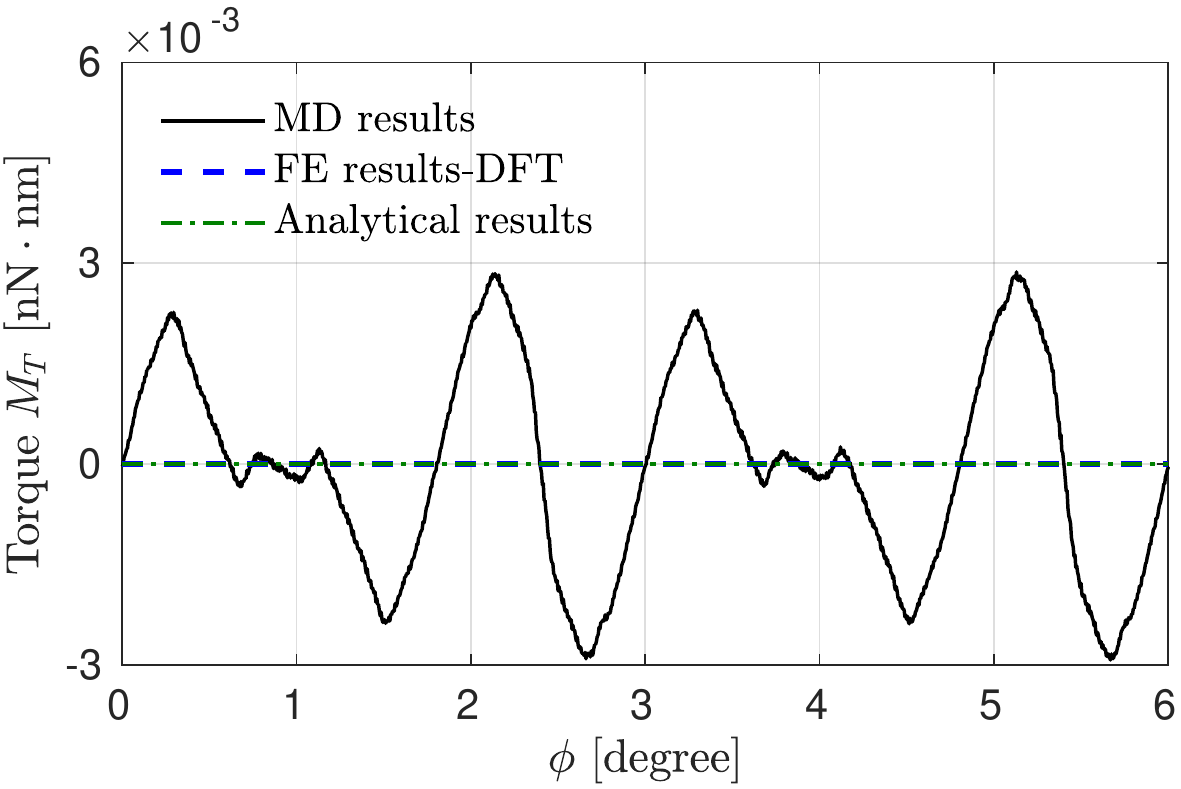}}
\put(-4.5,0){\includegraphics[height=48mm]{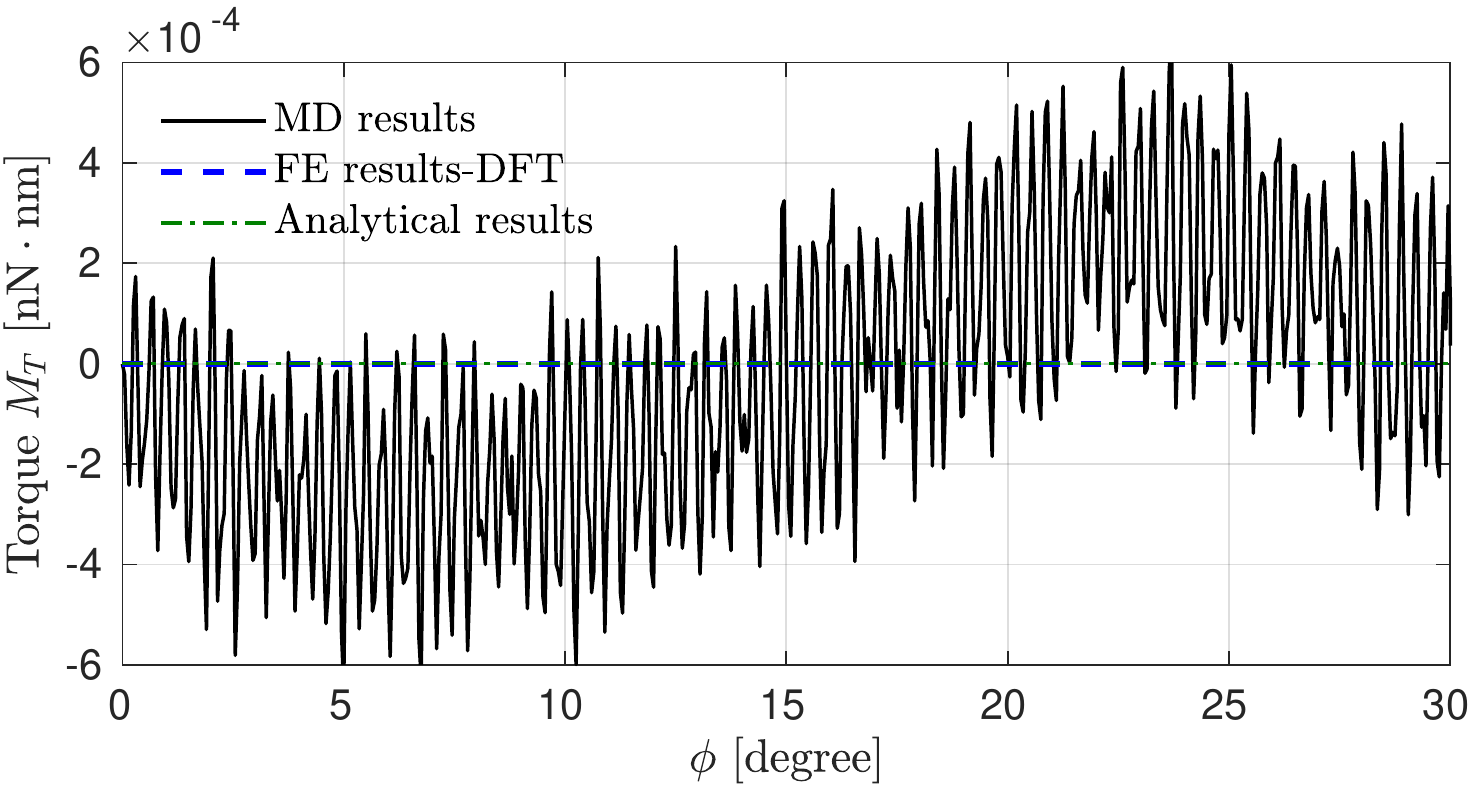}}
\put(-7.7,5){(a)}
\put(0.3,5){(b)}
\put(-3.5,0){(c)}
\end{picture}
\caption{CNT within CNT twisting: Variation of the resisting torque for (a)~Case 1, (b)~Case 2, and (c)~Case 3 with rotation angle $\phi$. } 
\label{cnt_twist_force}
\end{center}
\end{figure}
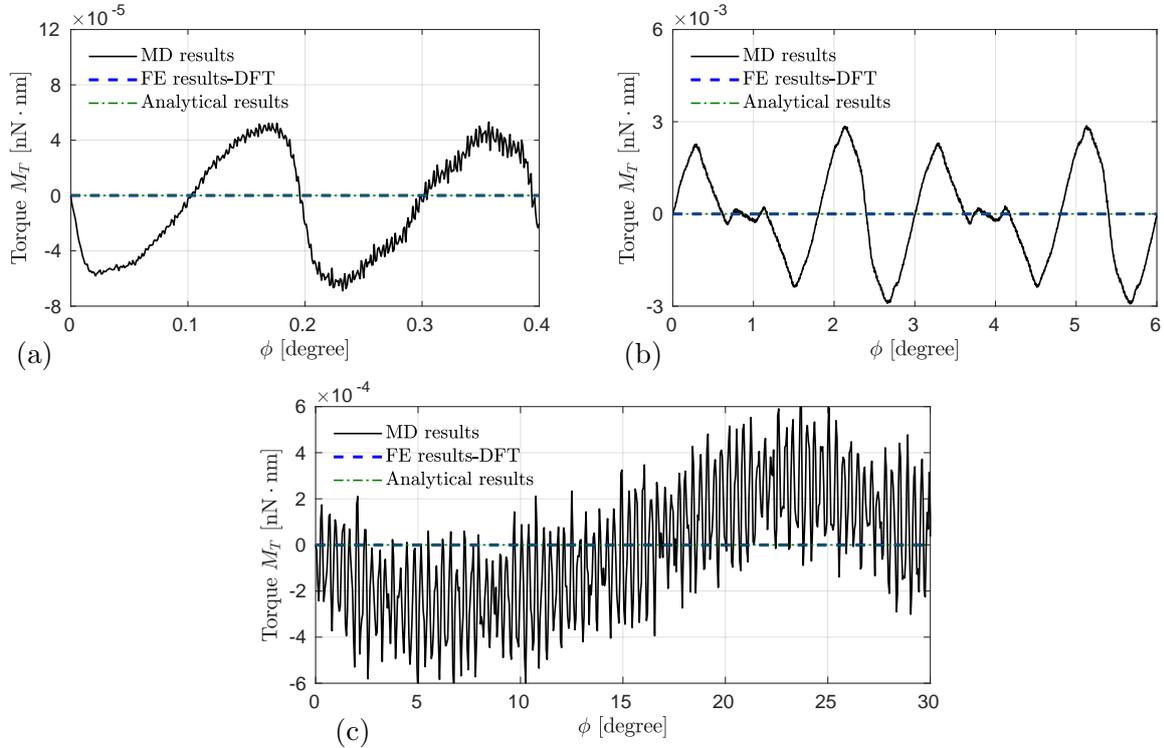

The comparison of the resisting torque as a function of the rotation angle of the inner CNT between FE, MD, and analytical expression is shown in Fig.~\ref{cnt_twist_force}.
The FE and analytical torques are zero, while for MD, the amplitude of the torques, defined as $M_\mrT^{\text{max}}-M_\mrT^{\text{min}}$, are $1.7051 \cdot 10^{-4}$, $6.6415 \cdot 10^{-3}$, and $1.3077 \cdot 10^{-3}$ $\text{nN}\cdot\text{nm}$ for Cases 1, 2, and 3, respectively. As in the case of pull-out, the periodicity and amplitude of the resisting torques of CNT within CNT twisting depend on the chirality of the CNTs. The periodicity of the resisting torque determined from the MD simulations agrees well with the theoretical calculations. 

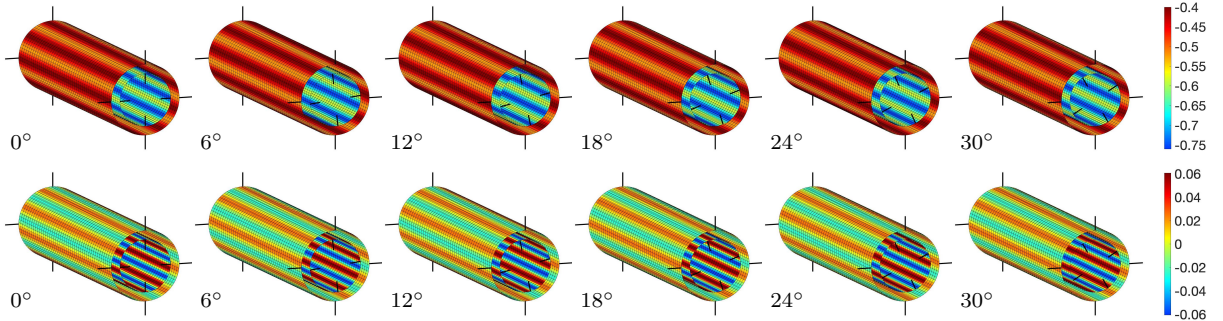
\begin{figure}[h]
\begin{center} \unitlength1cm
\begin{picture}(0,4.2)
\put(-7.95,1.75){\includegraphics[height=25mm]{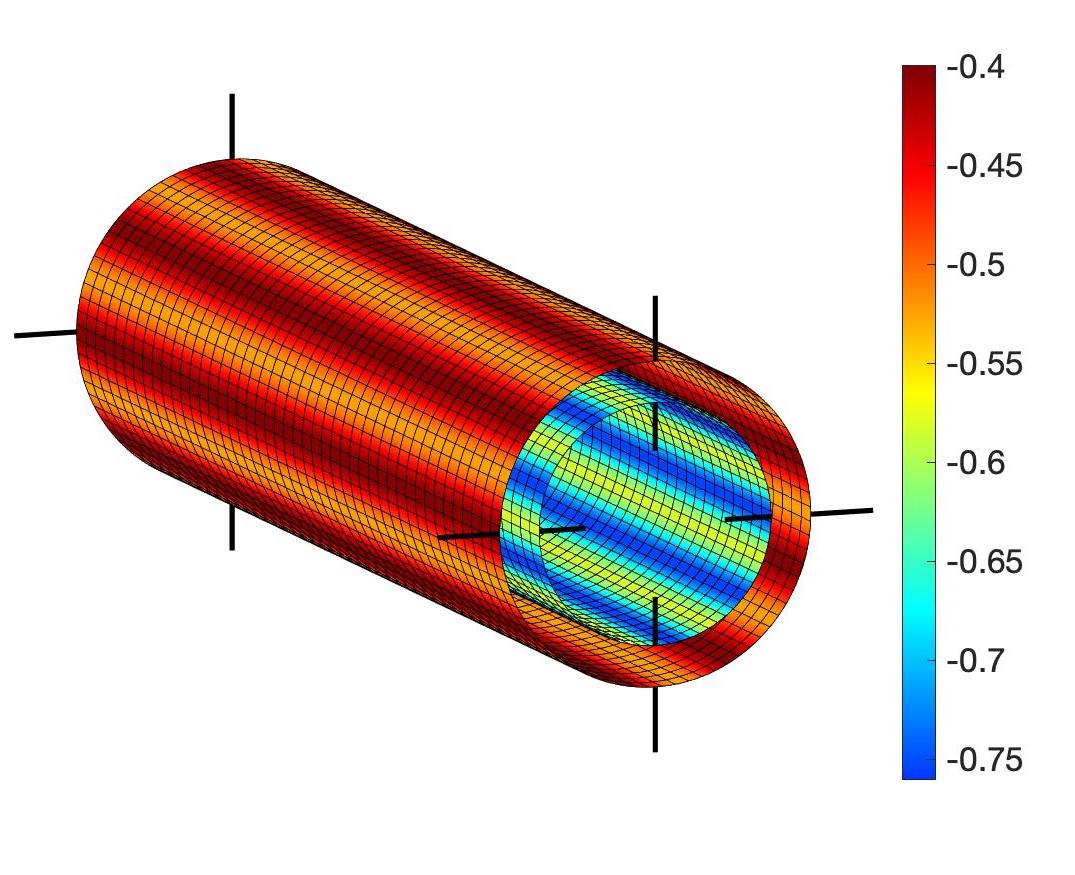}}
\put(-5.45,1.75){\includegraphics[height=25mm]{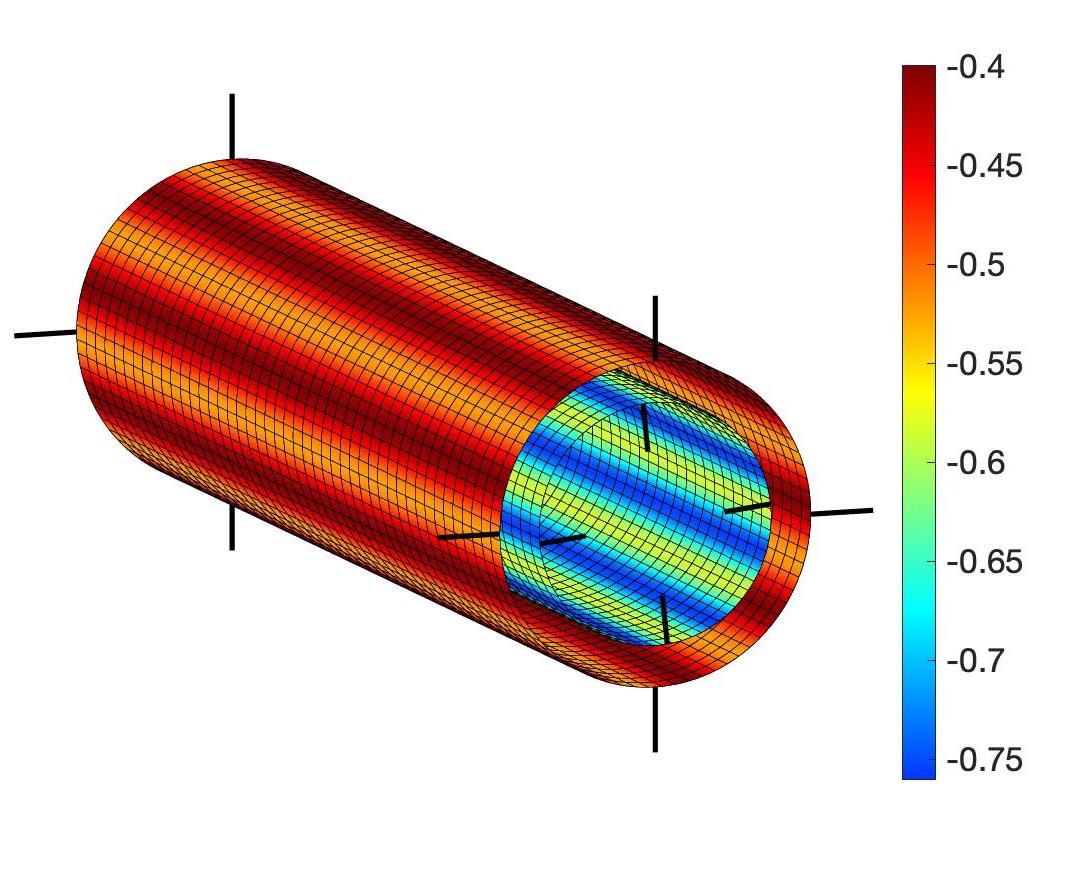}}
\put(-2.95,1.75){\includegraphics[height=25mm]{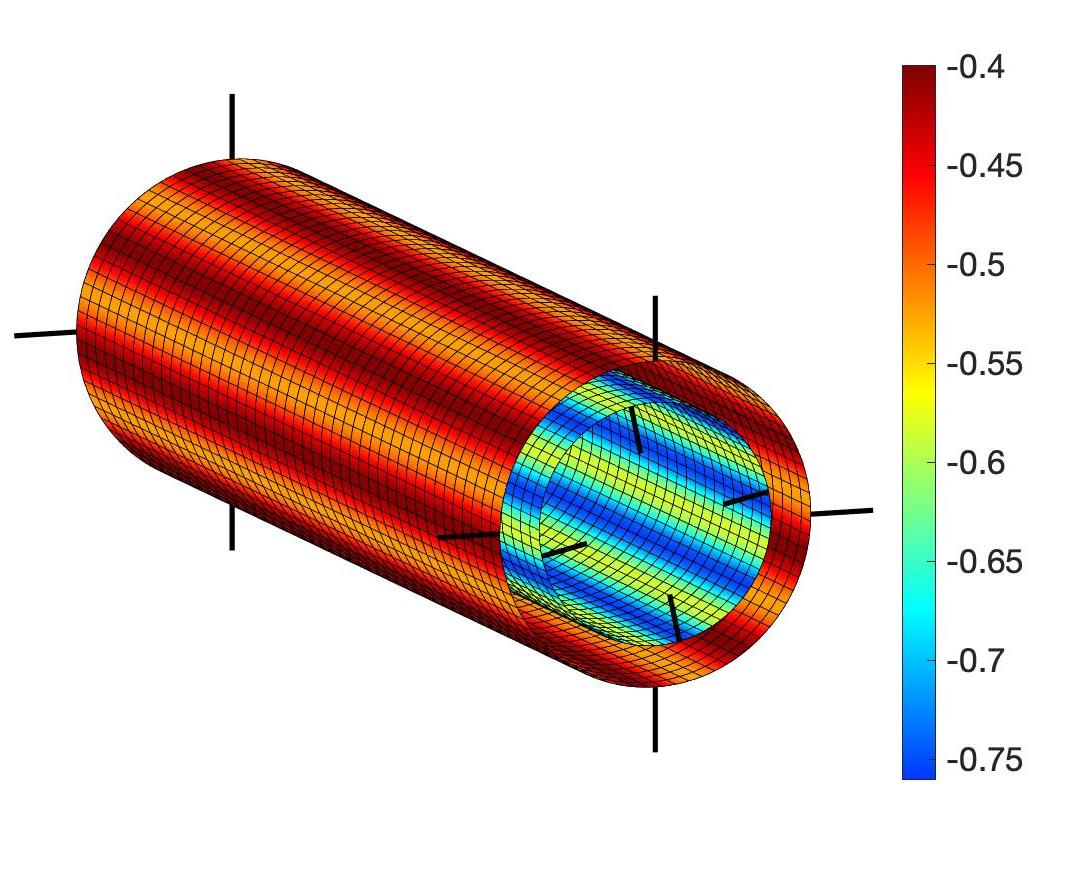}}
\put(-0.45,1.75){\includegraphics[height=25mm]{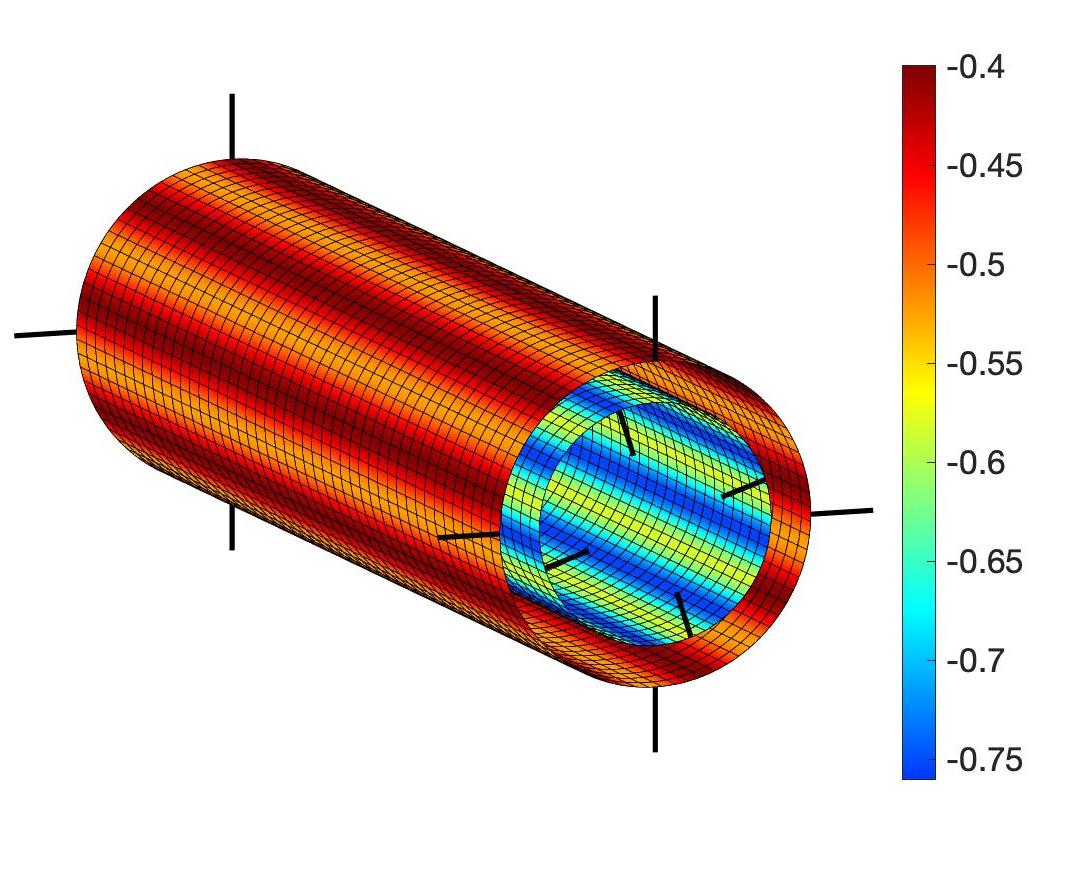}}
\put(2.05,1.75){\includegraphics[height=25mm]{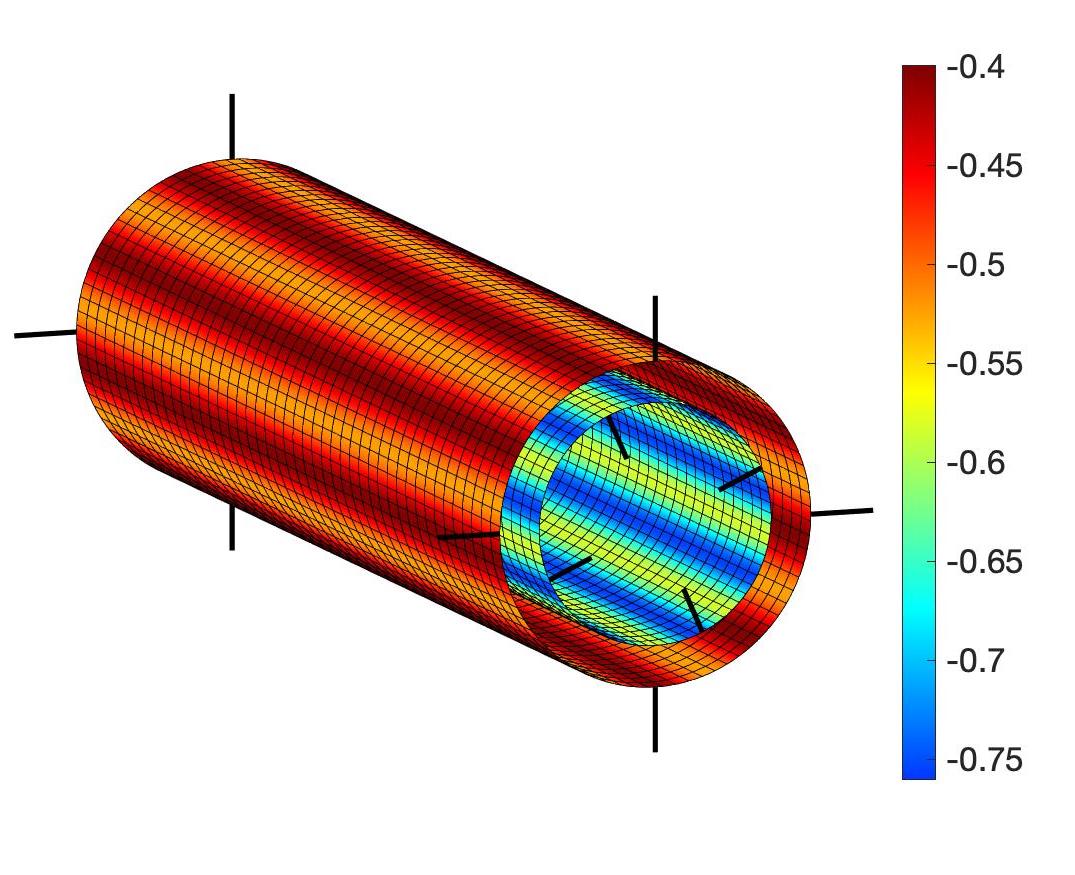}}
\put(4.55,1.75){\includegraphics[height=25mm]{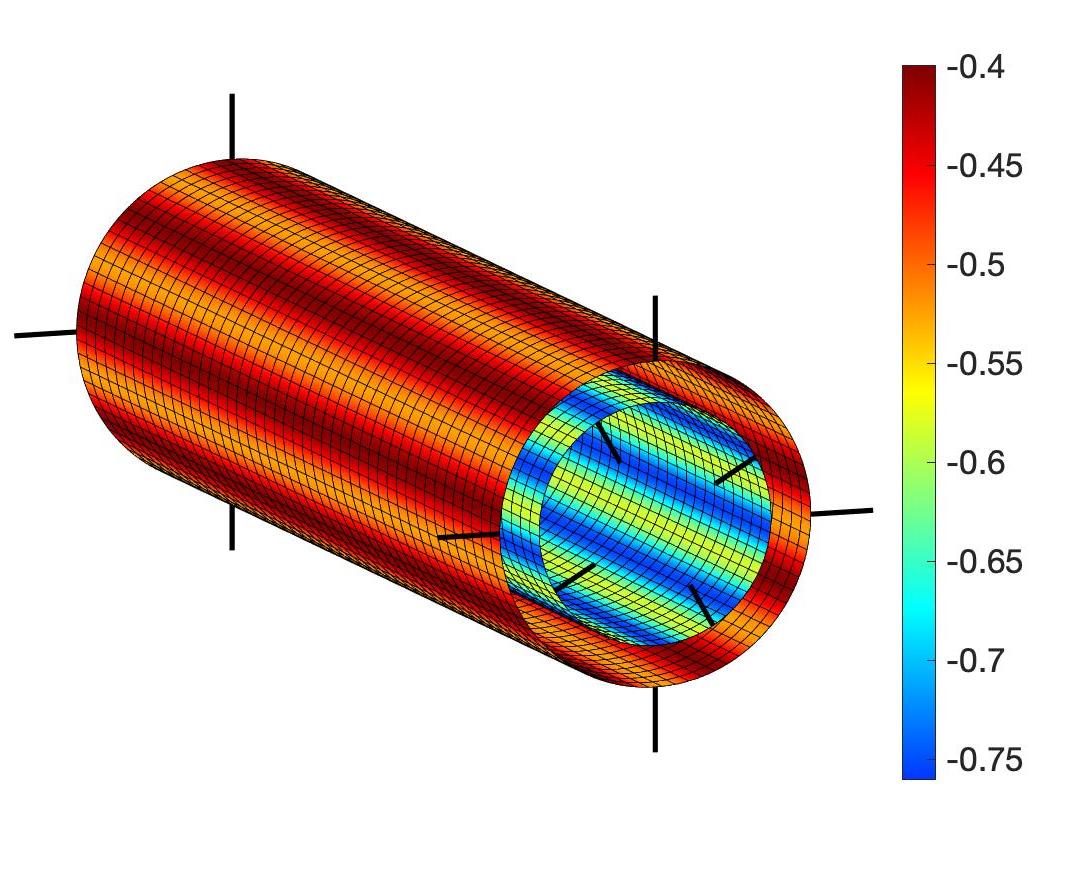}}
\put(-7.95,-.45){\includegraphics[height=25mm]{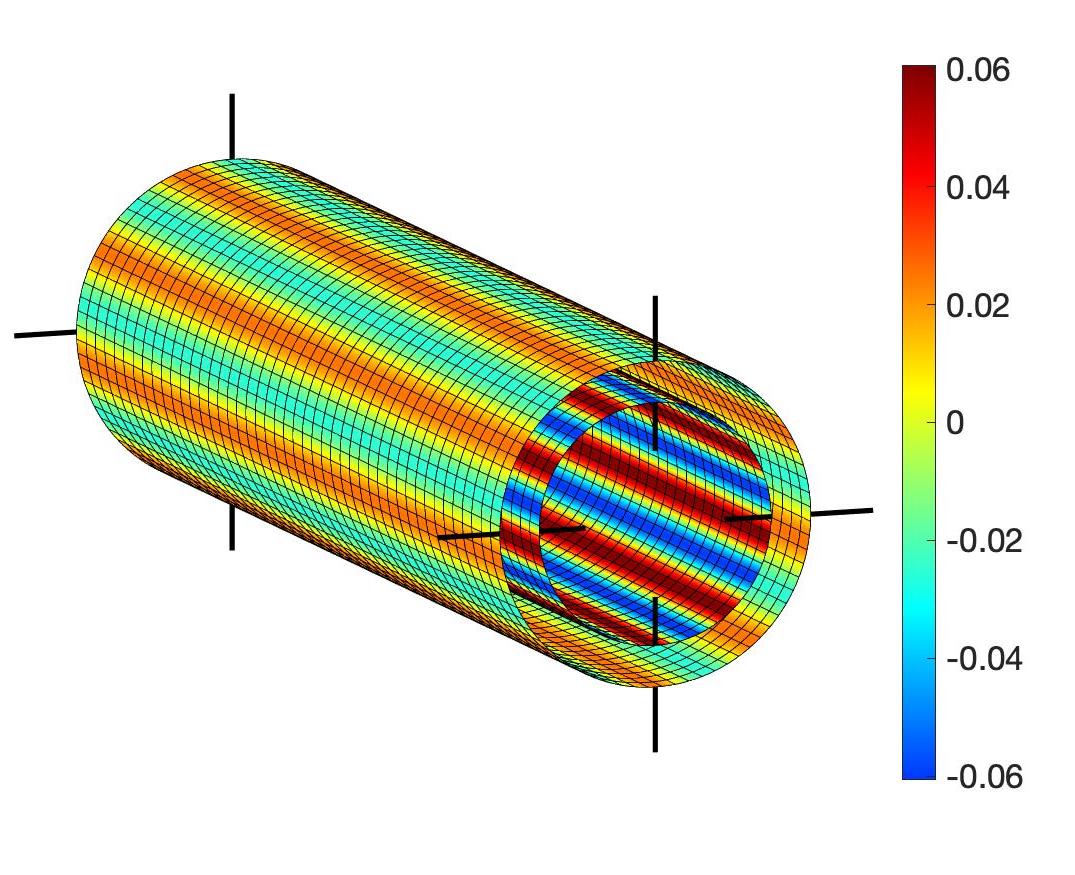}}
\put(-5.45,-.45){\includegraphics[height=25mm]{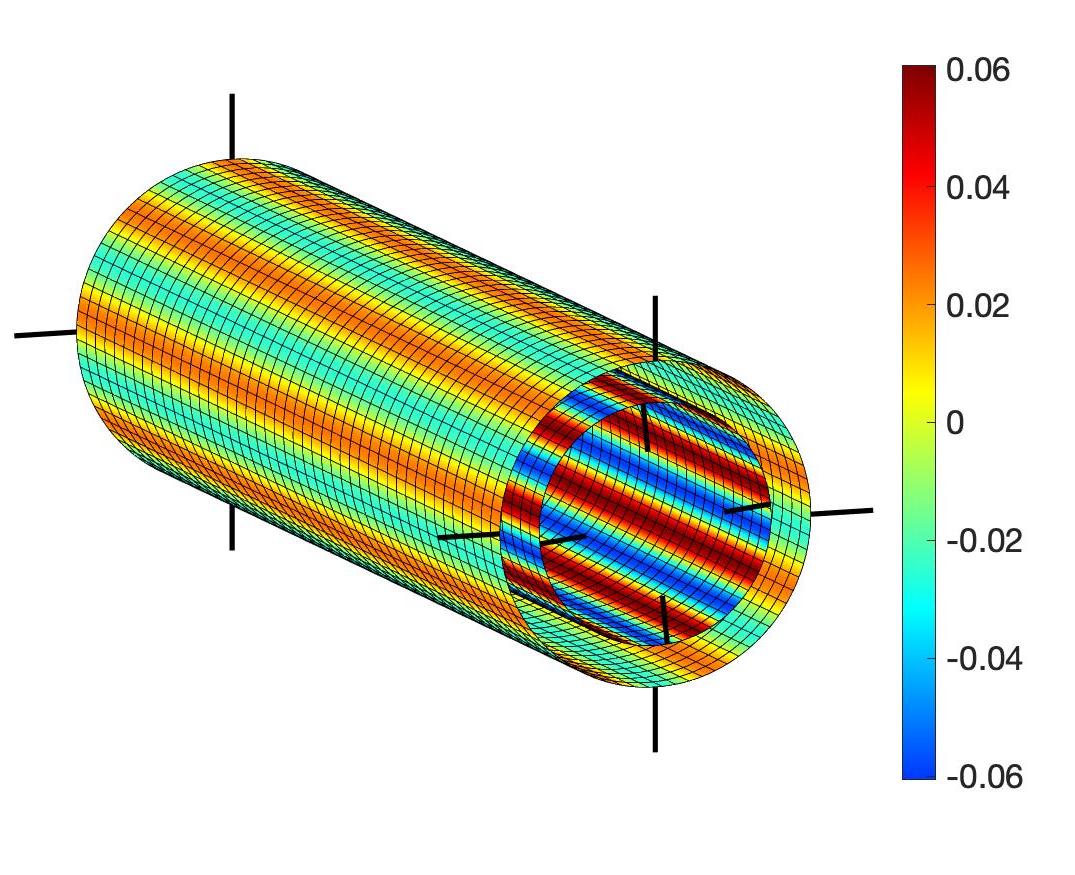}}
\put(-2.95,-.45){\includegraphics[height=25mm]{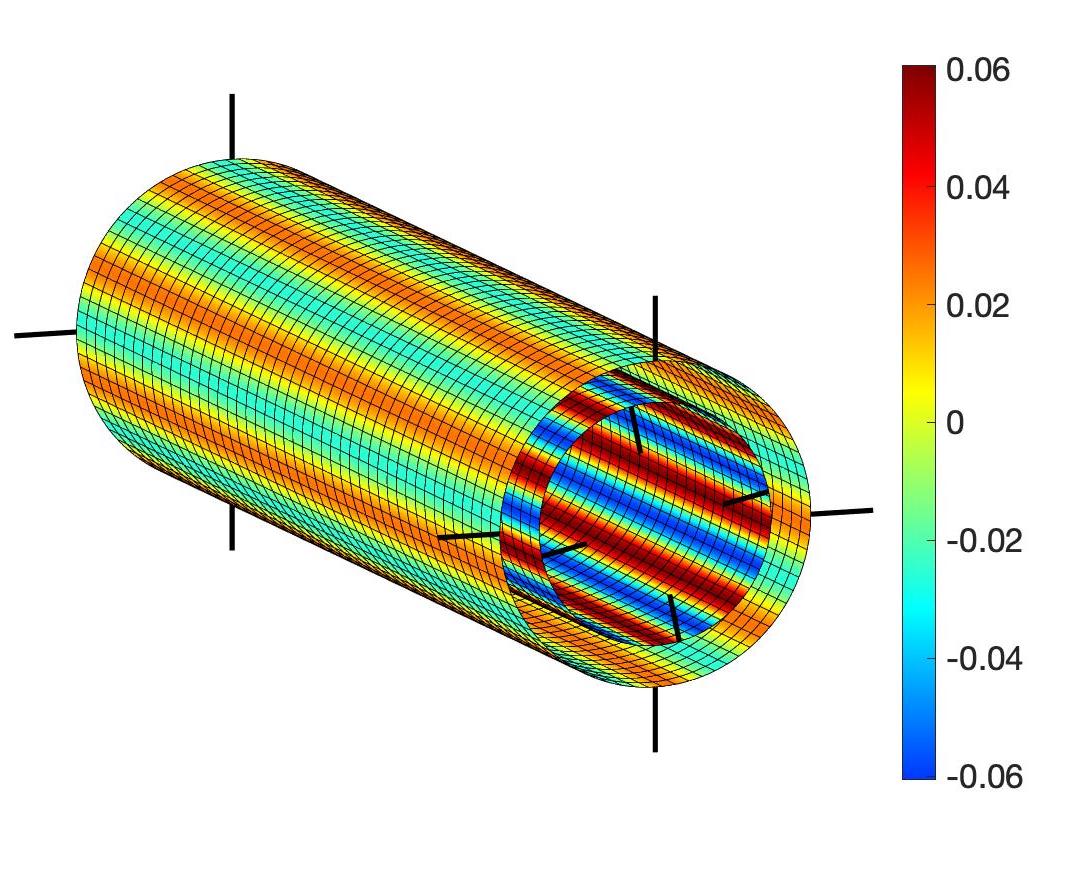}}
\put(-0.45,-.45){\includegraphics[height=25mm]{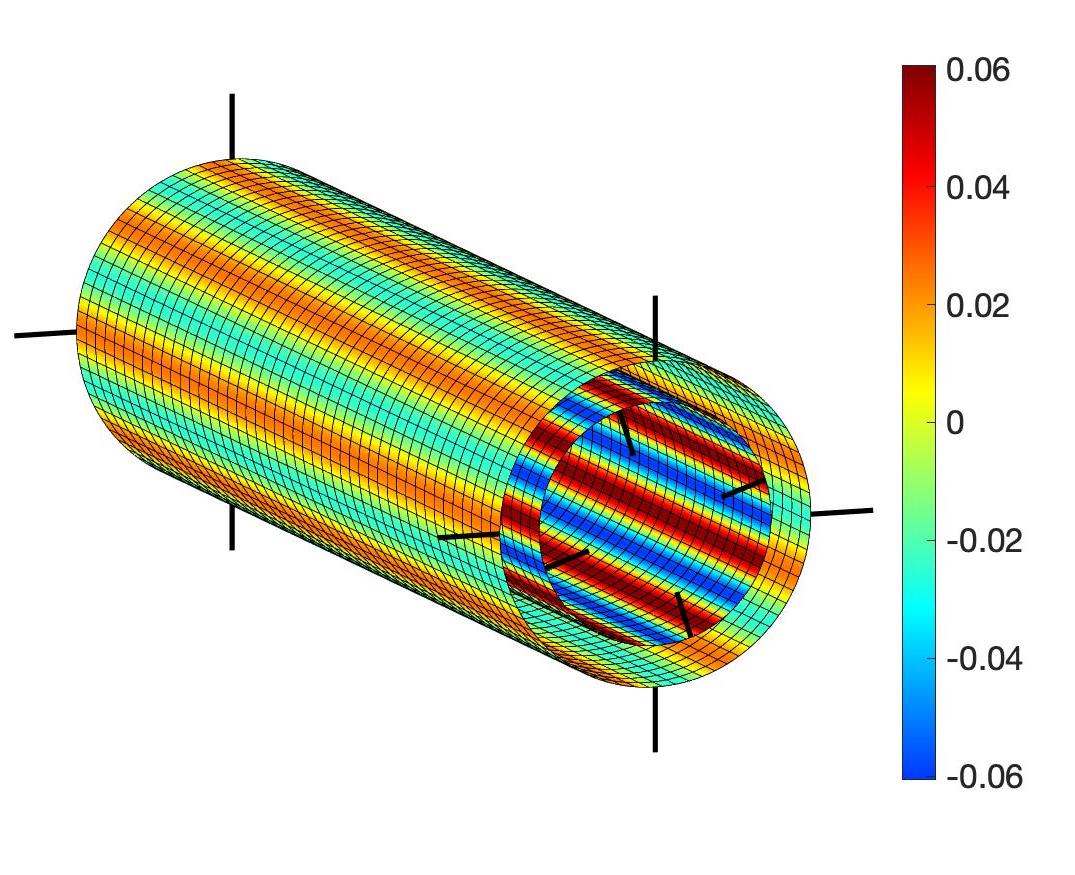}}
\put(2.05,-.45){\includegraphics[height=25mm]{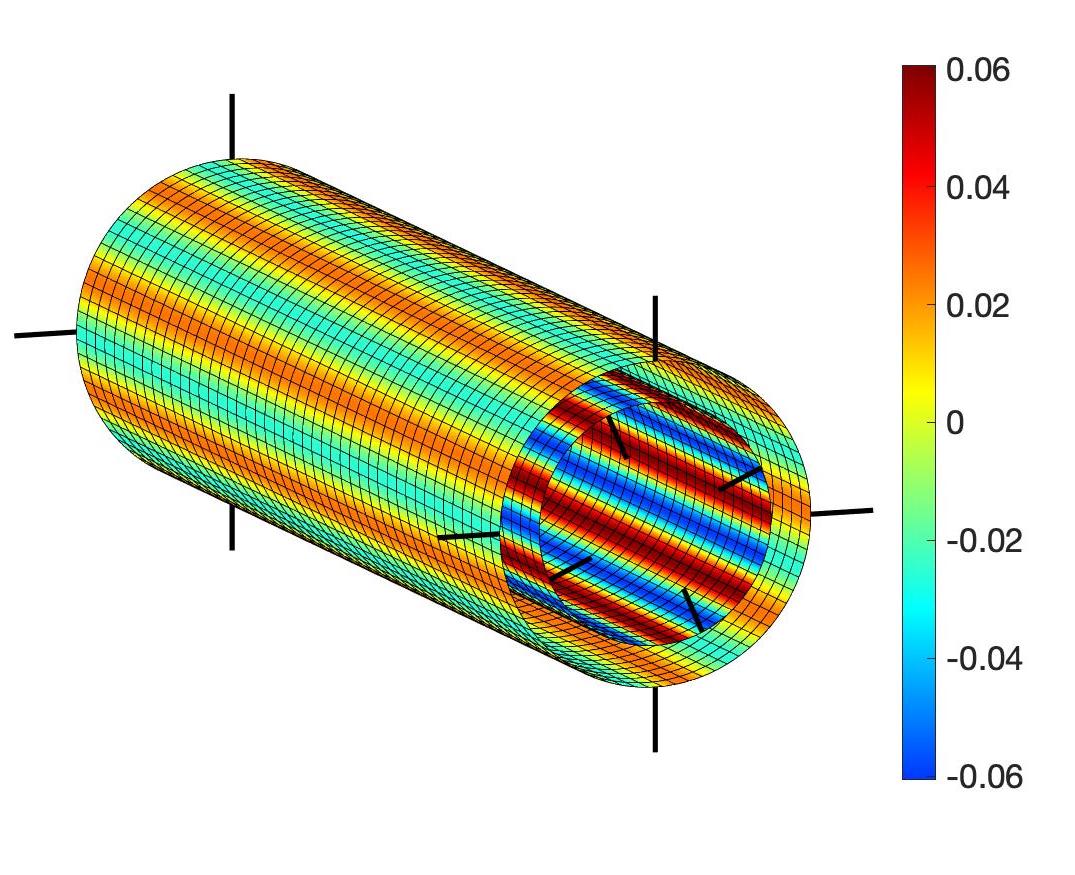}}
\put(4.55,-.45){\includegraphics[height=25mm]{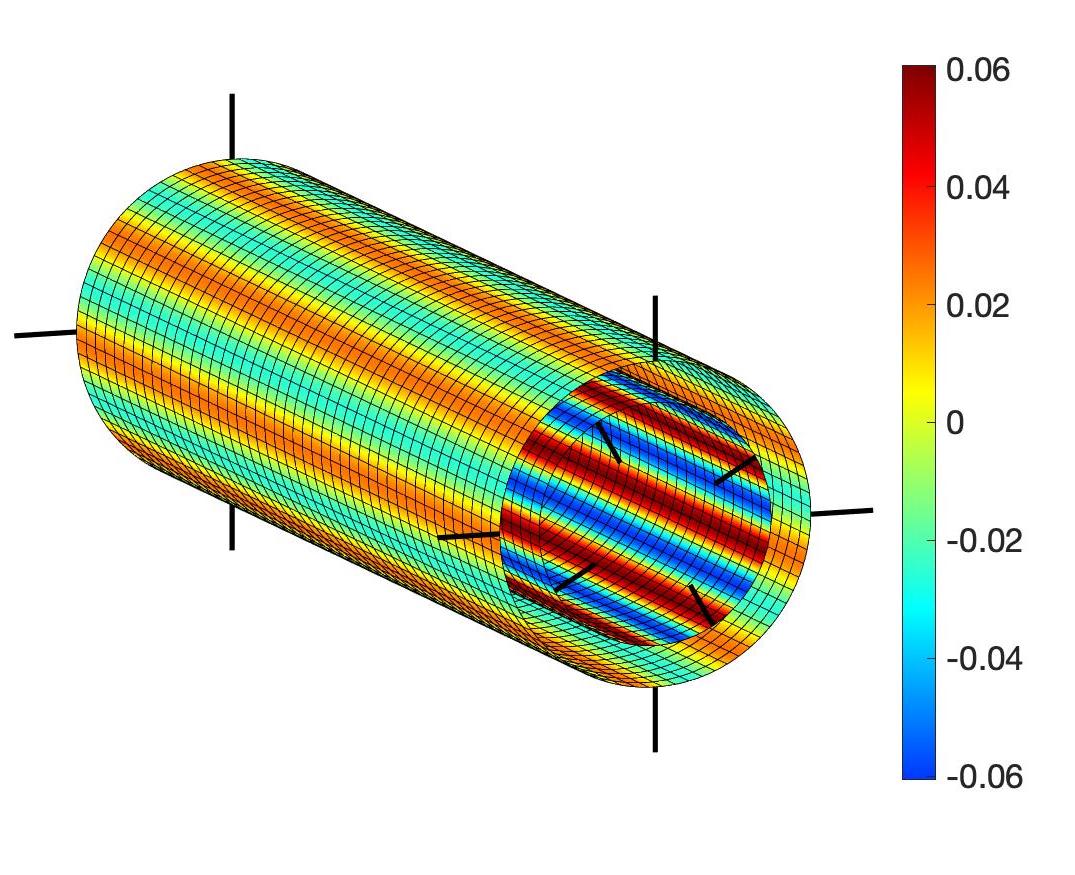}}
\put(7.1,1.85){\includegraphics[height=23mm]{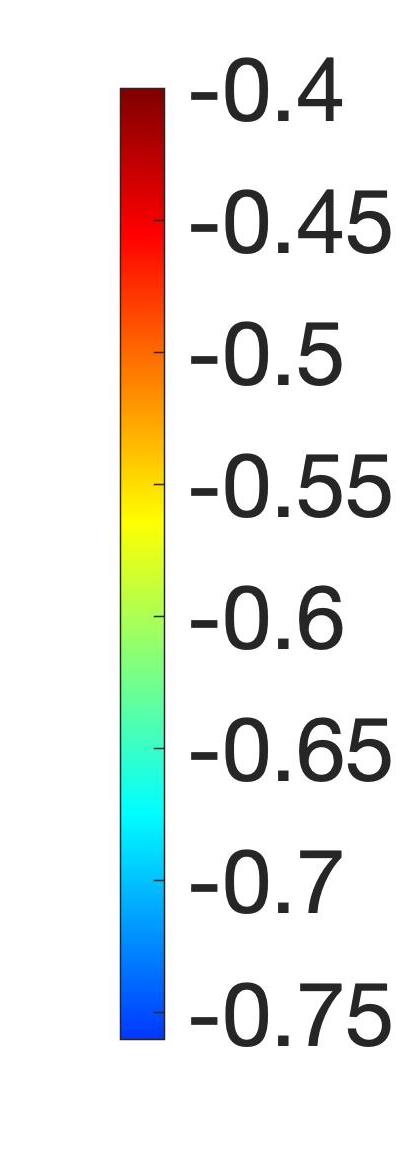}}
\put(7.1,-.35){\includegraphics[height=23mm]{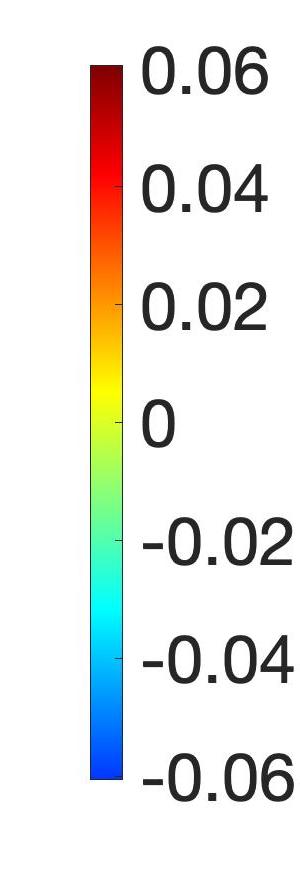}}
\put(-7.85,2.1){\scriptsize{$0^\circ$}}
\put(-5.35,2.1){\scriptsize{$6^\circ$}}
\put(-2.85,2.1){\scriptsize{$12^\circ$}}
\put(-0.35,2.1){\scriptsize{$18^\circ$}}
\put(2.15,2.1){\scriptsize{$24^\circ$}}
\put(4.65,2.1){\scriptsize{$30^\circ$}}
\put(-7.85,0){\scriptsize{$0^\circ$}}
\put(-5.35,0){\scriptsize{$6^\circ$}}
\put(-2.85,0){\scriptsize{$12^\circ$}}
\put(-0.35,0){\scriptsize{$18^\circ$}}
\put(2.15,0){\scriptsize{$24^\circ$}}
\put(4.65,0){\scriptsize{$30^\circ$}}
\end{picture}
\caption{Twisting CNT(26,0) inside CNT(35,0) (Case 1): Color plot of contact pressure $p$ in [GPa] (top row) and circumferential traction $t^2$ in [GPa] (bottom row) at $0^\circ$, $6^\circ$, $12^\circ$, $18^\circ$, $24^\circ$ and $30^\circ$ twist of the inner CNT, from left to right. 
The inner CNT is twisted counterclockwise.
The pressure and traction patterns move counter-clockwise on both CNTs, and they are faster than the twisting rate.
}
\label{f:twist2}
\end{center}
\end{figure}
\begin{figure}[h]
\begin{center} \unitlength1cm
\begin{picture}(0,4.2)
\put(-7.95,1.75){\includegraphics[height=25mm]{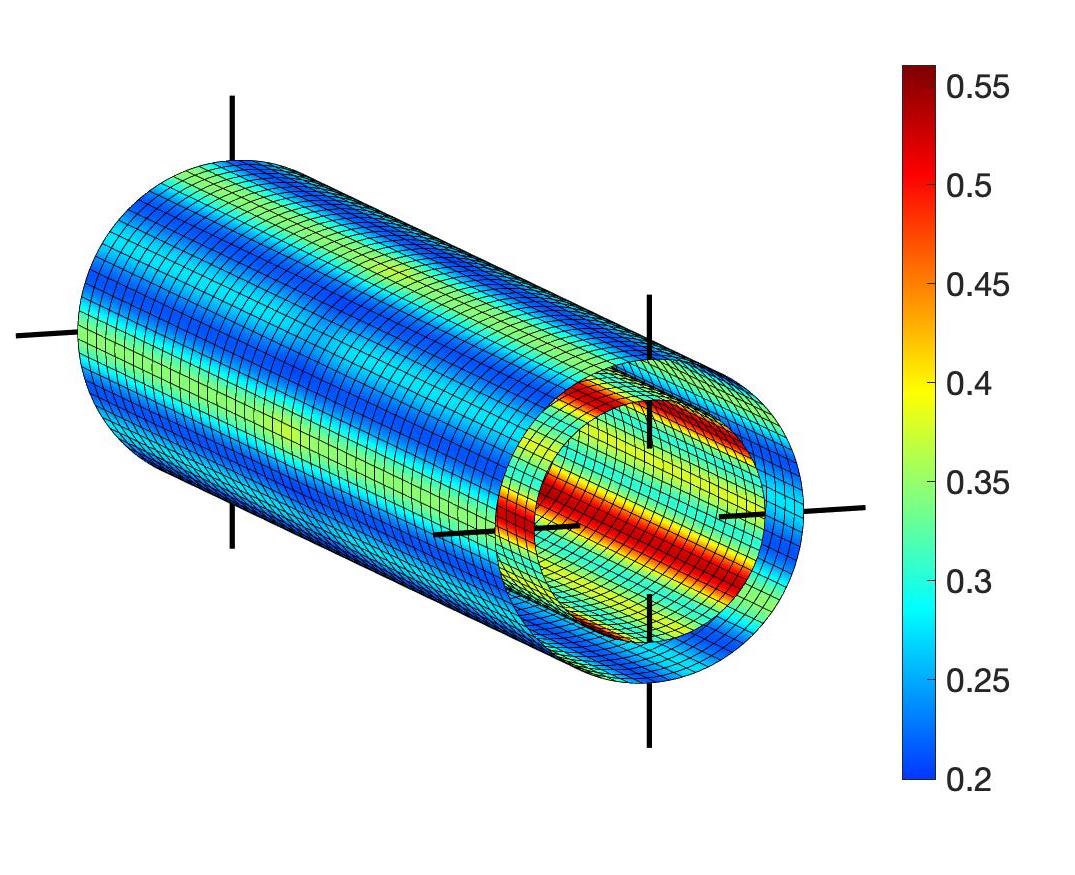}}
\put(-5.45,1.75){\includegraphics[height=25mm]{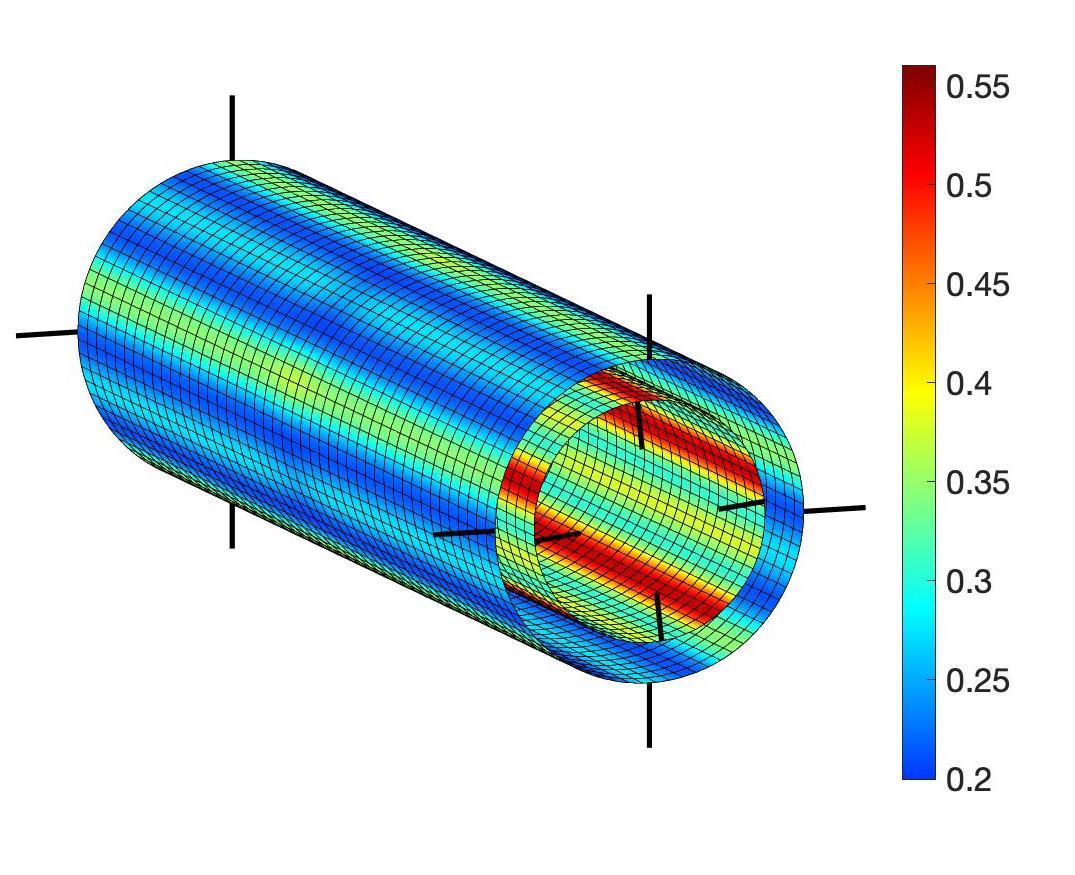}}
\put(-2.95,1.75){\includegraphics[height=25mm]{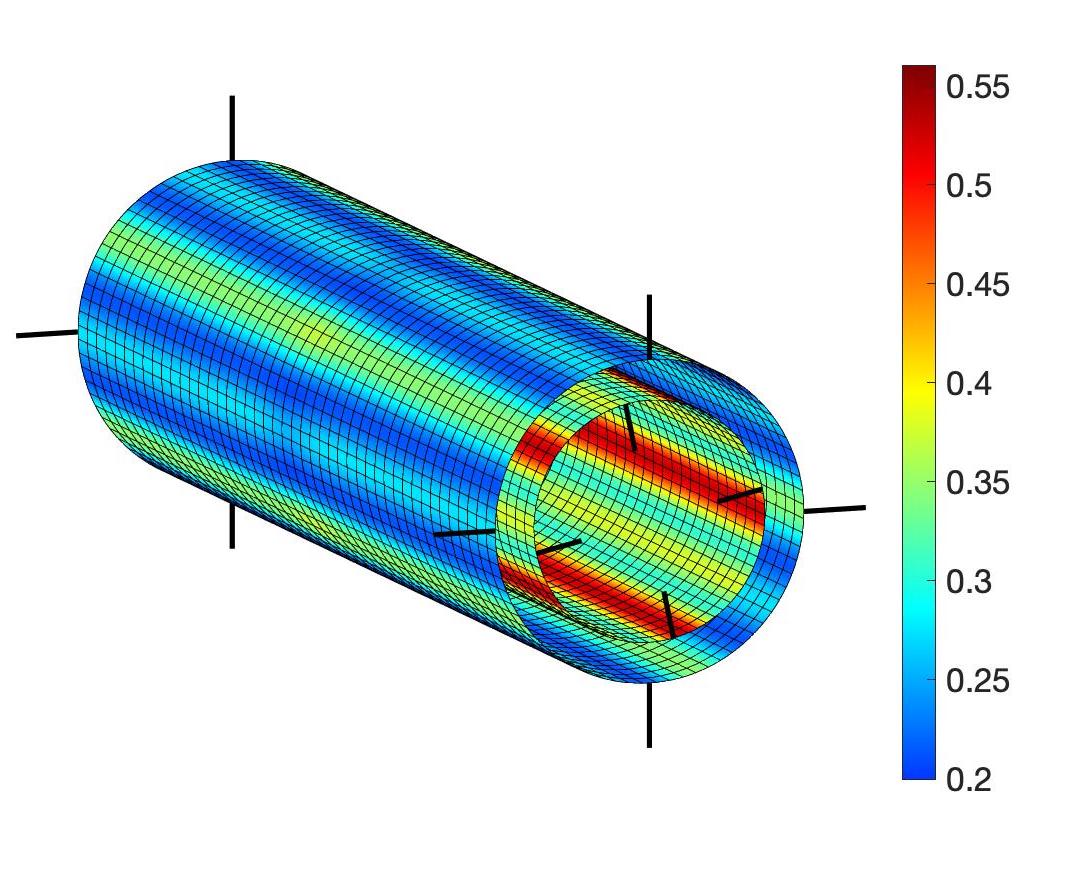}}
\put(-0.45,1.75){\includegraphics[height=25mm]{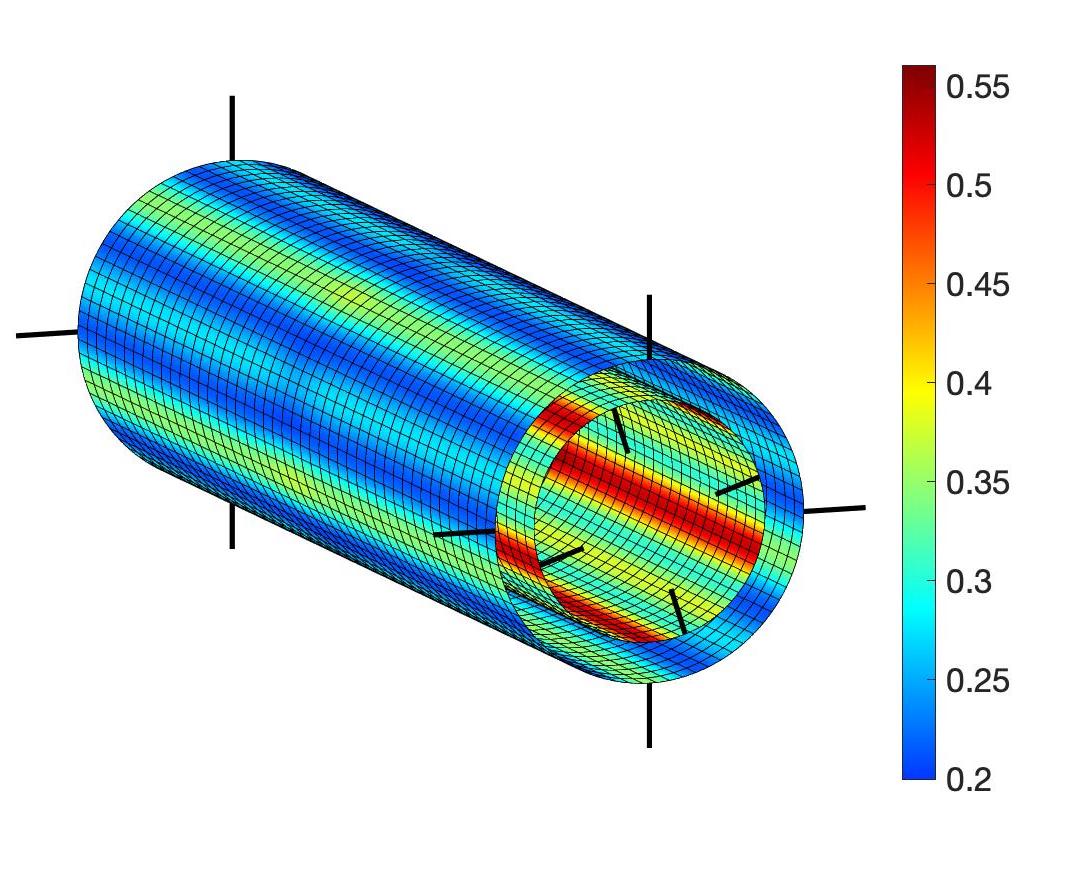}}
\put(2.05,1.75){\includegraphics[height=25mm]{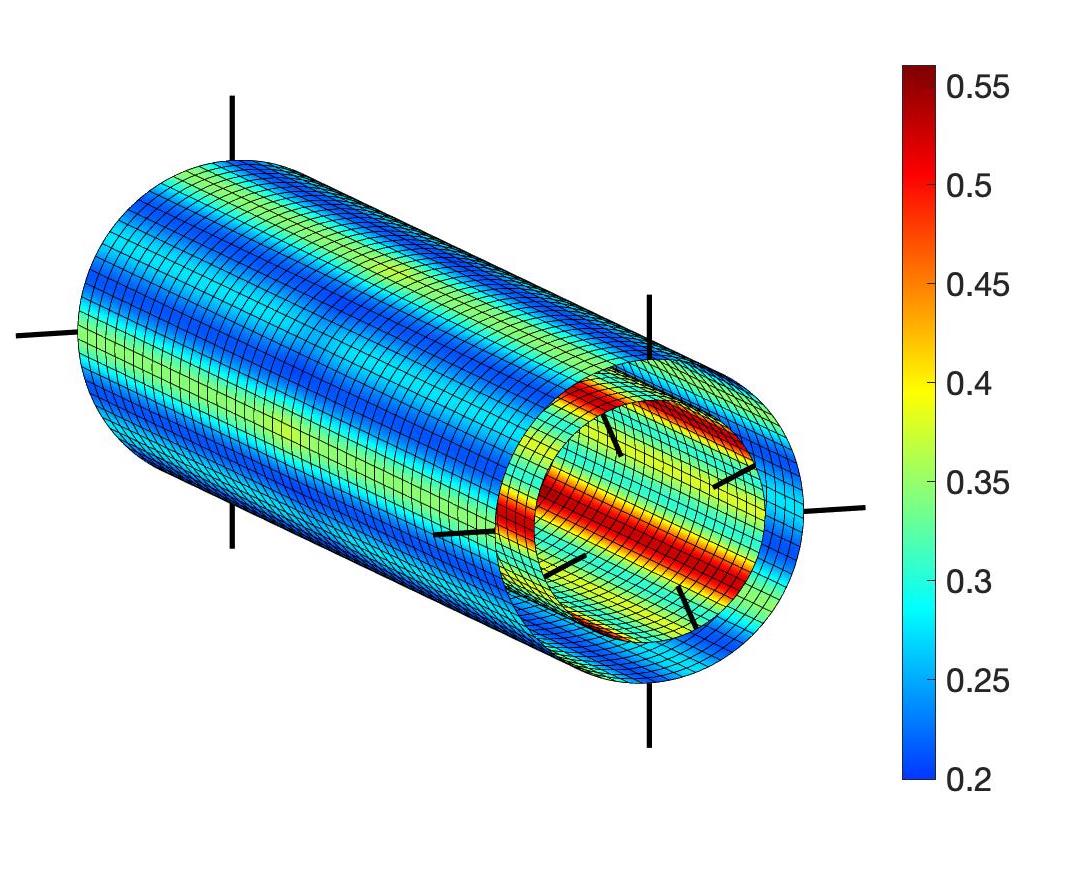}}
\put(4.55,1.75){\includegraphics[height=25mm]{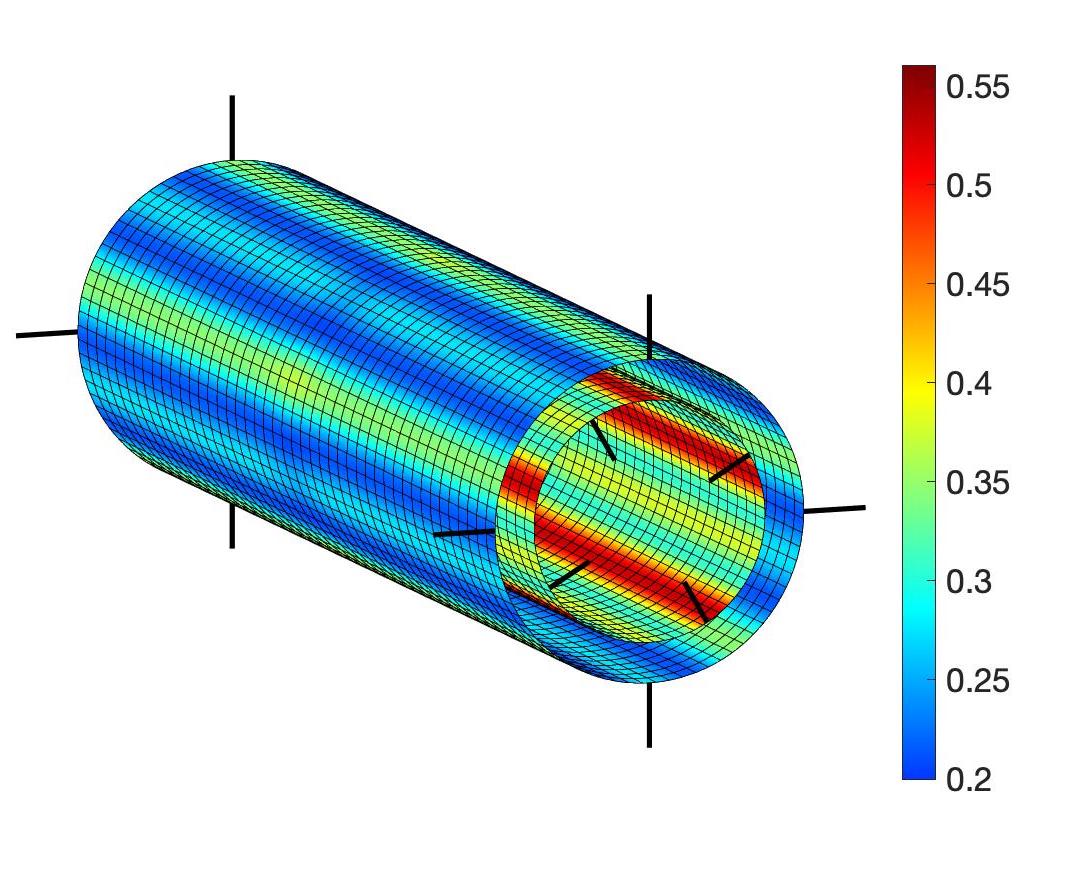}}
\put(-7.95,-.45){\includegraphics[height=25mm]{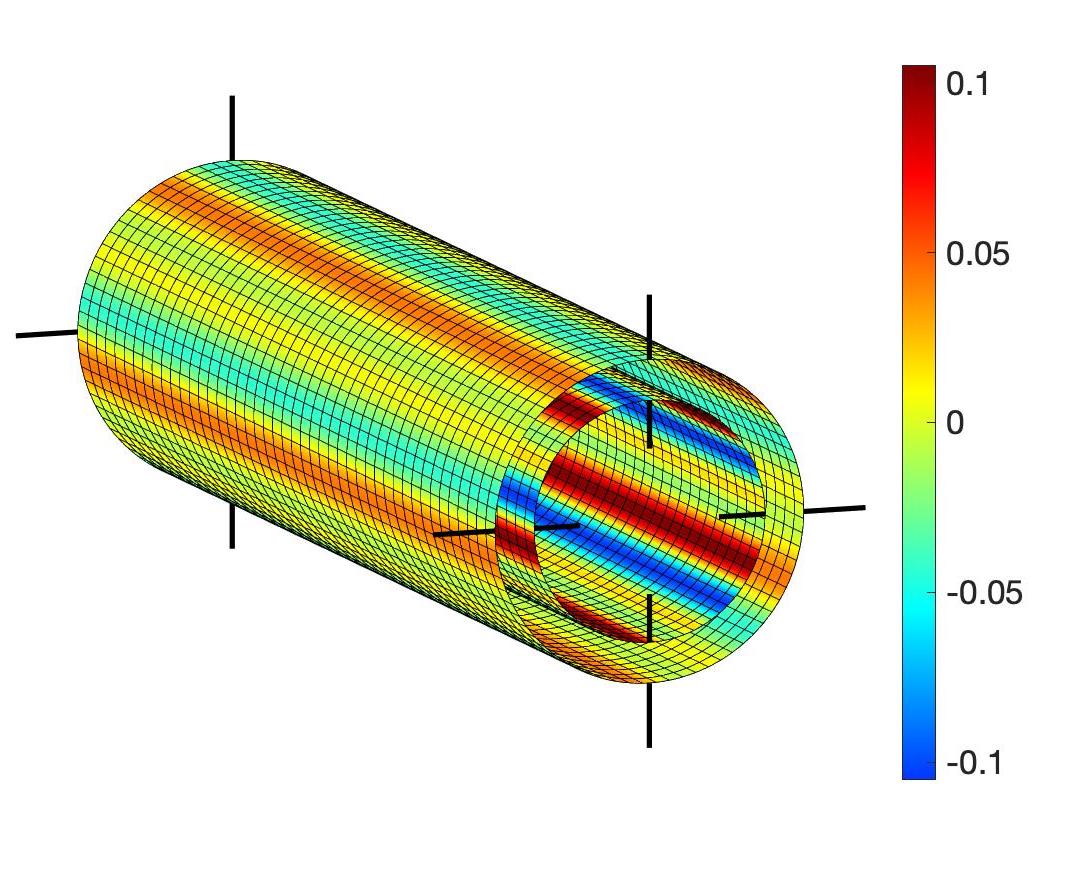}}
\put(-5.45,-.45){\includegraphics[height=25mm]{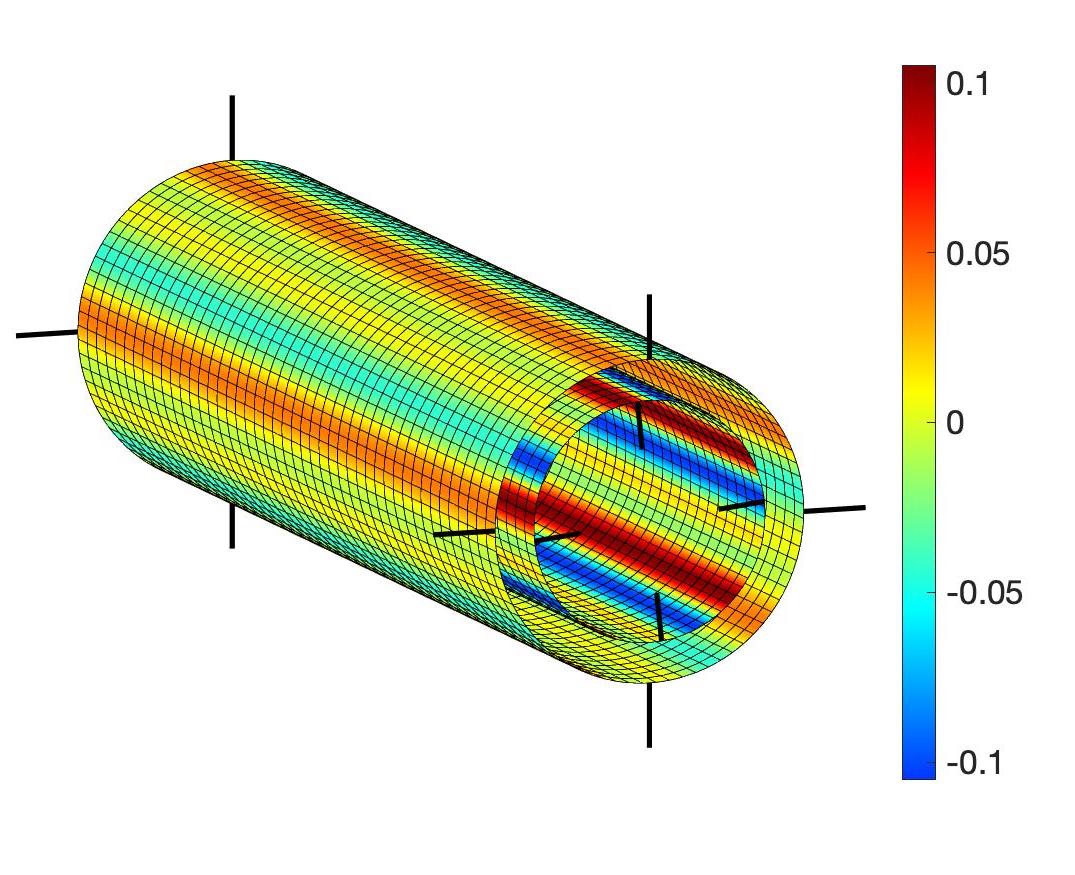}}
\put(-2.95,-.45){\includegraphics[height=25mm]{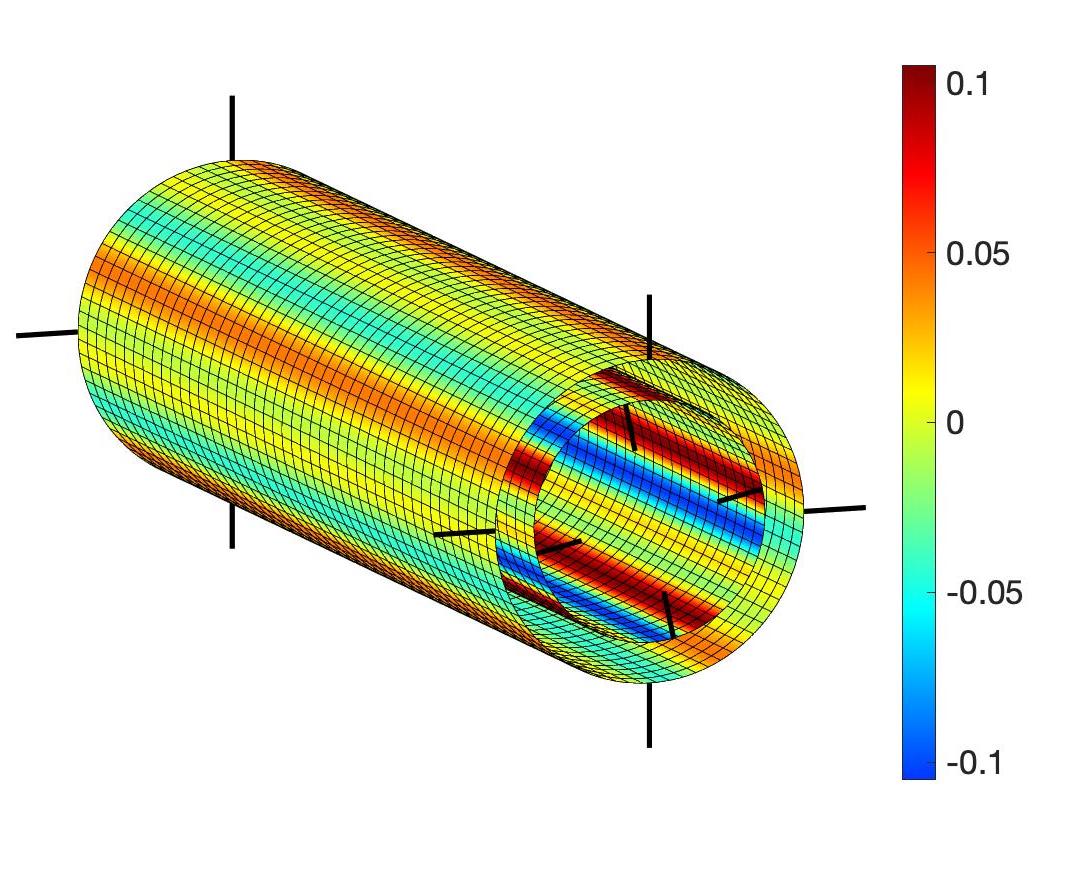}}
\put(-0.45,-.45){\includegraphics[height=25mm]{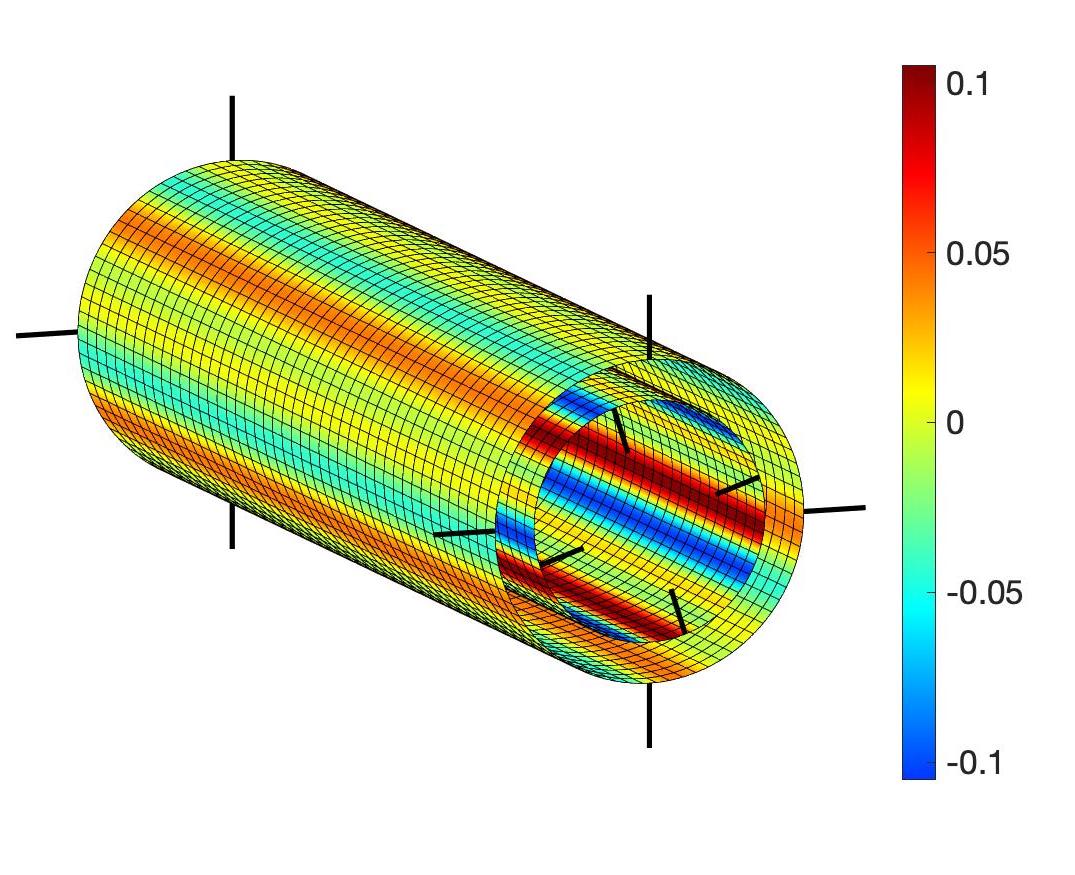}}
\put(2.05,-.45){\includegraphics[height=25mm]{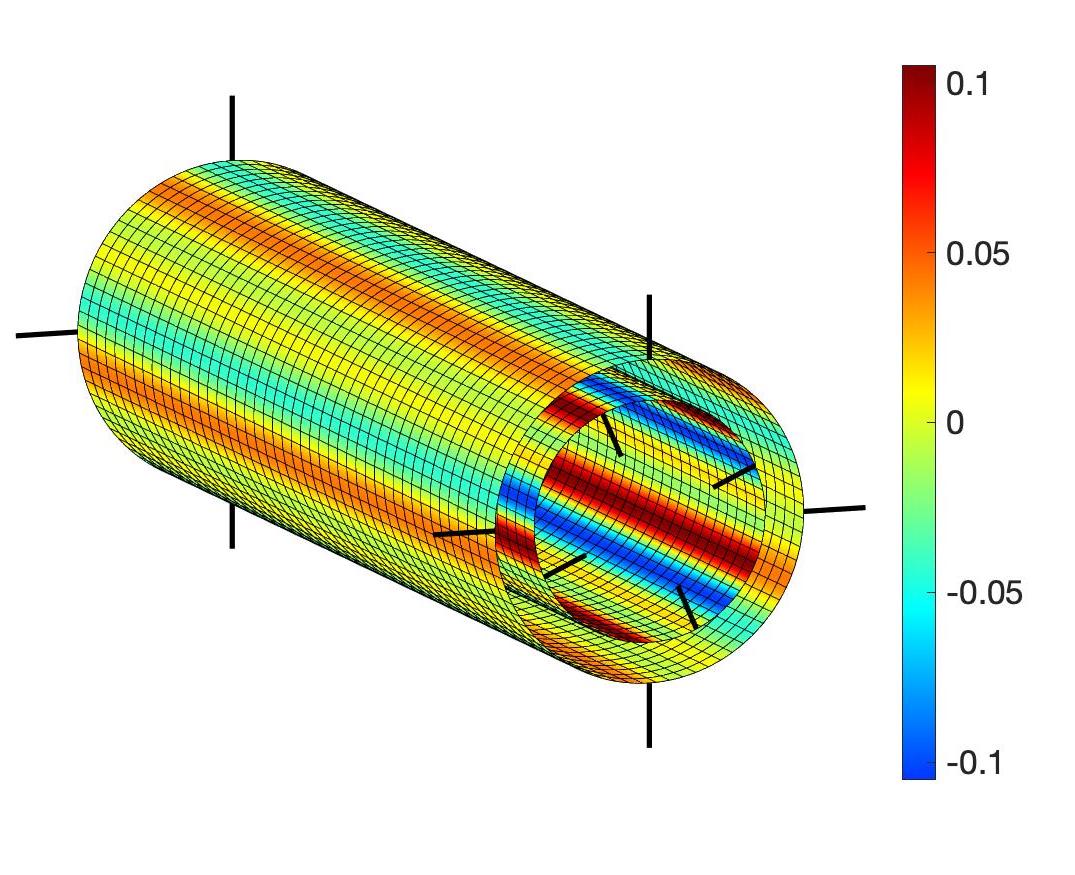}}
\put(4.55,-.45){\includegraphics[height=25mm]{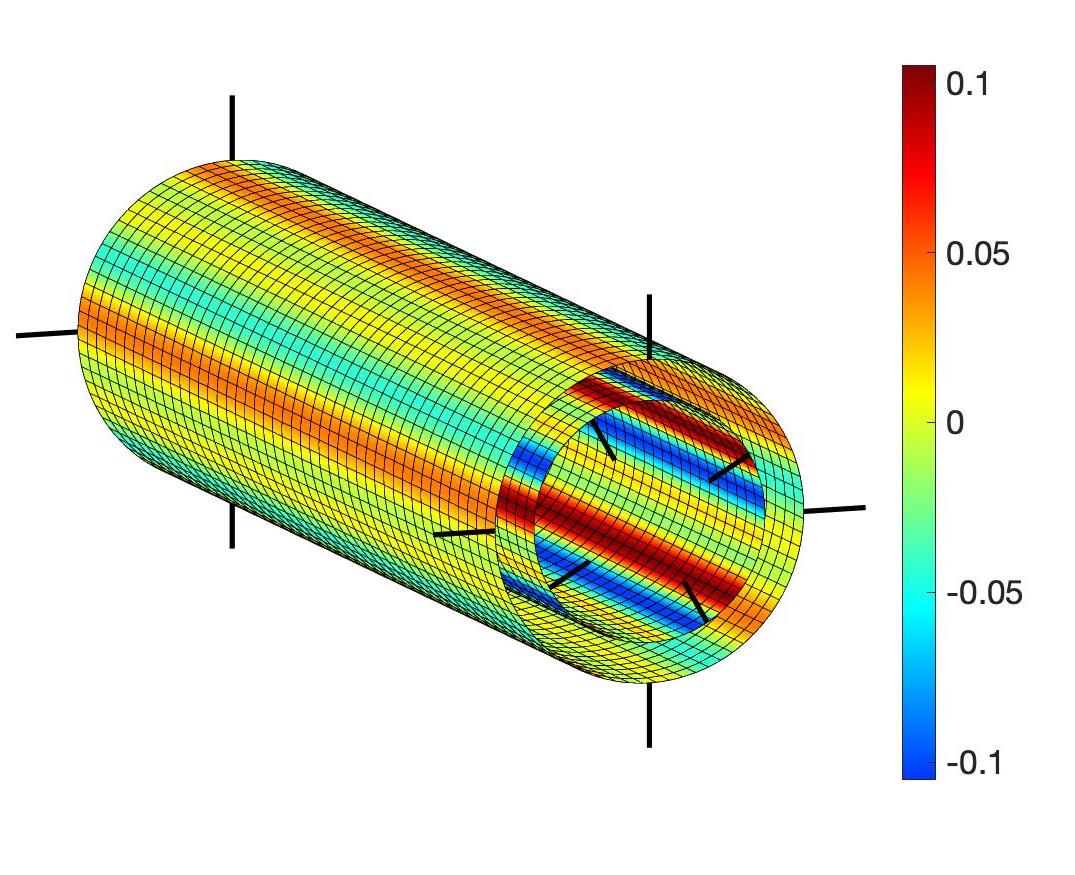}}
\put(7.1,1.85){\includegraphics[height=23mm]{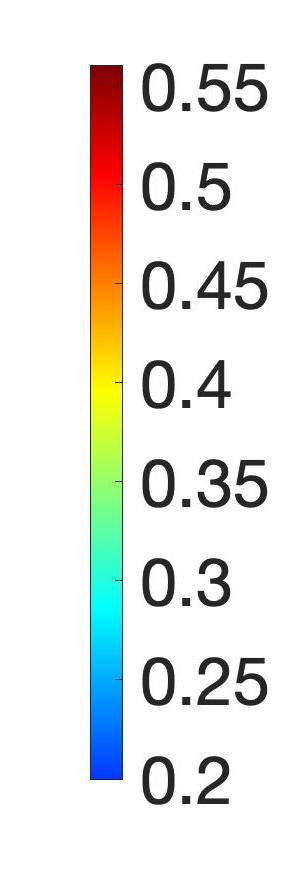}}
\put(7.1,-.35){\includegraphics[height=23mm]{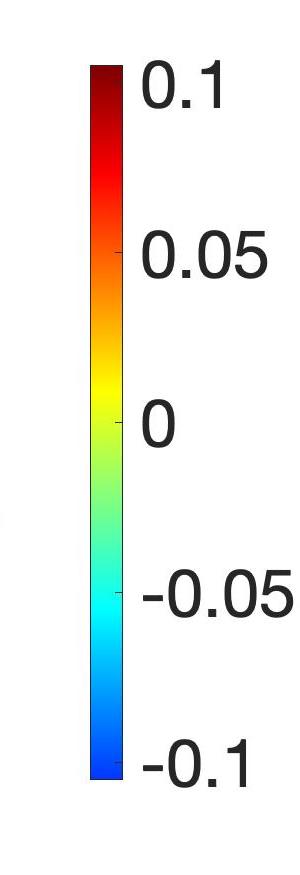}}
\put(-7.85,2.1){\scriptsize{$0^\circ$}}
\put(-5.35,2.1){\scriptsize{$6^\circ$}}
\put(-2.85,2.1){\scriptsize{$12^\circ$}}
\put(-0.35,2.1){\scriptsize{$18^\circ$}}
\put(2.15,2.1){\scriptsize{$24^\circ$}}
\put(4.65,2.1){\scriptsize{$30^\circ$}}
\put(-7.85,0){\scriptsize{$0^\circ$}}
\put(-5.35,0){\scriptsize{$6^\circ$}}
\put(-2.85,0){\scriptsize{$12^\circ$}}
\put(-0.35,0){\scriptsize{$18^\circ$}}
\put(2.15,0){\scriptsize{$24^\circ$}}
\put(4.65,0){\scriptsize{$30^\circ$}}
\end{picture}
\caption{Twisting CNT(15,15) inside CNT(20,20) (Case 2): Color plot of contact pressure $p$ in [GPa] (top row) and circumferential traction $t^2$ in [GPa] (bottom row) at $0^\circ$, $6^\circ$, $12^\circ$, $18^\circ$, $24^\circ$ and $30^\circ$ twist of the inner CNT, from left to right. 
The inner CNT is twisted counterclockwise.
The pressure and traction patterns move counter-clockwise on both CNTs, and they are faster than the twisting rate: three times faster on the outer CNT and four times faster on the inner CNT.
}
\label{f:twist1}
\end{center}
\end{figure}
\begin{figure}[!htbp]
\begin{center} \unitlength1cm
\begin{picture}(0,4.2)
\put(-7.95,1.75){\includegraphics[height=25mm]{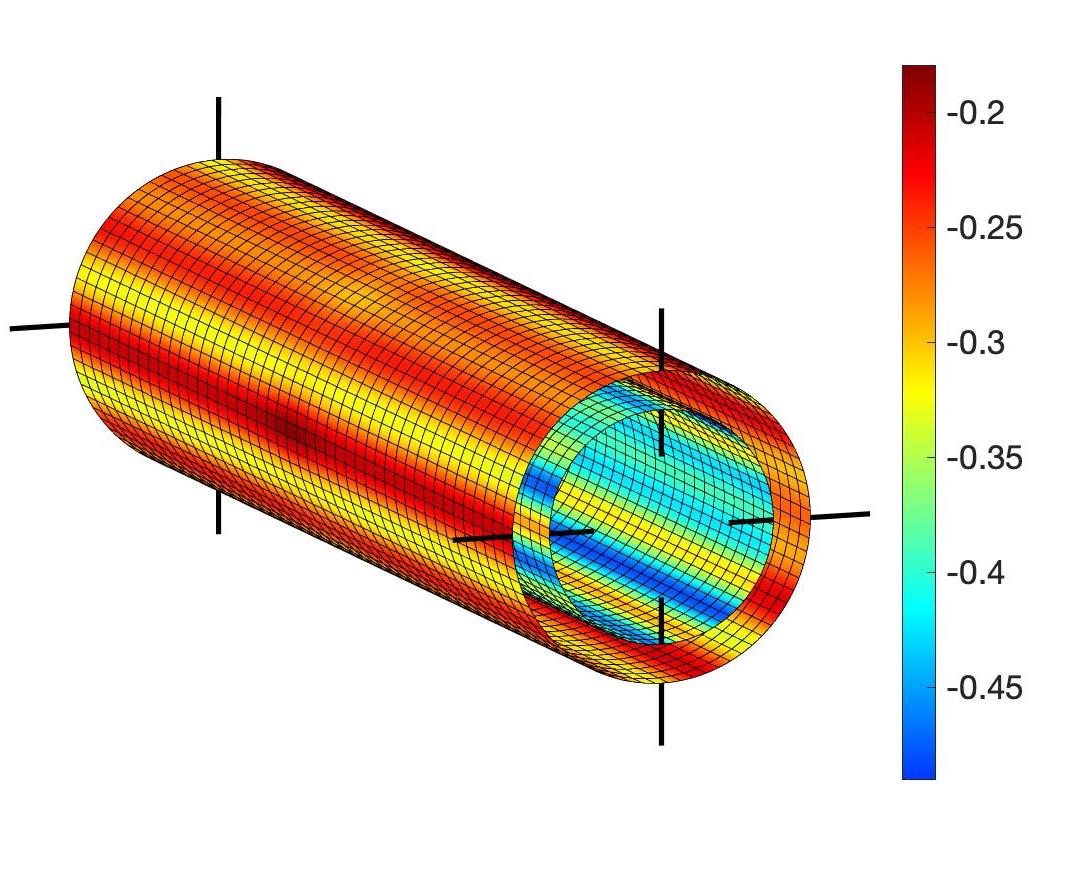}}
\put(-5.45,1.75){\includegraphics[height=25mm]{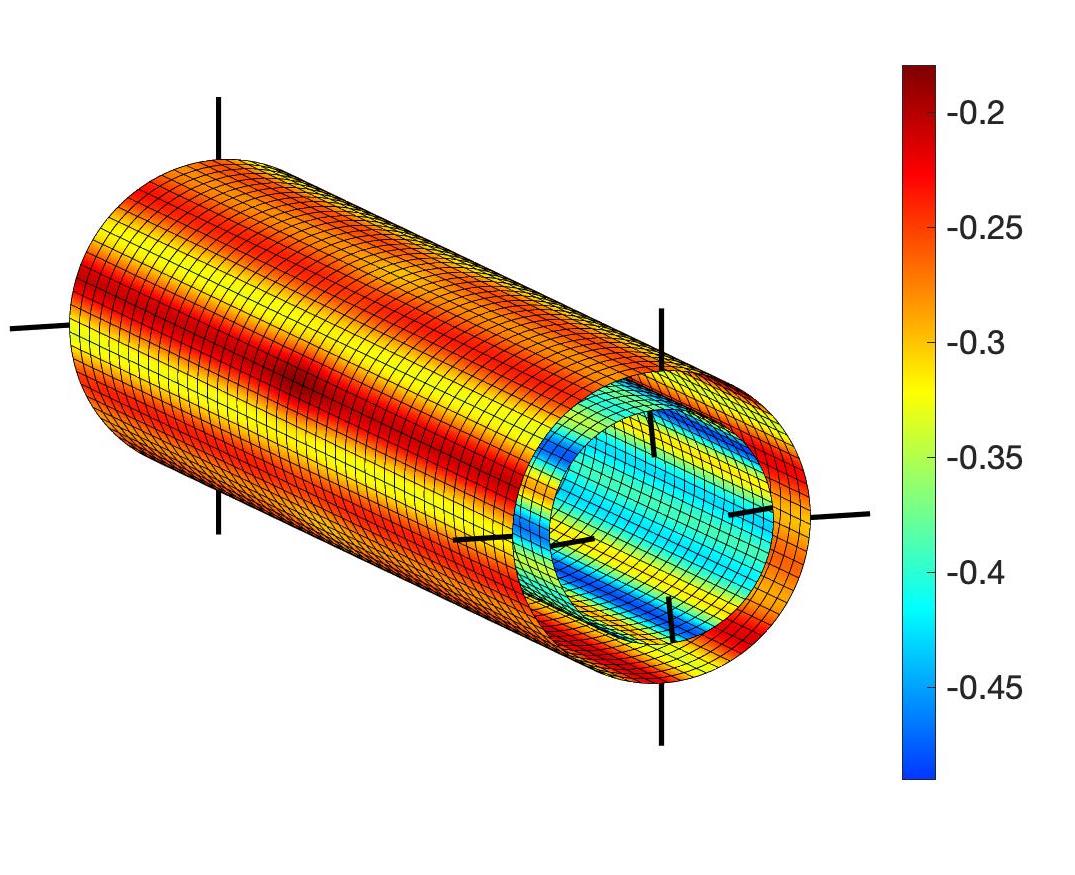}}
\put(-2.95,1.75){\includegraphics[height=25mm]{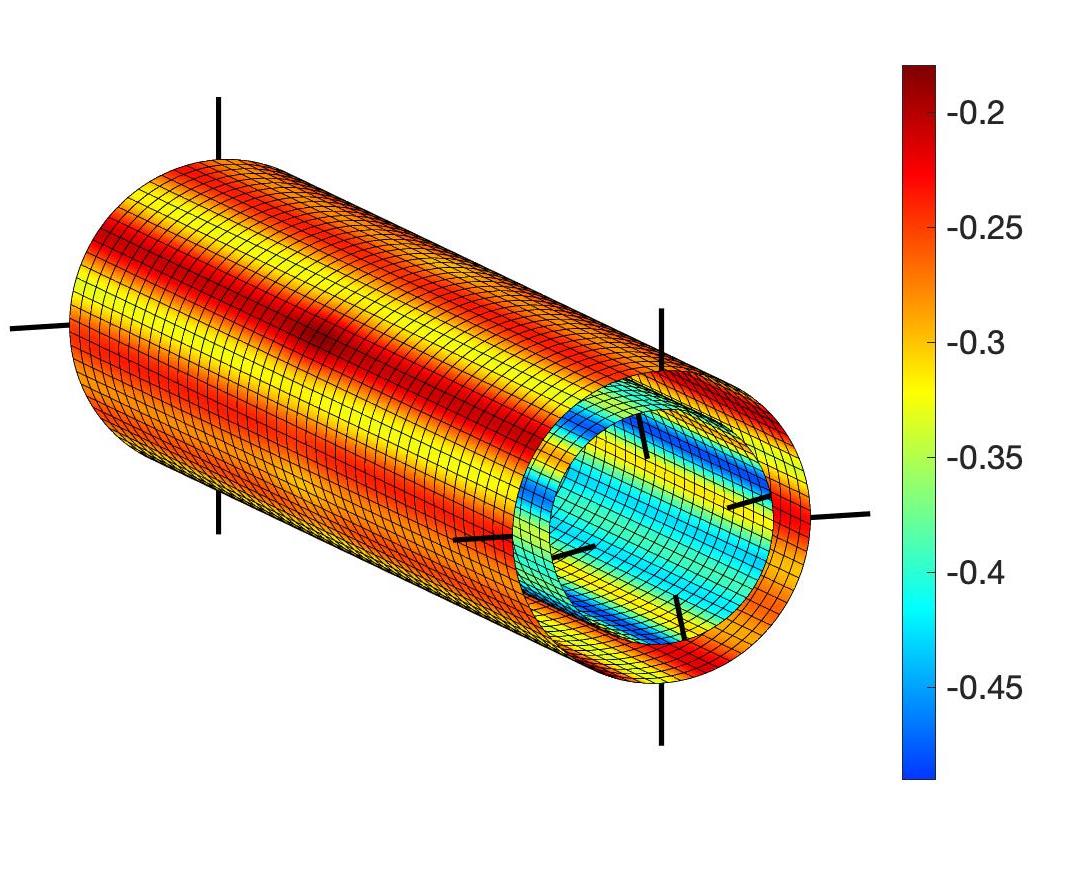}}
\put(-0.45,1.75){\includegraphics[height=25mm]{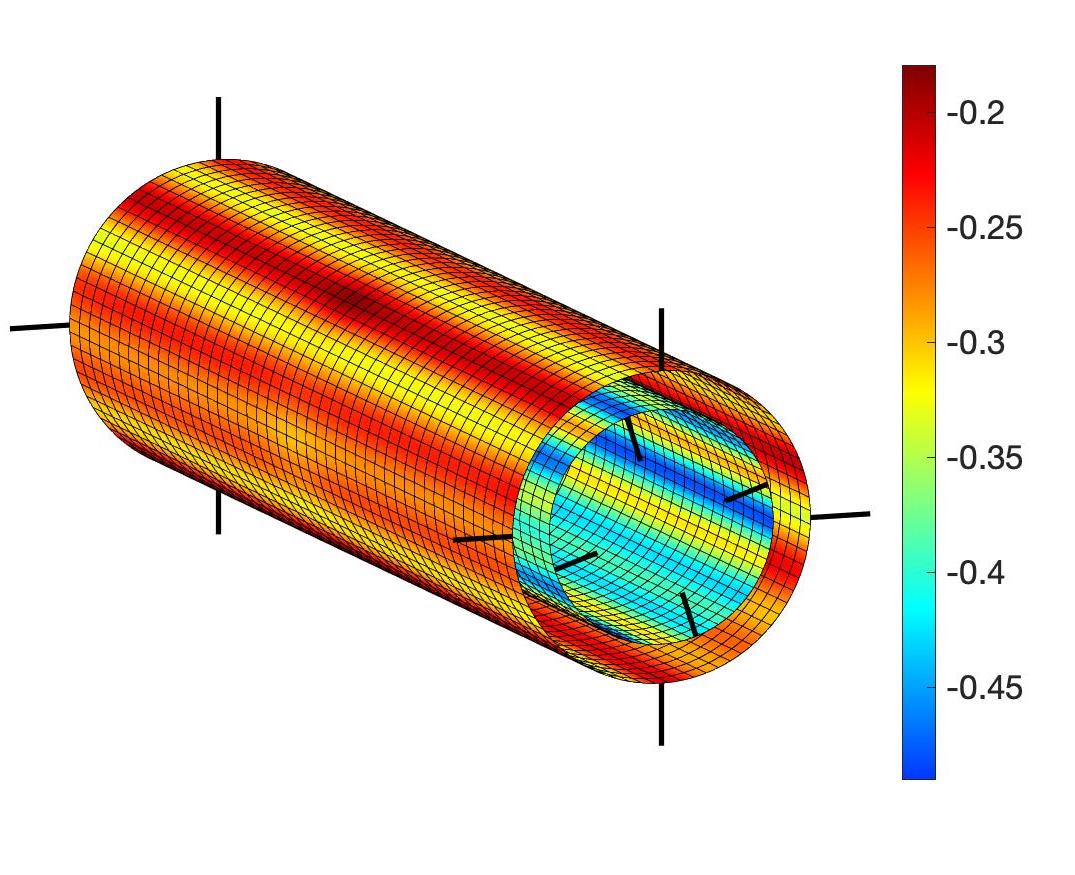}}
\put(2.05,1.75){\includegraphics[height=25mm]{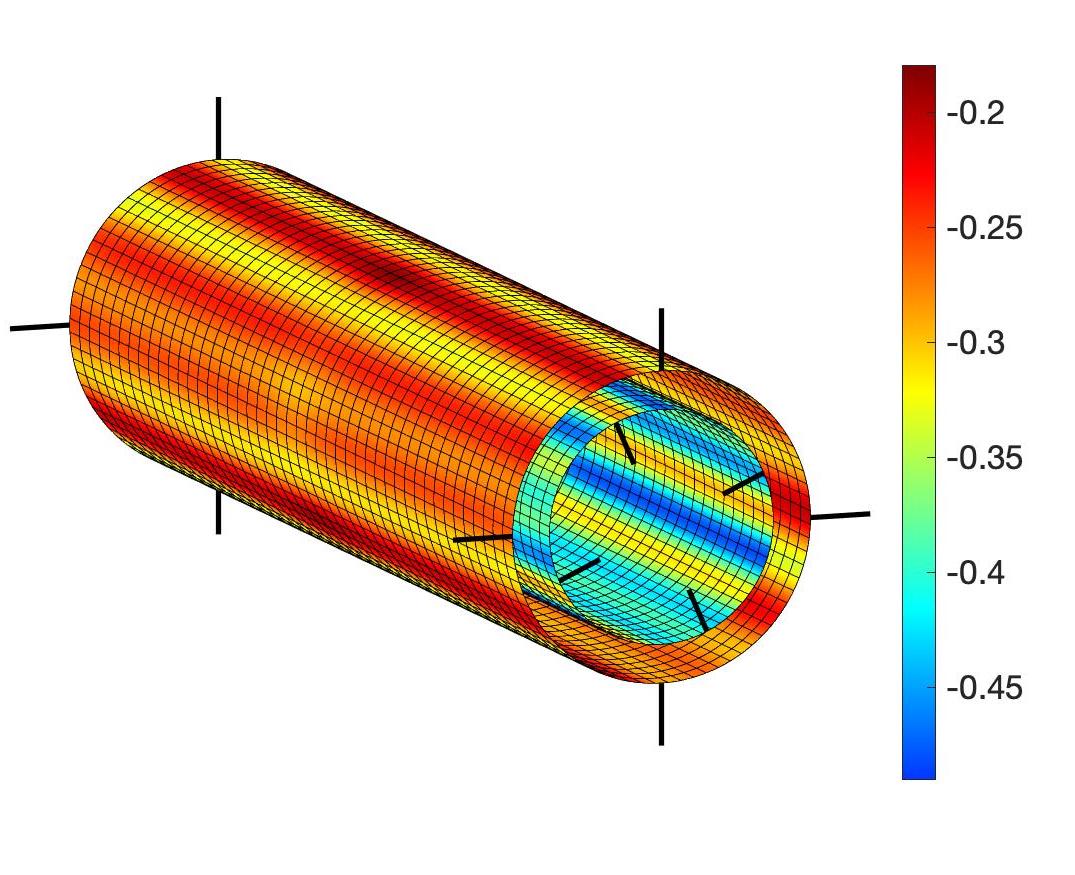}}
\put(4.55,1.75){\includegraphics[height=25mm]{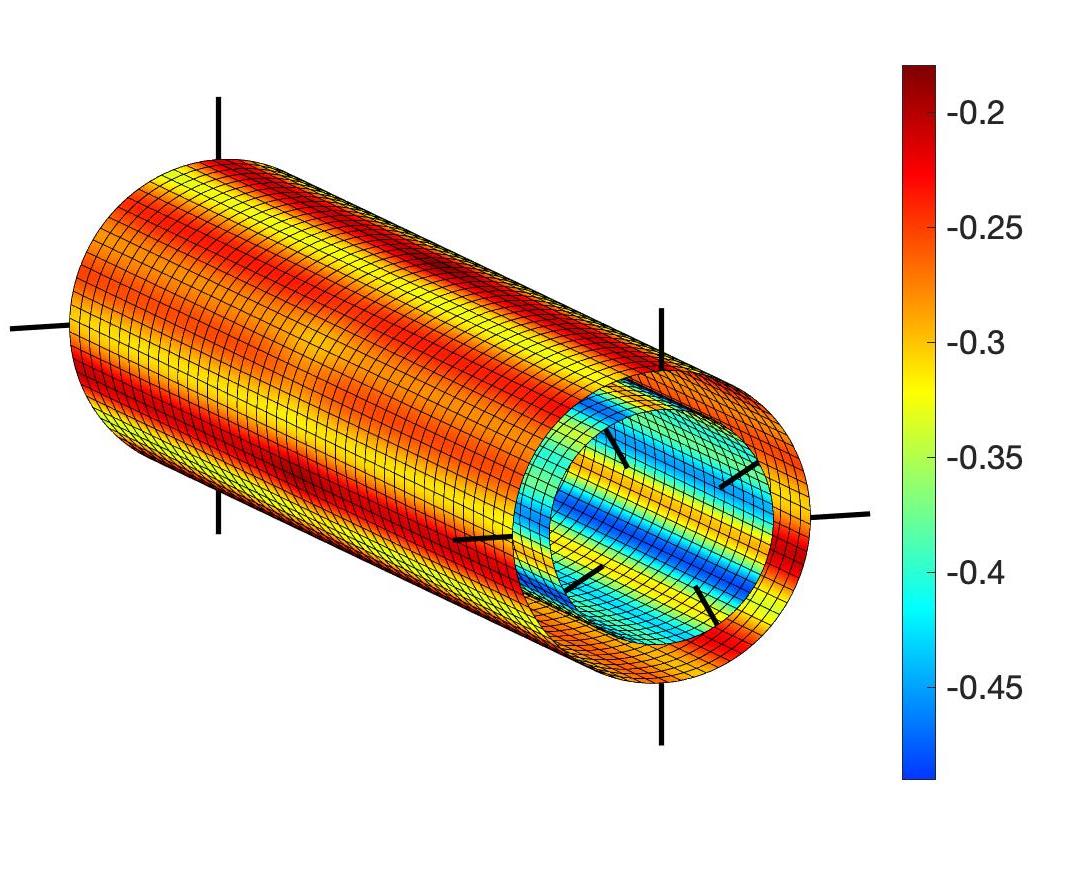}}
\put(-7.95,-.45){\includegraphics[height=25mm]{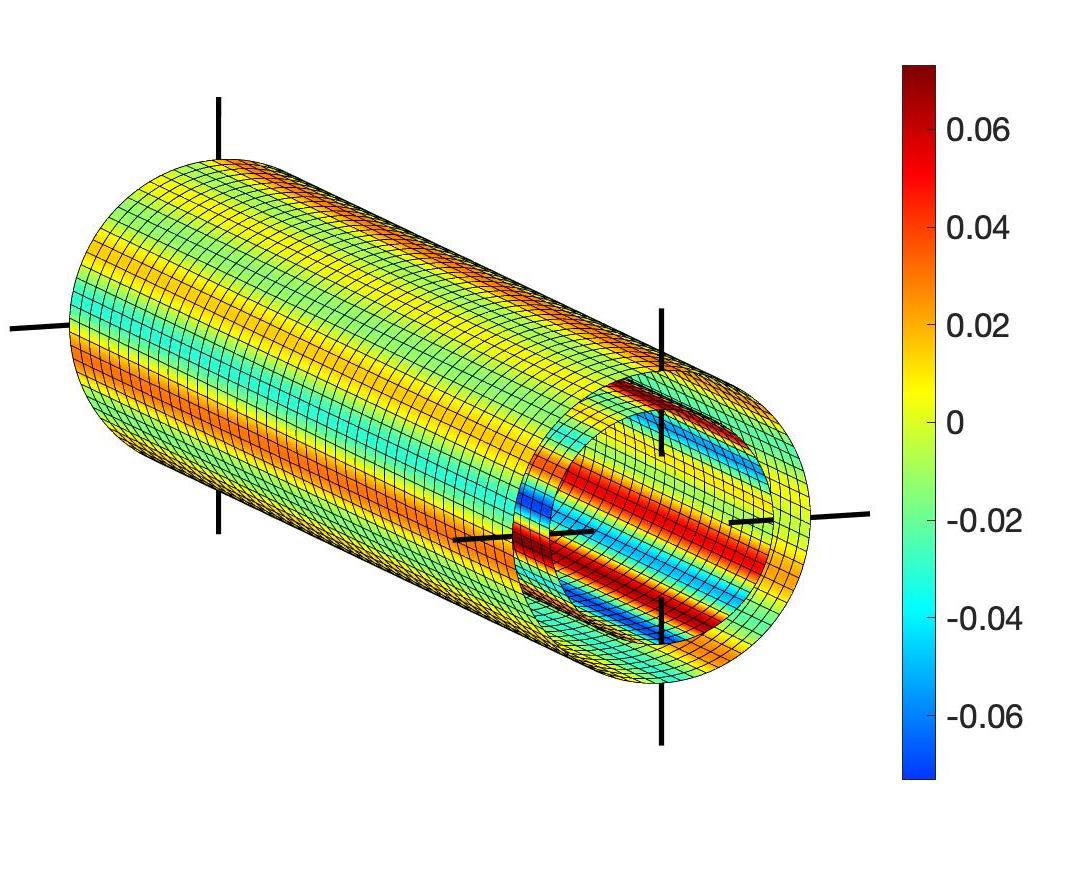}}
\put(-5.45,-.45){\includegraphics[height=25mm]{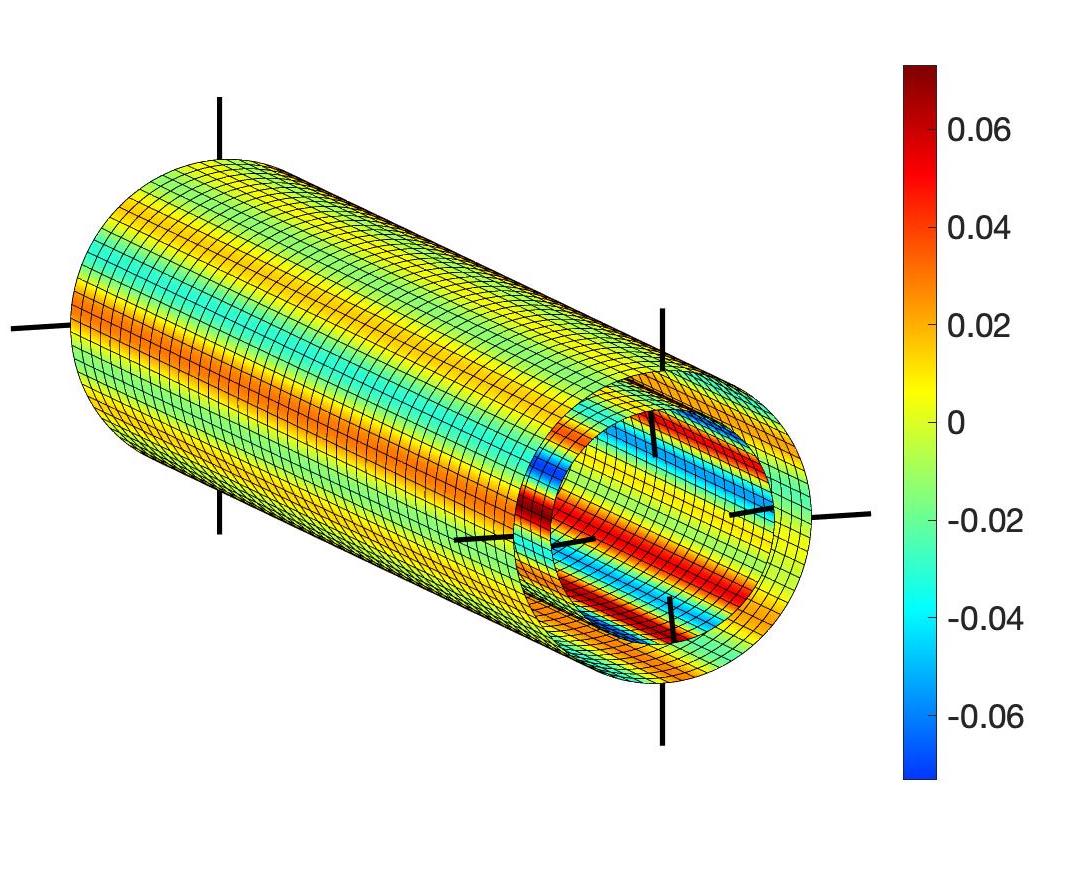}}
\put(-2.95,-.45){\includegraphics[height=25mm]{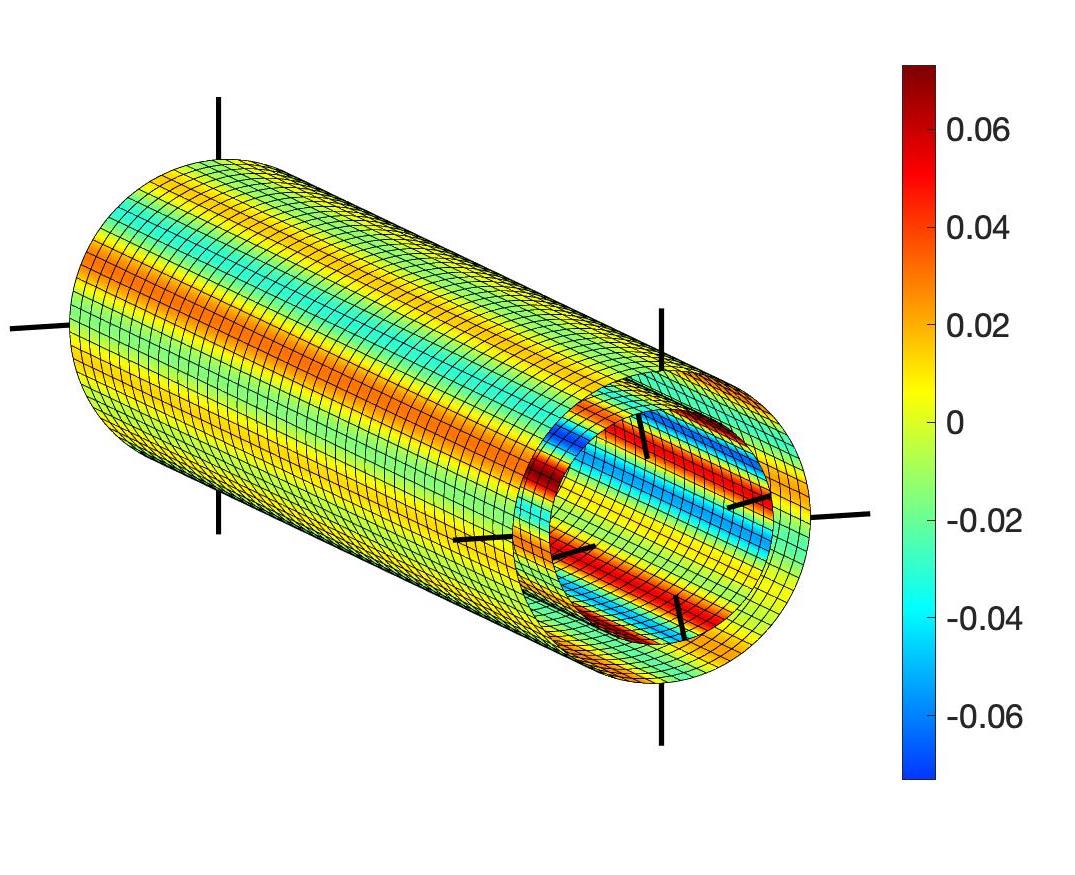}}
\put(-0.45,-.45){\includegraphics[height=25mm]{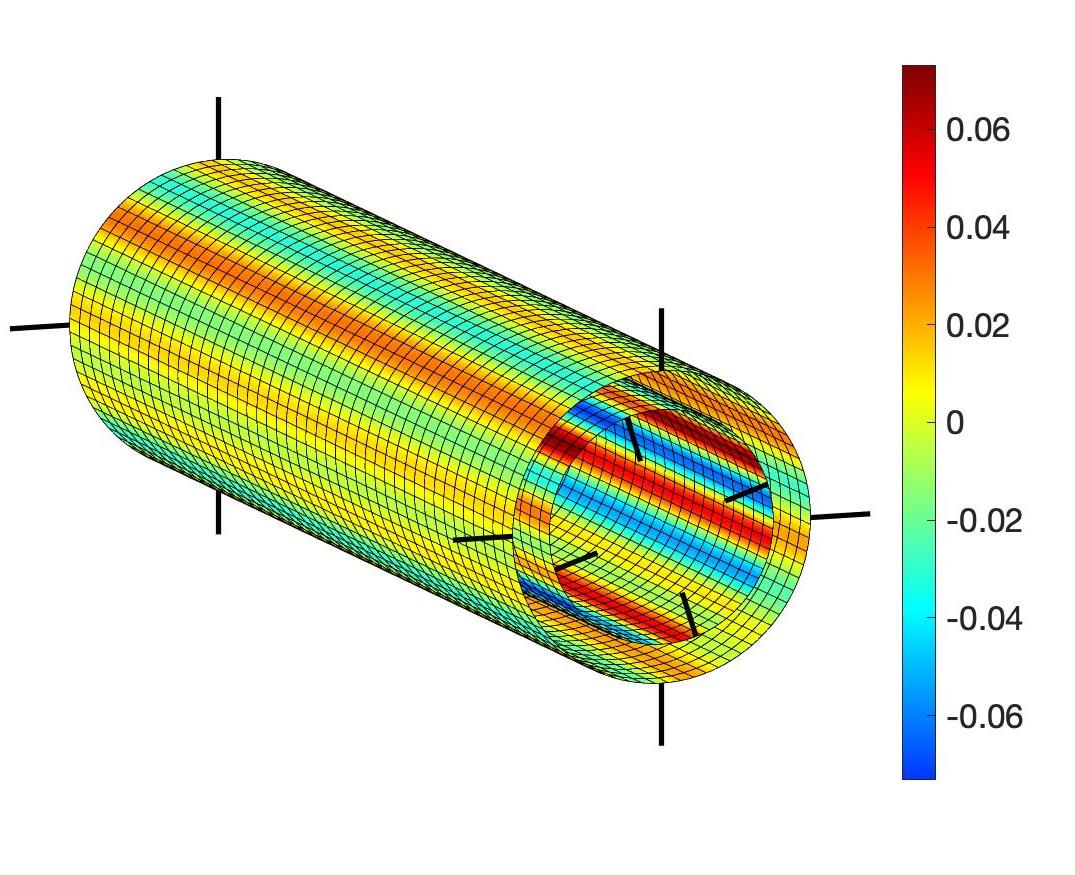}}
\put(2.05,-.45){\includegraphics[height=25mm]{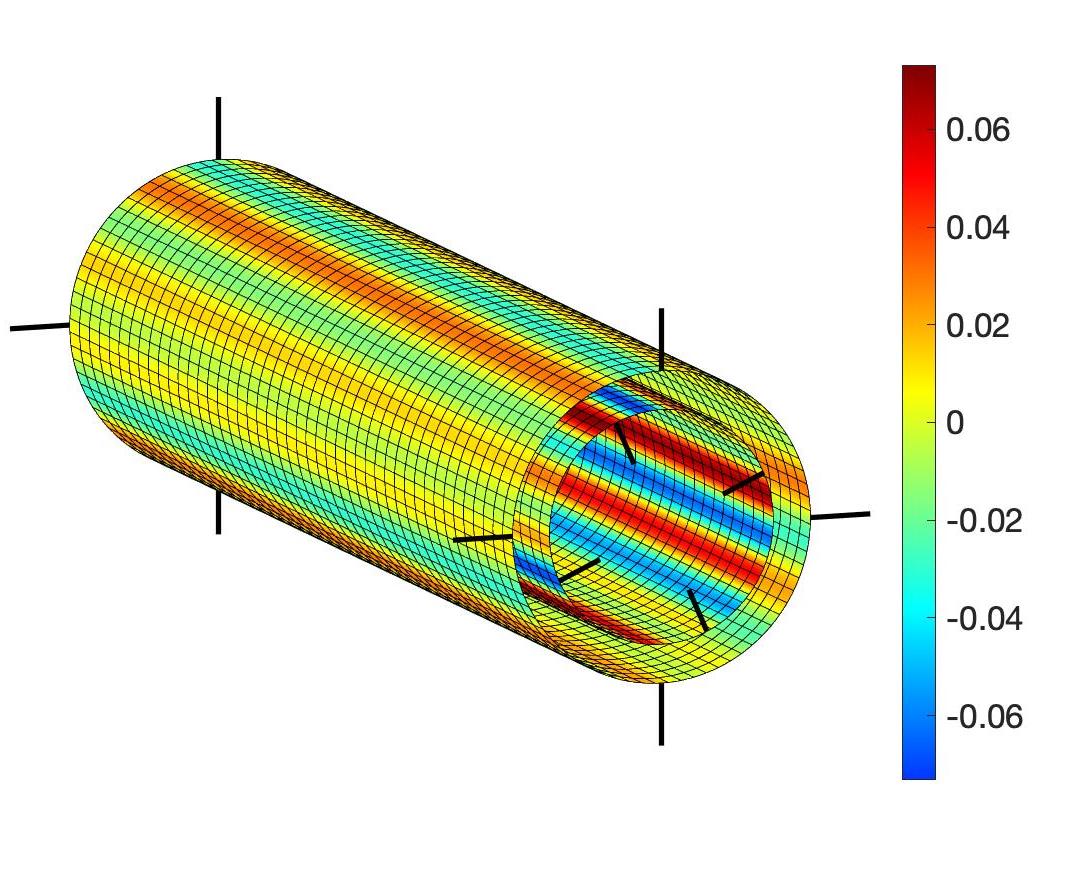}}
\put(4.55,-.45){\includegraphics[height=25mm]{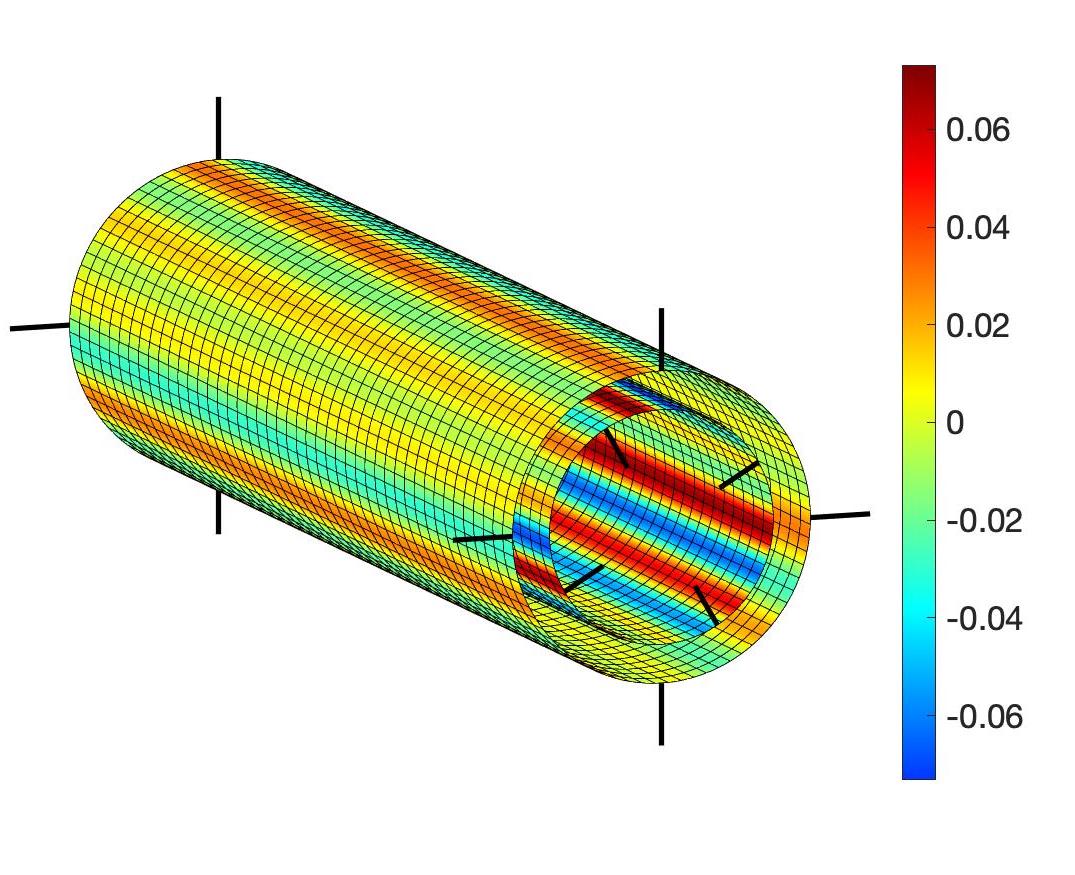}}
\put(7.1,1.85){\includegraphics[height=23mm]{CNT_pcc-2.jpg}}
\put(7.1,-.35){\includegraphics[height=23mm]{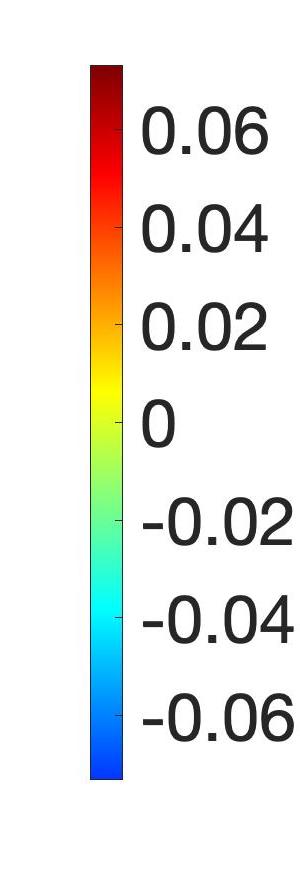}}
\put(-7.85,2.1){\scriptsize{$0^\circ$}}
\put(-5.35,2.1){\scriptsize{$6^\circ$}}
\put(-2.85,2.1){\scriptsize{$12^\circ$}}
\put(-0.35,2.1){\scriptsize{$18^\circ$}}
\put(2.15,2.1){\scriptsize{$24^\circ$}}
\put(4.65,2.1){\scriptsize{$30^\circ$}}
\put(-7.85,0){\scriptsize{$0^\circ$}}
\put(-5.35,0){\scriptsize{$6^\circ$}}
\put(-2.85,0){\scriptsize{$12^\circ$}}
\put(-0.35,0){\scriptsize{$18^\circ$}}
\put(2.15,0){\scriptsize{$24^\circ$}}
\put(4.65,0){\scriptsize{$30^\circ$}}
\end{picture}
\caption{Twisting CNT(21,9) inside CNT(28,12) (Case 3): Color plot of contact pressure $p$ in [GPa] (top row) and circumferential traction $t^2$ in [GPa] (bottom row) at $0^\circ$, $6^\circ$, $12^\circ$, $18^\circ$, $24^\circ$ and $30^\circ$ twist of the inner CNT, from left to right. 
The inner CNT is twisted counterclockwise.
The pressure and traction patterns move counter-clockwise on both CNTs, and they are faster than the twisting rate.
}
\label{f:twist3}
\end{center}
\end{figure}
Figs.~\ref{f:twist2}, \ref{f:twist1}, and \ref{f:twist3} show the contact pressures and circumferential tractions during twisting for the three cases determined from the FE simulations using the DFT parameters. In all three cases, due to the circumferential interference the contact forces and pressures vary in circumferential direction.

\subsubsection{Twisting summary}
As seen in Fig.~\ref{cnt_twist_force}, the torques from FE and analytical integration vanish. The MD torques are non-zero, but they are much smaller than what could have been expected from the pull-out forces observed in Fig.~\ref{cnt_pull_force}: Multiplying the pull-out force amplitudes by the DWCNT radii gives 2.052 nN-nm, 0.102 nN-nm and 0.035 nN-nm for the three cases, which is much higher than the torques observed for Cases 1 and 2. The reason for the difference between MD and FE torques is expected to lie in the insufficient MD boundary conditions and/or the approximate FE contact master surface treatment noted earlier. {The analytical result are based on the assumption that all deformations are negligible. As a consequence, their interaction integrates to zero.}

\section{Conclusion}\label{conclusion}
This work proposes a new continuum contact model to describe the interlayer interactions of curved graphene sheets in continuum formulations such as the finite element method. The interaction between two flat graphene layers shows non-dissipative sliding behavior when the separation gap between the two layers is larger than 0.29 nm. Thus, the interaction energy can be modeled using a surface potential that is then calibrated for various separation gaps between the sheets. The calibrated continuum model captures the sliding anisotropy of bilayer graphene for general sliding distances, both for the interaction potential and the resulting contact traction. The proposed continuum model is then implemented in a curvilinear finite element shell formulation to study the interactions of DWCNTs. Zigzag CNTs, whose axis is along the armchair direction, show maximum resistance to sliding, while the minimum is for chiral CNTs. The periodicity of pull-out forces and torques also {depends} on the chirality of DWCNTs. The FE simulations capture {these} CNT pull-out and twisting interactions sufficiently well.
\section*{Acknowledgements}
Financial support from the German Research Foundation (DFG) through grant SA1822/8-1 is gratefully acknowledged. The authors also thank Reza Ghaffari and Thang X. Duong for their support. 
\appendix
\section{Comparison of interlayer interaction potentials}\label{s:contact_pressure}\label{comp_pot}
Here, we {describe the interatomic potentials used and} compare the calibration results of Sec.~\ref{s:CCex} with those obtained from different interaction potentials available in the literature, such as the  Kolmogorov and Crespi \citep{Kolmogorov2005} and Lebedeva et al. \citep{Lebedeva2011_01} potentials.\\
 {The REBO potential is given by 
\eqb{l}
\ds E_{\text{REBO}} = \ds \sum_I \sum_{J=I+1}\big[E_{\text{R}}(r_{IJ})+b_{IJ}\,E_{\text{A}}(r_{IJ})\big]\,,
\label{eq:rebo}\eqe
where $r_{IJ} $ is the distance between the pair of atoms $I$ and $J$, and $b_{IJ}$ is an empirical bond-order term. $E_{\text{R}}$ and $E_{\text{A}}$ are, respectively, the repulsive and attractive terms taken from Stuart et al. \citep{Stuart2000}. The LJ term is given by
\eqb{l} 
\ds E_{\text{LJ}} = \ds 4 \epsilon_c \left[ \left(\frac{\sigma_c}{r_{IJ}}\right)^{12} - \left(\frac{\sigma_c}{r_{IJ}}\right)^6 \right]\,,
\label{eq:vdw}\eqe
where $\sigma_c = 3.4$ \AA \ and $\epsilon_c = 2.8437$ meV are the LJ parameters for carbon. In the REBO+LJ potential, the REBO part describes the short-range interactions, whereas the LJ part describes the nonbonded vdW interactions (see Stuart et al. \citep{Stuart2000} for details and the potential parameters).} 
The KC potential is given by  \citep{Kolmogorov2005} 
\eqb{l} 
\begin{split}
\ds E_{\text{KC}} &= \ds e^{-\tilde{\lambda}(r_{IJ}-z_0)}\left[C+f(\rho_{IJ})+f(\rho_{JI}) \right]-A\left(\frac{z_0}{r_{IJ}} \right)^{6}\,, \\
\ds \rho_{IJ}^2 &= \ds r_{IJ}^2-(\bn_I\cdot\br_{IJ})^2\,, \\
\ds \rho_{JI}^2 &= \ds r_{IJ}^2-(\bn_J\cdot\br_{IJ})^2\,, 
\text{and} \\
\ds f(\rho) &= \ds e^{-(\rho/\delta)^2}\sum_{n=0}^2 C_{2n}(\rho/\delta)^{2n}\,,
\end{split}
\label{eq:vdw_KC}\eqe
where the vector $\bn_k\,(k = I, J)$ is normal to the sp$^2$ plane in the vicinity of atom $k$, and $z_0 = 3.33$\AA, $C_0 = 21.84$meV, $C_2 = 12.06$meV, $C_4 = 4.711$meV, $C =6.678 \cdot 10^{-4}$meV, $\delta = 0.7718$\AA, $\tilde{\lambda} = 3.143$\AA$^{-1}$, and $A = 12.66$meV are the potential constants, taken from  Ouyang et al.\citep{Ouyang2018}.

The Lebedeva potential function is given by \citep{Lebedeva2011_01} 
\eqb{l} 
\ds E_{\text{Lebedeva}} = \ds B\,e^{-\alpha(r_{IJ}-z_0)}+C(1+D_1\,\rho_{IJ}^2+D_2\,\rho_{IJ}^4)e^{-\tilde{\lambda}_1\,\rho_{IJ}^2-\tilde{\lambda}_2(z_{IJ}^2-z_0^2)}-A\left(\frac{z_0}{r_{IJ}} \right)^{6}\,,
\label{eq:vdw_Leb}\eqe
where $A = 10.510$meV, $B = 11.652$meV, $C = 35.883$meV, $z_0 = 3.34$\AA, $\alpha = 4.16$\AA$^{-1}$, $D_1 = -0.86232$\AA$^{-2}$, $D_2 = 0.1005$\AA$^{-4}$, $\tilde{\lambda}_1 = 0.487$\AA$^{-2}$, and $\tilde{\lambda}_2 = 0.46445$\AA$^{-2}$ are the potential constants \citep{Lebedeva2011_01}.  

\begin{figure}[h]
\begin{center} \unitlength1cm
\begin{picture}(0,5.2)
\put(-8,0){\includegraphics[height=50mm]{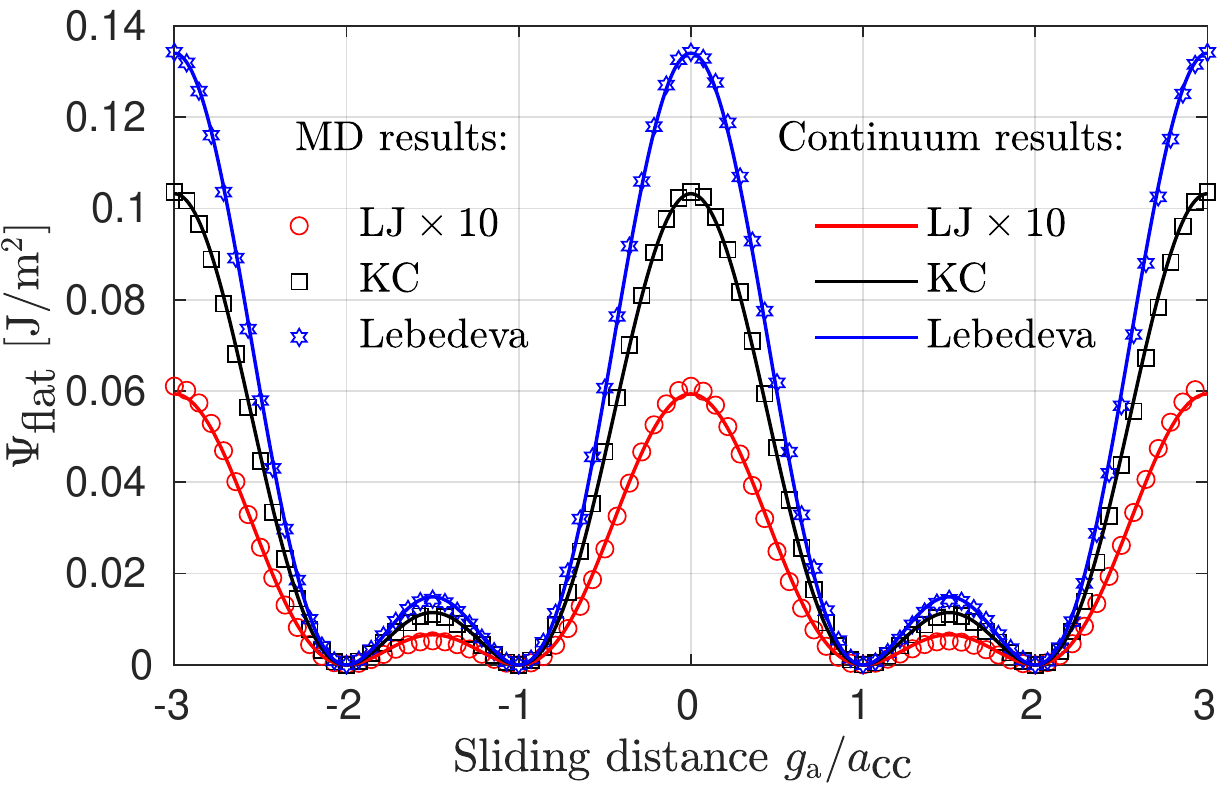}}
\put(0.2,0){\includegraphics[height=50mm]{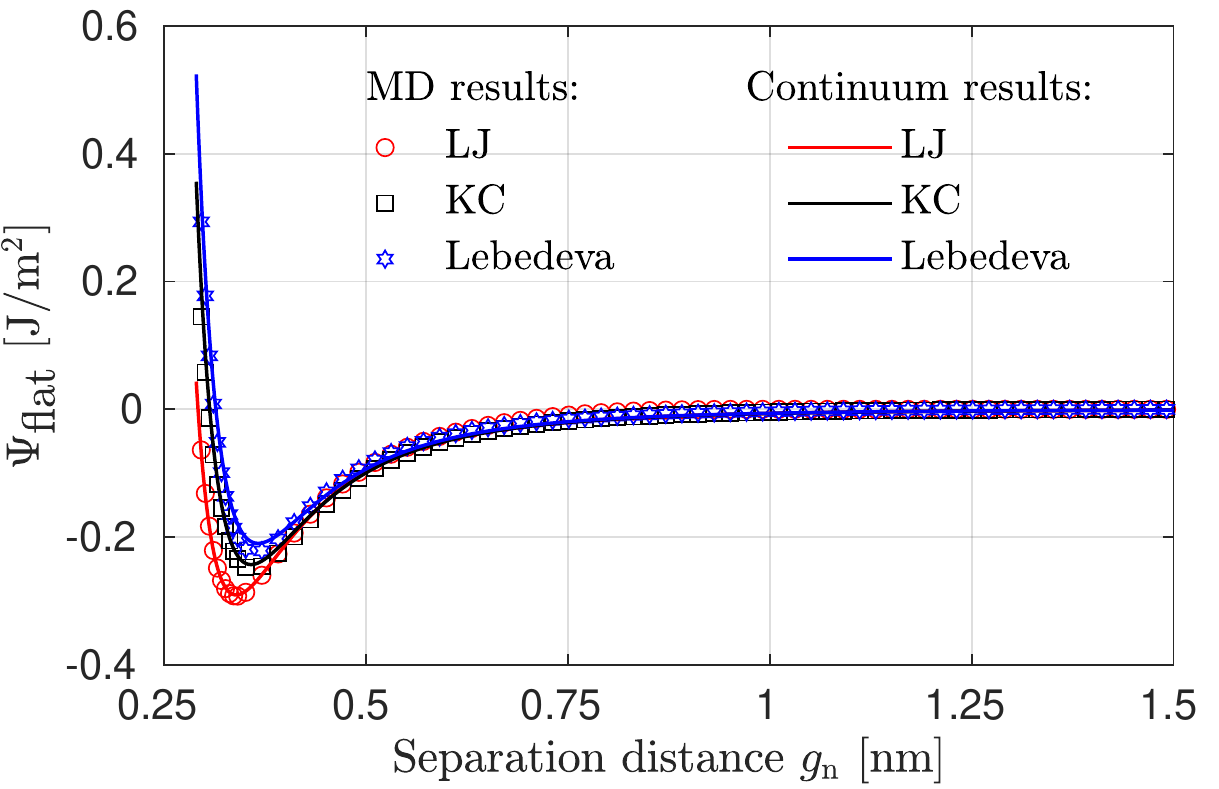}}
\put(-7.7,0){(a)}
\put(.3,0){(b)}
\end{picture}
\caption{Comparison of the interaction energy for {(a)~sliding along the armchair path at $g_\mathrm{n} = 0.3366 $ nm and (b)} for $g_\mathrm{a} = 0 $ and $g_\mathrm{z} = 0 $ (AA stacking) as a function of separation distance $g_\mathrm{n}$. {For better comparability, the results in (a) are plotted relative to the global minimum and the LJ results are scaled by a factor of 10. } } 
\label{fig:energy_comp}
\end{center}
\end{figure}

In the MD simulations, the interlayer interactions are now defined using these potentials. {Fig.~\ref{fig:energy_comp} shows that their normal contact and tangential sliding behavior are qualitatively the same. Therefore, the} same potential ansatz functions for $\Psi_{\mathrm{flat}}$, $\bar\Psi_\mrt$, $\Psi_1$ and $\Psi_2$ are used and the same procedure described in Sec.~\ref{s:CCex} is followed to determine the constants in Eqs.~\eqref{e:psi1} and \eqref{e:psi2}. These values are listed in Tab.~\ref{fitting_param}. 
\begin{table}[!htbp]
\centering
\begin{tabular}{|c|c|c|c|c|}
  \hline 
  Potential & $p_{01}$ [nN/nm$^2$] & $g_{01}$ [nm]  & $p_{02}$ [nN/nm$^2$] & $g_{02}$ [nm] \\[1mm] \hline 
LJ  & 5.8646 & 0.3376 & $4.404 \cdot 10^6$ & $1.875 \cdot 10^{-2}$ \\[.5mm]
KC  & 5.6448 & 0.3410 & $3.306 \cdot 10^4$ & $3.140 \cdot 10^{-2}$ \\[.5mm]
Lebedeva  & 4.9625 & 0.3460 & $3.985 \cdot 10^4$ & $3.160 \cdot 10^{-2}$ \\[.5mm]
   \hline
\end{tabular}
\caption{Fitting constants of Eq.~\eqref{e:psi1} and Eq.~\eqref{e:psi2} for different interaction potentials.}
\label{fitting_param}
\end{table}

The potential relief characteristics, such as the relative energy between AA and AB stacking ($\Delta \Psi_{\mathrm{flat}}^{\text{AA}}$), the relative energy between AB and SP stacking ($\Delta \Psi_{\mathrm{flat}}^{\text{SP}}$), and equilibrium distances of different stackings are listed in Tab.~\ref{pot_comp}.
\begin{table}[!htbp]
\centering
\begin{tabular}{|c|c|c|c|c|c|c|}
  \hline 
  Potential & \thead{$\Psi_{\mathrm{flat}}^{\text{AB}}$ \\ $\text{\big[meV/atom\big]}$} & \thead{$\Delta \Psi_{\mathrm{flat}}^{\text{AA}}$ \\ $\text{\big[meV/atom\big]}$} & \thead{$\Delta \Psi_{\mathrm{flat}}^{\text{SP}}$ \\ $\text{\big[meV/atom\big]}$} & \thead{$d_0^\text{AA}$ \\ $\text{\big[nm\big]}$} & \thead{$d_0^\text{AB}$\\ $\text{\big[nm\big]}$} & \thead{$d_0^\text{SP}$ \\ $\text{\big[nm\big]}$} \\[0mm] \hline 
LJ & -45.60 & 1.00 & 0.095 & 0.3394 & 0.3366 & 0.3370 \\[.5mm]
KC & -48.90 & 15.40 & 1.600 & 0.3580 & 0.3307 & 0.3343 \\[.5mm]
Lebedeva & -47.20 & 19.90 & 2.100 & 0.3667 & 0.3326 & 0.3373 \\[.5mm]
DFT & -50.60 & 19.50 & 2.070 & - & 0.3325 & - \\[0.5mm]
   \hline
\end{tabular}
\caption{Interaction energy of the AB stacking ($\Psi_{\mathrm{flat}}^{\text{AB}}$), relative energies between AA and AB stacking ($\Delta \Psi_{\mathrm{flat}}^{\text{AA}}$) and between SP and AB stacking ($\Delta \Psi_{\mathrm{flat}}^{\text{SP}}$) at equilibrium separation distance of the AB stacking, and equilibrium distances ($d_0$)  obtained from different interaction potentials. The DFT results are from Lebedeva et al. \citep{Lebedeva2011_01}.}
\label{pot_comp}
\end{table}
The separation distance is set to the equilibrium distance of the AB stacking. The potential relief characteristics obtained from the Lebedava potential agree better with the DFT data \citep{Lebedeva2011_01} than the other two interaction potentials.
{As noted already, the interaction behavior is qualitatively the same. Therefore, according to Eq.~\eqref{eq:tan_trac}, the tangential tractions only differ by the factors $\Psi_2^{\text{KC}}/\Psi_2^{\text{LJ}}$ and $\Psi_2^{\text{Lebedeva}}/\Psi_2^{\text{LJ}}$ for the KC and Lebedeva interaction potentials, respectively. }
\section{Elastic constants for graphene} \label{ele_prop}
\begin{figure}[!htbp]
\centering
  \includegraphics[width=0.5\columnwidth]{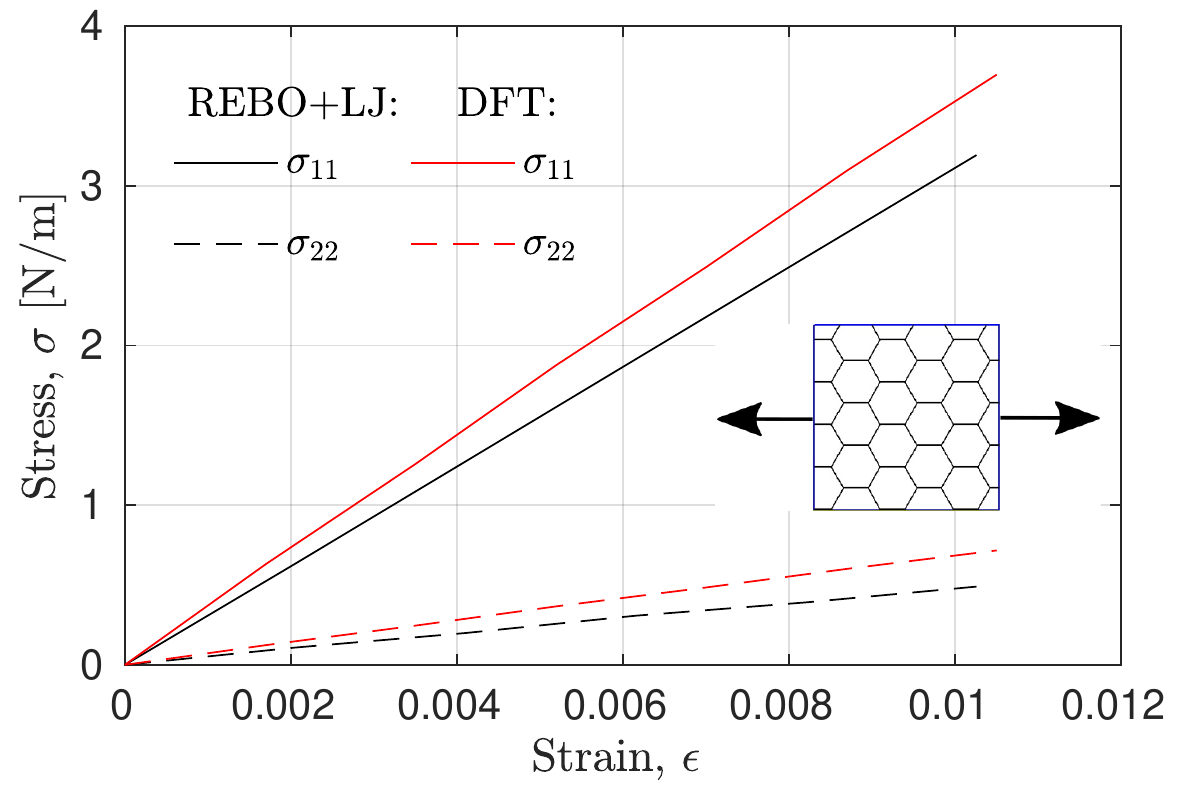}%
\caption{Variation of the stress along the stretch direction, $\sigma_{11}$, and perpendicular to the stretch direction, $\sigma_{22}$, with the strain.     \label{gn_prop}}
\end{figure}

Here, we calculate the elastic properties of a single layer graphene sheet (SLGS). The SLGS is stretched along the armchair direction to calculate these elastic properties while applying constraints to the lateral edge atoms. The stress along the stretch direction ($\sigma_{11}$) and perpendicular to the stretch direction ($\sigma_{22}$) is shown in Fig.~\ref{gn_prop}.
\vspace{5mm}
Within the small strain regime, the SLGS behaves like an isotropic material \citep{Reddy2006}, described by Hooke's law 
\begin{equation}
 \displaystyle \sigma_{ij} = \displaystyle 2\,\mu\,\varepsilon_{ij} + \lambda\,\delta_{ij}\,\varepsilon_{kk}~,
\end{equation}
where $\sigma_{ij}$ and $\varepsilon_{ij}$ are the stress and strain components, respectively, $\delta_{ij}$ is the Kronecker delta, and $\lambda$ and $\mu$ are the Lam\'e constants. The 2D Young's modulus (${E_{2\textnormal{D}}}$) and Poisson's ratio ($\nu$) are then determined through $\mu = {E_{2\textnormal{D}}}/{(2(1+\nu))}$ and $ \lambda =  {2\,\mu\,\nu}/{(1-\nu)}$.
\begin{table}[!htbp]
\centering
\begin{tabular}{|c|c|c|c|c|c|}
  \hline 
  Method/Potential & $\lambda$ (N/m) & $\mu$ (N/m) & $E_{\text{2D}}$ (N/m) & $\nu$ & $c_b$ (nN-nm) \\[0mm] \hline 
REBO+LJ & 48.27 & 131.42 & 304.89 & 0.16 & 0.354 \\[0.5mm]
DFT \citep{Shirazian2018_01}  & 69.01 & 143.05 & 340.46 & 0.19 & 0.242 \\[0.5mm]
   \hline
\end{tabular}
\caption{Elastic properties of SLGS obtained from MD and DFT simulations.}
\label{coil_param}
\end{table}
The bending stiffness $c_b$ is calculated by computing the potential energy of relaxed CNTs of different radii with respect to the ground state graphene sheet \citep{Lu_2009}. The potential energies as a function of curvatures are fitted by quadratic functions. The bending stiffness is then obtained taking double derivatives with respect to the curvature.  

The elastic properties determined from the REBO+LJ potential and DFT simulations available in the literature \citep{Shirazian2018_01} are given in Tab.~\ref{coil_param}. The REBO+LJ potential underestimates the 2D Young's modulus and Poisson's ratio by $\sim 10.5\%$ and $\sim 15.8\%$ compared to DFT, respectively. On the other hand, it overestimates the bending stiffness by $46.3\%$ compared to DFT.
\section{Contact linearization}


For a rigid master surface, the linearization of contact traction $\bt_\mrs$ from Sec.~\ref{s_cont_tra} is characterized by the increment
\eqb{l}
\Delta\bt_\mrs = \ds\pa{\bt_\mrs}{\bx_\mrs}\Delta\bx_\mrs\,. 
\eqe
Applying the product rule to Eq.~\eqref{e:ts} gives
\eqb{l}
\ds\pa{\bt_\mrs}{\bx_\mrs} = \, \ds \bn_\mrp\otimes\pa{p}{\bx_\mrs} + p\,\pa{\bn_\mrp}{\bx_\mrs} 
+ \ba^\mrp_\gamma\otimes\pa{t^\gamma}{\bx_\mrs} + t^\gamma \pa{\ba^\mrp_\gamma}{\bx_\mrs}\,,
\label{e:dtsdxs0}\eqe
where
\eqb{lll}
\ds\pa{p}{\bx_\mrs} \is \, \ds\pa{p}{\gn} \pa{\gn}{\bx_\mrs} + \pa{p}{g^\delta_{\mrc\mrc}} \pa{g^\delta_{\mrc\mrc}}{\bx_\mrs}\,, \\[7mm]
\ds\pa{t^\gamma}{\bx_\mrs} \is \ds\pa{t^\gamma}{\gn} \pa{\gn}{\bx_\mrs} + \pa{t^\gamma}{g^\delta_{\mrc\mrc}} \pa{g^\delta_{\mrc\mrc}}{\bx_\mrs}
\eqe
follow from the chain rule.
From this, Eq.~\eqref{e:dgndx} and Eq.~\eqref{e:dgcdx} then follows
\eqb{lllll}
\bn_\mrp\otimes\ds\pa{p}{\bx_\mrs}
\is \, \ds\pa{p}{\gn}\,\bn_\mrp\otimes\bn_\mrp \plus \ds\pa{p}{g^\delta_{\mrc\mrc}}\,Q^{\delta\mu}_\mrc\,\bn_\mrp\otimes\ba^\mrp_\mu\,, \\[7mm]
\ds\ba^\mrp_\gamma\otimes\pa{t^\gamma}{\bx_\mrs} 
\is \, \ds\pa{t^\gamma}{\gn}\,\ba^\mrp_\gamma\otimes\bn_\mrp \plus \ds\pa{t^\gamma}{g^\delta_{\mrc\mrc}}\,Q^{\delta\mu}_\mrc\,\ba^\mrp_\gamma\otimes\ba^\mrp_\mu\,.
\label{e:dtsdxs1}\eqe
Further
\eqb{l}
\ds\pa{\bn_\mrp}{\bx_\mrs} = \, \ds \pm\ds\frac{1}{R_\mrm\pm\gn}\,\ba^\mrp_2\otimes\ba^\mrp_2 
\label{e:dndx}\eqe
\citep{spbc}, and
\eqb{l}
\ds\pa{\ba^\mrp_1}{\bx_\mrs} = \, \ds \mathbf{0}\,,\quad
\ds\pa{\ba^\mrp_2}{\bx_\mrs} = \, \ds \mp\ds\frac{1}{R_\mrm\pm\gn}\,\bn_\mrp\otimes\ba^\mrp_2\,,
\label{e:dadx}\eqe
which follow from Eq.~\eqref{e:ap2}, Eq.~\eqref{e:dxidxs} and the chain rule 
\eqb{l}
\ds\pa{\ba^\mrp_\alpha}{\bx_\mrs} = \, \ds \pa{\ba^\mrp_\alpha}{\xi^\gamma_\mrp}\otimes\pa{\xi^\gamma_\mrp}{\bx_\mrs}\,.
\eqe
Equations \eqref{e:dndx} and \eqref{e:dadx} can also be written as
\eqb{lll}
\ds\pa{\bn_\mrp}{\bx_\mrs} \is \, \ds M^{\gamma\mu}\,\ba^\mrp_\gamma\otimes\ba^\mrp_\mu\,, \\[7mm]
\ds\pa{\ba^\mrp_\gamma}{\bx_\mrs} \is \, \ds -M_\gamma^\mu\,\bn_\mrp\otimes\ba^\mrp_\mu\,,
\label{e:dtsdxs2}\eqe
with
\eqb{l}
[M^{\alpha\beta}] = -[M_\alpha^\beta] = \left[\begin{matrix}
0 & 0 \\[3mm] 
0 & \ds\frac{\pm1}{R_\mrm\pm\gn}
\end{matrix}\right]\,.
\eqe
Inserting Eq.~\eqref{e:dtsdxs1} and Eq.~\eqref{e:dtsdxs2} into Eq.~\eqref{e:dtsdxs0} then yields the gradient
\eqb{lllll}
\ds\pa{\bt_\mrs}{\bx_\mrs}
\is \, \ds\pa{p}{\gn} \,\bn_\mrp\otimes\bn_\mrp 
\plus \bigg(\ds\pa{p}{g^\delta_{\mrc\mrc}}\,Q^{\delta\mu}_\mrc-t^\gamma\,M_\gamma^\mu\bigg)\,\bn_\mrp\otimes\ba^\mrp_\mu\,, \\[7mm]
\plus \ds\pa{t^\gamma}{\gn} \,\ba^\mrp_\gamma\otimes\bn_\mrp 
\plus \bigg(\ds\pa{t^\gamma}{g^\delta_{\mrc\mrc}}\,Q^{\delta\mu}_\mrc + p\,M^{\gamma\mu}\bigg)\, \ba^\mrp_\gamma\otimes\ba^\mrp_\mu\,,
\eqe
or
\eqb{lll}
\ds\pa{\bt_\mrs}{\bx_\mrs}
\is \, \ds C_{\mrn\mrn}\,\bn_\mrp\otimes\bn_\mrp + C_{\mrn\mra}^\alpha\,\bn_\mrp\otimes\ba^\mrp_\alpha
+ C_{\mra\mrn}^\alpha\,\ba^\mrp_\alpha\otimes\bn_\mrp + C_{\mra\mra}^{\alpha\beta}\,\ba^\mrp_\alpha\otimes\ba^\mrp_\beta\,,
\label{e:dtsdxs}\eqe
with
\eqb{lll}
C_{\mrn\mrn} \dis \ds\pa{p}{\gn}\,, \\[7mm]
C_{\mrn\mra}^\alpha \dis \ds\pa{p}{g^\delta_{\mrc\mrc}}\,Q^{\delta\alpha}_\mrc-t^\gamma\,M_\gamma^\alpha\,, \\[7mm]
C_{\mra\mrn}^\alpha \dis \ds\pa{t^\alpha}{\gn} \,, \\[7mm] 
C_{\mra\mra}^{\alpha\beta} \dis \ds\pa{t^\alpha}{g^\delta_{\mrc\mrc}}\,Q^{\delta\beta}_\mrc + p\,M^{\alpha\beta}\,.
\label{e:Cnn}\eqe
Eqs.~\eqref{e:t}-\eqref{e:tc2} yield
\eqb{lll}
\ds\pa{p}{\gn} \is \, \ds \Sm\,\big(p_1' + p_2'\,\bar\Psi_\mrt \big)\,, \\[7mm]
\ds\pa{p}{g^\delta_{\mrc\mrc}} \is \, \ds -\Sm\,\,p_2\,\bar t^\mrc_\delta\,,
\label{e:dp}\eqe
with $p'_i := \partial p_i/\partial\gn$, $i=1,2$, and
\eqb{lll}
\ds\pa{t^\gamma}{\gn} \is \, \ds -\Sm\,p_2\,\bar t_\alpha^\mrc\,Q^{\alpha\gamma}_\mrc\,, \\[7mm]
\ds\pa{t^\gamma}{g^\delta_{\mrc\mrc}} \is \, \ds \Sm\,\Psi_2\,Q^{\gamma\alpha}_{\mrc\mrT}\,\ds\pa{\bar t^\mrc_\alpha}{g^\delta_{\mrc\mrc}}\,.
\label{e:dt}\eqe
Since 
\eqb{l}
\ds\pa{\bar t^\mrc_\gamma}{g^\delta_{\mrc\mrc}} = \, \ds -\paqq{\bar\Psi_\mrt}{g^\gamma_{\mrc\mrc}}{g^\delta_{\mrc\mrc}} = \pa{\bar t^\mrc_\delta}{g^\gamma_{\mrc\mrc}}\,,
\eqe
Eqs.~\eqref{e:Cnn}-\eqref{e:dt} result in
\eqb{lll}
C_{\mrn\mrn} \is \ds \, \Sm\,\big( p_1' + p_2'\,\bar\Psi_\mrt\big) \,, \\[5mm]	
C_{\mrn\mra}^\alpha \is \, \ds 
	-\Sm\,p_2\,\bar t_\gamma^\mrc\,Q^{\gamma\alpha}_\mrc \,, \\[5mm]	
C_{\mra\mrn}^\alpha \is \, C_{\mrn\mra}^\alpha\,, \\[5mm] 
C_{\mra\mra}^{\alpha\beta} \is \, \ds 
	\Sm\,\Psi_2\,Q^{\alpha\gamma}_{\mrc\mrT}\,\ds\pa{\bar t^\mrc_\gamma}{g^\delta_{\mrc\mrc}}\,Q^{\delta\beta}_\mrc 
	+ p\,M^{\alpha\beta} \,.
\eqe
Since $C_{\mra\mrn}^\alpha = C_{\mrn\mra}^\alpha$ and $C_{\mra\mra}^{\alpha\beta} = C_{\mra\mra}^{\beta\alpha} $, the tangent is fully symmetric as it should be.
\section{Analytical expressions for CNT pull-out and twisting}\label{s_analytical_exp}
The axial pull-out force along $\be_1$ follows from integrating the traction $t^1$ over the slave CNT surface spanned by 
$\xi^1_{\mrp0}\in[-L,\,L]/2$ and $\xi^2_{\mrp0}\in[-\pi,\,\pi]R_\mrs$, i.e.
\eqb{l}
P = -\ds \int_{-\frac{L}{2}}^{\frac{L}{2}}\int_{-\pi}^\pi t^1(g_\mra,g_\mrz)\,R_\mrs\,\dif\bar\xi^{\,2}_{\mrp0}\,\dif\xi^1_{\mrp0}\,.
\label{e:P}\eqe
The axial twisting moment (along $\be_1$) follows from integrating the moment $R_\mrs\,t^2$ over the slave CNT surface, i.e.
\eqb{l}
M_\mrT = -\ds \int_{-\frac{L}{2}}^{\frac{L}{2}}\int_{-\pi}^\pi t^2(g_\mra,g_\mrz)\,R_\mrs^2\,\dif\bar\xi^{\,2}_{\mrp0}\,\dif\xi^1_{\mrp0}\,.
\label{e:M_T}\eqe
The minus signs appear since $P$ and $M_\mrT$ resists the tractions $t^1$ and $t^2$. 
According to Eqs.~\eqref{e:t}, \eqref{e:tc2}, \eqref{e:tc3}, \eqref{e:Qc}, (\ref{e:gQ}.2) and (\ref{e:dxidxs}.2) the latter are given by
\eqb{lll}
t^1 
\is \, \ds  \Sm \,\Psi_2\,\big(\bar t_\mra\cos\theta-\bar t_\mrz\sin\theta\,\big)\,,\\[5mm]
t^2 
\is \, \ds\Sm\,\Psi_2\,\big(\bar t_\mra\sin\theta+\bar t_\mrz\cos\theta\,\big)\frac{R_\mrm}{R_\mrm\pm\gn}\,.
\label{e:t2}\eqe
For rigid CNTs, the slave and master radii $R_\mrs$ and $R_\mrm$ are constant, and $R_\mrm\pm\gn=R_\mrs$.
Expressions \eqref{e:P}-\eqref{e:t2} can then be integrated analytically as is shown below.
As noted in Sec.~\ref{s:curv}, integral type \eqref{e:P} integrates equivalently over the inner CNT surface, the outer CNT surface, or the mid-surface, since
$\bar S = \bar R/R_\mrs$, where $\bar R$ is the average radius. 
Only the sign of $P$ 
differs on the outer and inner surface due to the sign difference of $t^1$ on those surfaces.
Integral type \eqref{e:M_T} integrates differently over both surfaces due to the factor $R_\mrm$ in front that is different on both surfaces.
Equilibrium can therefore only be satisfied if $M_T$ integrates to zero (for rigid CNTs).
\subsection{Pull-out of CNT(26,0) from within CNT(35,0)}\label{s:260P}
In this case the cylinder axis is aligned with the armchair direction ($\cos\theta=1$, $\sin\theta=0$), such that $t^1 = \Sm\,\Psi_2\,\bar t_\mra$ according to Eq.~(\ref{e:t2}.1). 
The initial gap is $G_\mrn = R_\mathrm{out} - R_\mathrm{in} = 9\ell_\mrz/(2\pi)$ and the length is denoted $L = L_\mra$.
%
For rigid cylinders with $u := \xi^1_\mrp - \bar\xi^1_{\mrp0}$, $\phi := \bar\xi^{\,2}_\mrp - \bar\xi^{\,2}_{\mrp0}$ and $\bar\xi^{\,2}_{\mrp0} \in[-\pi,\,\pi]$, the axial and circumferential gaps now becomes
\eqb{l}
g_\mra = u\,,\quad 
g_\mrz = \phi\,R_\mrm \mp\ds\frac{9\ell_\mrz}{2\pi}\bar\xi^{\,2}_{\mrp0}
\label{e:gz2}\eqe 
according to Eq.~\eqref{e:g1} and Eq.~\eqref{e:g2}, respectively.
The rear term of $g_\mrz$ lies in the interval $[-9,\,9]\ell_\mrz/2$.
That is, $g_\mrz$ spans exactly 9 periods of the interaction potential, irrespective of rotation angle $\phi$.
Therefore the rear term in Eq.~(\ref{e:btaz}.1) integrates to zero in Eq.~\eqref{e:P},
while the front term leads to the analytical pull-out force (for all $\phi$)
\eqb{l}
P_\mra(u) = -\ds \Sm\,\Psi_2\,L_\mra\,\bar R\,\int_{-\pi}^\pi \bar t_\mra(g_\mra,g_\mrz)\,\dif\bar\xi^{\,2}_{\mrp0} 
= -P_\mathrm{max} \,\sin\frac{4\pi\,u}{\ell_\mra}\,,
\eqe
with the amplitude
\eqb{l}
P_\mathrm{max} = \ds\frac{8\pi^2}{\ell_\mra}\,\Psi_2\,L_\mra\,\bar R\,.
\eqe
For $2\pi\bar R = 30.5\ell_\mrz$, $\ell_\mrz = \sqrt{3}a_{\mrc\mrc}$ and $L_\mra = 24\ell_\mra$ follows 
$P_\mathrm{max} = 2928\sqrt{3}\pi\,\Psi_2\,a_{\mrc\mrc}$.
The value $a_{\mrc\mrc} = 0.1397\,$nm then gives $G_\mrn = 0.3466\,$nm and $\Psi_2 = 7.7435\cdot10^{-4}\,$nN/nm and $P_\mathrm{max} = 1.7235\,$nN, which is the result shown in Fig.~\ref{cnt_pull_force}a.
\subsection{Pull-out of CNT(15,15) from within CNT(20,20)}\label{s:1515P}
In this case the cylinder axis is aligned with the zigzag direction ($\cos\theta=0$, $\sin\theta=1$), such that $t^1 = -\Sm\,\Psi_2\,\bar t_\mrz$ according to Eq.~(\ref{e:t2}.1). 
The initial gap is $G_\mrn = R_\mathrm{out} - R_\mathrm{in} = 5\ell_\mra/(2\pi)$ and the length is denoted $L = L_\mrz$.
%
For rigid cylinders, $\xi^1_\mrp - \xi^1_{\mrp0}$ corresponds to their relative axial motion $u$, while $\bar\xi^{\,2}_\mrp - \bar\xi^{\,2}_{\mrp0}$ corresponds to their relative rotation angle $\phi$.
Both are constant across the surface.
Further, $\bar\xi^{\,2}_{\mrp0} \in[-\pi,\,\pi]$. 
The axial and circumferential gaps thus become
\eqb{l}
g_\mrz = \ds u\,,\quad 
g_\mra = \ds \phi\,R_\mrm \mp\ds\frac{5\ell_\mra}{2\pi}\bar\xi^{\,2}_{\mrp0}
\label{e:ga1}\eqe 
according to Eq.~\eqref{e:g1} and Eq.~\eqref{e:g2}, respectively.
The rear term of $g_\mra$ lies in the interval $[-5,\,5]\ell_\mra/2$.
That is, $g_\mra$ spans exactly 5 periods of the interaction potential, irrespective of $\phi$.
Therefore the analytical pull-out force
\eqb{l}
P_\mrz(u,\phi) 
= \ds\Psi_2\,L_\mrz\,\bar R\,\int_{-\pi}^\pi \bar t_\mrz(g_\mra,g_\mrz)\,\dif\bar\xi^{\,2}_{\mrp0}
\eqe
from Eq.~\eqref{e:P} integrates to zero for all $u$ and $\phi$ according to Eq.~(\ref{e:btaz}.2).
\subsection{Pull-out of CNT(21,9) from within CNT(28,12)}\label{s:219P}
In this case $\cos\theta=17/(2\bar c)$ and $\sin\theta=3\sqrt{3}/(2\bar c)$, with $\bar c := \sqrt{79}$, such that 
\eqb{l}
t^1 = \ds\frac{\Sm\,\Psi_2}{2\bar c}\Big(17\,\bar t_\mra-3\sqrt{3}\,\bar t_\mrz\Big)\,,
\eqe
according to Eq.~(\ref{e:t2}.1). 
Inserting Eq.~\eqref{e:btaz} then leads to
\eqb{l}
t^1 = 
\ds\frac{2\pi\,\Sm\,\Psi_2}{\ell_u}\Big(17\sin2\hat g_\mra + 13\sin\big(\hat g_\mra - \hat g_\mrz\big) + 4\sin\big(\hat g_\mra + \hat g_\mrz\big) \Big)\,,
\label{e:t1P3}\eqe
where
\eqb{llllrll}
\hat g_\mra \dis \ds\frac{2\pi\,g_\mra}{\ell_\mra} \is 17\pi\ds\frac{u}{\ell_u} 
+ \frac{3\pi\,R_\mrm}{\ell_\phi}\phi \mp \ds\frac{3}{2}\bar\xi^{\,2}_{\mrp0}\,, \\[7mm]
\hat g_\mrz \dis \ds\frac{2\pi\,g_\mrz}{\ell_\mrz} \is -9\pi\ds\frac{u}{\ell_u} 
+ \frac{17\pi\,R_\mrm}{\ell_\phi}\phi \mp \ds\frac{17}{2}\bar\xi^{\,2}_{\mrp0}
\label{e:hatg}\eqe
follows from Eqs.~\eqref{e:gaz}, \eqref{e:g1}, \eqref{e:g2} and $G_\mrn = \sqrt{3}\,\bar c\,a_\mathrm{cc}/(2\pi)$ with $u := \xi^1_\mrp - \xi^1_{\mrp0}$, $\phi := \bar\xi^{\,2}_\mrp - \bar\xi^{\,2}_{\mrp0}$ 
and $\ell_u := \bar c\,\ell_\mra$, $\ell_\phi := \bar c\,\ell_\mrz$. 
The front terms in Eq.~\eqref{e:hatg} do not change the fact that the three sine-terms in Eq.~\eqref{e:t1P3} contain exactly 3, 7 and 10 full periods within $\bar\xi^{\,2}_{\mrp0} \in[-\pi,\,\pi]$, respectively.
Integral Eq.~\eqref{e:P} therefore vanishes, and the pull-out force becomes zero for all $u$ and $\phi$.
\subsection{Twisting CNT(26,0) inside CNT(35,0)}
In this case the cylinder axis is aligned with the armchair direction ($\cos\theta=1$, $\sin\theta=0$). 
The circumferential traction on the slave surface is therefore $t^2 = \Sm\,\Psi_2\,\bar t_\mrz\, R_\mrm/R_\mrs$ according to Eq.~(\ref{e:t2}.2).
As noted in Sec.~\ref{s:260P}, the initial gap is $G_\mrn = 9\ell_\mrz/(2\pi)$ 
such that $g_\mrz$ spans exactly 9 periods of the interaction potential, see Eq.~\eqref{e:gz2}.
Therefore the analytical twisting moment
\eqb{l}
M_\mrT(\phi,u) = -\ds\Psi_2\,L_\mrz\,\bar R\,R_\mrm\,\int_{-\pi}^\pi \bar t_\mrz(g_\mra,g_\mrz)\,\dif\bar\xi^{\,2}_\mrp
\eqe
from Eq.~\eqref{e:M_T} integrates to zero for all $\phi$ and $u$ according to Eq.~(\ref{e:btaz}.2).
\subsection{Twisting CNT(15,15) inside CNT(20,20)}\label{s:1515M}
In this case the cylinder axis is aligned with the zigzag direction ($\cos\theta=0$, $\sin\theta=1$). 
The circumferential traction on the slave surface is therefore $t^2 = \Sm\,\Psi_2\,\bar t_\mra\, R_\mrm/R_\mrs$, with $\Sm = \bar R/R_\mrs$, according to Eq.~(\ref{e:t2}.2).
As noted in Sec.~\ref{s:1515P}, the initial gap is $G_\mrn = 5\ell_\mra/(2\pi)$ 
such that $g_\mra$ spans exactly 5 periods of the interaction potential, see Eq.~\eqref{e:ga1}.
Therefore the analytical twisting moment
\eqb{l}
\displaystyle M_\mrT(\phi,u) = -\ds\Psi_2\,L_\mrz\,\bar R\,R_\mrm\,\int_{-\pi}^\pi \bar t_\mra(g_\mra,g_\mrz)\,\dif\bar\xi^{\,2}_\mrp
\eqe
from Eq.~\eqref{e:M_T} integrates to zero for all $\phi$ and $u$ according to Eq.~(\ref{e:btaz}.1).
\subsection{Twisting CNT(21,9) inside CNT(28,12)}
In this case $\cos\theta=17/(2\bar c)$ and $\sin\theta=3\sqrt{3}/(2\bar c)$, $\bar c = \sqrt{79}$, such that 
\eqb{l}
\displaystyle t^2 = \ds\frac{\Sm\,\Psi_2\,R_\mrm}{2\bar c\,R_\mrs}\Big(3\sqrt{3}\,\bar t_\mra + 17\,\bar t_\mrz\Big)\,,
\label{e:t2M3}\eqe
according to Eq.~(\ref{e:t2}.2). 
Inserting Eq.~\eqref{e:btaz} then leads to
\eqb{l}
\displaystyle t^2 = \ds\frac{2\pi\,\Sm\,\Psi_2\,R_\mrm}{\ell_\phi\,R_\mrs}\Big(3\sin2\hat g_\mra - 7\sin\big(\hat g_\mra - \hat g_\mrz\big) + 10\sin\big(\hat g_\mra + \hat g_\mrz\big) \Big)\,,
\label{e:t2M3}\eqe
with $\hat g_\mra$ and $\hat g_\mrz$ given in Eq.~\eqref{e:hatg}.
From Eq.~\eqref{e:hatg} again follows that the three sine-terms in Eq.~\eqref{e:t2M3} contain exactly 3, 7 and 10 full periods within $\bar\xi^{\,2}_{\mrp0} \in[-\pi,\,\pi]$, respectively.
Integral Eq.~\eqref{e:M_T} therefore vanishes, and the twisting moment again becomes zero for all $\phi$ and $u$.

\bibliographystyle{apalike}
\bibliography{achemso}

\end{document}

%% file: neco_updated.tex
\newcommand{\bitm}{\begin{itemize}}
\newcommand{\eitm}{\end{itemize}}
\newcommand{\bnumr}{\begin{enumerate}}
\newcommand{\enumr}{\end{enumerate}}

\newcommand{\mrT}{\mathrm{T}}

\newcommand {\eqb}[1]{\begin{equation}\begin{array}{#1}}
\newcommand {\eqe}{\end{array}\end{equation}}

\newcommand {\esb}[1]{\begin{equation*}\begin{array}{#1}}
\newcommand {\ese}{\end{array}\end{equation*}}
\newcommand {\ds}{\displaystyle}

\newcommand {\pa}[2]{\frac{\partial{#1}}{\partial{#2}}}

\newcommand {\paqq}[3]{\frac{\partial^2{#1}}{\partial{#2}\,\partial{#3}}}
\newcommand {\back}{\! \! \!}
\newcommand {\is}{\back &=& \back}
\newcommand {\dis}{\back &:=& \back}

\newcommand {\ais}{\back &\approx& \back}

\newcommand {\plus}{\back &+& \back}

\newcommand {\norm}[1]{\|#1\|}

\newcommand {\sign}{\mathrm{sign}\,}

\newcommand {\dif}{\mathrm{d}}

\newcommand {\II}{{I\kern-.3em I}}
\newcommand {\III}{{I\kern-.3em I\kern-.3em I}}

\newcommand {\mra}{\mathrm{a}}
\newcommand {\mrb}{\mathrm{b}}
\newcommand {\mrc}{\mathrm{c}}
\newcommand {\mrd}{\mathrm{d}}
\newcommand {\mre}{\mathrm{e}}

\newcommand {\mri}{\mathrm{i}}

\newcommand {\mrm}{\mathrm{m}}
\newcommand {\mrn}{\mathrm{n}}
\newcommand {\mro}{\mathrm{o}}
\newcommand {\mrp}{\mathrm{p}}

\newcommand {\mrs}{\mathrm{s}}
\newcommand {\mrt}{\mathrm{t}}

\newcommand {\mrz}{\mathrm{z}}

\newcommand {\mf}{\mathbf{f}}

\newcommand {\mk}{\mathbf{k}}

\newcommand {\muu}{\mathbf{u}}

\newcommand {\mx}{\mathbf{x}}

\newcommand {\ba}{\boldsymbol{a}}

\newcommand {\be}{\boldsymbol{e}}

\newcommand {\bg}{\boldsymbol{g}}

\newcommand {\bn}{\boldsymbol{n}}

\newcommand {\br}{\boldsymbol{r}}

\newcommand {\bt}{\boldsymbol{t}}
\newcommand {\bu}{\boldsymbol{u}}

\newcommand {\bx}{\boldsymbol{x}}

\newcommand {\mN}{\mathbf{N}}

\newcommand {\mX}{\mathbf{X}}

\newcommand {\bF}{\boldsymbol{F}}

\newcommand {\bP}{\boldsymbol{P}}

\newcommand {\bX}{\boldsymbol{X}}

\newcommand {\eps}{\varepsilon}

\newcommand {\bone}{\mathbf{1}}

\newcommand {\IR}{{\rm\kern.24em
   \vrule width.02em height1.53ex depth-.05ex
   \kern-.3em R}}
\newcommand {\ic}{{\rm\kern.20em
   \vrule width.02em height1.0ex depth-.05ex
   \kern-.22em c}}
\newcommand {\ia}{{\rm\kern.20em
   \vrule width.02em height1.05ex depth-.0ex
   \kern-.25em a}}
\newcommand {\IC}{{\rm\kern.24em
   \vrule width.02em height1.4ex depth-.05ex
   \kern-.26em C}}
\newcommand {\ID}{{\rm\kern.34em
   \vrule width.02em height1.5ex depth-.05ex
   \kern-.36em D}}
\newcommand {\IS}{{\rm\kern.24em
   \vrule width.02em height1.6ex depth.05ex
   \kern-.26em S}}
\newcommand {\IT}{{\rm\kern.50em
   \vrule width.02em height1.55ex depth-.05ex
   \kern-.52em T}}

\newcommand {\IE}{{\rm\kern.24em
   \vrule width.02em height1.55ex depth-.05ex
   \kern-.33em E}}
\newcommand {\IEa}{{\rm\kern.24em
   \vrule width.02em height1.55ex depth-.05ex
   \kern-.33em E}^{1}_{ijkl}}
\newcommand {\IEb}{{\rm\kern.24em
   \vrule width.02em height1.55ex depth-.05ex
   \kern-.33em E}^{2}_{ijkl}}

\newcommand {\sS}{\mathcal{S}}

\newcommand {\Ass}[2]{\kern 0.9ex \vrule width0.45em height0.2ex depth0ex \kern -2.1ex \bigwedge_{#1}^{#2}}
\newcommand {\ASS}[2]{\kern 1.45ex \vrule width0.5em height0.2ex depth0ex \kern -2.65ex \bigwedge_{#1}^{#2}}




%% file: friction.bbl
\begin{thebibliography}{}

\bibitem[Afsharirad et~al., 2021]{AFSHARIRAD2021}
Afsharirad, F., Mousanezhad, S., Biglari, H., and Rahmani, O. (2021).
\newblock Molecular dynamics of axial interwall van der {W}aals force and
  mechanical vibration of double-walled carbon nanotubes.
\newblock {\em Mater. Today Comm.}, 28:102708.

\bibitem[Arciniega and Reddy, 2005]{arciniega05}
Arciniega, R.~A. and Reddy, J.~N. (2005).
\newblock Tensor-based finite element formulation for geometrically nonlinear
  analysis of shell structures.
\newblock {\em AIAA J.}, {43}(9):2024--2038.

\bibitem[Arroyo and Belytschko, 2004a]{Arroyo2004b}
Arroyo, M. and Belytschko, T. (2004a).
\newblock Finite crystal elasticity of carbon nanotubes based on the
  exponential {C}auchy-{B}orn rule.
\newblock {\em Phys. Rev. B}, 69:115415.

\bibitem[Arroyo and Belytschko, 2004b]{Arroyo2004a}
Arroyo, M. and Belytschko, T. (2004b).
\newblock Finite element methods for the non-linear mechanics of crystalline
  sheets and nanotubes.
\newblock {\em Int. J. Numer. Methods Eng.}, 59(3):419--456.

\bibitem[Arroyo and Belytschko, 2005]{arroyo2005}
Arroyo, M. and Belytschko, T. (2005).
\newblock Continuum mechanics modeling and simulation of carbon nanotubes.
\newblock {\em Meccanica}, { 40}:455--469.

\bibitem[Bae et~al., 2010]{Bae2010}
Bae, S., Kim, H., Lee, Y., Xu, X., Park, J.-S., Zheng, Y., Balakrishnan, J.,
  Lei, T., Ri~Kim, H., Song, Y.~I., et~al. (2010).
\newblock Roll-to-roll production of 30-inch graphene films for transparent
  electrodes.
\newblock {\em Nature Nanotech.}, 5(8):574--578.

\bibitem[Ba{\c{s}}ar and Ding, 1996]{basar96}
Ba{\c{s}}ar, Y. and Ding, Y. (1996).
\newblock Finite-element analysis of hyperelastic thin shells with large
  strains.
\newblock {\em Comput. Mech}, {18}(3):200--214.

\bibitem[Berman et~al., 2014]{Berman2014_01}
Berman, D., Erdemir, A., and Sumant, A.~V. (2014).
\newblock Graphene: A new emerging lubricant.
\newblock {\em Mater. Today}, { 17}(1):31--42.

\bibitem[Buczkowski and Kleiber, 1997]{buczkowski97}
Buczkowski, R. and Kleiber, M. (1997).
\newblock Elasto-plastic interface model for 3{D}-frictional orthotropic
  contact problems.
\newblock {\em Int. J. Numer. Methods Eng.}, 40(4):599--619.

\bibitem[Bunch et~al., 2007]{Bunch2007}
Bunch, J.~S., van~der Zande, A.~M., Verbridge, S.~S., Frank, I.~W., Tanenbaum,
  D.~M., Parpia, J.~M., Craighead, H.~G., and McEuen, P.~L. (2007).
\newblock Electromechanical resonators from graphene sheets.
\newblock {\em Science}, 315(5811):490--493.

\bibitem[Cao et~al., 2018]{Cao2018}
Cao, Y., Fatemi, V., Fang, S., Watanabe, K., Taniguchi, T., Kaxiras, E., and
  Jarillo-Herrero, P. (2018).
\newblock Unconventional superconductivity in magic-angle graphene
  superlattices.
\newblock {\em Nature}, 556(7699):43--50.

\bibitem[Cox, 1952]{Cox_1952}
Cox, H.~L. (1952).
\newblock The elasticity and strength of paper and other fibrous materials.
\newblock {\em Br. J. Appl. Phys.}, 3(3):72.

\bibitem[Dienwiebel et~al., 2005]{Dienwiebel2005}
Dienwiebel, M., Pradeep, N., Verhoeven, G.~S., Zandbergen, H.~W., and Frenken,
  J.~W. (2005).
\newblock Model experiments of superlubricity of graphite.
\newblock {\em Surf. Sci.}, 576(1-3):197--211.

\bibitem[Dienwiebel et~al., 2004]{Dienwiebel2004a}
Dienwiebel, M., Verhoeven, G.~S., Pradeep, N., Frenken, J. W.~M., Heimberg,
  J.~A., and Zandbergen, H.~W. (2004).
\newblock Superlubricity of graphite.
\newblock {\em Phys. Rev. Lett.}, 92:126101.

\bibitem[Dresselhaus et~al., 1995]{DRESSELHAUS1995}
Dresselhaus, M., Dresselhaus, G., and Saito, R. (1995).
\newblock Physics of carbon nanotubes.
\newblock {\em Carbon}, 33(7):883--891.

\bibitem[Duong et~al., 2017]{solidshell}
Duong, T.~X., Roohbakhshan, F., and Sauer, R.~A. (2017).
\newblock A new rotation-free isogeometric thin shell formulation and a
  corresponding continuity constraint for patch boundaries.
\newblock {\em Comput. Methods Appl. Mech. Eng.}, {316}:43--83.

\bibitem[Evans and Holian, 1985]{Evans1985}
Evans, D. and Holian, B. (1985).
\newblock The {N}ose-–{H}oover thermostat.
\newblock {\em J. Chem. Phys.}, 83(8):4069--4074.

\bibitem[Feng et~al., 2013]{Feng2013}
Feng, X., Kwon, S., Park, J.~Y., and Salmeron, M. (2013).
\newblock Superlubric sliding of graphene nanoflakes on graphene.
\newblock {\em ACS Nano}, 7(2):1718--1724.
\newblock PMID: 23327483.

\bibitem[Ghaffari et~al., 2017]{graphene}
Ghaffari, R., Duong, T.~X., and Sauer, R.~A. (2017).
\newblock A new shell formulation for graphene structures based on existing
  ab-initio data.
\newblock {\em Int. J. Solids Struc.}, {135}:37--60.

\bibitem[Girifalco et~al., 2000]{Girifalco_2000}
Girifalco, L.~A., Hodak, M., and Lee, R.~S. (2000).
\newblock Carbon nanotubes, buckyballs, ropes, and a universal graphitic
  potential.
\newblock {\em Phys. Rev. B}, 62:13104--13110.

\bibitem[Guo et~al., 2007]{Guo2007}
Guo, Y., Guo, W., and Chen, C. (2007).
\newblock Modifying atomic-scale friction between two graphene sheets: A
  molecular-force-field study.
\newblock {\em Phys. Rev. B}, 76:155429.

\bibitem[Hu et~al., 2022]{hu22}
Hu, L., Cong, Y., Renaud, C., and Feng, Z.-Q. (2022).
\newblock A bi-potential contact formulation of orthotropic adhesion between
  soft bodies.
\newblock {\em Comput. Mech.}, 69(4):931--945.

\bibitem[Jiang and Park, 2015]{Jiang2015_01}
Jiang, J.-W. and Park, H.~S. (2015).
\newblock A {Gaussian} treatment for the friction issue of {Lennard-Jones}
  potential in layered materials: Application to friction between graphene,
  {MoS$_2$}, and black phosphorus.
\newblock {\em J. Appl. Phys.}, 117(12):124304.

\bibitem[Jones, 1924]{Lj1924}
Jones, J. (1924).
\newblock On the determination of molecular fields. {II}. from the equation of
  state of a gas.
\newblock {\em Proc. R. Soc. A}, 106(738):463--477.

\bibitem[Jones and Papadopoulos, 2006]{jones06}
Jones, R.~E. and Papadopoulos, P. (2006).
\newblock Simulating anisotropic frictional response using smoothly
  interpolated traction fields.
\newblock {\em Comput. Methods Appl. Mech. Eng.}, 195(7):588--613.

\bibitem[Kolmogorov and Crespi, 2005]{Kolmogorov2005}
Kolmogorov, A.~N. and Crespi, V.~H. (2005).
\newblock Registry-dependent interlayer potential for graphitic systems.
\newblock {\em Phys. Rev. B}, 71:235415.

\bibitem[Konyukhov and Schweizerhof, 2006a]{Konyukhov2006_01}
Konyukhov, A. and Schweizerhof, K. (2006a).
\newblock Covariant description of contact interfaces considering anisotropy
  for adhesion and friction: Part 1. {Formulation} and analysis of the
  computational model.
\newblock {\em Comput. Methods Appl. Mech. Eng.}, {196}(1):103--117.

\bibitem[Konyukhov and Schweizerhof, 2006b]{Konyukhov2006_02}
Konyukhov, A. and Schweizerhof, K. (2006b).
\newblock Covariant description of contact interfaces considering anisotropy
  for adhesion and friction: Part 2. {Linearization}, finite element
  implementation and numerical analysis of the model.
\newblock {\em Comput. Methods Appl. Mech. Eng.}, {196}(1):289--303.

\bibitem[Koshino and Nam, 2020]{Koshino2020}
Koshino, M. and Nam, N. N.~T. (2020).
\newblock Effective continuum model for relaxed twisted bilayer graphene and
  moir\'e electron-phonon interaction.
\newblock {\em Phys. Rev. B}, { 101}:195425.

\bibitem[Kumar et~al., 2016]{Kumar2016}
Kumar, H., Dong, L., and Shenoy, V.~B. (2016).
\newblock Limits of coherency and strain transfer in flexible 2d van der waals
  heterostructures: formation of strain solitons and interlayer debonding.
\newblock {\em Sci. Rep.}, 6(1):1--8.

\bibitem[Laursen, 2002]{laursen}
Laursen, T.~A. (2002).
\newblock {\em Computational Contact and Impact Mechanics: Fundamentals of
  modeling interfacial phenomena in nonlinear finite element analysis}.
\newblock Springer, Berlin, Heidelberg.

\bibitem[Lebedeva et~al., 2010]{Lebedeva2010}
Lebedeva, I.~V., Knizhnik, A.~A., Popov, A.~M., Ershova, O.~V., Lozovik, Y.~E.,
  and Potapkin, B.~V. (2010).
\newblock Fast diffusion of a graphene flake on a graphene layer.
\newblock {\em Phys. Rev. B}, 82:155460.

\bibitem[Lebedeva et~al., 2011a]{Lebedeva2011}
Lebedeva, I.~V., Knizhnik, A.~A., Popov, A.~M., Ershova, O.~V., Lozovik, Y.~E.,
  and Potapkin, B.~V. (2011a).
\newblock Diffusion and drift of graphene flake on graphite surface.
\newblock {\em J. Chem. Phys.}, 134(10):104505.

\bibitem[Lebedeva et~al., 2011b]{Lebedeva2011_01}
Lebedeva, I.~V., Knizhnik, A.~A., Popov, A.~M., Lozovik, Y.~E., and Potapkin,
  B.~V. (2011b).
\newblock Interlayer interaction and relative vibrations of bilayer graphene.
\newblock {\em Phys. Chem. Chem. Phys.}, {13}:5687--5695.

\bibitem[Leven et~al., 2016]{Leven2016}
Leven, I., Maaravi, T., Azuri, I., Kronik, L., and Hod, O. (2016).
\newblock Interlayer potential for graphene/h-bn heterostructures.
\newblock {\em J. Chem. Theory Comput.}, { 12}(6):2896--2905.

\bibitem[Li and Kim, 2020]{LI2020}
Li, H. and Kim, W.~K. (2020).
\newblock A comparison study between the {L}ennard-{J}ones and {DRIP}
  potentials for friction of graphene layers.
\newblock {\em Comput. Mater. Sci.}, 180:109723.

\bibitem[Liu et~al., 2016]{Liu2016}
Liu, Y., Weiss, N.~O., Duan, X., Cheng, H.-C., Huang, Y., and Duan, X. (2016).
\newblock Van der {W}aals heterostructures and devices.
\newblock {\em Nature Rev. Mat.}, 1(9):1--17.

\bibitem[Liu, 2014]{Liu_2014}
Liu, Z. (2014).
\newblock The diversity of friction behavior between bi-layer graphenes.
\newblock {\em Nanotechnology}, 25(7):075703.

\bibitem[Lopes~dos Santos et~al., 2012]{Lopes2012}
Lopes~dos Santos, J. M.~B., Peres, N. M.~R., and Castro~Neto, A.~H. (2012).
\newblock Continuum model of the twisted graphene bilayer.
\newblock {\em Phys. Rev. B}, {86}:155449.

\bibitem[Lu et~al., 2009]{Lu_2009}
Lu, Q., Arroyo, M., and Huang, R. (2009).
\newblock Elastic bending modulus of monolayer graphene.
\newblock {\em J. Phys. D: Appl. Phys.}, 42(10):102002.

\bibitem[Maaravi et~al., 2017]{Maaravi2017}
Maaravi, T., Leven, I., Azuri, I., Kronik, L., and Hod, O. (2017).
\newblock Interlayer potential for homogeneous graphene and hexagonal boron
  nitride systems: Reparametrization for many-body dispersion effects.
\newblock {\em J. Phys. Chem. C}, { 121}(41):22826--22835.

\bibitem[Mele, 2010]{Mele2010}
Mele, E.~J. (2010).
\newblock Commensuration and interlayer coherence in twisted bilayer graphene.
\newblock {\em Phys. Rev. B}, { 81}:161405.

\bibitem[Mergel et~al., 2019]{mergel19}
Mergel, J.~C., Sahli, R., Scheibert, J., and Sauer, R.~A. (2019).
\newblock Continuum contact models for coupled adhesion and friction.
\newblock {\em J. Adhes.}, 95(12):1101--1133.

\bibitem[Mergel et~al., 2021]{mergel21}
Mergel, J.~C., Scheibert, J., and Sauer, R.~A. (2021).
\newblock Contact with coupled adhesion and friction: {C}omputational
  framework, applications, and new insights.
\newblock {\em J. Mech. Phys. Solids}, 146:104194.

\bibitem[Merkle, 1993]{Merkle_1993}
Merkle, R.~C. (1993).
\newblock A proof about molecular bearings.
\newblock {\em Nanotechnology}, 4(2):86.

\bibitem[Mokhalingam et~al., 2020]{Mokhalingam2020}
Mokhalingam, A., Ghaffari, R., Sauer, R.~A., and Gupta, S.~S. (2020).
\newblock Comparing quantum, molecular and continuum models for graphene at
  large deformations.
\newblock {\em Carbon}, {159}:478--494.

\bibitem[Morovati et~al., 2022]{Morovati2022}
Morovati, V., Xue, Z., Liechti, K.~M., and Huang, R. (2022).
\newblock Interlayer coupling and strain localization in small-twist-angle
  graphene flakes.
\newblock {\em Extreme Mech. Lett.}, { 55}:101829.

\bibitem[Novoselov et~al., 2004]{Novoselov2004}
Novoselov, K., Geim, A., Morozov, S., Jiang, D., Zhang, Y., and Dubonos~et al.,
  S. (2004).
\newblock Electric field effect in atomically thin carbon films.
\newblock {\em Science}, 306(5696):666--669.

\bibitem[Ouyang et~al., 2020]{Ouyang2020}
Ouyang, W., Azuri, I., Mandelli, D., Tkatchenko, A., Kronik, L., Urbakh, M.,
  and Hod, O. (2020).
\newblock Mechanical and tribological properties of layered materials under
  high pressure: Assessing the importance of many-body dispersion effects.
\newblock {\em J. Chem. Theory Comput.}, { 16}(1):666--676.

\bibitem[Ouyang et~al., 2018]{Ouyang2018}
Ouyang, W., Mandelli, D., Urbakh, M., and Hod, O. (2018).
\newblock Nanoserpents: Graphene nanoribbon motion on two-dimensional hexagonal
  materials.
\newblock {\em Nano Lett.}, { 18}(9):6009--6016.

\bibitem[Park and Kwak, 1994]{park94}
Park, J.~K. and Kwak, B.~M. (1994).
\newblock {Three-dimensional frictional contact analysis using the homotopy
  method}.
\newblock {\em J. Appl. Mech.}, 61(3):703--709.

\bibitem[Plimpton, 1995]{lammps}
Plimpton, S. (1995).
\newblock Fast parallel algorithms for short-range molecular dynamics.
\newblock {\em J. Comput. Phys.}, {117}(1):1--19.

\bibitem[Polak and Ribiere, 1969]{Polak1969}
Polak, E. and Ribiere, G. (1969).
\newblock Note sur la convergence de m\'ethodes de directions conjugu\'ees.
\newblock {\em ESAIM: Math. Model. Num. - Mod\'elisation Math\'ematique et
  Analyse Num\'erique}, 3(R1):35--43.

\bibitem[Popov et~al., 2011]{Popov2011}
Popov, A.~M., Lebedeva, I.~V., Knizhnik, A.~A., Lozovik, Y.~E., and Potapkin,
  B.~V. (2011).
\newblock Molecular dynamics simulation of the self-retracting motion of a
  graphene flake.
\newblock {\em Phys. Rev. B}, 84:245437.

\bibitem[Reddy et~al., 2006]{Reddy2006}
Reddy, C., Rajendran, S., and Liew, K. (2006).
\newblock Equilibrium configuration and continuum elastic properties of finite
  sized graphene.
\newblock {\em Nanotechnology}, 17(3):864--870.

\bibitem[Reguzzoni et~al., 2012]{Reguzzoni2012}
Reguzzoni, M., Fasolino, A., Molinari, E., and Righi, M.~C. (2012).
\newblock Potential energy surface for graphene on graphene: Ab initio
  derivation, analytical description, and microscopic interpretation.
\newblock {\em Phys. Rev. B}, 86:245434.

\bibitem[Rodr\'iguez-Tembleque and Abascal, 2013]{rodriguez13}
Rodr\'iguez-Tembleque, L. and Abascal, R. (2013).
\newblock Fast {FE}--{BEM} algorithms for orthotropic frictional contact.
\newblock {\em Int. J. Numer. Methods Eng.}, 94(7):687--707.

\bibitem[Rodr\'iguez-Tembleque et~al., 2012]{RODRIGUEZ20121}
Rodr\'iguez-Tembleque, L., Abascal, R., and Aliabadi, M.~H. (2012).
\newblock Anisotropic wear framework for 3d contact and rolling problems.
\newblock {\em Comput. Methods Appl. Mech. Eng.}, 241:1--19.

\bibitem[Sauer, 2006]{sauer-phd}
Sauer, R.~A. (2006).
\newblock {\em An atomic interaction based continuum model for computational
  multiscale contact mechanics}.
\newblock PhD thesis, University of California, Berkeley, USA.

\bibitem[Sauer, 2016]{sauer16}
Sauer, R.~A. (2016).
\newblock A frictional sliding algorithm for liquid droplets.
\newblock {\em Comput. Mech.}, 58(6):937--956.

\bibitem[Sauer and {De~Lorenzis}, 2013]{spbc}
Sauer, R.~A. and {De~Lorenzis}, L. (2013).
\newblock A computational contact formulation based on surface potentials.
\newblock {\em Comput. Methods Appl. Mech. Eng.}, {253}:369--395.

\bibitem[Sauer and {De~Lorenzis}, 2015]{spbf}
Sauer, R.~A. and {De~Lorenzis}, L. (2015).
\newblock An unbiased computational contact formulation for {3D} friction.
\newblock {\em Int. J. Numer. Meth. Engrg.}, {101}(4):251--280.

\bibitem[Sauer and Wriggers, 2009]{sauer09b}
Sauer, R.~A. and Wriggers, P. (2009).
\newblock Formulation and analysis of a {3D} finite element implementation for
  adhesive contact at the nanoscale.
\newblock {\em Comput. Methods Appl. Mech. Eng.}, {198}:3871--3883.

\bibitem[Shen et~al., 2020]{shen2020}
Shen, C., Chu, Y., Wu, Q., Li, N., Wang, S., Zhao, Y., Tang, J., Liu, J., Tian,
  J., Watanabe, K., et~al. (2020).
\newblock Correlated states in twisted double bilayer graphene.
\newblock {\em Nature Phys.}, 16(5):520--525.

\bibitem[Shirazian et~al., 2018]{Shirazian2018_01}
Shirazian, F., Ghaffari, R., Hu, M., and Sauer, R.~A. (2018).
\newblock Hyperelastic material modeling of graphene based on density
  functional calculations.
\newblock {\em Proc. Appl. Math. Mech.}, 18(1):e201800419.

\bibitem[Stuart et~al., 2000]{Stuart2000}
Stuart, S., Tutein, A., and Harrison, J. (2000).
\newblock A reactive potential for hydrocarbons with intermolecular
  interactions.
\newblock {\em J. Chem. Phys.}, 112(14):6472--6486.

\bibitem[Stupkiewicz et~al., 2014]{stupkiewicz14}
Stupkiewicz, S., Lewandowski, M.~J., and Lengiewicz, J. (2014).
\newblock Micromechanical analysis of friction anisotropy in rough elastic
  contacts.
\newblock {\em Int. J. Solids Struct.}, 51(23):3931--3943.

\bibitem[Sun et~al., 2018]{Sun2018}
Sun, J., Zhang, Y., Lu, Z., Li, Q., Xue, Q., Du, S., Pu, J., and Wang, L.
  (2018).
\newblock Superlubricity enabled by pressure-induced friction collapse.
\newblock {\em J. Phys. Chem. Lett.}, 9(10):2554--2559.
\newblock PMID: 29714483.

\bibitem[Swope et~al., 1982]{Swope1982}
Swope, W.~C., Andersen, H.~C., Berens, P.~H., and Wilson, K.~R. (1982).
\newblock {A computer simulation method for the calculation of equilibrium
  constants for the formation of physical clusters of molecules: Application to
  small water clusters}.
\newblock {\em J. Chem. Phys.}, 76(1):637--649.

\bibitem[Temizer, 2014]{temizer14}
Temizer, I. (2014).
\newblock Computational homogenization of soft matter friction: {I}sogeometric
  framework and elastic boundary layers.
\newblock {\em Int. J. Numer. Methods Eng.}, 100(13):953--981.

\bibitem[Tomlinson, 1929]{Tomlinson1929_01}
Tomlinson, G.~A. (1929).
\newblock A molecular theory of friction.
\newblock {\em Lond. Edinb. Dublin Philos. Mag. J. Sci.}, 7(46):905--939.

\bibitem[Verhoeven et~al., 2004]{Ver2004}
Verhoeven, G.~S., Dienwiebel, M., and Frenken, J. W.~M. (2004).
\newblock Model calculations of superlubricity of graphite.
\newblock {\em Phys. Rev. B}, 70:165418.

\bibitem[Wang et~al., 2017a]{Wang2017}
Wang, G., Dai, Z., Wang, Y., Tan, P., Liu, L., Xu, Z., Wei, Y., Huang, R., and
  Zhang, Z. (2017a).
\newblock Measuring interlayer shear stress in bilayer graphene.
\newblock {\em Phys. Rev. Lett.}, 119:036101.

\bibitem[Wang et~al., 2017b]{Wang2017b}
Wang, S., Chen, Y., Ma, Y., Wang, Z., and Zhang, J. (2017b).
\newblock Size effect on interlayer shear between graphene sheets.
\newblock {\em J. Appl. Phys.}, 122(7):074301.

\bibitem[Wen et~al., 2018]{Wen2018}
Wen, M., Carr, S., Fang, S., Kaxiras, E., and Tadmor, E.~B. (2018).
\newblock Dihedral-angle-corrected registry-dependent interlayer potential for
  multilayer graphene structures.
\newblock {\em Phys. Rev. B}, 98:235404.

\bibitem[Wriggers, 2006]{wriggers-contact}
Wriggers, P. (2006).
\newblock {\em Computational Contact Mechanics}.
\newblock Springer, Berlin, Heidelberg, 2$^{\text{nd}}$ edition.

\bibitem[Xu et~al., 2011]{Xu_2011}
Xu, L., Ma, T.-B., Hu, Y.-Z., and Wang, H. (2011).
\newblock Vanishing stick{\textendash}slip friction in few-layer graphenes: the
  thickness effect.
\newblock {\em Nanotechnology}, 22(28):285708.

\bibitem[Xue et~al., 2022]{Xue2022}
Xue, Z., Chen, G., Wang, C., and Huang, R. (2022).
\newblock Peeling and sliding of graphene nanoribbons with periodic van der
  {W}aals interactions.
\newblock {\em J. Mech. Phys. Solids}, 158:104698.

\bibitem[Zhang et~al., 2015]{ZHANG2015}
Zhang, H., Guo, Z., Gao, H., and Chang, T. (2015).
\newblock Stiffness-dependent interlayer friction of graphene.
\newblock {\em Carbon}, 94:60--66.

\bibitem[Zheng et~al., 2008]{Zheng2008}
Zheng, Q., Jiang, B., Liu, S., Weng, Y., Lu, L., Xue, Q., Zhu, J., Jiang, Q.,
  Wang, S., and Peng, L. (2008).
\newblock Self-retracting motion of graphite microflakes.
\newblock {\em Phys. Rev. Lett.}, 100:067205.

\bibitem[Zmitrowicz, 1981]{Zmitrowicz1981}
Zmitrowicz, A. (1981).
\newblock A theoretical model of anisotropic dry friction.
\newblock {\em Wear}, 73(1):9--39.

\bibitem[Zmitrowicz, 1989]{Zmitrowicz1989}
Zmitrowicz, A. (1989).
\newblock Mathematical descriptions of anisotropic friction.
\newblock {\em Int. J. Solids Struct.}, 25(8):837--862.

\bibitem[Zmitrowicz, 1992]{Zmitrowicz1992}
Zmitrowicz, A. (1992).
\newblock A constitutive modelling of centrosymmetric and non-centrosymmetric
  anisotropic friction.
\newblock {\em Int. J. Solids Struct.}, 29(23):3025--3043.

\end{thebibliography}
